\documentclass[%
 reprint,
 amsmath,amssymb,
 aps,
]{revtex4-2}

\usepackage{caption}
\usepackage{subcaption}
\usepackage{graphicx}
\usepackage{dcolumn}
\usepackage{bm}
\usepackage{hyperref}
\usepackage{xcolor}
\usepackage{aas_macros}

\newcommand{\mpcoh}{h^{-1}{\rm Mpc}}
\newcommand{\msunoh}{h^{-1}{\rm M}_\odot}

\begin{document}

\preprint{APS/123-QED}

\title{Examining Lyman-alpha Emitters through MillenniumTNG in anticipation of DESI-II}

\author{Jyotsna Ravi}
\email{jyotsna.ravi@berkeley.edu}
\affiliation{%
Department of Physics, University of California, Berkeley, CA 94720, USA
}%

\author{Boryana Hadzhiyska}%
\email{boryanah@berkeley.edu}
\affiliation{%
Department of Physics, University of California, Berkeley, CA 94720, USA
}%

\author{Martin J. White}%
\affiliation{%
Department of Physics, University of California, Berkeley, CA 94720, USA
}%

\author{Lars Hernquist}
\affiliation{%
Center for Astrophysics | Harvard \& Smithsonian, 60 Garden Street, Cambridge, MA 02138, USA
}%

\author{Sownak Bose}
\affiliation{%
Institute for Computational Cosmology, Department of Physics, University of Durham, South Road, Durham, DH1 3LE, UK
}%

\date{\today}

\begin{abstract}
The new frontier of cosmological experiments will be aimed at targeting galaxies at higher redshifts. The goal of this study is to conduct a timely analysis of the high-redshift star-forming galaxy populations, which will be informative in designing next-generation experiments and their extragalactic targets. We use the hydrodynamical simulation MillenniumTNG (MTNG) to model Lyman-alpha Emitting (LAE) galaxies to extract key properties such as their clustering and occupation statistics. We define LAEs through an empirical relation between star formation rate (SFR) and Lyman-alpha flux. We also explore two other definitions, finding that imposing an additional cut on the maximum stellar mass of the galaxy sample, which approximates the effect of a low escape fraction at high halo mass, leads to a 5-10\% decrease of the linear bias of the population. As expected, we find that the HOD mass parameters rapidly (and logarithmically) decrease with increasing number density. Additionally, the HOD parameter $\sigma$ also decreases with number density, implying that the SFR-halo mass relationship becomes tighter for low-luminosity objects. Surprisingly, the non-linear clustering, estimated by the parameter $r_0$, is fixed at fixed number density across the two redshifts we study, whereas the linear bias parameter varies with redshift as $b(z) \propto (1 + z)$, suggesting that our LAE samples are relatively stable and long-lived. Finally, we study the amount of galaxy assembly bias present at $z = 2, \ 3$ and find that while at $z = 2$ it is roughly $\lesssim$10\%, at $z = 3$ it decreases significantly to $\lesssim$5\%. This suggests that assembly bias effects become less important at high $z$ likely due to the lower number of cumulative two-halo interactions (mergers, splashback, stripping, etc.). While our study is based on a single full-physics simulation, we expect our results to reflect qualitatively the properties of LAEs in the Universe. We demonstrate that our findings are in good agreement with previous results in the literature using both observations and simulations.
\end{abstract}

\maketitle


\section{Introduction}

Thanks to the influx of high-volume, high-quality observational data in the next decade, photometric and spectroscopic galaxy surveys such as the Dark Energy Survey \citep[DES,][]{2018PhRvD..98d3526A,2022PhRvD.105b3520A}, the Dark Energy Spectroscopic Instrument  \citep[DESI,][]{2016arXiv161100036D}, Euclid \citep{2011arXiv1110.3193L}, the Vera Rubin Observatory \citep[Rubin][]{2009arXiv0912.0201L} and Nancy Grace Roman Space Telescope \citep[\textit{Roman,}][]{2015arXiv150303757S,2021MNRAS.507.1746E} will provide an extraordinary opportunity to stress test the current cosmological paradigm and improve our models of galaxy formation and evolution. The `golden standard' of spectroscopic galaxy surveys is the analysis of the baryon acoustic oscillations (BAO) peak \citep{2005ApJ...633..560E}, which enables the extraction of cosmological information from the large-scale clustering of galaxies. However, an alternative path to unraveling the long-standing mysteries of dark matter, dark energy, gravity, and neutrinos lies in the decade scale between the BAO scale and the non-linear scale, which upcoming surveys will measure with exquisite precision.

The small-scale signal encodes the relationship between galaxies and the underlying dark matter distribution, which can help us refine our understanding of galaxy formation and evolution and also quantify the impact of small-scale physics on the large scales used in cosmological analysis. Most standard methods for modeling the small-scale galaxy distribution rely on the tenet that galaxy formation requires a gravitationally-bound dark matter halo or subhalo to accumulate and condense gas \citep{2000MNRAS.318.1144P,2000MNRAS.318..203S,2000MNRAS.311..793B,2001ApJ...550L.129W,2002ApJ...575..587B,2002PhR...372....1C,2009ApJ...695..900Y}.

The most prominent examples of such galaxy-(sub)halo models are the halo occupation distribution model \citep[HOD,][]{2002ApJ...575..587B, 2002PhR...372....1C,  2004MNRAS.350.1153Y, 2005ApJ...633..791Z} and the subhalo abundance matching model \citep[SHAM,][]{2006ApJ...647..201C, 2010ApJ...717..379B, 2014ApJ...783..118R, 2016MNRAS.459.3040G, 2016MNRAS.460.3100C, 2022MNRAS.509.1614F}. The HOD prescription takes as input the average number of central and satellite galaxies in a given halo as a function of halo mass and outputs a summary statistic of interest or a galaxy catalog. On the other hand, SHAM connects galaxies to dark matter subhalos assuming a (typically) monotonic relation between a galaxy property such as stellar mass or star formation rate and a subhalo dark-matter property such as maximum circular velocity. Both methods come with well-known shortcomings when fitting observational and simulated data \citep[e.g.,][]{Norberg, Zehavi, 2017MNRAS.467.3024L, 2019ApJ...887...17Z,2020MNRAS.493.5506H, 2021MNRAS.508..175C}. For reviews on modeling the galaxy-halo connection and dark-matter simulations, see \citet{2018ARA&A..56..435W,2022LRCA....8....1A}

One of the main limitations of these empirical models is their handling of the effect of `galaxy assembly bias' \citep{Gao2005}.
Galaxy assembly bias refers to a manifestation of a discrepancy between the actual distribution of galaxies and one inferred from dark matter halos using their present-day mass alone \citep[e.g.,][]{2007MNRAS.374.1303C}. Additional halo properties, for example halo formation time, local environment, concentration, triaxiality, spin, or velocity dispersion need to be considered to describe the clustering correctly. The standard implementation of the popular HOD model does not consider halo properties apart from mass, and hence completely neglects galaxy assembly bias. Similarly, the baseline SHAM model does not take baryonic effects such as tidal stripping and disruption into consideration, which affect subhalos in $N$-body and hydro simulations differently and may thus distort the subhalo properties in the presence of baryons \citep[see e.g.,][]{2021MNRAS.501.1603H}.

As ongoing experiments exhaust the potential of these measurements at low redshift, the next-generation of planned experiments will target the high-redshift Universe. One of these new-frontier experiments is DESI-II, the successor of DESI, which will provide clean low-noise measurements of large-scale observables deep into the past of our Universe. As measurements of the clustering at high redshifts have not been featured extensively in precision cosmology studies, various details about the galaxy-halo connection of the targeted galaxy populations warrant careful theoretical modeling. Luckily, recent progress in the realm of hydrodynamical simulations, which provide a plausible picture of the real Universe, offer an exciting venue for obtaining high-fidelity models of the high-redshift Universe.

Hydrodynamical simulations \citep[e.g.,][]{2014MNRAS.444.1518V,2014Natur.509..177V,2014MNRAS.445..175G,2015MNRAS.446..521S,2019ComAC...6....2N,2022MNRAS.511.4005K} simulate the dark matter component along with the gas and stars. In this type of simulations, baryonic and galaxy processes are tracked by a combination of fluid equations and subgrid models, which renders them too expensive to run in the volumes needed to analyze modern galaxy surveys. Nonetheless, they can still provide invaluable insight into how and where galaxies form in relation to their dark-matter hosts. In this study, we employ the new full-physics simulation MillenniumTNG (MTNG), which has comparable resolution to the largest simulation from the IllustrisTNG suite, TNG300-1, but offers a factor of $\sim$15 improvement in the volume \citep[for a review of the IllustrisTNG project, see][]{2019ComAC...6....2N}. The improved statistics open up the possibility of selecting galaxy samples with lower number densities, making MTNG an ideal testing ground for developing and validating various theoretical tools in anticipation of the future high-redshift galaxy surveys.

The Lyman-alpha (Ly$\alpha$) emission line is commonly observed in the spectra of high-redshift galaxies \citep[e.g.,][]{2003ApJ...588...65S,2005ApJ...620L...1O,2006Natur.440.1145H}. Detections of Ly$\alpha$ emitting galaxies (LAEs) in large numbers at $z = 2–6$ using narrow-band filters have opened up a new window for probing cosmology at earlier epochs. In this paper, we address the long-standing question of modeling the large-scale properties of LAEs at $z = 2$ and 3, a regime that is of interest to future wide-field surveys. Our goal is to conduct a timely analysis of the high-redshift star-forming galaxy populations, which we hope will be informative in designing the next-generation experiments and their extragalactic targets. While our study is based on the MTNG740 simulation, we expect our results to reflect qualitatively the properties of the real Universe. An overview of the full simulation suite of MillenniumTNG and an analysis of its matter clustering and halo statistics is given in \citet{Aguayo2022}, while \citet{Pakmor2022} provide a detailed description of the hydro simulation together with an examination of the properties of its galaxy clusters. Further introductory papers present analyses of high-redshift galaxies \citep{Kannan2022},  weak gravitational lensing \citep{Ferlito2022}, intrinsic alignment \citep{Delgado2022}, galaxy clustering \citep{Bose2022}, cosmological inference from galaxy clustering \citep{Contreras2022}, one- and two-halo term analysis \citep{2023MNRAS.524.2507H,2023MNRAS.524.2524H}, and semi-analytic galaxies on the light cone \citep{Barrera2022}.

This paper is organized as follows. In Section~\ref{sec:methods}, we describe the MTNG740 hydrodynamical simulation and provide details about our LAE selection procedure as well as basic definitions of the large-scale statistics used in our study. In Section~\ref{sec:results}, we discuss our main findings pertaining to the galaxy-halo modeling, linear bias, redshift-space clustering and assembly bias properties of our LAE samples. Finally, in Section~\ref{sec:conclusions}, we summarize our main results and put our work into perspective, providing a qualitative comparison with similar studies performed in simulations and observations.

\section{Methods}
\label{sec:methods}

\subsection{The MillenniumTNG simulations}
\label{sec:mtng}

The aim of the MillenniumTNG project is to provide a set of numerical simulations that make accurate predictions about the complex interaction between galaxy processes and large-scale structure in sufficiently large volumes. While the full simulation suite consists of several hydrodynamical and $N$-body simulations at various resolutions and box sizes, here we employ only one of these products, namely, the largest available full-physics box, \textsc{MTNG740}. This simulation contains $2 \times 4320^3$ resolution elements in a comoving volume of $(0.125\,h^{-3} {\rm Gpc})^3$, which corresponds to a mass resolution of $2.1 \times 10^7 \, \msunoh$ in the baryons and $1.1 \times 10^8 \, \msunoh$ in the dark matter. Throughout this paper, we refer to this simulation as MTNG for simplicity. 

The physics and cosmology model of MTNG echoes that of IllustrisTNG \citep{2017MNRAS.465.3291W,2018MNRAS.473.4077P,2018MNRAS.475..648P,2018MNRAS.475..624N,2018MNRAS.477.1206N,2018MNRAS.480.5113M,2018MNRAS.475..676S, 2019MNRAS.490.3234N, 2019MNRAS.490.3196P}, with a resolution comparable to but slightly lower than that of the largest IllustrisTNG box, TNG300-1. MTNG also uses the same hydrodynamical moving-mesh code \textsc{AREPO} \citep[][]{2010MNRAS.401..791S}, a main feature of which is the use of a Voronoi tessellation for the construction of the computational mesh. Haloes (groups) in MTNG are identified by applying the standard friends-of-friends \citep[FoF,][]{1985ApJ...292..371D} algorithm to the dark matter particles, adopting a linking length of $b = 0.2$ (in units of the mean interparticle distance). Gravitationally bound substructures in halos are identified with the {\small SUBFIND-HBT} algorithm described in \citet{Springel2021}.

We derive the two quantities halo mass and halo radius as the total mass (and corresponding radius) enclosed in a sphere around the halo center with mean density $\Delta_c$ times the critical density of the Universe. $\Delta_c$ is defined using the generalized solution of the collapse of a spherical top-hat perturbation in a low-density universe and fit by the polynomial function \citep{1998ApJ...495...80B}:
\begin{equation}
  \Delta_c(z) = 18 \pi^2 + 82 x - 39 x^2 \, ,
  \label{eq:Delta_c}
\end{equation} 
where $x = \Omega_m(z) - 1$, and $\Omega_m(z)$ is the matter energy density at redshift $z$.

\subsection{Halo occupation distribution}

One way to quantify the connection between galaxies and the underlying dark-matter distribution is through the halo occupation distribution (HOD). Studying the HOD of a given galaxy population can provide us with vital insight on cosmic formation and evolution. The HOD is usually studied separately for the central and satellite populations. The central galaxies are typically bright galaxies located in the center of their respective dark matter halo. Thus, for a given halo, there will either be zero or one central galaxy. In contrast, satellite galaxies orbit around the central galaxy. A given halo can have anywhere between 0 and $\sim$100 satellite galaxies. The HOD predicts the average number of central and satellite galaxies, respectively, as a function of halo mass. 

The distribution of central and satellite galaxies for magnitude-limited and red galaxies can be modeled using the formalism of \citet{Zheng:2004id}:
\begin{equation} \label{eq:ncen}
\langle N_{\rm cen}(M_h) \rangle=\frac{1}{2}\left[1 + \textrm{erf}(\frac{\textrm{log}M_h - \textrm{log}M_{\rm min}}{\sigma_{\textrm{log}M}})\right]
\end{equation}
\begin{equation} \label{eq:nsat}
\langle N_{\rm sat}(M_h) \rangle =\left(\frac{M_h-M_{\rm cut}}{M_1}\right)^\alpha
\end{equation}
Where $M_{\rm min}$ roughly represents the mass threshold at which halos can form a central galaxy, $\sigma_{\log M}$ denotes how quickly the central galaxy slope transitions between zero and one, $M_{\rm cut}$ is the cut-off mass at which halos can host satellite galaxies, $M_1$ is a normalization factor, and $\alpha$ represents the slope of the satellite galaxy curve. In our analysis we use $M_h \equiv M_{\rm tophat}$ as our mass proxy (see Eq.~\ref{eq:Delta_c}). 

In this study, we are interested in modeling Lyman-alpha emitting galaxies (LAEs), the HOD of which is not well constrained. We will show that the formalism of \citet{Zheng:2004id} works well for the most simplistic method of selecting those galaxies in MTNG.

\subsection{Galaxy selection}
\label{sec:selection}

In this section, we summarize the different methods we employ in this study to select LAEs.

Since LAEs often contain young and massive stars, a good proxy for selecting these galaxies from simulations is the star formation rate (SFR). However, there are several other factors that can influence their identification, including the escape fraction, equivalent width and radiative transfer effects \citep{2020ARA&A..58..617O}. In addition, because Ly$\alpha$ scatters through the ISM
and IGM, it is difficult to determine the amount of extinction, whereas SFR derived from the UV flux density is likely more stable. For this reason, we use the relation from \citet{2010MNRAS.401.2343D} to approximate the Ly$\alpha$ SFR:
\begin{equation} \label{eq:M}
\mathcal{M} \equiv \frac{\rm SFR(UV)}{\rm
SFR(Ly\alpha{\rm )}}  ,
\end{equation}
where $\mathcal{M}$ is a random variable drawn from the logarithmic probability distribution function:
\begin{eqnarray}
P(\mathcal{M})d\mathcal{M}=\frac{1}{\sigma\sqrt{2\pi}}{\rm exp}\Big{[}
  -\frac{1}{2}\Big{(}\frac{[\log
      \mathcal{M}]-x}{\sigma}\Big{)}^2\Big{]}\frac{d\mathcal{M}}{\mathcal{M}\ln
  10},
\label{eq:pdf}
\end{eqnarray}
where $x = 0.04$ and $\sigma = 0.35$ for $z \leq 3$. We can express the Lyman-alpha luminosity ($L_{\rm Ly\alpha}$) as:
\begin{equation}
     L_{\rm Ly\alpha} = \frac{1.1 \times 10^{42} \textrm{erg} \ s^{-1}}{\mathcal{M}}\frac{\rm SFR}{{\rm M}_\odot {\rm yr}^{-1}}
     \label{eq:L_lya}
\end{equation}
The observable quantity through which LAEs are typically selected is the Lyman-alpha flux, $f_{\rm Ly\alpha}$, defined as: 
\begin{equation} \label{eq:f_lya}
f_{\rm Ly\alpha} = \frac{L_{\rm Ly\alpha}}{4\pi d_L^2}
\end{equation}
where $d_L$ is the luminosity distance to an object at some particular redshift. In this work, we will consider LAEs at $z = 2$ and 3. As future high-redshift observatories such as DESI-II begin to prepare for the target selection of LAEs, in this work, we explore a range of $f_{\rm Ly\alpha}$ threshold values in an effort to inform these future efforts.

\begin{figure}
\includegraphics[width=\columnwidth]{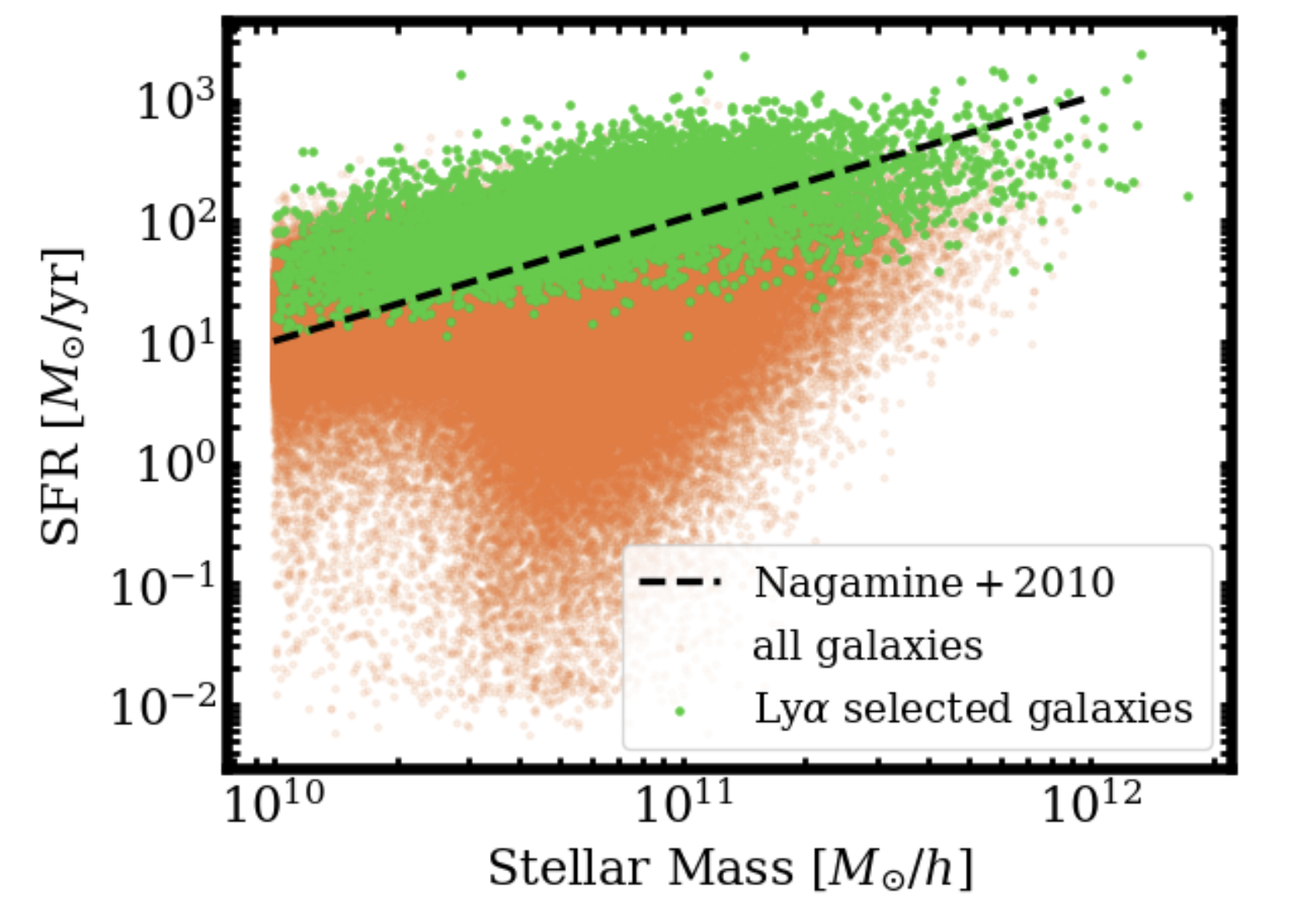}
\caption{\label{fig:selection}
Scatter plot showing the SFR and stellar mass for all galaxies in orange and for our \textbf{default} selection of LAEs in green (see Section~\ref{sec:selection}). We see that the LAEs occupy the upper region of this scatter plot, corresponding to the objects with highest SFR. This agrees with our intuitive expectation for the LAEs as a star-forming sample. Our selection includes predominantly objects with high stellar masses. We also show in dashed black the mean SFR-stellar mass fit for LAEs from \citet{Nagamine}, finding it to be in very good agreement with our LAEs. 
}
\label{fig:scatter}
\end{figure}

LAEs may be selected by an additional cut in rest-frame equivalent width (REW). As some future missions are considering making such a selection, we also examine how a cut in REW affects the LAE clustering and occupation statistics. We obtain a proxy to the REW as follows:
\begin{equation} 
\label{eq:rew}
{\rm REW} = \mathcal{M} \left(\frac{{\rm REW}_c}{C}\right) \simeq 85\AA\ \mathcal{M} ,
\end{equation}
where $C = 0.89$, ${\rm REW}_c = 76${\AA}, and $\mathcal{M}$ is defined in Eqs.~\ref{eq:M}, \ref{eq:pdf}. 
We note that ideally REW would be tied to the internal properties of the galaxy, however such detailed modeling is beyond the scope of this work, and we leave a more realistic modeling of that effect for future investigations that utilize radiative transfer simulations. Finally, since our galaxies also lack dust, we test the effect of applying an upper threshold cut in stellar mass, which mimics the effect of lower escape fractions at higher halo masses \citep{2014MNRAS.442.1805I}. To summarize, we construct three different samples defined below:
\begin{itemize}
    \item \textbf{Default}: Our default sample consists of applying a flux cut, $f_{\rm Ly\alpha}$. In particular, future surveys such as DESI-II will likely apply a cut in the range of $5 \times 10^{-17}$ and $1 \times 10^{-16}$ ${\rm erg} \ s^{-1} \ {\rm cm}^{-2}$. The corresponding number densities are $1.42 \times 10^{-2}$ and $9.6 \times 10^{-3} h^{3}{\rm Mpc}^{-3}$, respectively. Our default sample is the mid-point flux threshold of $f_{\rm Ly\alpha} > 7.5 \times 10^{-17}$, corresponding to a number density of $0.0114$ $h^{3}{\rm Mpc}^{-3}$ at redshift of $z=2$ and $0.00515$ $h^{3}{\rm Mpc}^{-3}$ at $z=3$. In later sections, we study how the parameters change as we vary this threshold. 
    \item \textbf{Default $+$ REW}: We next construct a sample, which in addition to the default $f_{\rm Ly\alpha}$ cut, has an REW cut of ${\rm REW} > 20${\AA}, which is equivalent to imposing a cut of $\mathcal{M}>0.23$. We effectively remove 39572 galaxies and obtain a corresponding number density of $n_{\rm gal} = 0.0110$ $h^{3}{\rm Mpc}^{-3}$. Given our model for the equivalent width (see Eq.~\ref{eq:pdf}), this process is equivalent to a `random downsampling' of objects with probability $P(\mathcal{M}>0.23) \approx 97\%$.
    \item \textbf{Default $+$ stellar mass}: Finally, we study the effect of applying a stellar mass maximum threshold instead of an REW cut, such that the number of galaxies remains the same as in the previous selection, \textbf{default $+$ REW}. This corresponds to a stellar mass cut of $M_\ast < 3.75 \times 10^{10} \, \msunoh$ and a number density of  $n_{\rm gal} = 0.0110$ $h^{3}{\rm Mpc}^{-3}$.
\end{itemize}

 We show a scatter plot of SFR and stellar mass at $z = 2$ in Fig.~\ref{fig:scatter}, which illustrates the \textbf{default} galaxy selection ($f_{\rm Ly\alpha} > 7.5 \times 10^{-17}$ ${\rm erg} \ s^{-1} \ {\rm cm}^{-2}$). We find that it occupies the upper region in the plot, which corresponds to highly star-forming galaxies, as expected. Due to the correlation of these two quantities, we see that our approximate cut on SFR also predominantly selects galaxies with large stellar masses. We also show a comparison to the mean SFR-stellar mass curve for LAEs from \citet{Nagamine} and find that it is in very good agreement with the galaxies we have identified as LAEs in MTNG. 

\begin{figure}
\includegraphics[width=\columnwidth]{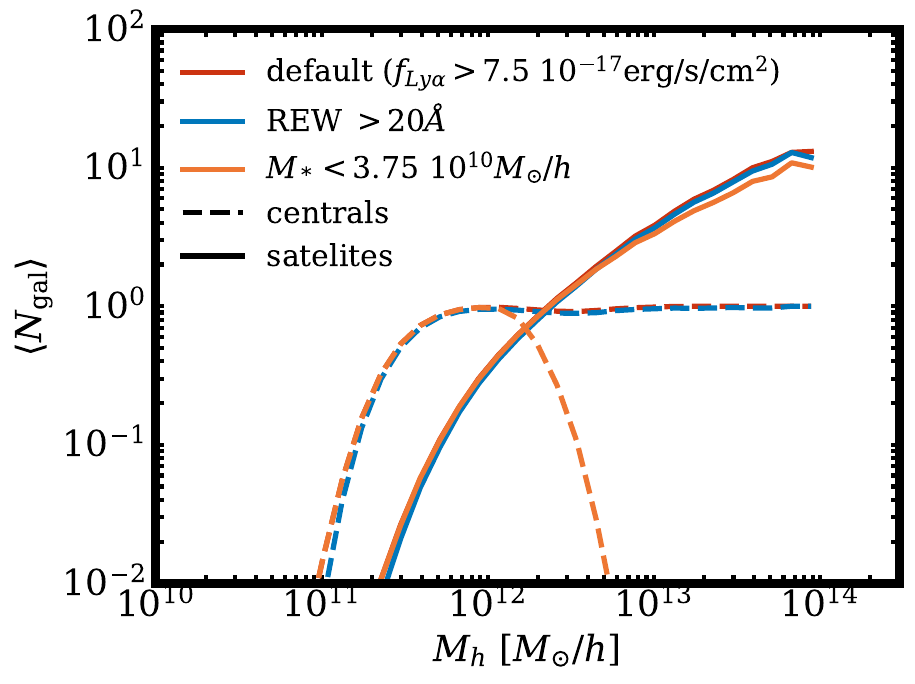}
\caption{Halo occupation distribution for the three LAE selection schemes detailed in Section~\ref{sec:selection}. The \textbf{default} (red) and \textbf{default $+$ REW} (blue) samples are virtually identical (since the latter is a downsampled version of the former). We note that a slight shift to higher halo masses is seen in the latter sample, as halos with scarce occupation numbers get eliminated from the sample through the process of random sampling. The HOD of the \textbf{default $+$ stellar mass} sample (orange) is not well approximated by the \citet{Zheng:2004id} formalism, as halos with high halo masses are removed from the sample. This mimics the decrease of photon escape fraction with increasing halo mass.
}
\label{fig:hod}
\end{figure}

In Fig.~\ref{fig:hod}, we examine the occupation statistics of our three samples at $z = 2$. We see that the HOD of the \textbf{default} and the \textbf{default $+$ REW} samples are very similar to each other and well-approximated by the \citet{Zheng:2004id} formalism. Since the REW cut effectively downsamples the number of galaxies in the selection, the main impact on the HOD is to slightly lower the occupation distribution of the centrals and satellites. On the other hand, the maximum stellar mass selection has noticeably different central occupation distribution. Namely, the HOD of the centrals declines steeply past $M_h \gtrsim 10^{12} \, \msunoh$, since the majority of stellar massive galaxies live in massive halos. While there is some obscuration of high-mass halos, it is likely that our rather simplistic treatment leads to a more exaggerated effect. The \textbf{default $+$ stellar mass} sample can thus give us a rough estimate of the maximum effect of dust obscuration on the HOD and clustering of LAEs. We plan to examine other strategies in future work.

\subsection{Clustering statistics}

\begin{figure}
\includegraphics[width=\columnwidth]{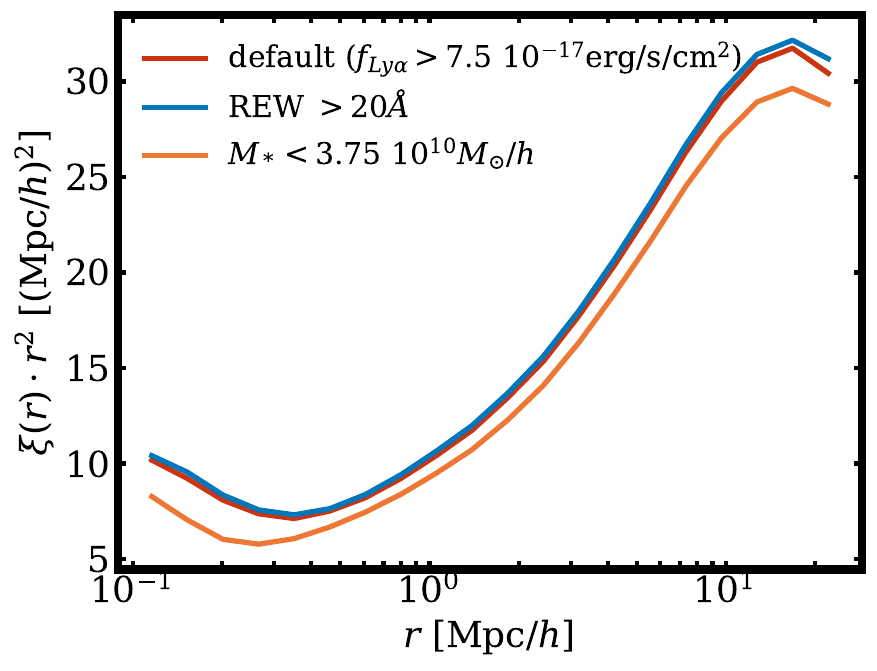}
\caption{Real-space correlation function for the three LAE selection schemes detailed in Section~\ref{sec:selection}. As expected from Fig.~\ref{fig:hod}, the clustering statistics of the \textbf{default} (red) and \textbf{default $+$ REW} (blue) samples are largely identical, with the latter sample being a tiny bit more clustered (reflecting the slight shift to higher halo masses relative to the \textbf{default} sample, as seen in its HOD). The \textbf{default $+$ stellar mass} sample (orange), on the other hand, has overall lower clustering by about 5-10\% due to the decrease in mean halo mass, and as a result, its one-halo term is also pushed to smaller scales.
}
\label{fig:corrfunc}
\end{figure}

Another quantity that allows us to study LAEs and extract information about their large-scale distribution is the correlation function. In a cubic box with periodic boundary conditions, we can obtain the correlation function in real space using the natural estimator as:
\begin{equation} 
\label{eq:xi}
\xi(r) = \left[\frac{DD(r)}{RR(r)}\right] - 1
\end{equation}
$DD(r)$ is the normalized number of galaxy pairs in a spherical shell at a distance $r$, and $RR(r)$ is the normalized number of pairs in the same spherical shell for a randomly generated galaxy sample. We can split the obtained correlation function into a `one-halo' and a `two-halo' piece, where the `one-halo' piece receives contributions from within the halo (i.e., central-satellite and satellite-satellite pairs in the same halo), whereas the `two-halo' piece comes from central-central, central-satellite, satellite-satellite in two distinct halos.

In Fig.~\ref{fig:corrfunc}, we display the correlation function for the three LAE samples defined in Section~\ref{sec:selection}. Similarly to Fig.~\ref{fig:hod}, we find that the \textbf{default} and the \textbf{default $+$ REW} samples result in almost indistinguishable correlation functions. The small difference that we see can be attributed to the slight shift of central and satellite occupations towards higher halo masses (and thus higher linear bias) when we apply the REW cut. The \textbf{default $+$ stellar mass} sample, on the other hand, has two distinctive features. On one hand, the amplitude (linear bias) is lower on large scales by roughly 5-10\% due to the elimination of some of the most massive (and thus most clustered) halos when we impose the stellar mass cut. On the other, the one-halo term is more reduced and shifted towards smaller values of the pair distance $r$, as our maximum stellar mass cut has 
reduced the mean halo mass and thus halo radius to smaller values, curtailing the range of the one-halo term (which scales as the mean virial radius of the host halos).


Using the galaxy correlation function, we define the galaxy bias in real space as
\begin{equation} 
\label{eq:bias}
b(r) \equiv \sqrt{\frac{\xi(r)}{\xi_{\rm matter}(r)}} \, ,
\end{equation}
where $\xi_{\rm matter}(r)$ is the non-linear correlation function of the matter distribution, which in our case, is obtained from the downsampled dark matter particles in the dark-matter-only counterpart to MTNG-740. 
Over a range of scales near the non-linear scale, the correlation function $\xi(r)$ can be modeled by a power law with a slope of $\gamma \approx 1.8$. In this work, we will hold this slope fixed and can thus approximately express the clustering on quasi-linear scales as:
\begin{equation} \label{eq:r0}
\xi(r) = \left(\frac{r}{r_0}\right)^{-1.8},
\end{equation}
where $r_0$ is a characteristic amplitude, which is effectively also a measure of the linear bias.
Throughout this work, we quote the linear bias as the value of $b(r)$ at $r = 10 \, \mpcoh$ and fit $r_0$ by using Eq.~\ref{eq:r0} as our model via \texttt{curvefit}. The large-scale bias should hold on large, linear scales while the power-law form holds in the non-linear regime. Our choice of scale is thus a compromise between these two regimes.


\subsection{Redshift space distortions}

\begin{figure}[h]
\includegraphics[width=\columnwidth]{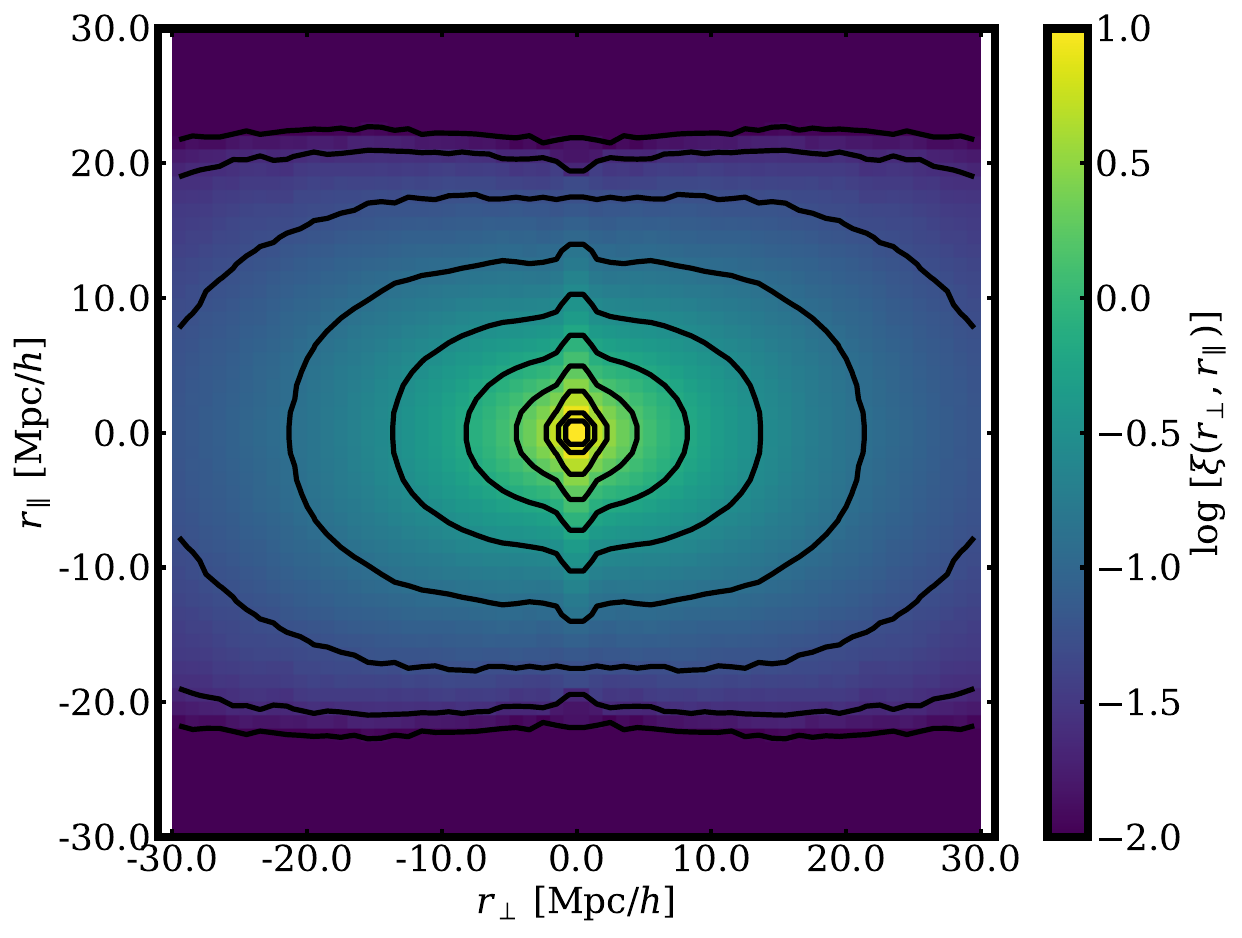}
\caption{`Butterfly' plot showing the redshift-space correlation function, $\xi(r_\perp, r_\parallel)$ as a function of transverse ($r_\perp$) and parallel to the line-of-sight ($r_\parallel$) distance of the galaxy pairs for LAE-like galaxies at $z = 2$. We can clearly see the Finger-of-God effect on small scales ($r \sim 1 \ \mpcoh$), causing an elongation, and the Kaiser (infall) effect on moderate scales ($r \sim 10 \ \mpcoh$), causing a squashing along the line-of-sight.
}
\label{fig:butterfly}
\end{figure}

In reality, when measuring the 3D distribution of galaxies, we always observe them in redshift-space,
where the redshift information of a given galaxy contains both a Hubble flow as well as a peculiar velocity term. The observed redshift of a galaxy can be expressed as the sum of two components:
\begin{equation}
  z_{\rm obs} = z_{\rm cosmo} + \frac{v_{\rm los}}{ac}
  \label{eq:z}
\end{equation}
where $a = 1/(1+z_{\rm cosmo})$ is the scale factor, $c$ is the speed of light, $v_{\rm los}$ is the component of the `peculiar velocity' along the
line-of-sight (LOS) and $z_{\rm cosmo}$ is the cosmological redshift. This second term results from the proper motion of the galaxies and contributes a shift to their apparent comoving positions, given by
\begin{equation}
  \Delta s = \frac{v_{\rm los}}{aH(a)},
\end{equation}
where $H(a)$ is the Hubble parameter at $a$. We refer to this distortion in the galaxy density field along the LOS as redshift space distortions (RSDs).

The RSD effects distort the observed coordinates in the LOS direction, so we can consider the clustering signal, $\xi$, as a function of two variables: the LOS separation, $r_\parallel$, and the separation transverse to the LOS, $r_\perp$. We show the clustering signal of our \textbf{default} sample (see Sec.~\ref{sec:selection}) in Fig. \ref{fig:butterfly},
which has two main features: `squashing' in the LOS direction on large scales ($\sim$$4 \, \mpcoh$) due to the reduction in apparent separation between pairs of galaxies, and `stretching' on small scales (i.e., at $r_\perp \lesssim 1\, \mpcoh$) due to the thermal motions of the galaxies within their dark matter halo hosts. We refer to these as the Kaiser and the Finger-of-God effects, which are caused by the infall of galaxies into clusters and the virial motions of galaxies within the halo,
respectively \citep[see e.g.,][]{2011MNRAS.417.1913R}. In subsequent sections, we examine the amplitude of the LAE redshift-space effects. We look at these redshift distortion effects for \textbf{default + REW} and \textbf{default $+$ stellar mass} selection processes, which are outlined in section IIC. While Fig. \ref{fig:butterfly}, 
gives us a nice qualitative understanding of these `squashing' and `stretching' effects, it is easier to understand this phenomena through plotting the Legendre multipoles:
\begin{equation}
\xi_{\ell}(r) = \frac{2\ell+1}{2} \int d\mu \ \xi(r, \mu) L_{\ell}(\mu),
\label{eq:xi_ell}
\end{equation}
due to the fact that the majority of redshift space information can be accessed by measuring the first three even multipoles ($\ell = 0,2,4$). Here, $r$ is the redshift space separation, $r^2 = r_\perp^2 + r_\parallel^2$, $\mu = r_\parallel/r$ denotes the cosine angle of the galaxy pair with respect to the line of sight, and $L_\ell$ is the Legendre polynomial of order $\ell$.

\begin{figure}
\includegraphics[width=\columnwidth]{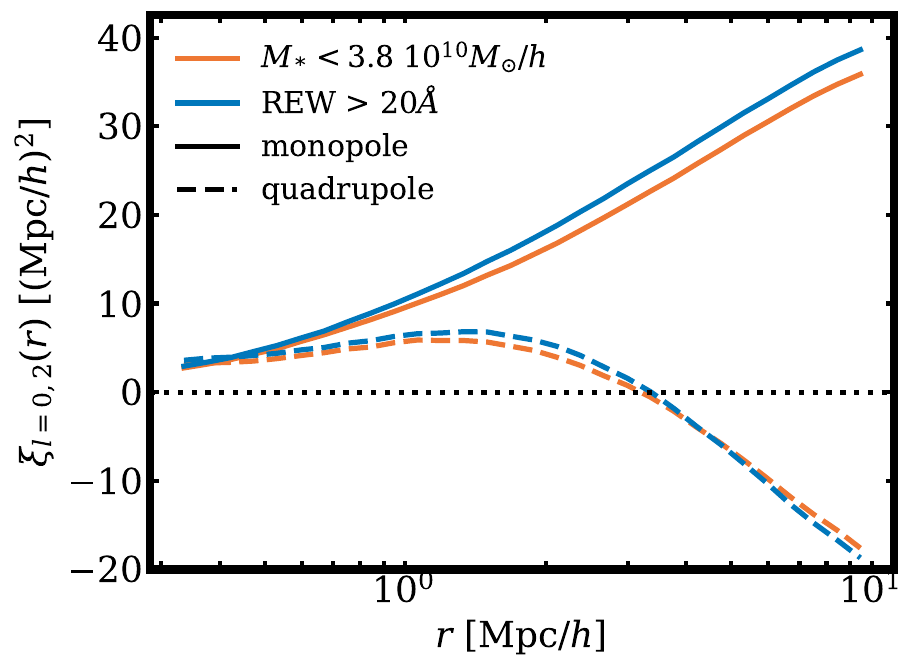}
\caption{\label{fig:xi_ell}
Correlation function multipoles $\xi_{\ell=0,2}$ (see Eq.~\ref{eq:xi_ell}) of the \textbf{default $+$ REW} (blue) and \textbf{default $+$ stellar mass} (orange) samples at $z = 2$ (see Section~\ref{sec:selection} for definitions of the samples). The \textbf{default} sample is not shown, as it is virtually indistinguishable from the \textbf{default $+$ REW} sample, as expected. Due to the lower mean halo mass of the \textbf{default $+$ stellar mass} and thus lower virial velocities of the galaxies in a halo, we see that the small-scale quadrupole (dashed) has a lower amplitude. On large scales, the quadrupole signal of the two samples is very similar, while
echoing Fig.~\ref{fig:corrfunc}, we see that the isotropic clustering (solid) of the \textbf{default $+$ stellar mass} sample is more suppressed, suggesting that that sample exhibits a larger amount of anisotropy. 
}
\end{figure}

The correlation function multipoles for our three samples (see Section~\ref{sec:selection}) at $z = 2$ shown in Fig.~\ref{fig:xi_ell}. The monopole signal ($\ell = 0$) measures the isotropic clustering, while the quadrupole signal ($\ell = 2$) measures the non-isotropic signal due to LOS (Finger-of-God and Kaiser) effects.
For both the \textbf{default} and \textbf{default $+$ REW} samples, we find the quadrupole crosses zero at $r = 3-4 \, \mpcoh$. This feature is important to note as this crossover happens at a much lower $r$ value in comparison to luminous red galaxies (LRGs), where the crossover happens at around $r = 8 \, \mpcoh$ \citep{2011MNRAS.417.1913R}. This difference suggests that the Finger-of-God effects for LAEs are weaker. Between the two selection processes, we see that the \textbf{default $+$ stellar mass} sample has a slightly lower crossing point and a slightly lower amplitude, meaning the Finger-of-God effects are slightly weaker for that sample. We surmise that this difference is the result of the lower mean halo mass of the \textbf{default $+$ stellar mass} sample, which implies smaller virial velocities of the galaxies and thus smaller Finger-of-God effects. On large scales, we see that the quadrupole signal for both samples is pretty similar, whereas the monopole is smaller for the \textbf{default $+$ stellar mass} sample, suggesting that the amount of anistropicity for that sample is larger. This is generally expected for lower-mass samples.

\section{Results}
\label{sec:results}

\subsection{HOD parameters}

\begin{figure*}
\includegraphics[width=\textwidth]{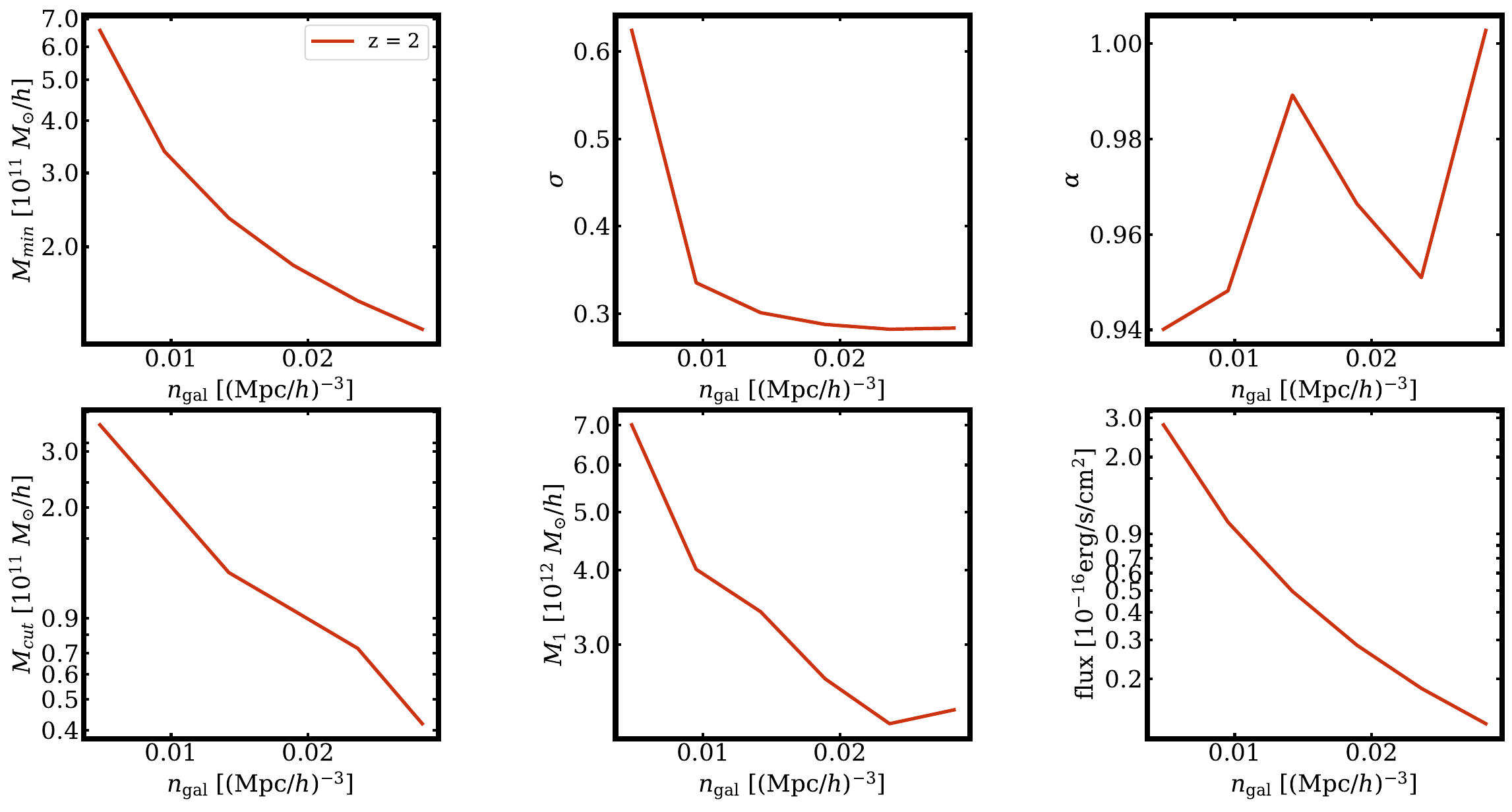}
\caption{\label{fig:hod_num_z2} 
Standard HOD parameters (from \citet{Zheng:2004id}) and Ly$\alpha$ flux ($f_{\rm Ly\alpha}$) cuts (varying between $5 \times 10^{-17} \div 2$ and $1 \times 10^{-16} \times 2$ ${\rm erg} \ {\rm s}^{-1} \ {\rm cm}^{-2}$) at $z = 2$ as a function of number density. Overall, we find that the mass-based parameters decrease with increasing number density (since the mean halo mass also decreases). The parameter $\sigma$, dictating the rate of transitioning from zero to one of the central occupations, decreases with number density, suggesting that the relation between SFR and halo mass becomes tighter for low-luminosity objects.
}
\end{figure*}

\begin{figure*}
\includegraphics[width=\textwidth]{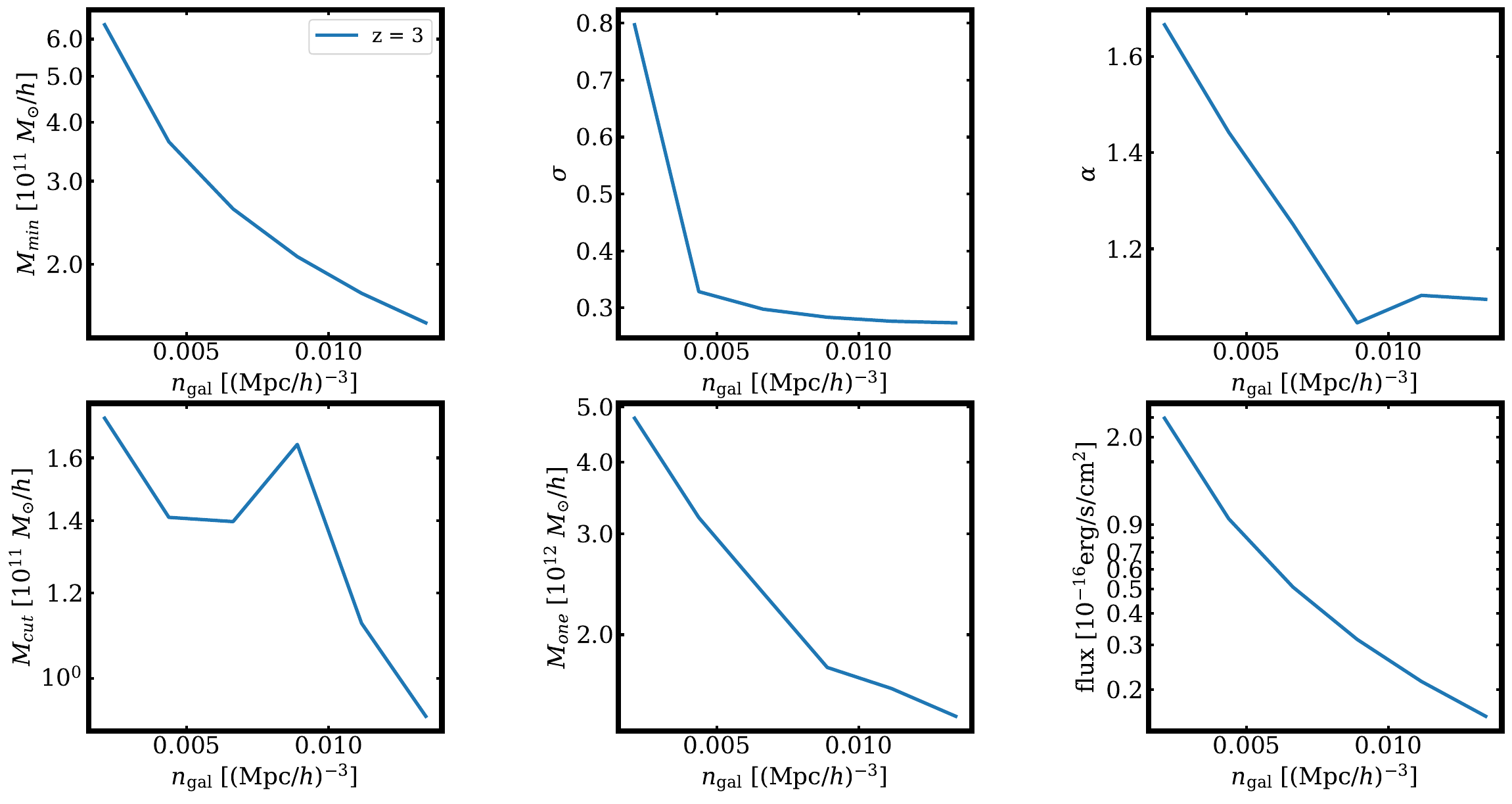}
\caption{\label{fig:hod_num_z3}
Same as Fig.~\ref{fig:hod_num_z2}, but at $z = 3$. We find similar trends to $z = 2$, but the fits appear to be noisier for the $M_{\rm cut}$ parameter. Unlike Fig.~\ref{fig:hod_num_z2}, we find that the power-law slope parameter $\alpha$, controlling the satellite occupation, varies more widely (from 1.7 to 1.1), suggesting that one needs to adopt a broader prior for that parameter in the actual small-scale clustering analysis of upcoming surveys.
}
\end{figure*}

In Fig.~\ref{fig:hod_num_z2} and Fig.~\ref{fig:hod_num_z3}, we show the variation of the five standard HOD parameters from the \citet{Zheng:2004id} formalism with number density, which have been fit via \texttt{scipy}'s \texttt{curvefit} function. In the last panel (lower right) in each figure, we also show the relation between number density and Ly$\alpha$ flux. Overall, we find a rapid (logarithmic) decrease with number density for all three mass-based HOD parameters ($M_1$, $M_{\rm cut}$, and $M_{\rm min}$). Qualitatively, that is expected, as when we go to lower luminosities (fluxes), we include more low-mass halos, whose biases are also lower (see Table~\ref{tab:num_dens} and the following section on clustering). These findings hold true in the case of $z = 3$, though we find that the fits for the $M_{\rm cut}$ parameter are a bit more noisy. Interestingly, the parameter $\sigma$, controlling the rate at which the HOD of the centrals transitions from zero to one, also drops very rapidly with number density. This suggests that the SFR ($f_{\rm Ly\alpha} \propto {\rm SFR}$) and halo mass relation becomes tighter, as we include more low-luminosity galaxies. Another striking feature is the behavior of the parameter $\alpha$, which determines the power-law slope of the satellite occupation distribution. While at $z = 2$, it takes a stable value of $\alpha \approx 0.97$ (note $\alpha \sim 1$ for magnitude-limited and stellar-mass-selected samples), at $z = 3$ it varies between 1.7 and 1.1. Upon close inspection, we find that these differences are due to the fact that lower-density probes do not exhibit the typical broken power-law behavior expected of satellites in the \citet{Zheng:2004id} HOD model. These high values for $\alpha$ imply that when performing fits to data from upcoming surveys, one might need to free the prior on that parameter and allow the analysis chains to explore a broader parameter space. 

\subsection{Galaxy clustering}

\begin{figure*}
\includegraphics[width=\textwidth]{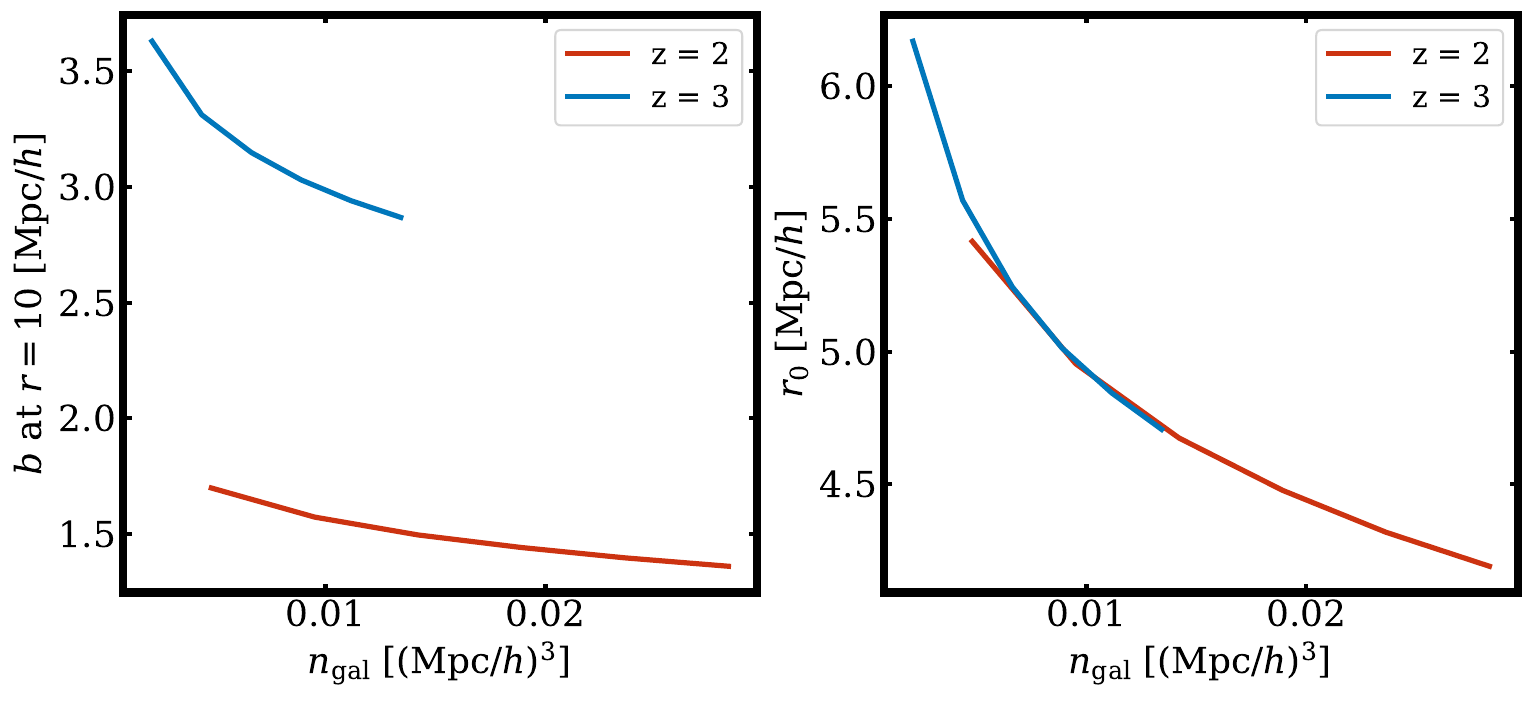}
\caption{\label{fig:bias_r0} 
Linear bias $b$ and `non-linear bias' $r_0$ as a function of number density at $z = 2$ and 3. Similarly to the HOD mass-based parameters, the bias values decrease with increasing number density due to the lower mean halo mass. The fact that $r_0$ is approximately redshift-independent illustrates that the clustering is `stable' at fixed number density for these two redshifts.  This suggests that the bias should grow as $\propto (1+z)$ to counteract the drop in the matter clustering. Such behavior is expected of a massive, long-lived population that is undergoing little merging. We note that a more realistic LAE selection method (e.g., involving radiative transfer, scattering and escape fraction effects) might change some of this behavior by e.g., preferentially selecting younger galaxies.
}
\end{figure*}

Similarly to the previous section where we studied how the standard HOD parameters vary as a function of number density, here we extend the analysis to some clustering statistics such as the linear bias, $b$ (Eq.~\ref{eq:bias}), and the power-law amplitude, $r_0$ (Eq.~\ref{eq:r0}). As before, the number densities shown correspond to a range of flux cuts between $5 \times 10^{-17} \div 2$ and $1 \times 10^{-16} \times 2$ ${\rm erg} \ s^{-1} \ {\rm cm}^{-2}$ at redshifts $z = 2, \ 3$. The results are shown in Fig.~\ref{fig:bias_r0}. As expected, both bias parameters decrease as a function of number density, since the high-number density samples include a larger fraction of low-mass halos, which trace regions of lower density. Furthermore, we see that the linear bias is larger at $z = 3$ than at $z = 2$. This is because at higher redshifts, high-bias objects are even rarer, resulting in a larger linear bias. We see a very similar trend in $r_0$ values: $r_0$ decreases as a function of number density and is larger at higher redshifts. However, unlike the case of the linear bias parameter, the $r_0$ values overlap at $z = 2$ and $z = 3$ at fixed number density.

\begin{table*}
\begin{tabular}{||c c c c c c c c c||} 
 \hline
 $n_{\rm gal}$ & $M_{\rm min}$ & $\sigma$ & $M_{\rm cut}$ & $M_{1}$ & $\alpha$ & $f_{\rm Ly\alpha}$ & $b$ & $r_0$ \\ [0.5ex] 
 [$h^3$Mpc$^{-3}$] & [$10^{11}$ $\msunoh$] &  -- & [$10^{11}$ $\msunoh$] & [$10^{12}$ $\msunoh$] & -- & [$10^{-16}$ s$^{-1}$ erg/cm$^2$]& -- & [$\mpcoh$] \\ [0.5ex]
 \hline\hline
 & & & & $z = 2$ & & & &\\
 \hline
 0.00481 & 6.57 & 0.62 & 3.64 & 7.00 & 0.94 & 2.79 & 1.71 & 5.43\\ 
 \hline
 0.00951 & 3.38 & 0.34 & 2.14 & 4.01 & 0.95 & 1.02 & 1.58 & 4.96\\
 \hline
 0.0142 & 2.34 & 0.30 & 1.25 & 3.41 & 0.99 & 0.50 & 1.50 & 4.67\\
 \hline
 0.0189 & 1.81 & 0.29 & 0.95 & 2.63 & 0.97 & 0.28 & 1.44 & 4.47\\
 \hline
 0.0236 & 1.49 & 0.28 & 0.72 & 2.21 & 0.95 & 0.18 & 1.40 & 4.32\\
 \hline
 0.0283 & 1.27 & 0.28 & 0.42 & 2.33 & 1.00 & 0.13 & 1.36 & 4.19\\ 
\hline\hline
 & & & & $z = 3$ & & & &\\
 \hline
 0.00211 & 6.43 & 0.79 & 1.74 & 4.77 & 1.67 & 2.38 & 3.63 & 6.17\\ 
 \hline
 0.00437 & 3.65 & 0.33 & 1.41 & 3.20 & 1.44 & 0.95 & 3.31 & 5.57\\
 \hline
 0.00663 & 2.62 & 0.30 & 1.40 & 2.37 & 1.25 & 0.50 & 3.15 & 5.24\\
 \hline
 0.00889 & 2.08 & 0.28 & 1.65 & 1.75 & 1.05 & 0.32 & 3.03 & 5.01\\
 \hline
 0.0112 & 1.74 & 0.28 & 1.12 & 1.61 & 1.10 & 0.21 & 2.94 & 4.84\\
 \hline
 0.0134 & 1.50 & 0.27 & 0.92 & 1.44 & 1.10 & 0.16 & 2.87 & 4.70\\ 
 \hline
\end{tabular}
\caption{\label{tab:num_dens}
Best-fit values of the HOD parameters from the \citet{Zheng:2004id} model, minimum Ly$\alpha$ flux ($f_{\rm Ly\alpha}$) cuts, linear bias ($b$), and `non-linear bias' ($r_0$) for several number densities ($n_{\rm gal}$) at $z = 2$ for the \textbf{default} LAE sample (see Section~\ref{sec:selection}). The number densities vary in accordance with the imposed flux cuts in the range of $5 \times 10^{-17} \div 2$ and $1 \times 10^{-16} \times 2$ ${\rm erg} \ s^{-1} $.}
\end{table*}

We provide an intuitive explanation for why that might be the case as follows. At these redshifts, the star-forming population of galaxies (which we select to be our \textbf{default} LAE-like sample) also has a large stellar mass (see Fig.~\ref{fig:selection}). By tracing the merger history of the LAE progenitors, we find a large overlap ($\gtrsim 90\%$) between the objects identified as LAEs at $z = 2$ and $z = 3$, suggesting that these populations are relatively stable and long-lived. As a result, we can approximately express their linear bias as being proportional to $b(z) \propto (1+z)\propto D^{-1}(z)$. Indeed, provided that this is true, the `non-linear bias', for which $r_0$ is a good proxy, is expected to remain stable at fixed number density in order to counteract the drop in the matter clustering, which is exactly what we find. We note that it is possible that a more realistic selection of LAEs (e.g., including radiative transfer effects) would feature a younger and less stable sample of star-forming galaxies (as in the case of our extreme \textbf{default $+$ stellar mass} sample), and we leave this for future work.


\subsection{Assembly bias signature}

Galaxy assembly bias originates from two effects: halo assembly bias and halo occupation variation. The former manifests itself as a difference in the halo clustering among halos of the same mass that differ by some secondary property \cite[e.g., formation time, concentration, spin, see also][]{2007MNRAS.377L...5G}, while the latter comes from the dependence of the halo occupancy (i.e., the number of galaxies per halo) on properties of the host halo other than its mass \citep[e.g.,][]{2018ApJ...853...84Z,2018MNRAS.480.3978A}. In this section, we perform the standard `shuffling test' to estimate the effect of occupation variation. The reason the latter is of particular interest to future surveys is that modern analysis of the small-scale signal often makes use of cosmological $N$-body simulations, which do not implement any baryonic processes but trace the dark-matter component, providing accurate statistics of its virialized structures (i.e., halos) and thus halo assembly bias \citep[see e.g.,][]{2022MNRAS.510.3301Y}. In particular, often small-scale analysis is based on emulators built on an extensive suite of $N$-body simulations \citep[e.g.,][]{2021MNRAS.508.4017M}, spanning a number of cosmologies, which enables the extraction of cosmological constraints by marginalizing over the parameters of the (sub)halo model. As such, these emulators already encapsulate halo assembly bias effects, as the halo statistics are dynamically tracked in the simulation \footnote{A caveat in that statement is the dependence of halo properties on the halo finding algorithm and simulation resolution}. The shuffling procedure, on the other hand, directly probes the connection between the luminous component and its dark matter host beyond the clustering properties of the halos.

Below, we outline the steps of our shuffling test. Broadly, its aim is to estimate how large of an effect is hidden in the halo properties beyond halo mass. This is done by preserving the halo occupation dependence on mass while randomly shuffling the galaxy occupations at fixed halo mass. This procedure mimics the standard implementation of the baseline HOD model, as it preserves the mean occupation number as a function of mass. At the same time, it makes no assumptions regarding the shape of the HOD or the statistical distribution of the central and satellite occupations (typically taken to be binomial and Poisson, respectively). In detail, we first identify the halo hosts of our LAE-like galaxies, which provides us with an occupancy number for each halo. We next define logarithmic mass bins between $\log M_{\rm min} = 12$ and $\log M_{\rm max} = 15$ of width $\Delta \log M = 0.1$, in units of $\msunoh$. We test that the width of the bins does not affect the final result. In each mass bin,
\begin{itemize}
    \item[1.] We randomly shuffle the galaxy occupations of the halos belonging to that bin.
    \item[2.] As our interest is in the effect on quasilinear scales ($r \sim 10 \, \mpcoh$), we place the galaxies in the shuffled sample at the center of their host halo. Alternatively, we could adopt a simple one-halo-term prescription or simply preserve the spatial distribution of the halos with respect to the halo center when performing the shuffling. However, even so the one- to two-halo term transition would not be ideally matched as more simplistic recipes disregard the non-spherical shape of the halo and its substructure. We leave a more dedicated study of the one-halo term of LAEs for future analysis. 
    \item[3.] Measure a large-scale statistics of choice such as the two-point clustering. The correlation function can be calculated using the natural estimator (see Eq.~\ref{eq:xi}). To obtain the mean correlation function, ratio of shuffled to unshuffled clustering, and corresponding errors for the full box, we calculate the mean and jackknife errors of the correlation functions for the 125 subsamples:
    \begin{equation}
        \overline{\xi(r)}=\frac{1}{n}\sum_{i=1}^{n} \overline{\xi(r)}_i
    \end{equation}
    \begin{equation}
    {\rm Var}(\overline{\xi(r)})=\frac{n-1}{n} \sum_{i=1}^{n} (\overline{\xi(r)}_i - \overline{\xi(r)})^2 ,
    \end{equation}
    where $n=125$ and $\overline{\xi(r)}_i$ is the correlation function value at $r$ for subsample $i$ (i.e. excluding the galaxies residing within volume element $i$ in the correlation function computation).
\end{itemize}

Previous studies have shown that the effect of using the full-physics rather than the dark-matter-only counterpart of the simulation is negligible \citep{2020MNRAS.493.5506H}, as the mass properties of the halos are largely unaffected by the presence of baryonic physics (note that this does not hold true for the subhalo properties).

\begin{figure}
\includegraphics[width=\columnwidth]{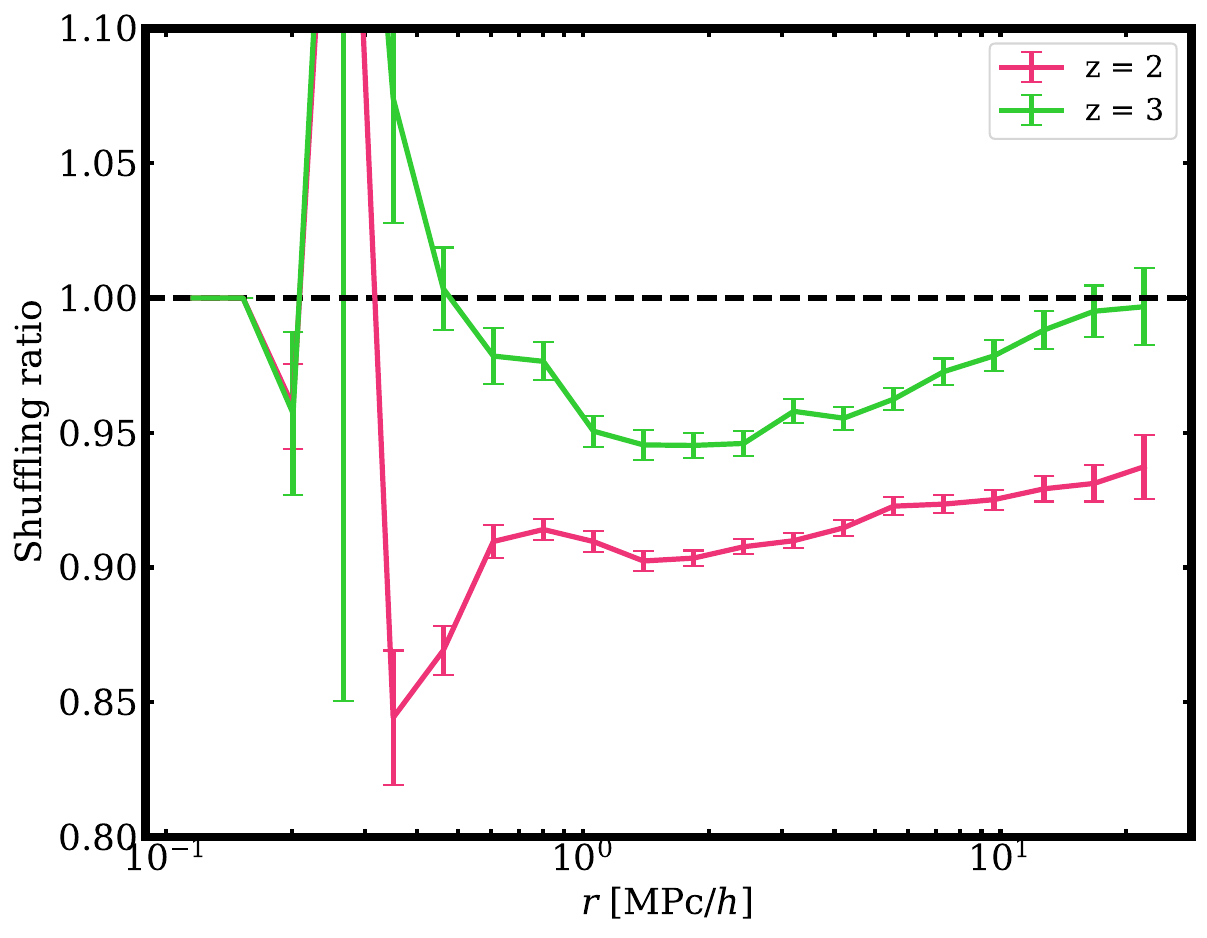}
\caption{Clustering ratio, defined as $\xi_{\rm shuff}(r)/\xi_{\rm orig}(r)$ between a randomly shuffled sample (at constant halo mass) and the original sample at $z = 2$ and $z = 3$. This test shows the amount of galaxy assembly bias present in the LAE sample. We see that on average, this effect is about 8\% at $r \gtrsim 1 \ \mpcoh$ for the galaxies at $z = 2$, and about $\sim$2.5\% for the LAEs at $z = 3$, suggesting that assembly bias affects this galaxy population a bit less at high redshifts, likely due to the lower number of two-halo interactions (mergers, splashback, stripping). The sharp feature at $r \sim 0.3 \ \mpcoh$ is due to our satellite placement procedure (we put all galaxies in a given halo at its center), which leads to discontinuities in transitioning between the one- and two-halo, but is confined to these scales and does not affect the larger scales, $r \gtrsim 0.6 \ \mpcoh$.}
\label{fig:gab}
\end{figure}

In Fig.~\ref{fig:gab}, we show the ratio of the shuffled sample to the original (unshuffled) sample, i.e. $\xi_{\rm shuff}(r)/\xi_{\rm orig}(r)$. The results are shown for our \textbf{default} selection of LAEs (see Section~\ref{sec:selection}) at the two redshifts of interest, $z = 2$ and 3. We see that on average, the deviation from one is about 8\% at $r \gtrsim 1 \, \mpcoh$ for the galaxies at $z = 2$, and about $\sim$3\% for the LAEs at $z = 3$, demonstrating that assembly bias affects this galaxy population less at the higher redshift compared with the lower one. The discrepancy at $z = 2$ is reminiscent of that of LRGs and ELGs found in previous studies \citep[e.g.,][]{2021MNRAS.502.3599H}. We interpret the finding at $z = 3$ as being the consequence of astrophysical and numerical effects. The numerical effects are related to the well-known fact that in highly clustered regions, spatial halo finding algorithms such as the FoF algorithm (used to define halos in MTNG) struggle with deblending halos, which can lead to biases in the reported halo masses: for example, an intermediate-sized halo on the outskirts of a larger halo may have its mass underestimated and hence have fewer galaxies ascribed to it by a mass-only HOD prescription compared with the number it would have received if its mass had been correctly identified. That issue is of course exacerbated at lower redshifts, for which we are more likely to find large galaxy clusters and halos in close proximity \citep{2011ApJS..195....4M}.

We also expect that galaxy formation and evolution is impacted by environmental and tidal effects, which can lead to splashback effects, quenching or triggering of star formation, merging of halos, and stripping of the dark matter envelope of small infalling halos \citep[see e.g.,][for a similar study of red and blue galaxies]{2023MNRAS.524.2524H}. These astrophysical considerations suggest that one needs to take into account additional properties of the halo beyond its mass in order to correctly predict the clustering of LAE galaxies. These effects are less pronounced at $z = 3$ both due to the fact that the cumulative number of halo interactions is lower and also because being less massive at this redshift, halos are less likely to have active AGNs, which cause quenching of the star-forming galaxies. We leave for the future a more comprehensive study of the properties the help reconcile the observed discrepancy with the mass-only prediction. After identifying these properties, it is important to incorporate a dependence on them in the galaxy-halo models used to analyze observations in order to get an unbiased inference on halo mass and cosmological parameters \citep[see e.g.,][for a similar study of red and blue galaxies]{2023MNRAS.524.2507H}.

\section{Discussion}
\label{sec:disc}

LAEs have been an object of study for both observers and simulators. In this section, to put our findings into context, we discuss several relevant works using either simulations or observations.

\begin{figure*}
\includegraphics[width=0.48\textwidth]{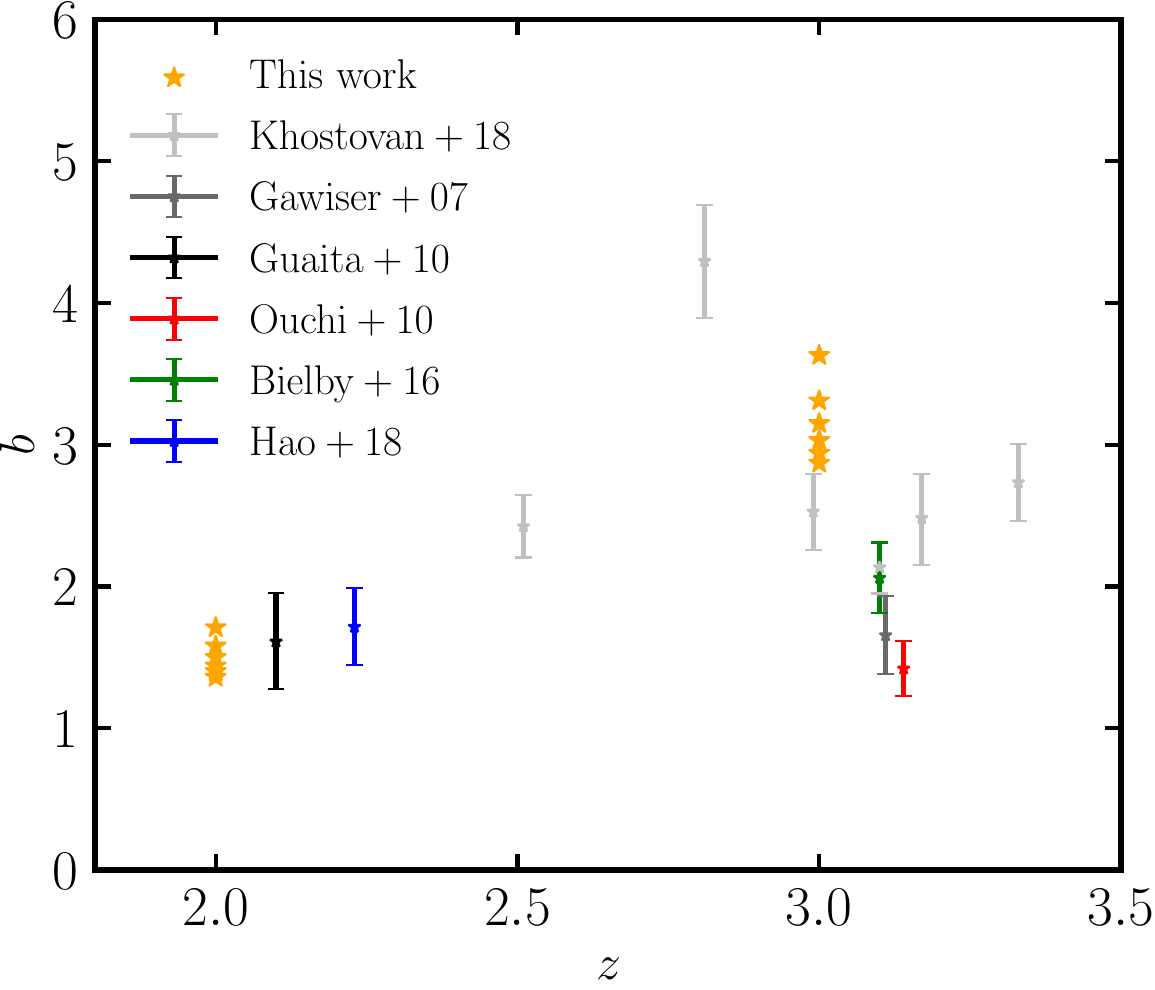}
\includegraphics[width=0.48\textwidth]{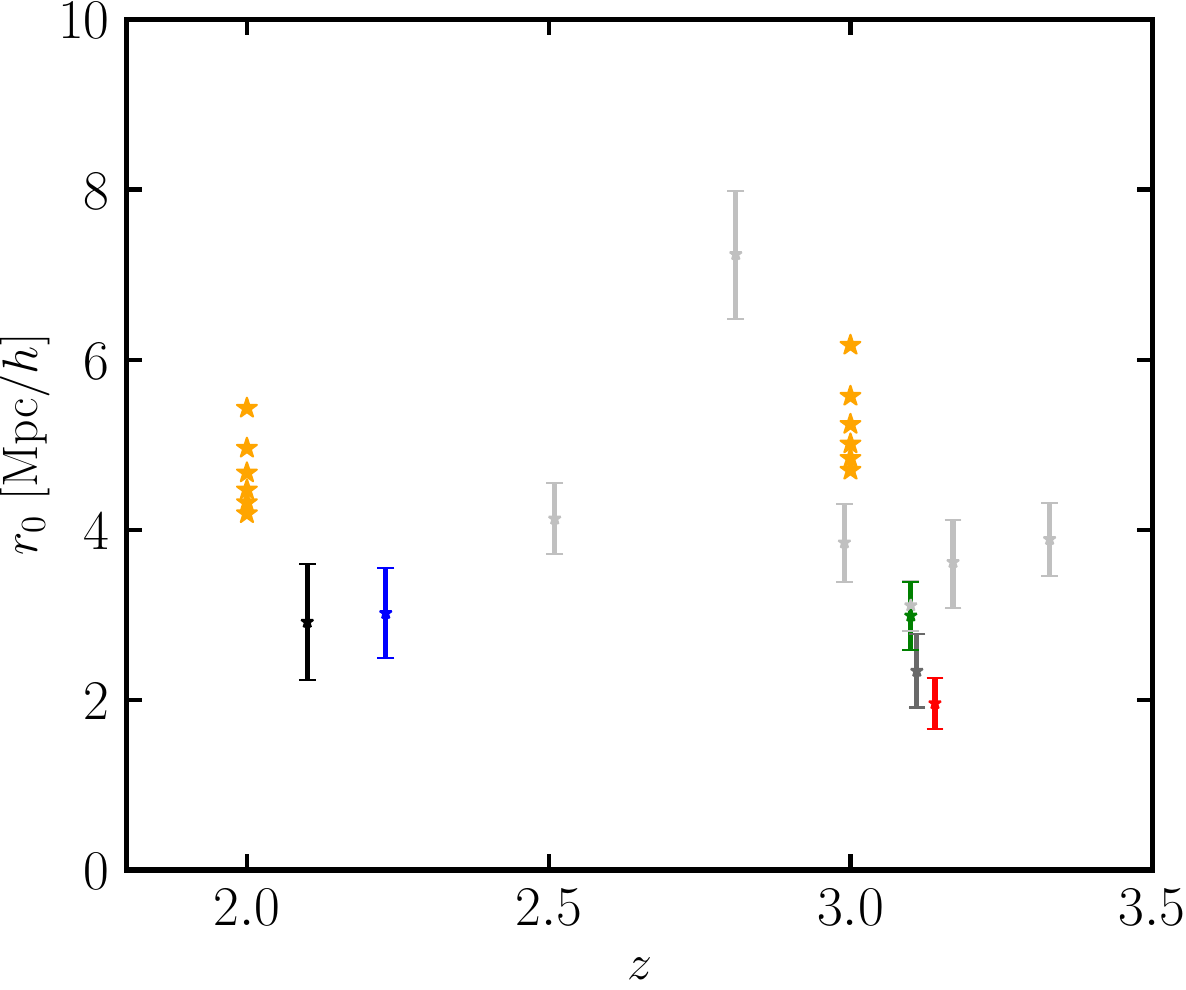}%
\caption{\label{fig:b_r0_lit} 
Comparison between the linear bias $b$ (left panel) and clustering parameter $r_0$ (right panel) obtained in this study compared with measurements from observations \citep{Gawiser,Guaita,Ouchi,Bielby,Hao,Khostovan}. We find overall good agreement between our inferred $b$ and $r_0$ and the observational probes. However, our values for $b$ at $z = 3$ are a bit higher than \citet{Bielby,Ouchi} and \citet{Gawiser}, but are comparable to the \citet{Khostovan} ones. Similarly, our values of $r_0$ are a bit higher, especially at the low-redshift end. These differences could be attributed to e.g., our lack of an escape fraction model, which might have led us to select a slightly less clustered sample.
}
\end{figure*}

The small-scale clustering of LAEs at $z \sim 2 - 3$ has been measured by various groups in the last couple of decades \citep{Gawiser,Guaita,Ouchi,Bielby,Hao,Khostovan,2023ApJ...951..119R,2024ApJ...962...36L}. In Fig.~\ref{fig:b_r0_lit}, we show a comparison between the linear bias $b$ and clustering parameter $r_0$ obtained in this study and the observational constraints on these parameters. We find overall good agreement between our inferred bias at $z = 2$ and $z = 3$ and the observational probes. In particular, similarly to \citet{Guaita} and \citet{Hao}, we find that the bias at $z \sim 2$ ranges between $b = 1$ and 2, while at $z \sim 3$ it increases to values around 3. Our values for $b$ at $z = 3$ are a bit higher than \citet{Bielby,Ouchi} and \citet{Gawiser}, but are comparable to the \citet{Khostovan} findings. Similarly, we find that our values of $r_0$ are roughly consistent with observations, albeit again a bit higher, especially at the low-redshift end. While this quantity $r_0$ is sensitive to the scales used in the fit, the difference (both for $b$ and $r_0$) could also be attributed to the fact that our simulation lacks an escape fraction model and other radiative transfer effects, which if accounted for, might have led us to select a slightly less massive (less clustered) sample.

Several groups have also identified LAEs using hydrodynamical simulations and semi-analytic models and studied their clustering properties and galaxy-halo connection \citep[e.g.,][]{2007MNRAS.381...75M,2018MNRAS.479.2564W,2019MNRAS.485.1350W,Nagamine,Garel}. In Fig.~\ref{fig:scatter}, we showed that our SFR-stellar mass relation for the LAEs agrees very well with the \citet{Nagamine} fit. In addition, Fig. 8 and Table 2 of \citet{Nagamine} presents the bias $b$ and clustering parameter $r_0$ at $z = 3$ for two different scenarios: `stochastic’ and `escape fraction’. We find good agreement with their values of about $b \approx 3$ for their `stochastic scenario’ and consistent values for $r_0$ for both scenarios ($r_0 \approx 3-5 \ \mpcoh$). Additionally, we compare our HOD to Fig. 11 in \citet{Garel} and find it to be consistent with their mid-luminosity sample, which roughly matches the luminosity range of our sample as well. We attribute differences on the high-mass end to the fact that we do not incorporate dust attenuation, escape fraction, EW modeling and resonant scattering, which their semi-analytic model does feature. A key finding of that paper is the similitude between their LAE and LBG sample, since both populations target star-forming galaxies with bright UV emission. The authors attribute the remaining differences to: ``EW selections, UV detection limits, and a decreasing Lya-to-UV escape fraction ratio in high SFR galaxies’’. We leave a comparative study of LAEs and LBGs for future work.

\section{Conclusions}
\label{sec:conclusions}
Future cosmological experiments such as DESI-II will likely target strongly emitting galaxies at high redshifts ($z \gtrsim 2$). In this study, we use the hydrodynamical simulation MillenniumTNG (MTNG) to study Lyman-alpha Emitting (LAE) galaxies at $z = 2$ and 3, with the hope that our analysis will inform these planned galaxy surveys. In particular, we are interested in determining key galaxy-halo properties of LAEs such as their clustering and occupation statistics. 

In Section~\ref{sec:selection}, we define LAEs in three different ways. Our \textbf{default} sample is obtained through an empirical relation linking star formation rate (SFR) and Lyman-alpha flux (see Fig.~\ref{fig:selection}). We also explore a sample that imposes an additional REW cut (\textbf{default $+$ REW}), which behaves qualitatively almost identically to the default case. Finally, we study a sample that combines a Ly$\alpha$ flux cut with a stellar mass maximum threshold (\textbf{default $+$ stellar mass}). This last definition approximates the effect of a steady lowering of the escape fraction of photons with increasing halo mass. This is reflected on the halo occupation of the centrals as seen in Fig.~\ref{fig:hod} as well as on the correlation function (shown in Fig.~\ref{fig:corrfunc}), which sees a 5-10\% decrease of the clustering amplitude at $r \approx 10 \ \mpcoh$. Furthermore, the redshift clustering of the \textbf{default} LAEs suggests that the Finger-of-God effects are weaker compared with a typical low-redshift red population, with a transition scale of $r \approx 4 \ \mpcoh$ (see Fig.~\ref{fig:butterfly} and Fig.~\ref{fig:xi_ell}).

In Fig.~\ref{fig:hod_num_z2} and Fig.~\ref{fig:hod_num_z3}, we show the evolution of HOD parameters as a function of number density (or equivalently, Ly$\alpha$ flux) at the two redshifts of interest. We find that the HOD mass parameters rapidly (and logarithmically) decrease with increasing number density, due to the overall lowering of the mean halo mass of the population. Additionally, the HOD parameter $\sigma$ also decreases with number density, implying that the SFR-halo mass relationship becomes tighter for low-luminosity objects. At $z = 3$, the power-law slope parameter $\alpha$ takes a surprisingly large range of values (1.1 to 1.7).

In Fig.~\ref{fig:bias_r0}, we showcase the linear and quasi-linear `bias' parameters $b$ and $r_0$, respectively, as a function of number density at both redshifts. Interestingly, the non-linear clustering parameter $r_0$ appears to be fixed at fixed number density and constant at both epochs. On the other hand, the linear bias parameter $b$ varies with redshift as $b(z) \propto (1 + z)$ in order to cancel the drop in the matter clustering with increasing redshift. This suggests that the LAE samples defined in this work are relatively stable and long-lived. Indeed, we find that a large fraction of the LAE progenitors at $z = 2$ have a large overlap with the sample at $z = 3$.

In Fig.~\ref{fig:gab}, we examine the amount of galaxy assembly bias at $z = 2$ and 3 through a shuffling test. We report that at $z = 2$ the discrepancy between the shuffling and original sample is roughly $\lesssim$10\%, suggesting that a cosmological analysis of the small-scale clustering that ignores the effect of halo properties beyond halo mass could potentially lead to substantial ($\sim$10\%) biases on the inferred cosmological parameters (e.g., $\sigma_8$ and mean halo mass). At $z = 3$, the assembly bias effect decreases to about 5\%, which implies that assembly bias effects become less important at high $z$. We conjecture that this is the case due to the lower cumulative number of two-halo interactions (mergers, splashback, stripping, etc.) between massive galaxies at higher redshifts. 

While our study is based on a single full-physics simulation (MTNG), we expect our results to reflect accurately the qualitative behavior of LAEs in the Universe. In subsequent work, we plan to investigate the effect of incorporating radiative transfer physics on this population of LAEs, on their astrophysical and cosmological properties \citep{2011ApJ...726...38Z,2018A&A...614A..31B,2021ApJ...912L..24M,2024arXiv240113166E}.


\begin{acknowledgments}
We wish to acknowledge the support of the PI$^2$ program for supporting J.R. through the summer of 2023. B.H. is grateful to the Miller Institute for their generous support.
\end{acknowledgments}

\nocite{*}

\bibliography{apssamp}

\providecommand{\noopsort}[1]{}\providecommand{\singleletter}[1]{#1}%
\begin{thebibliography}{102}%
\makeatletter
\providecommand \@ifxundefined [1]{%
 \@ifx{#1\undefined}
}%
\providecommand \@ifnum [1]{%
 \ifnum #1\expandafter \@firstoftwo
 \else \expandafter \@secondoftwo
 \fi
}%
\providecommand \@ifx [1]{%
 \ifx #1\expandafter \@firstoftwo
 \else \expandafter \@secondoftwo
 \fi
}%
\providecommand \natexlab [1]{#1}%
\providecommand \enquote  [1]{``#1''}%
\providecommand \bibnamefont  [1]{#1}%
\providecommand \bibfnamefont [1]{#1}%
\providecommand \citenamefont [1]{#1}%
\providecommand \href@noop [0]{\@secondoftwo}%
\providecommand \href [0]{\begingroup \@sanitize@url \@href}%
\providecommand \@href[1]{\@@startlink{#1}\@@href}%
\providecommand \@@href[1]{\endgroup#1\@@endlink}%
\providecommand \@sanitize@url [0]{\catcode `\\12\catcode `\$12\catcode
  `\&12\catcode `\#12\catcode `\^12\catcode `\_12\catcode `\%12\relax}%
\providecommand \@@startlink[1]{}%
\providecommand \@@endlink[0]{}%
\providecommand \url  [0]{\begingroup\@sanitize@url \@url }%
\providecommand \@url [1]{\endgroup\@href {#1}{\urlprefix }}%
\providecommand \urlprefix  [0]{URL }%
\providecommand \Eprint [0]{\href }%
\providecommand \doibase [0]{https://doi.org/}%
\providecommand \selectlanguage [0]{\@gobble}%
\providecommand \bibinfo  [0]{\@secondoftwo}%
\providecommand \bibfield  [0]{\@secondoftwo}%
\providecommand \translation [1]{[#1]}%
\providecommand \BibitemOpen [0]{}%
\providecommand \bibitemStop [0]{}%
\providecommand \bibitemNoStop [0]{.\EOS\space}%
\providecommand \EOS [0]{\spacefactor3000\relax}%
\providecommand \BibitemShut  [1]{\csname bibitem#1\endcsname}%
\let\auto@bib@innerbib\@empty
\bibitem [{\citenamefont {{Abbott}}\ \emph {et~al.}(2018)\citenamefont
  {{Abbott}}, \citenamefont {{Abdalla}}, \citenamefont {{Alarcon}},
  \citenamefont {{Aleksi{\'c}}}, \citenamefont {{Allam}}, \citenamefont
  {{Allen}}, \citenamefont {{Amara}}, \citenamefont {{Annis}}, \citenamefont
  {{Asorey}}, \citenamefont {{Avila}}, \citenamefont {{Bacon}}, \citenamefont
  {{Balbinot}}, \citenamefont {{Banerji}}, \citenamefont {{Banik}},
  \citenamefont {{Barkhouse}}, \citenamefont {{Baumer}}, \citenamefont
  {{Baxter}}, \citenamefont {{Bechtol}}, \citenamefont {{Becker}},
  \citenamefont {{Benoit-L{\'e}vy}}, \citenamefont {{Benson}}, \citenamefont
  {{Bernstein}}, \citenamefont {{Bertin}}, \citenamefont {{Blazek}},
  \citenamefont {{Bridle}}, \citenamefont {{Brooks}}, \citenamefont {{Brout}},
  \citenamefont {{Buckley-Geer}}, \citenamefont {{Burke}}, \citenamefont
  {{Busha}}, \citenamefont {{Campos}}, \citenamefont {{Capozzi}}, \citenamefont
  {{Carnero Rosell}}, \citenamefont {{Carrasco Kind}}, \citenamefont
  {{Carretero}}, \citenamefont {{Castander}}, \citenamefont {{Cawthon}},
  \citenamefont {{Chang}}, \citenamefont {{Chen}}, \citenamefont {{Childress}},
  \citenamefont {{Choi}}, \citenamefont {{Conselice}}, \citenamefont
  {{Crittenden}}, \citenamefont {{Crocce}}, \citenamefont {{Cunha}},
  \citenamefont {{D'Andrea}}, \citenamefont {{da Costa}}, \citenamefont
  {{Das}}, \citenamefont {{Davis}}, \citenamefont {{Davis}}, \citenamefont {{De
  Vicente}}, \citenamefont {{DePoy}}, \citenamefont {{DeRose}}, \citenamefont
  {{Desai}}, \citenamefont {{Diehl}}, \citenamefont {{Dietrich}}, \citenamefont
  {{Dodelson}}, \citenamefont {{Doel}}, \citenamefont {{Drlica-Wagner}},
  \citenamefont {{Eifler}}, \citenamefont {{Elliott}}, \citenamefont
  {{Elsner}}, \citenamefont {{Elvin-Poole}}, \citenamefont {{Estrada}},
  \citenamefont {{Evrard}}, \citenamefont {{Fang}}, \citenamefont
  {{Fernandez}}, \citenamefont {{Fert{\'e}}}, \citenamefont {{Finley}},
  \citenamefont {{Flaugher}}, \citenamefont {{Fosalba}}, \citenamefont
  {{Friedrich}}, \citenamefont {{Frieman}}, \citenamefont
  {{Garc{\'\i}a-Bellido}}, \citenamefont {{Garcia-Fernandez}}, \citenamefont
  {{Gatti}}, \citenamefont {{Gaztanaga}}, \citenamefont {{Gerdes}},
  \citenamefont {{Giannantonio}}, \citenamefont {{Gill}}, \citenamefont
  {{Glazebrook}}, \citenamefont {{Goldstein}}, \citenamefont {{Gruen}},
  \citenamefont {{Gruendl}}, \citenamefont {{Gschwend}}, \citenamefont
  {{Gutierrez}}, \citenamefont {{Hamilton}}, \citenamefont {{Hartley}},
  \citenamefont {{Hinton}}, \citenamefont {{Honscheid}}, \citenamefont
  {{Hoyle}}, \citenamefont {{Huterer}}, \citenamefont {{Jain}}, \citenamefont
  {{James}}, \citenamefont {{Jarvis}}, \citenamefont {{Jeltema}}, \citenamefont
  {{Johnson}}, \citenamefont {{Johnson}}, \citenamefont {{Kacprzak}},
  \citenamefont {{Kent}}, \citenamefont {{Kim}}, \citenamefont {{King}},
  \citenamefont {{Kirk}}, \citenamefont {{Kokron}}, \citenamefont {{Kovacs}},
  \citenamefont {{Krause}}, \citenamefont {{Krawiec}}, \citenamefont
  {{Kremin}}, \citenamefont {{Kuehn}}, \citenamefont {{Kuhlmann}},
  \citenamefont {{Kuropatkin}}, \citenamefont {{Lacasa}}, \citenamefont
  {{Lahav}}, \citenamefont {{Li}}, \citenamefont {{Liddle}}, \citenamefont
  {{Lidman}}, \citenamefont {{Lima}}, \citenamefont {{Lin}}, \citenamefont
  {{MacCrann}}, \citenamefont {{Maia}}, \citenamefont {{Makler}}, \citenamefont
  {{Manera}}, \citenamefont {{March}}, \citenamefont {{Marshall}},
  \citenamefont {{Martini}}, \citenamefont {{McMahon}}, \citenamefont
  {{Melchior}}, \citenamefont {{Menanteau}}, \citenamefont {{Miquel}},
  \citenamefont {{Miranda}}, \citenamefont {{Mudd}}, \citenamefont {{Muir}},
  \citenamefont {{M{\"o}ller}}, \citenamefont {{Neilsen}}, \citenamefont
  {{Nichol}}, \citenamefont {{Nord}}, \citenamefont {{Nugent}}, \citenamefont
  {{Ogando}}, \citenamefont {{Palmese}}, \citenamefont {{Peacock}},
  \citenamefont {{Peiris}}, \citenamefont {{Peoples}}, \citenamefont
  {{Percival}}, \citenamefont {{Petravick}}, \citenamefont {{Plazas}},
  \citenamefont {{Porredon}}, \citenamefont {{Prat}}, \citenamefont {{Pujol}},
  \citenamefont {{Rau}}, \citenamefont {{Refregier}}, \citenamefont {{Ricker}},
  \citenamefont {{Roe}}, \citenamefont {{Rollins}}, \citenamefont {{Romer}},
  \citenamefont {{Roodman}}, \citenamefont {{Rosenfeld}}, \citenamefont
  {{Ross}}, \citenamefont {{Rozo}}, \citenamefont {{Rykoff}}, \citenamefont
  {{Sako}}, \citenamefont {{Salvador}}, \citenamefont {{Samuroff}},
  \citenamefont {{S{\'a}nchez}}, \citenamefont {{Sanchez}}, \citenamefont
  {{Santiago}}, \citenamefont {{Scarpine}}, \citenamefont {{Schindler}},
  \citenamefont {{Scolnic}}, \citenamefont {{Secco}}, \citenamefont
  {{Serrano}}, \citenamefont {{Sevilla-Noarbe}}, \citenamefont {{Sheldon}},
  \citenamefont {{Smith}}, \citenamefont {{Smith}}, \citenamefont {{Smith}},
  \citenamefont {{Soares-Santos}}, \citenamefont {{Sobreira}}, \citenamefont
  {{Suchyta}}, \citenamefont {{Tarle}}, \citenamefont {{Thomas}}, \citenamefont
  {{Troxel}}, \citenamefont {{Tucker}}, \citenamefont {{Tucker}}, \citenamefont
  {{Uddin}}, \citenamefont {{Varga}}, \citenamefont {{Vielzeuf}}, \citenamefont
  {{Vikram}}, \citenamefont {{Vivas}}, \citenamefont {{Walker}}, \citenamefont
  {{Wang}}, \citenamefont {{Wechsler}}, \citenamefont {{Weller}}, \citenamefont
  {{Wester}}, \citenamefont {{Wolf}}, \citenamefont {{Yanny}}, \citenamefont
  {{Yuan}}, \citenamefont {{Zenteno}}, \citenamefont {{Zhang}}, \citenamefont
  {{Zhang}}, \citenamefont {{Zuntz}},\ and\ \citenamefont {{Dark Energy Survey
  Collaboration}}}]{2018PhRvD..98d3526A}%
  \BibitemOpen
  \bibfield  {author} {\bibinfo {author} {\bibfnamefont {T.~M.~C.}\
  \bibnamefont {{Abbott}}}, \bibinfo {author} {\bibfnamefont {F.~B.}\
  \bibnamefont {{Abdalla}}}, \bibinfo {author} {\bibfnamefont {A.}~\bibnamefont
  {{Alarcon}}}, \bibinfo {author} {\bibfnamefont {J.}~\bibnamefont
  {{Aleksi{\'c}}}}, \bibinfo {author} {\bibfnamefont {S.}~\bibnamefont
  {{Allam}}}, \bibinfo {author} {\bibfnamefont {S.}~\bibnamefont {{Allen}}},
  \bibinfo {author} {\bibfnamefont {A.}~\bibnamefont {{Amara}}}, \bibinfo
  {author} {\bibfnamefont {J.}~\bibnamefont {{Annis}}}, \bibinfo {author}
  {\bibfnamefont {J.}~\bibnamefont {{Asorey}}}, \bibinfo {author}
  {\bibfnamefont {S.}~\bibnamefont {{Avila}}}, \bibinfo {author} {\bibfnamefont
  {D.}~\bibnamefont {{Bacon}}}, \bibinfo {author} {\bibfnamefont
  {E.}~\bibnamefont {{Balbinot}}}, \bibinfo {author} {\bibfnamefont
  {M.}~\bibnamefont {{Banerji}}}, \bibinfo {author} {\bibfnamefont
  {N.}~\bibnamefont {{Banik}}}, \bibinfo {author} {\bibfnamefont
  {W.}~\bibnamefont {{Barkhouse}}}, \bibinfo {author} {\bibfnamefont
  {M.}~\bibnamefont {{Baumer}}}, \bibinfo {author} {\bibfnamefont
  {E.}~\bibnamefont {{Baxter}}}, \bibinfo {author} {\bibfnamefont
  {K.}~\bibnamefont {{Bechtol}}}, \bibinfo {author} {\bibfnamefont {M.~R.}\
  \bibnamefont {{Becker}}}, \bibinfo {author} {\bibfnamefont {A.}~\bibnamefont
  {{Benoit-L{\'e}vy}}}, \bibinfo {author} {\bibfnamefont {B.~A.}\ \bibnamefont
  {{Benson}}}, \bibinfo {author} {\bibfnamefont {G.~M.}\ \bibnamefont
  {{Bernstein}}}, \bibinfo {author} {\bibfnamefont {E.}~\bibnamefont
  {{Bertin}}}, \bibinfo {author} {\bibfnamefont {J.}~\bibnamefont {{Blazek}}},
  \bibinfo {author} {\bibfnamefont {S.~L.}\ \bibnamefont {{Bridle}}}, \bibinfo
  {author} {\bibfnamefont {D.}~\bibnamefont {{Brooks}}}, \bibinfo {author}
  {\bibfnamefont {D.}~\bibnamefont {{Brout}}}, \bibinfo {author} {\bibfnamefont
  {E.}~\bibnamefont {{Buckley-Geer}}}, \bibinfo {author} {\bibfnamefont
  {D.~L.}\ \bibnamefont {{Burke}}}, \bibinfo {author} {\bibfnamefont {M.~T.}\
  \bibnamefont {{Busha}}}, \bibinfo {author} {\bibfnamefont {A.}~\bibnamefont
  {{Campos}}}, \bibinfo {author} {\bibfnamefont {D.}~\bibnamefont {{Capozzi}}},
  \bibinfo {author} {\bibfnamefont {A.}~\bibnamefont {{Carnero Rosell}}},
  \bibinfo {author} {\bibfnamefont {M.}~\bibnamefont {{Carrasco Kind}}},
  \bibinfo {author} {\bibfnamefont {J.}~\bibnamefont {{Carretero}}}, \bibinfo
  {author} {\bibfnamefont {F.~J.}\ \bibnamefont {{Castander}}}, \bibinfo
  {author} {\bibfnamefont {R.}~\bibnamefont {{Cawthon}}}, \bibinfo {author}
  {\bibfnamefont {C.}~\bibnamefont {{Chang}}}, \bibinfo {author} {\bibfnamefont
  {N.}~\bibnamefont {{Chen}}}, \bibinfo {author} {\bibfnamefont
  {M.}~\bibnamefont {{Childress}}}, \bibinfo {author} {\bibfnamefont
  {A.}~\bibnamefont {{Choi}}}, \bibinfo {author} {\bibfnamefont
  {C.}~\bibnamefont {{Conselice}}}, \bibinfo {author} {\bibfnamefont
  {R.}~\bibnamefont {{Crittenden}}}, \bibinfo {author} {\bibfnamefont
  {M.}~\bibnamefont {{Crocce}}}, \bibinfo {author} {\bibfnamefont {C.~E.}\
  \bibnamefont {{Cunha}}}, \bibinfo {author} {\bibfnamefont {C.~B.}\
  \bibnamefont {{D'Andrea}}}, \bibinfo {author} {\bibfnamefont {L.~N.}\
  \bibnamefont {{da Costa}}}, \bibinfo {author} {\bibfnamefont
  {R.}~\bibnamefont {{Das}}}, \bibinfo {author} {\bibfnamefont {T.~M.}\
  \bibnamefont {{Davis}}}, \bibinfo {author} {\bibfnamefont {C.}~\bibnamefont
  {{Davis}}}, \bibinfo {author} {\bibfnamefont {J.}~\bibnamefont {{De
  Vicente}}}, \bibinfo {author} {\bibfnamefont {D.~L.}\ \bibnamefont
  {{DePoy}}}, \bibinfo {author} {\bibfnamefont {J.}~\bibnamefont {{DeRose}}},
  \bibinfo {author} {\bibfnamefont {S.}~\bibnamefont {{Desai}}}, \bibinfo
  {author} {\bibfnamefont {H.~T.}\ \bibnamefont {{Diehl}}}, \bibinfo {author}
  {\bibfnamefont {J.~P.}\ \bibnamefont {{Dietrich}}}, \bibinfo {author}
  {\bibfnamefont {S.}~\bibnamefont {{Dodelson}}}, \bibinfo {author}
  {\bibfnamefont {P.}~\bibnamefont {{Doel}}}, \bibinfo {author} {\bibfnamefont
  {A.}~\bibnamefont {{Drlica-Wagner}}}, \bibinfo {author} {\bibfnamefont
  {T.~F.}\ \bibnamefont {{Eifler}}}, \bibinfo {author} {\bibfnamefont {A.~E.}\
  \bibnamefont {{Elliott}}}, \bibinfo {author} {\bibfnamefont {F.}~\bibnamefont
  {{Elsner}}}, \bibinfo {author} {\bibfnamefont {J.}~\bibnamefont
  {{Elvin-Poole}}}, \bibinfo {author} {\bibfnamefont {J.}~\bibnamefont
  {{Estrada}}}, \bibinfo {author} {\bibfnamefont {A.~E.}\ \bibnamefont
  {{Evrard}}}, \bibinfo {author} {\bibfnamefont {Y.}~\bibnamefont {{Fang}}},
  \bibinfo {author} {\bibfnamefont {E.}~\bibnamefont {{Fernandez}}}, \bibinfo
  {author} {\bibfnamefont {A.}~\bibnamefont {{Fert{\'e}}}}, \bibinfo {author}
  {\bibfnamefont {D.~A.}\ \bibnamefont {{Finley}}}, \bibinfo {author}
  {\bibfnamefont {B.}~\bibnamefont {{Flaugher}}}, \bibinfo {author}
  {\bibfnamefont {P.}~\bibnamefont {{Fosalba}}}, \bibinfo {author}
  {\bibfnamefont {O.}~\bibnamefont {{Friedrich}}}, \bibinfo {author}
  {\bibfnamefont {J.}~\bibnamefont {{Frieman}}}, \bibinfo {author}
  {\bibfnamefont {J.}~\bibnamefont {{Garc{\'\i}a-Bellido}}}, \bibinfo {author}
  {\bibfnamefont {M.}~\bibnamefont {{Garcia-Fernandez}}}, \bibinfo {author}
  {\bibfnamefont {M.}~\bibnamefont {{Gatti}}}, \bibinfo {author} {\bibfnamefont
  {E.}~\bibnamefont {{Gaztanaga}}}, \bibinfo {author} {\bibfnamefont {D.~W.}\
  \bibnamefont {{Gerdes}}}, \bibinfo {author} {\bibfnamefont {T.}~\bibnamefont
  {{Giannantonio}}}, \bibinfo {author} {\bibfnamefont {M.~S.~S.}\ \bibnamefont
  {{Gill}}}, \bibinfo {author} {\bibfnamefont {K.}~\bibnamefont
  {{Glazebrook}}}, \bibinfo {author} {\bibfnamefont {D.~A.}\ \bibnamefont
  {{Goldstein}}}, \bibinfo {author} {\bibfnamefont {D.}~\bibnamefont
  {{Gruen}}}, \bibinfo {author} {\bibfnamefont {R.~A.}\ \bibnamefont
  {{Gruendl}}}, \bibinfo {author} {\bibfnamefont {J.}~\bibnamefont
  {{Gschwend}}}, \bibinfo {author} {\bibfnamefont {G.}~\bibnamefont
  {{Gutierrez}}}, \bibinfo {author} {\bibfnamefont {S.}~\bibnamefont
  {{Hamilton}}}, \bibinfo {author} {\bibfnamefont {W.~G.}\ \bibnamefont
  {{Hartley}}}, \bibinfo {author} {\bibfnamefont {S.~R.}\ \bibnamefont
  {{Hinton}}}, \bibinfo {author} {\bibfnamefont {K.}~\bibnamefont
  {{Honscheid}}}, \bibinfo {author} {\bibfnamefont {B.}~\bibnamefont
  {{Hoyle}}}, \bibinfo {author} {\bibfnamefont {D.}~\bibnamefont {{Huterer}}},
  \bibinfo {author} {\bibfnamefont {B.}~\bibnamefont {{Jain}}}, \bibinfo
  {author} {\bibfnamefont {D.~J.}\ \bibnamefont {{James}}}, \bibinfo {author}
  {\bibfnamefont {M.}~\bibnamefont {{Jarvis}}}, \bibinfo {author}
  {\bibfnamefont {T.}~\bibnamefont {{Jeltema}}}, \bibinfo {author}
  {\bibfnamefont {M.~D.}\ \bibnamefont {{Johnson}}}, \bibinfo {author}
  {\bibfnamefont {M.~W.~G.}\ \bibnamefont {{Johnson}}}, \bibinfo {author}
  {\bibfnamefont {T.}~\bibnamefont {{Kacprzak}}}, \bibinfo {author}
  {\bibfnamefont {S.}~\bibnamefont {{Kent}}}, \bibinfo {author} {\bibfnamefont
  {A.~G.}\ \bibnamefont {{Kim}}}, \bibinfo {author} {\bibfnamefont
  {A.}~\bibnamefont {{King}}}, \bibinfo {author} {\bibfnamefont
  {D.}~\bibnamefont {{Kirk}}}, \bibinfo {author} {\bibfnamefont
  {N.}~\bibnamefont {{Kokron}}}, \bibinfo {author} {\bibfnamefont
  {A.}~\bibnamefont {{Kovacs}}}, \bibinfo {author} {\bibfnamefont
  {E.}~\bibnamefont {{Krause}}}, \bibinfo {author} {\bibfnamefont
  {C.}~\bibnamefont {{Krawiec}}}, \bibinfo {author} {\bibfnamefont
  {A.}~\bibnamefont {{Kremin}}}, \bibinfo {author} {\bibfnamefont
  {K.}~\bibnamefont {{Kuehn}}}, \bibinfo {author} {\bibfnamefont
  {S.}~\bibnamefont {{Kuhlmann}}}, \bibinfo {author} {\bibfnamefont
  {N.}~\bibnamefont {{Kuropatkin}}}, \bibinfo {author} {\bibfnamefont
  {F.}~\bibnamefont {{Lacasa}}}, \bibinfo {author} {\bibfnamefont
  {O.}~\bibnamefont {{Lahav}}}, \bibinfo {author} {\bibfnamefont {T.~S.}\
  \bibnamefont {{Li}}}, \bibinfo {author} {\bibfnamefont {A.~R.}\ \bibnamefont
  {{Liddle}}}, \bibinfo {author} {\bibfnamefont {C.}~\bibnamefont {{Lidman}}},
  \bibinfo {author} {\bibfnamefont {M.}~\bibnamefont {{Lima}}}, \bibinfo
  {author} {\bibfnamefont {H.}~\bibnamefont {{Lin}}}, \bibinfo {author}
  {\bibfnamefont {N.}~\bibnamefont {{MacCrann}}}, \bibinfo {author}
  {\bibfnamefont {M.~A.~G.}\ \bibnamefont {{Maia}}}, \bibinfo {author}
  {\bibfnamefont {M.}~\bibnamefont {{Makler}}}, \bibinfo {author}
  {\bibfnamefont {M.}~\bibnamefont {{Manera}}}, \bibinfo {author}
  {\bibfnamefont {M.}~\bibnamefont {{March}}}, \bibinfo {author} {\bibfnamefont
  {J.~L.}\ \bibnamefont {{Marshall}}}, \bibinfo {author} {\bibfnamefont
  {P.}~\bibnamefont {{Martini}}}, \bibinfo {author} {\bibfnamefont {R.~G.}\
  \bibnamefont {{McMahon}}}, \bibinfo {author} {\bibfnamefont {P.}~\bibnamefont
  {{Melchior}}}, \bibinfo {author} {\bibfnamefont {F.}~\bibnamefont
  {{Menanteau}}}, \bibinfo {author} {\bibfnamefont {R.}~\bibnamefont
  {{Miquel}}}, \bibinfo {author} {\bibfnamefont {V.}~\bibnamefont {{Miranda}}},
  \bibinfo {author} {\bibfnamefont {D.}~\bibnamefont {{Mudd}}}, \bibinfo
  {author} {\bibfnamefont {J.}~\bibnamefont {{Muir}}}, \bibinfo {author}
  {\bibfnamefont {A.}~\bibnamefont {{M{\"o}ller}}}, \bibinfo {author}
  {\bibfnamefont {E.}~\bibnamefont {{Neilsen}}}, \bibinfo {author}
  {\bibfnamefont {R.~C.}\ \bibnamefont {{Nichol}}}, \bibinfo {author}
  {\bibfnamefont {B.}~\bibnamefont {{Nord}}}, \bibinfo {author} {\bibfnamefont
  {P.}~\bibnamefont {{Nugent}}}, \bibinfo {author} {\bibfnamefont {R.~L.~C.}\
  \bibnamefont {{Ogando}}}, \bibinfo {author} {\bibfnamefont {A.}~\bibnamefont
  {{Palmese}}}, \bibinfo {author} {\bibfnamefont {J.}~\bibnamefont
  {{Peacock}}}, \bibinfo {author} {\bibfnamefont {H.~V.}\ \bibnamefont
  {{Peiris}}}, \bibinfo {author} {\bibfnamefont {J.}~\bibnamefont {{Peoples}}},
  \bibinfo {author} {\bibfnamefont {W.~J.}\ \bibnamefont {{Percival}}},
  \bibinfo {author} {\bibfnamefont {D.}~\bibnamefont {{Petravick}}}, \bibinfo
  {author} {\bibfnamefont {A.~A.}\ \bibnamefont {{Plazas}}}, \bibinfo {author}
  {\bibfnamefont {A.}~\bibnamefont {{Porredon}}}, \bibinfo {author}
  {\bibfnamefont {J.}~\bibnamefont {{Prat}}}, \bibinfo {author} {\bibfnamefont
  {A.}~\bibnamefont {{Pujol}}}, \bibinfo {author} {\bibfnamefont {M.~M.}\
  \bibnamefont {{Rau}}}, \bibinfo {author} {\bibfnamefont {A.}~\bibnamefont
  {{Refregier}}}, \bibinfo {author} {\bibfnamefont {P.~M.}\ \bibnamefont
  {{Ricker}}}, \bibinfo {author} {\bibfnamefont {N.}~\bibnamefont {{Roe}}},
  \bibinfo {author} {\bibfnamefont {R.~P.}\ \bibnamefont {{Rollins}}}, \bibinfo
  {author} {\bibfnamefont {A.~K.}\ \bibnamefont {{Romer}}}, \bibinfo {author}
  {\bibfnamefont {A.}~\bibnamefont {{Roodman}}}, \bibinfo {author}
  {\bibfnamefont {R.}~\bibnamefont {{Rosenfeld}}}, \bibinfo {author}
  {\bibfnamefont {A.~J.}\ \bibnamefont {{Ross}}}, \bibinfo {author}
  {\bibfnamefont {E.}~\bibnamefont {{Rozo}}}, \bibinfo {author} {\bibfnamefont
  {E.~S.}\ \bibnamefont {{Rykoff}}}, \bibinfo {author} {\bibfnamefont
  {M.}~\bibnamefont {{Sako}}}, \bibinfo {author} {\bibfnamefont {A.~I.}\
  \bibnamefont {{Salvador}}}, \bibinfo {author} {\bibfnamefont
  {S.}~\bibnamefont {{Samuroff}}}, \bibinfo {author} {\bibfnamefont
  {C.}~\bibnamefont {{S{\'a}nchez}}}, \bibinfo {author} {\bibfnamefont
  {E.}~\bibnamefont {{Sanchez}}}, \bibinfo {author} {\bibfnamefont
  {B.}~\bibnamefont {{Santiago}}}, \bibinfo {author} {\bibfnamefont
  {V.}~\bibnamefont {{Scarpine}}}, \bibinfo {author} {\bibfnamefont
  {R.}~\bibnamefont {{Schindler}}}, \bibinfo {author} {\bibfnamefont
  {D.}~\bibnamefont {{Scolnic}}}, \bibinfo {author} {\bibfnamefont {L.~F.}\
  \bibnamefont {{Secco}}}, \bibinfo {author} {\bibfnamefont {S.}~\bibnamefont
  {{Serrano}}}, \bibinfo {author} {\bibfnamefont {I.}~\bibnamefont
  {{Sevilla-Noarbe}}}, \bibinfo {author} {\bibfnamefont {E.}~\bibnamefont
  {{Sheldon}}}, \bibinfo {author} {\bibfnamefont {R.~C.}\ \bibnamefont
  {{Smith}}}, \bibinfo {author} {\bibfnamefont {M.}~\bibnamefont {{Smith}}},
  \bibinfo {author} {\bibfnamefont {J.}~\bibnamefont {{Smith}}}, \bibinfo
  {author} {\bibfnamefont {M.}~\bibnamefont {{Soares-Santos}}}, \bibinfo
  {author} {\bibfnamefont {F.}~\bibnamefont {{Sobreira}}}, \bibinfo {author}
  {\bibfnamefont {E.}~\bibnamefont {{Suchyta}}}, \bibinfo {author}
  {\bibfnamefont {G.}~\bibnamefont {{Tarle}}}, \bibinfo {author} {\bibfnamefont
  {D.}~\bibnamefont {{Thomas}}}, \bibinfo {author} {\bibfnamefont {M.~A.}\
  \bibnamefont {{Troxel}}}, \bibinfo {author} {\bibfnamefont {D.~L.}\
  \bibnamefont {{Tucker}}}, \bibinfo {author} {\bibfnamefont {B.~E.}\
  \bibnamefont {{Tucker}}}, \bibinfo {author} {\bibfnamefont {S.~A.}\
  \bibnamefont {{Uddin}}}, \bibinfo {author} {\bibfnamefont {T.~N.}\
  \bibnamefont {{Varga}}}, \bibinfo {author} {\bibfnamefont {P.}~\bibnamefont
  {{Vielzeuf}}}, \bibinfo {author} {\bibfnamefont {V.}~\bibnamefont
  {{Vikram}}}, \bibinfo {author} {\bibfnamefont {A.~K.}\ \bibnamefont
  {{Vivas}}}, \bibinfo {author} {\bibfnamefont {A.~R.}\ \bibnamefont
  {{Walker}}}, \bibinfo {author} {\bibfnamefont {M.}~\bibnamefont {{Wang}}},
  \bibinfo {author} {\bibfnamefont {R.~H.}\ \bibnamefont {{Wechsler}}},
  \bibinfo {author} {\bibfnamefont {J.}~\bibnamefont {{Weller}}}, \bibinfo
  {author} {\bibfnamefont {W.}~\bibnamefont {{Wester}}}, \bibinfo {author}
  {\bibfnamefont {R.~C.}\ \bibnamefont {{Wolf}}}, \bibinfo {author}
  {\bibfnamefont {B.}~\bibnamefont {{Yanny}}}, \bibinfo {author} {\bibfnamefont
  {F.}~\bibnamefont {{Yuan}}}, \bibinfo {author} {\bibfnamefont
  {A.}~\bibnamefont {{Zenteno}}}, \bibinfo {author} {\bibfnamefont
  {B.}~\bibnamefont {{Zhang}}}, \bibinfo {author} {\bibfnamefont
  {Y.}~\bibnamefont {{Zhang}}}, \bibinfo {author} {\bibfnamefont
  {J.}~\bibnamefont {{Zuntz}}},\ and\ \bibinfo {author} {\bibnamefont {{Dark
  Energy Survey Collaboration}}},\ }\bibfield  {title} {\bibinfo {title} {{Dark
  Energy Survey year 1 results: Cosmological constraints from galaxy clustering
  and weak lensing}},\ }\href {https://doi.org/10.1103/PhysRevD.98.043526}
  {\bibfield  {journal} {\bibinfo  {journal} {\prd}\ }\textbf {\bibinfo
  {volume} {98}},\ \bibinfo {eid} {043526} (\bibinfo {year} {2018})},\ \Eprint
  {https://arxiv.org/abs/1708.01530} {arXiv:1708.01530 [astro-ph.CO]}
  \BibitemShut {NoStop}%
\bibitem [{\citenamefont {{Abbott}}\ \emph {et~al.}(2022)\citenamefont
  {{Abbott}}, \citenamefont {{Aguena}}, \citenamefont {{Alarcon}},
  \citenamefont {{Allam}}, \citenamefont {{Alves}}, \citenamefont {{Amon}},
  \citenamefont {{Andrade-Oliveira}}, \citenamefont {{Annis}}, \citenamefont
  {{Avila}}, \citenamefont {{Bacon}}, \citenamefont {{Baxter}}, \citenamefont
  {{Bechtol}}, \citenamefont {{Becker}}, \citenamefont {{Bernstein}},
  \citenamefont {{Bhargava}}, \citenamefont {{Birrer}}, \citenamefont
  {{Blazek}}, \citenamefont {{Brandao-Souza}}, \citenamefont {{Bridle}},
  \citenamefont {{Brooks}}, \citenamefont {{Buckley-Geer}}, \citenamefont
  {{Burke}}, \citenamefont {{Camacho}}, \citenamefont {{Campos}}, \citenamefont
  {{Carnero Rosell}}, \citenamefont {{Carrasco Kind}}, \citenamefont
  {{Carretero}}, \citenamefont {{Castander}}, \citenamefont {{Cawthon}},
  \citenamefont {{Chang}}, \citenamefont {{Chen}}, \citenamefont {{Chen}},
  \citenamefont {{Choi}}, \citenamefont {{Conselice}}, \citenamefont
  {{Cordero}}, \citenamefont {{Costanzi}}, \citenamefont {{Crocce}},
  \citenamefont {{da Costa}}, \citenamefont {{da Silva Pereira}}, \citenamefont
  {{Davis}}, \citenamefont {{Davis}}, \citenamefont {{De Vicente}},
  \citenamefont {{DeRose}}, \citenamefont {{Desai}}, \citenamefont {{Di
  Valentino}}, \citenamefont {{Diehl}}, \citenamefont {{Dietrich}},
  \citenamefont {{Dodelson}}, \citenamefont {{Doel}}, \citenamefont {{Doux}},
  \citenamefont {{Drlica-Wagner}}, \citenamefont {{Eckert}}, \citenamefont
  {{Eifler}}, \citenamefont {{Elsner}}, \citenamefont {{Elvin-Poole}},
  \citenamefont {{Everett}}, \citenamefont {{Evrard}}, \citenamefont {{Fang}},
  \citenamefont {{Farahi}}, \citenamefont {{Fernandez}}, \citenamefont
  {{Ferrero}}, \citenamefont {{Fert{\'e}}}, \citenamefont {{Fosalba}},
  \citenamefont {{Friedrich}}, \citenamefont {{Frieman}}, \citenamefont
  {{Garc{\'\i}a-Bellido}}, \citenamefont {{Gatti}}, \citenamefont
  {{Gaztanaga}}, \citenamefont {{Gerdes}}, \citenamefont {{Giannantonio}},
  \citenamefont {{Giannini}}, \citenamefont {{Gruen}}, \citenamefont
  {{Gruendl}}, \citenamefont {{Gschwend}}, \citenamefont {{Gutierrez}},
  \citenamefont {{Harrison}}, \citenamefont {{Hartley}}, \citenamefont
  {{Herner}}, \citenamefont {{Hinton}}, \citenamefont {{Hollowood}},
  \citenamefont {{Honscheid}}, \citenamefont {{Hoyle}}, \citenamefont {{Huff}},
  \citenamefont {{Huterer}}, \citenamefont {{Jain}}, \citenamefont {{James}},
  \citenamefont {{Jarvis}}, \citenamefont {{Jeffrey}}, \citenamefont
  {{Jeltema}}, \citenamefont {{Kovacs}}, \citenamefont {{Krause}},
  \citenamefont {{Kron}}, \citenamefont {{Kuehn}}, \citenamefont
  {{Kuropatkin}}, \citenamefont {{Lahav}}, \citenamefont {{Leget}},
  \citenamefont {{Lemos}}, \citenamefont {{Liddle}}, \citenamefont {{Lidman}},
  \citenamefont {{Lima}}, \citenamefont {{Lin}}, \citenamefont {{MacCrann}},
  \citenamefont {{Maia}}, \citenamefont {{Marshall}}, \citenamefont
  {{Martini}}, \citenamefont {{McCullough}}, \citenamefont {{Melchior}},
  \citenamefont {{Mena-Fern{\'a}ndez}}, \citenamefont {{Menanteau}},
  \citenamefont {{Miquel}}, \citenamefont {{Mohr}}, \citenamefont {{Morgan}},
  \citenamefont {{Muir}}, \citenamefont {{Myles}}, \citenamefont {{Nadathur}},
  \citenamefont {{Navarro-Alsina}}, \citenamefont {{Nichol}}, \citenamefont
  {{Ogando}}, \citenamefont {{Omori}}, \citenamefont {{Palmese}}, \citenamefont
  {{Pandey}}, \citenamefont {{Park}}, \citenamefont {{Paz-Chinch{\'o}n}},
  \citenamefont {{Petravick}}, \citenamefont {{Pieres}}, \citenamefont {{Plazas
  Malag{\'o}n}}, \citenamefont {{Porredon}}, \citenamefont {{Prat}},
  \citenamefont {{Raveri}}, \citenamefont {{Rodriguez-Monroy}}, \citenamefont
  {{Rollins}}, \citenamefont {{Romer}}, \citenamefont {{Roodman}},
  \citenamefont {{Rosenfeld}}, \citenamefont {{Ross}}, \citenamefont
  {{Rykoff}}, \citenamefont {{Samuroff}}, \citenamefont {{S{\'a}nchez}},
  \citenamefont {{Sanchez}}, \citenamefont {{Sanchez}}, \citenamefont {{Sanchez
  Cid}}, \citenamefont {{Scarpine}}, \citenamefont {{Schubnell}}, \citenamefont
  {{Scolnic}}, \citenamefont {{Secco}}, \citenamefont {{Serrano}},
  \citenamefont {{Sevilla-Noarbe}}, \citenamefont {{Sheldon}}, \citenamefont
  {{Shin}}, \citenamefont {{Smith}}, \citenamefont {{Soares-Santos}},
  \citenamefont {{Suchyta}}, \citenamefont {{Swanson}}, \citenamefont
  {{Tabbutt}}, \citenamefont {{Tarle}}, \citenamefont {{Thomas}}, \citenamefont
  {{To}}, \citenamefont {{Troja}}, \citenamefont {{Troxel}}, \citenamefont
  {{Tucker}}, \citenamefont {{Tutusaus}}, \citenamefont {{Varga}},
  \citenamefont {{Walker}}, \citenamefont {{Weaverdyck}}, \citenamefont
  {{Wechsler}}, \citenamefont {{Weller}}, \citenamefont {{Yanny}},
  \citenamefont {{Yin}}, \citenamefont {{Zhang}}, \citenamefont {{Zuntz}},\
  and\ \citenamefont {{DES Collaboration}}}]{2022PhRvD.105b3520A}%
  \BibitemOpen
  \bibfield  {author} {\bibinfo {author} {\bibfnamefont {T.~M.~C.}\
  \bibnamefont {{Abbott}}}, \bibinfo {author} {\bibfnamefont {M.}~\bibnamefont
  {{Aguena}}}, \bibinfo {author} {\bibfnamefont {A.}~\bibnamefont {{Alarcon}}},
  \bibinfo {author} {\bibfnamefont {S.}~\bibnamefont {{Allam}}}, \bibinfo
  {author} {\bibfnamefont {O.}~\bibnamefont {{Alves}}}, \bibinfo {author}
  {\bibfnamefont {A.}~\bibnamefont {{Amon}}}, \bibinfo {author} {\bibfnamefont
  {F.}~\bibnamefont {{Andrade-Oliveira}}}, \bibinfo {author} {\bibfnamefont
  {J.}~\bibnamefont {{Annis}}}, \bibinfo {author} {\bibfnamefont
  {S.}~\bibnamefont {{Avila}}}, \bibinfo {author} {\bibfnamefont
  {D.}~\bibnamefont {{Bacon}}}, \bibinfo {author} {\bibfnamefont
  {E.}~\bibnamefont {{Baxter}}}, \bibinfo {author} {\bibfnamefont
  {K.}~\bibnamefont {{Bechtol}}}, \bibinfo {author} {\bibfnamefont {M.~R.}\
  \bibnamefont {{Becker}}}, \bibinfo {author} {\bibfnamefont {G.~M.}\
  \bibnamefont {{Bernstein}}}, \bibinfo {author} {\bibfnamefont
  {S.}~\bibnamefont {{Bhargava}}}, \bibinfo {author} {\bibfnamefont
  {S.}~\bibnamefont {{Birrer}}}, \bibinfo {author} {\bibfnamefont
  {J.}~\bibnamefont {{Blazek}}}, \bibinfo {author} {\bibfnamefont
  {A.}~\bibnamefont {{Brandao-Souza}}}, \bibinfo {author} {\bibfnamefont
  {S.~L.}\ \bibnamefont {{Bridle}}}, \bibinfo {author} {\bibfnamefont
  {D.}~\bibnamefont {{Brooks}}}, \bibinfo {author} {\bibfnamefont
  {E.}~\bibnamefont {{Buckley-Geer}}}, \bibinfo {author} {\bibfnamefont
  {D.~L.}\ \bibnamefont {{Burke}}}, \bibinfo {author} {\bibfnamefont
  {H.}~\bibnamefont {{Camacho}}}, \bibinfo {author} {\bibfnamefont
  {A.}~\bibnamefont {{Campos}}}, \bibinfo {author} {\bibfnamefont
  {A.}~\bibnamefont {{Carnero Rosell}}}, \bibinfo {author} {\bibfnamefont
  {M.}~\bibnamefont {{Carrasco Kind}}}, \bibinfo {author} {\bibfnamefont
  {J.}~\bibnamefont {{Carretero}}}, \bibinfo {author} {\bibfnamefont {F.~J.}\
  \bibnamefont {{Castander}}}, \bibinfo {author} {\bibfnamefont
  {R.}~\bibnamefont {{Cawthon}}}, \bibinfo {author} {\bibfnamefont
  {C.}~\bibnamefont {{Chang}}}, \bibinfo {author} {\bibfnamefont
  {A.}~\bibnamefont {{Chen}}}, \bibinfo {author} {\bibfnamefont
  {R.}~\bibnamefont {{Chen}}}, \bibinfo {author} {\bibfnamefont
  {A.}~\bibnamefont {{Choi}}}, \bibinfo {author} {\bibfnamefont
  {C.}~\bibnamefont {{Conselice}}}, \bibinfo {author} {\bibfnamefont
  {J.}~\bibnamefont {{Cordero}}}, \bibinfo {author} {\bibfnamefont
  {M.}~\bibnamefont {{Costanzi}}}, \bibinfo {author} {\bibfnamefont
  {M.}~\bibnamefont {{Crocce}}}, \bibinfo {author} {\bibfnamefont {L.~N.}\
  \bibnamefont {{da Costa}}}, \bibinfo {author} {\bibfnamefont {M.~E.}\
  \bibnamefont {{da Silva Pereira}}}, \bibinfo {author} {\bibfnamefont
  {C.}~\bibnamefont {{Davis}}}, \bibinfo {author} {\bibfnamefont {T.~M.}\
  \bibnamefont {{Davis}}}, \bibinfo {author} {\bibfnamefont {J.}~\bibnamefont
  {{De Vicente}}}, \bibinfo {author} {\bibfnamefont {J.}~\bibnamefont
  {{DeRose}}}, \bibinfo {author} {\bibfnamefont {S.}~\bibnamefont {{Desai}}},
  \bibinfo {author} {\bibfnamefont {E.}~\bibnamefont {{Di Valentino}}},
  \bibinfo {author} {\bibfnamefont {H.~T.}\ \bibnamefont {{Diehl}}}, \bibinfo
  {author} {\bibfnamefont {J.~P.}\ \bibnamefont {{Dietrich}}}, \bibinfo
  {author} {\bibfnamefont {S.}~\bibnamefont {{Dodelson}}}, \bibinfo {author}
  {\bibfnamefont {P.}~\bibnamefont {{Doel}}}, \bibinfo {author} {\bibfnamefont
  {C.}~\bibnamefont {{Doux}}}, \bibinfo {author} {\bibfnamefont
  {A.}~\bibnamefont {{Drlica-Wagner}}}, \bibinfo {author} {\bibfnamefont
  {K.}~\bibnamefont {{Eckert}}}, \bibinfo {author} {\bibfnamefont {T.~F.}\
  \bibnamefont {{Eifler}}}, \bibinfo {author} {\bibfnamefont {F.}~\bibnamefont
  {{Elsner}}}, \bibinfo {author} {\bibfnamefont {J.}~\bibnamefont
  {{Elvin-Poole}}}, \bibinfo {author} {\bibfnamefont {S.}~\bibnamefont
  {{Everett}}}, \bibinfo {author} {\bibfnamefont {A.~E.}\ \bibnamefont
  {{Evrard}}}, \bibinfo {author} {\bibfnamefont {X.}~\bibnamefont {{Fang}}},
  \bibinfo {author} {\bibfnamefont {A.}~\bibnamefont {{Farahi}}}, \bibinfo
  {author} {\bibfnamefont {E.}~\bibnamefont {{Fernandez}}}, \bibinfo {author}
  {\bibfnamefont {I.}~\bibnamefont {{Ferrero}}}, \bibinfo {author}
  {\bibfnamefont {A.}~\bibnamefont {{Fert{\'e}}}}, \bibinfo {author}
  {\bibfnamefont {P.}~\bibnamefont {{Fosalba}}}, \bibinfo {author}
  {\bibfnamefont {O.}~\bibnamefont {{Friedrich}}}, \bibinfo {author}
  {\bibfnamefont {J.}~\bibnamefont {{Frieman}}}, \bibinfo {author}
  {\bibfnamefont {J.}~\bibnamefont {{Garc{\'\i}a-Bellido}}}, \bibinfo {author}
  {\bibfnamefont {M.}~\bibnamefont {{Gatti}}}, \bibinfo {author} {\bibfnamefont
  {E.}~\bibnamefont {{Gaztanaga}}}, \bibinfo {author} {\bibfnamefont {D.~W.}\
  \bibnamefont {{Gerdes}}}, \bibinfo {author} {\bibfnamefont {T.}~\bibnamefont
  {{Giannantonio}}}, \bibinfo {author} {\bibfnamefont {G.}~\bibnamefont
  {{Giannini}}}, \bibinfo {author} {\bibfnamefont {D.}~\bibnamefont {{Gruen}}},
  \bibinfo {author} {\bibfnamefont {R.~A.}\ \bibnamefont {{Gruendl}}}, \bibinfo
  {author} {\bibfnamefont {J.}~\bibnamefont {{Gschwend}}}, \bibinfo {author}
  {\bibfnamefont {G.}~\bibnamefont {{Gutierrez}}}, \bibinfo {author}
  {\bibfnamefont {I.}~\bibnamefont {{Harrison}}}, \bibinfo {author}
  {\bibfnamefont {W.~G.}\ \bibnamefont {{Hartley}}}, \bibinfo {author}
  {\bibfnamefont {K.}~\bibnamefont {{Herner}}}, \bibinfo {author}
  {\bibfnamefont {S.~R.}\ \bibnamefont {{Hinton}}}, \bibinfo {author}
  {\bibfnamefont {D.~L.}\ \bibnamefont {{Hollowood}}}, \bibinfo {author}
  {\bibfnamefont {K.}~\bibnamefont {{Honscheid}}}, \bibinfo {author}
  {\bibfnamefont {B.}~\bibnamefont {{Hoyle}}}, \bibinfo {author} {\bibfnamefont
  {E.~M.}\ \bibnamefont {{Huff}}}, \bibinfo {author} {\bibfnamefont
  {D.}~\bibnamefont {{Huterer}}}, \bibinfo {author} {\bibfnamefont
  {B.}~\bibnamefont {{Jain}}}, \bibinfo {author} {\bibfnamefont {D.~J.}\
  \bibnamefont {{James}}}, \bibinfo {author} {\bibfnamefont {M.}~\bibnamefont
  {{Jarvis}}}, \bibinfo {author} {\bibfnamefont {N.}~\bibnamefont {{Jeffrey}}},
  \bibinfo {author} {\bibfnamefont {T.}~\bibnamefont {{Jeltema}}}, \bibinfo
  {author} {\bibfnamefont {A.}~\bibnamefont {{Kovacs}}}, \bibinfo {author}
  {\bibfnamefont {E.}~\bibnamefont {{Krause}}}, \bibinfo {author}
  {\bibfnamefont {R.}~\bibnamefont {{Kron}}}, \bibinfo {author} {\bibfnamefont
  {K.}~\bibnamefont {{Kuehn}}}, \bibinfo {author} {\bibfnamefont
  {N.}~\bibnamefont {{Kuropatkin}}}, \bibinfo {author} {\bibfnamefont
  {O.}~\bibnamefont {{Lahav}}}, \bibinfo {author} {\bibfnamefont {P.~F.}\
  \bibnamefont {{Leget}}}, \bibinfo {author} {\bibfnamefont {P.}~\bibnamefont
  {{Lemos}}}, \bibinfo {author} {\bibfnamefont {A.~R.}\ \bibnamefont
  {{Liddle}}}, \bibinfo {author} {\bibfnamefont {C.}~\bibnamefont {{Lidman}}},
  \bibinfo {author} {\bibfnamefont {M.}~\bibnamefont {{Lima}}}, \bibinfo
  {author} {\bibfnamefont {H.}~\bibnamefont {{Lin}}}, \bibinfo {author}
  {\bibfnamefont {N.}~\bibnamefont {{MacCrann}}}, \bibinfo {author}
  {\bibfnamefont {M.~A.~G.}\ \bibnamefont {{Maia}}}, \bibinfo {author}
  {\bibfnamefont {J.~L.}\ \bibnamefont {{Marshall}}}, \bibinfo {author}
  {\bibfnamefont {P.}~\bibnamefont {{Martini}}}, \bibinfo {author}
  {\bibfnamefont {J.}~\bibnamefont {{McCullough}}}, \bibinfo {author}
  {\bibfnamefont {P.}~\bibnamefont {{Melchior}}}, \bibinfo {author}
  {\bibfnamefont {J.}~\bibnamefont {{Mena-Fern{\'a}ndez}}}, \bibinfo {author}
  {\bibfnamefont {F.}~\bibnamefont {{Menanteau}}}, \bibinfo {author}
  {\bibfnamefont {R.}~\bibnamefont {{Miquel}}}, \bibinfo {author}
  {\bibfnamefont {J.~J.}\ \bibnamefont {{Mohr}}}, \bibinfo {author}
  {\bibfnamefont {R.}~\bibnamefont {{Morgan}}}, \bibinfo {author}
  {\bibfnamefont {J.}~\bibnamefont {{Muir}}}, \bibinfo {author} {\bibfnamefont
  {J.}~\bibnamefont {{Myles}}}, \bibinfo {author} {\bibfnamefont
  {S.}~\bibnamefont {{Nadathur}}}, \bibinfo {author} {\bibfnamefont
  {A.}~\bibnamefont {{Navarro-Alsina}}}, \bibinfo {author} {\bibfnamefont
  {R.~C.}\ \bibnamefont {{Nichol}}}, \bibinfo {author} {\bibfnamefont
  {R.~L.~C.}\ \bibnamefont {{Ogando}}}, \bibinfo {author} {\bibfnamefont
  {Y.}~\bibnamefont {{Omori}}}, \bibinfo {author} {\bibfnamefont
  {A.}~\bibnamefont {{Palmese}}}, \bibinfo {author} {\bibfnamefont
  {S.}~\bibnamefont {{Pandey}}}, \bibinfo {author} {\bibfnamefont
  {Y.}~\bibnamefont {{Park}}}, \bibinfo {author} {\bibfnamefont
  {F.}~\bibnamefont {{Paz-Chinch{\'o}n}}}, \bibinfo {author} {\bibfnamefont
  {D.}~\bibnamefont {{Petravick}}}, \bibinfo {author} {\bibfnamefont
  {A.}~\bibnamefont {{Pieres}}}, \bibinfo {author} {\bibfnamefont {A.~A.}\
  \bibnamefont {{Plazas Malag{\'o}n}}}, \bibinfo {author} {\bibfnamefont
  {A.}~\bibnamefont {{Porredon}}}, \bibinfo {author} {\bibfnamefont
  {J.}~\bibnamefont {{Prat}}}, \bibinfo {author} {\bibfnamefont
  {M.}~\bibnamefont {{Raveri}}}, \bibinfo {author} {\bibfnamefont
  {M.}~\bibnamefont {{Rodriguez-Monroy}}}, \bibinfo {author} {\bibfnamefont
  {R.~P.}\ \bibnamefont {{Rollins}}}, \bibinfo {author} {\bibfnamefont {A.~K.}\
  \bibnamefont {{Romer}}}, \bibinfo {author} {\bibfnamefont {A.}~\bibnamefont
  {{Roodman}}}, \bibinfo {author} {\bibfnamefont {R.}~\bibnamefont
  {{Rosenfeld}}}, \bibinfo {author} {\bibfnamefont {A.~J.}\ \bibnamefont
  {{Ross}}}, \bibinfo {author} {\bibfnamefont {E.~S.}\ \bibnamefont
  {{Rykoff}}}, \bibinfo {author} {\bibfnamefont {S.}~\bibnamefont
  {{Samuroff}}}, \bibinfo {author} {\bibfnamefont {C.}~\bibnamefont
  {{S{\'a}nchez}}}, \bibinfo {author} {\bibfnamefont {E.}~\bibnamefont
  {{Sanchez}}}, \bibinfo {author} {\bibfnamefont {J.}~\bibnamefont
  {{Sanchez}}}, \bibinfo {author} {\bibfnamefont {D.}~\bibnamefont {{Sanchez
  Cid}}}, \bibinfo {author} {\bibfnamefont {V.}~\bibnamefont {{Scarpine}}},
  \bibinfo {author} {\bibfnamefont {M.}~\bibnamefont {{Schubnell}}}, \bibinfo
  {author} {\bibfnamefont {D.}~\bibnamefont {{Scolnic}}}, \bibinfo {author}
  {\bibfnamefont {L.~F.}\ \bibnamefont {{Secco}}}, \bibinfo {author}
  {\bibfnamefont {S.}~\bibnamefont {{Serrano}}}, \bibinfo {author}
  {\bibfnamefont {I.}~\bibnamefont {{Sevilla-Noarbe}}}, \bibinfo {author}
  {\bibfnamefont {E.}~\bibnamefont {{Sheldon}}}, \bibinfo {author}
  {\bibfnamefont {T.}~\bibnamefont {{Shin}}}, \bibinfo {author} {\bibfnamefont
  {M.}~\bibnamefont {{Smith}}}, \bibinfo {author} {\bibfnamefont
  {M.}~\bibnamefont {{Soares-Santos}}}, \bibinfo {author} {\bibfnamefont
  {E.}~\bibnamefont {{Suchyta}}}, \bibinfo {author} {\bibfnamefont {M.~E.~C.}\
  \bibnamefont {{Swanson}}}, \bibinfo {author} {\bibfnamefont {M.}~\bibnamefont
  {{Tabbutt}}}, \bibinfo {author} {\bibfnamefont {G.}~\bibnamefont {{Tarle}}},
  \bibinfo {author} {\bibfnamefont {D.}~\bibnamefont {{Thomas}}}, \bibinfo
  {author} {\bibfnamefont {C.}~\bibnamefont {{To}}}, \bibinfo {author}
  {\bibfnamefont {A.}~\bibnamefont {{Troja}}}, \bibinfo {author} {\bibfnamefont
  {M.~A.}\ \bibnamefont {{Troxel}}}, \bibinfo {author} {\bibfnamefont {D.~L.}\
  \bibnamefont {{Tucker}}}, \bibinfo {author} {\bibfnamefont {I.}~\bibnamefont
  {{Tutusaus}}}, \bibinfo {author} {\bibfnamefont {T.~N.}\ \bibnamefont
  {{Varga}}}, \bibinfo {author} {\bibfnamefont {A.~R.}\ \bibnamefont
  {{Walker}}}, \bibinfo {author} {\bibfnamefont {N.}~\bibnamefont
  {{Weaverdyck}}}, \bibinfo {author} {\bibfnamefont {R.}~\bibnamefont
  {{Wechsler}}}, \bibinfo {author} {\bibfnamefont {J.}~\bibnamefont
  {{Weller}}}, \bibinfo {author} {\bibfnamefont {B.}~\bibnamefont {{Yanny}}},
  \bibinfo {author} {\bibfnamefont {B.}~\bibnamefont {{Yin}}}, \bibinfo
  {author} {\bibfnamefont {Y.}~\bibnamefont {{Zhang}}}, \bibinfo {author}
  {\bibfnamefont {J.}~\bibnamefont {{Zuntz}}},\ and\ \bibinfo {author}
  {\bibnamefont {{DES Collaboration}}},\ }\bibfield  {title} {\bibinfo {title}
  {{Dark Energy Survey Year 3 results: Cosmological constraints from galaxy
  clustering and weak lensing}},\ }\href
  {https://doi.org/10.1103/PhysRevD.105.023520} {\bibfield  {journal} {\bibinfo
   {journal} {\prd}\ }\textbf {\bibinfo {volume} {105}},\ \bibinfo {eid}
  {023520} (\bibinfo {year} {2022})},\ \Eprint
  {https://arxiv.org/abs/2105.13549} {arXiv:2105.13549 [astro-ph.CO]}
  \BibitemShut {NoStop}%
\bibitem [{\citenamefont {{DESI Collaboration}}\ \emph
  {et~al.}(2016)\citenamefont {{DESI Collaboration}}, \citenamefont
  {{Aghamousa}}, \citenamefont {{Aguilar}}, \citenamefont {{Ahlen}},
  \citenamefont {{Alam}}, \citenamefont {{Allen}}, \citenamefont {{Allende
  Prieto}}, \citenamefont {{Annis}}, \citenamefont {{Bailey}}, \citenamefont
  {{Balland}}, \citenamefont {{Ballester}}, \citenamefont {{Baltay}},
  \citenamefont {{Beaufore}}, \citenamefont {{Bebek}}, \citenamefont {{Beers}},
  \citenamefont {{Bell}}, \citenamefont {{Bernal}}, \citenamefont {{Besuner}},
  \citenamefont {{Beutler}}, \citenamefont {{Blake}}, \citenamefont
  {{Bleuler}}, \citenamefont {{Blomqvist}}, \citenamefont {{Blum}},
  \citenamefont {{Bolton}}, \citenamefont {{Briceno}}, \citenamefont
  {{Brooks}}, \citenamefont {{Brownstein}}, \citenamefont {{Buckley-Geer}},
  \citenamefont {{Burden}}, \citenamefont {{Burtin}}, \citenamefont {{Busca}},
  \citenamefont {{Cahn}}, \citenamefont {{Cai}}, \citenamefont {{Cardiel-Sas}},
  \citenamefont {{Carlberg}}, \citenamefont {{Carton}}, \citenamefont
  {{Casas}}, \citenamefont {{Castander}}, \citenamefont {{Cervantes-Cota}},
  \citenamefont {{Claybaugh}}, \citenamefont {{Close}}, \citenamefont
  {{Coker}}, \citenamefont {{Cole}}, \citenamefont {{Comparat}}, \citenamefont
  {{Cooper}}, \citenamefont {{Cousinou}}, \citenamefont {{Crocce}},
  \citenamefont {{Cuby}}, \citenamefont {{Cunningham}}, \citenamefont
  {{Davis}}, \citenamefont {{Dawson}}, \citenamefont {{de la Macorra}},
  \citenamefont {{De Vicente}}, \citenamefont {{Delubac}}, \citenamefont
  {{Derwent}}, \citenamefont {{Dey}}, \citenamefont {{Dhungana}}, \citenamefont
  {{Ding}}, \citenamefont {{Doel}}, \citenamefont {{Duan}}, \citenamefont
  {{Ealet}}, \citenamefont {{Edelstein}}, \citenamefont {{Eftekharzadeh}},
  \citenamefont {{Eisenstein}}, \citenamefont {{Elliott}}, \citenamefont
  {{Escoffier}}, \citenamefont {{Evatt}}, \citenamefont {{Fagrelius}},
  \citenamefont {{Fan}}, \citenamefont {{Fanning}}, \citenamefont {{Farahi}},
  \citenamefont {{Farihi}}, \citenamefont {{Favole}}, \citenamefont {{Feng}},
  \citenamefont {{Fernandez}}, \citenamefont {{Findlay}}, \citenamefont
  {{Finkbeiner}}, \citenamefont {{Fitzpatrick}}, \citenamefont {{Flaugher}},
  \citenamefont {{Flender}}, \citenamefont {{Font-Ribera}}, \citenamefont
  {{Forero-Romero}}, \citenamefont {{Fosalba}}, \citenamefont {{Frenk}},
  \citenamefont {{Fumagalli}}, \citenamefont {{Gaensicke}}, \citenamefont
  {{Gallo}}, \citenamefont {{Garcia-Bellido}}, \citenamefont {{Gaztanaga}},
  \citenamefont {{Pietro Gentile Fusillo}}, \citenamefont {{Gerard}},
  \citenamefont {{Gershkovich}}, \citenamefont {{Giannantonio}}, \citenamefont
  {{Gillet}}, \citenamefont {{Gonzalez-de-Rivera}}, \citenamefont
  {{Gonzalez-Perez}}, \citenamefont {{Gott}}, \citenamefont {{Graur}},
  \citenamefont {{Gutierrez}}, \citenamefont {{Guy}}, \citenamefont {{Habib}},
  \citenamefont {{Heetderks}}, \citenamefont {{Heetderks}}, \citenamefont
  {{Heitmann}}, \citenamefont {{Hellwing}}, \citenamefont {{Herrera}},
  \citenamefont {{Ho}}, \citenamefont {{Holland}}, \citenamefont {{Honscheid}},
  \citenamefont {{Huff}}, \citenamefont {{Hutchinson}}, \citenamefont
  {{Huterer}}, \citenamefont {{Hwang}}, \citenamefont {{Illa Laguna}},
  \citenamefont {{Ishikawa}}, \citenamefont {{Jacobs}}, \citenamefont
  {{Jeffrey}}, \citenamefont {{Jelinsky}}, \citenamefont {{Jennings}},
  \citenamefont {{Jiang}}, \citenamefont {{Jimenez}}, \citenamefont
  {{Johnson}}, \citenamefont {{Joyce}}, \citenamefont {{Jullo}}, \citenamefont
  {{Juneau}}, \citenamefont {{Kama}}, \citenamefont {{Karcher}}, \citenamefont
  {{Karkar}}, \citenamefont {{Kehoe}}, \citenamefont {{Kennamer}},
  \citenamefont {{Kent}}, \citenamefont {{Kilbinger}}, \citenamefont {{Kim}},
  \citenamefont {{Kirkby}}, \citenamefont {{Kisner}}, \citenamefont
  {{Kitanidis}}, \citenamefont {{Kneib}}, \citenamefont {{Koposov}},
  \citenamefont {{Kovacs}}, \citenamefont {{Koyama}}, \citenamefont {{Kremin}},
  \citenamefont {{Kron}}, \citenamefont {{Kronig}}, \citenamefont
  {{Kueter-Young}}, \citenamefont {{Lacey}}, \citenamefont {{Lafever}},
  \citenamefont {{Lahav}}, \citenamefont {{Lambert}}, \citenamefont
  {{Lampton}}, \citenamefont {{Landriau}}, \citenamefont {{Lang}},
  \citenamefont {{Lauer}}, \citenamefont {{Le Goff}}, \citenamefont {{Le
  Guillou}}, \citenamefont {{Le Van Suu}}, \citenamefont {{Lee}}, \citenamefont
  {{Lee}}, \citenamefont {{Leitner}}, \citenamefont {{Lesser}}, \citenamefont
  {{Levi}}, \citenamefont {{L'Huillier}}, \citenamefont {{Li}}, \citenamefont
  {{Liang}}, \citenamefont {{Lin}}, \citenamefont {{Linder}}, \citenamefont
  {{Loebman}}, \citenamefont {{Luki{\'c}}}, \citenamefont {{Ma}}, \citenamefont
  {{MacCrann}}, \citenamefont {{Magneville}}, \citenamefont {{Makarem}},
  \citenamefont {{Manera}}, \citenamefont {{Manser}}, \citenamefont
  {{Marshall}}, \citenamefont {{Martini}}, \citenamefont {{Massey}},
  \citenamefont {{Matheson}}, \citenamefont {{McCauley}}, \citenamefont
  {{McDonald}}, \citenamefont {{McGreer}}, \citenamefont {{Meisner}},
  \citenamefont {{Metcalfe}}, \citenamefont {{Miller}}, \citenamefont
  {{Miquel}}, \citenamefont {{Moustakas}}, \citenamefont {{Myers}},
  \citenamefont {{Naik}}, \citenamefont {{Newman}}, \citenamefont {{Nichol}},
  \citenamefont {{Nicola}}, \citenamefont {{Nicolati da Costa}}, \citenamefont
  {{Nie}}, \citenamefont {{Niz}}, \citenamefont {{Norberg}}, \citenamefont
  {{Nord}}, \citenamefont {{Norman}}, \citenamefont {{Nugent}}, \citenamefont
  {{O'Brien}}, \citenamefont {{Oh}}, \citenamefont {{Olsen}}, \citenamefont
  {{Padilla}}, \citenamefont {{Padmanabhan}}, \citenamefont {{Padmanabhan}},
  \citenamefont {{Palanque-Delabrouille}}, \citenamefont {{Palmese}},
  \citenamefont {{Pappalardo}}, \citenamefont {{P{\^a}ris}}, \citenamefont
  {{Park}}, \citenamefont {{Patej}}, \citenamefont {{Peacock}}, \citenamefont
  {{Peiris}}, \citenamefont {{Peng}}, \citenamefont {{Percival}}, \citenamefont
  {{Perruchot}}, \citenamefont {{Pieri}}, \citenamefont {{Pogge}},
  \citenamefont {{Pollack}}, \citenamefont {{Poppett}}, \citenamefont
  {{Prada}}, \citenamefont {{Prakash}}, \citenamefont {{Probst}}, \citenamefont
  {{Rabinowitz}}, \citenamefont {{Raichoor}}, \citenamefont {{Ree}},
  \citenamefont {{Refregier}}, \citenamefont {{Regal}}, \citenamefont {{Reid}},
  \citenamefont {{Reil}}, \citenamefont {{Rezaie}}, \citenamefont {{Rockosi}},
  \citenamefont {{Roe}}, \citenamefont {{Ronayette}}, \citenamefont
  {{Roodman}}, \citenamefont {{Ross}}, \citenamefont {{Ross}}, \citenamefont
  {{Rossi}}, \citenamefont {{Rozo}}, \citenamefont {{Ruhlmann-Kleider}},
  \citenamefont {{Rykoff}}, \citenamefont {{Sabiu}}, \citenamefont
  {{Samushia}}, \citenamefont {{Sanchez}}, \citenamefont {{Sanchez}},
  \citenamefont {{Schlegel}}, \citenamefont {{Schneider}}, \citenamefont
  {{Schubnell}}, \citenamefont {{Secroun}}, \citenamefont {{Seljak}},
  \citenamefont {{Seo}}, \citenamefont {{Serrano}}, \citenamefont
  {{Shafieloo}}, \citenamefont {{Shan}}, \citenamefont {{Sharples}},
  \citenamefont {{Sholl}}, \citenamefont {{Shourt}}, \citenamefont {{Silber}},
  \citenamefont {{Silva}}, \citenamefont {{Sirk}}, \citenamefont {{Slosar}},
  \citenamefont {{Smith}}, \citenamefont {{Smoot}}, \citenamefont {{Som}},
  \citenamefont {{Song}}, \citenamefont {{Sprayberry}}, \citenamefont
  {{Staten}}, \citenamefont {{Stefanik}}, \citenamefont {{Tarle}},
  \citenamefont {{Sien Tie}}, \citenamefont {{Tinker}}, \citenamefont
  {{Tojeiro}}, \citenamefont {{Valdes}}, \citenamefont {{Valenzuela}},
  \citenamefont {{Valluri}}, \citenamefont {{Vargas-Magana}}, \citenamefont
  {{Verde}}, \citenamefont {{Walker}}, \citenamefont {{Wang}}, \citenamefont
  {{Wang}}, \citenamefont {{Weaver}}, \citenamefont {{Weaverdyck}},
  \citenamefont {{Wechsler}}, \citenamefont {{Weinberg}}, \citenamefont
  {{White}}, \citenamefont {{Yang}}, \citenamefont {{Yeche}}, \citenamefont
  {{Zhang}}, \citenamefont {{Zhao}}, \citenamefont {{Zheng}}, \citenamefont
  {{Zhou}}, \citenamefont {{Zhou}}, \citenamefont {{Zhu}}, \citenamefont
  {{Zou}},\ and\ \citenamefont {{Zu}}}]{2016arXiv161100036D}%
  \BibitemOpen
  \bibfield  {author} {\bibinfo {author} {\bibnamefont {{DESI Collaboration}}},
  \bibinfo {author} {\bibfnamefont {A.}~\bibnamefont {{Aghamousa}}}, \bibinfo
  {author} {\bibfnamefont {J.}~\bibnamefont {{Aguilar}}}, \bibinfo {author}
  {\bibfnamefont {S.}~\bibnamefont {{Ahlen}}}, \bibinfo {author} {\bibfnamefont
  {S.}~\bibnamefont {{Alam}}}, \bibinfo {author} {\bibfnamefont {L.~E.}\
  \bibnamefont {{Allen}}}, \bibinfo {author} {\bibfnamefont {C.}~\bibnamefont
  {{Allende Prieto}}}, \bibinfo {author} {\bibfnamefont {J.}~\bibnamefont
  {{Annis}}}, \bibinfo {author} {\bibfnamefont {S.}~\bibnamefont {{Bailey}}},
  \bibinfo {author} {\bibfnamefont {C.}~\bibnamefont {{Balland}}}, \bibinfo
  {author} {\bibfnamefont {O.}~\bibnamefont {{Ballester}}}, \bibinfo {author}
  {\bibfnamefont {C.}~\bibnamefont {{Baltay}}}, \bibinfo {author}
  {\bibfnamefont {L.}~\bibnamefont {{Beaufore}}}, \bibinfo {author}
  {\bibfnamefont {C.}~\bibnamefont {{Bebek}}}, \bibinfo {author} {\bibfnamefont
  {T.~C.}\ \bibnamefont {{Beers}}}, \bibinfo {author} {\bibfnamefont {E.~F.}\
  \bibnamefont {{Bell}}}, \bibinfo {author} {\bibfnamefont {J.~L.}\
  \bibnamefont {{Bernal}}}, \bibinfo {author} {\bibfnamefont {R.}~\bibnamefont
  {{Besuner}}}, \bibinfo {author} {\bibfnamefont {F.}~\bibnamefont
  {{Beutler}}}, \bibinfo {author} {\bibfnamefont {C.}~\bibnamefont {{Blake}}},
  \bibinfo {author} {\bibfnamefont {H.}~\bibnamefont {{Bleuler}}}, \bibinfo
  {author} {\bibfnamefont {M.}~\bibnamefont {{Blomqvist}}}, \bibinfo {author}
  {\bibfnamefont {R.}~\bibnamefont {{Blum}}}, \bibinfo {author} {\bibfnamefont
  {A.~S.}\ \bibnamefont {{Bolton}}}, \bibinfo {author} {\bibfnamefont
  {C.}~\bibnamefont {{Briceno}}}, \bibinfo {author} {\bibfnamefont
  {D.}~\bibnamefont {{Brooks}}}, \bibinfo {author} {\bibfnamefont {J.~R.}\
  \bibnamefont {{Brownstein}}}, \bibinfo {author} {\bibfnamefont
  {E.}~\bibnamefont {{Buckley-Geer}}}, \bibinfo {author} {\bibfnamefont
  {A.}~\bibnamefont {{Burden}}}, \bibinfo {author} {\bibfnamefont
  {E.}~\bibnamefont {{Burtin}}}, \bibinfo {author} {\bibfnamefont {N.~G.}\
  \bibnamefont {{Busca}}}, \bibinfo {author} {\bibfnamefont {R.~N.}\
  \bibnamefont {{Cahn}}}, \bibinfo {author} {\bibfnamefont {Y.-C.}\
  \bibnamefont {{Cai}}}, \bibinfo {author} {\bibfnamefont {L.}~\bibnamefont
  {{Cardiel-Sas}}}, \bibinfo {author} {\bibfnamefont {R.~G.}\ \bibnamefont
  {{Carlberg}}}, \bibinfo {author} {\bibfnamefont {P.-H.}\ \bibnamefont
  {{Carton}}}, \bibinfo {author} {\bibfnamefont {R.}~\bibnamefont {{Casas}}},
  \bibinfo {author} {\bibfnamefont {F.~J.}\ \bibnamefont {{Castander}}},
  \bibinfo {author} {\bibfnamefont {J.~L.}\ \bibnamefont {{Cervantes-Cota}}},
  \bibinfo {author} {\bibfnamefont {T.~M.}\ \bibnamefont {{Claybaugh}}},
  \bibinfo {author} {\bibfnamefont {M.}~\bibnamefont {{Close}}}, \bibinfo
  {author} {\bibfnamefont {C.~T.}\ \bibnamefont {{Coker}}}, \bibinfo {author}
  {\bibfnamefont {S.}~\bibnamefont {{Cole}}}, \bibinfo {author} {\bibfnamefont
  {J.}~\bibnamefont {{Comparat}}}, \bibinfo {author} {\bibfnamefont {A.~P.}\
  \bibnamefont {{Cooper}}}, \bibinfo {author} {\bibfnamefont {M.~C.}\
  \bibnamefont {{Cousinou}}}, \bibinfo {author} {\bibfnamefont
  {M.}~\bibnamefont {{Crocce}}}, \bibinfo {author} {\bibfnamefont {J.-G.}\
  \bibnamefont {{Cuby}}}, \bibinfo {author} {\bibfnamefont {D.~P.}\
  \bibnamefont {{Cunningham}}}, \bibinfo {author} {\bibfnamefont {T.~M.}\
  \bibnamefont {{Davis}}}, \bibinfo {author} {\bibfnamefont {K.~S.}\
  \bibnamefont {{Dawson}}}, \bibinfo {author} {\bibfnamefont {A.}~\bibnamefont
  {{de la Macorra}}}, \bibinfo {author} {\bibfnamefont {J.}~\bibnamefont {{De
  Vicente}}}, \bibinfo {author} {\bibfnamefont {T.}~\bibnamefont {{Delubac}}},
  \bibinfo {author} {\bibfnamefont {M.}~\bibnamefont {{Derwent}}}, \bibinfo
  {author} {\bibfnamefont {A.}~\bibnamefont {{Dey}}}, \bibinfo {author}
  {\bibfnamefont {G.}~\bibnamefont {{Dhungana}}}, \bibinfo {author}
  {\bibfnamefont {Z.}~\bibnamefont {{Ding}}}, \bibinfo {author} {\bibfnamefont
  {P.}~\bibnamefont {{Doel}}}, \bibinfo {author} {\bibfnamefont {Y.~T.}\
  \bibnamefont {{Duan}}}, \bibinfo {author} {\bibfnamefont {A.}~\bibnamefont
  {{Ealet}}}, \bibinfo {author} {\bibfnamefont {J.}~\bibnamefont
  {{Edelstein}}}, \bibinfo {author} {\bibfnamefont {S.}~\bibnamefont
  {{Eftekharzadeh}}}, \bibinfo {author} {\bibfnamefont {D.~J.}\ \bibnamefont
  {{Eisenstein}}}, \bibinfo {author} {\bibfnamefont {A.}~\bibnamefont
  {{Elliott}}}, \bibinfo {author} {\bibfnamefont {S.}~\bibnamefont
  {{Escoffier}}}, \bibinfo {author} {\bibfnamefont {M.}~\bibnamefont
  {{Evatt}}}, \bibinfo {author} {\bibfnamefont {P.}~\bibnamefont
  {{Fagrelius}}}, \bibinfo {author} {\bibfnamefont {X.}~\bibnamefont {{Fan}}},
  \bibinfo {author} {\bibfnamefont {K.}~\bibnamefont {{Fanning}}}, \bibinfo
  {author} {\bibfnamefont {A.}~\bibnamefont {{Farahi}}}, \bibinfo {author}
  {\bibfnamefont {J.}~\bibnamefont {{Farihi}}}, \bibinfo {author}
  {\bibfnamefont {G.}~\bibnamefont {{Favole}}}, \bibinfo {author}
  {\bibfnamefont {Y.}~\bibnamefont {{Feng}}}, \bibinfo {author} {\bibfnamefont
  {E.}~\bibnamefont {{Fernandez}}}, \bibinfo {author} {\bibfnamefont {J.~R.}\
  \bibnamefont {{Findlay}}}, \bibinfo {author} {\bibfnamefont {D.~P.}\
  \bibnamefont {{Finkbeiner}}}, \bibinfo {author} {\bibfnamefont {M.~J.}\
  \bibnamefont {{Fitzpatrick}}}, \bibinfo {author} {\bibfnamefont
  {B.}~\bibnamefont {{Flaugher}}}, \bibinfo {author} {\bibfnamefont
  {S.}~\bibnamefont {{Flender}}}, \bibinfo {author} {\bibfnamefont
  {A.}~\bibnamefont {{Font-Ribera}}}, \bibinfo {author} {\bibfnamefont {J.~E.}\
  \bibnamefont {{Forero-Romero}}}, \bibinfo {author} {\bibfnamefont
  {P.}~\bibnamefont {{Fosalba}}}, \bibinfo {author} {\bibfnamefont {C.~S.}\
  \bibnamefont {{Frenk}}}, \bibinfo {author} {\bibfnamefont {M.}~\bibnamefont
  {{Fumagalli}}}, \bibinfo {author} {\bibfnamefont {B.~T.}\ \bibnamefont
  {{Gaensicke}}}, \bibinfo {author} {\bibfnamefont {G.}~\bibnamefont
  {{Gallo}}}, \bibinfo {author} {\bibfnamefont {J.}~\bibnamefont
  {{Garcia-Bellido}}}, \bibinfo {author} {\bibfnamefont {E.}~\bibnamefont
  {{Gaztanaga}}}, \bibinfo {author} {\bibfnamefont {N.}~\bibnamefont {{Pietro
  Gentile Fusillo}}}, \bibinfo {author} {\bibfnamefont {T.}~\bibnamefont
  {{Gerard}}}, \bibinfo {author} {\bibfnamefont {I.}~\bibnamefont
  {{Gershkovich}}}, \bibinfo {author} {\bibfnamefont {T.}~\bibnamefont
  {{Giannantonio}}}, \bibinfo {author} {\bibfnamefont {D.}~\bibnamefont
  {{Gillet}}}, \bibinfo {author} {\bibfnamefont {G.}~\bibnamefont
  {{Gonzalez-de-Rivera}}}, \bibinfo {author} {\bibfnamefont {V.}~\bibnamefont
  {{Gonzalez-Perez}}}, \bibinfo {author} {\bibfnamefont {S.}~\bibnamefont
  {{Gott}}}, \bibinfo {author} {\bibfnamefont {O.}~\bibnamefont {{Graur}}},
  \bibinfo {author} {\bibfnamefont {G.}~\bibnamefont {{Gutierrez}}}, \bibinfo
  {author} {\bibfnamefont {J.}~\bibnamefont {{Guy}}}, \bibinfo {author}
  {\bibfnamefont {S.}~\bibnamefont {{Habib}}}, \bibinfo {author} {\bibfnamefont
  {H.}~\bibnamefont {{Heetderks}}}, \bibinfo {author} {\bibfnamefont
  {I.}~\bibnamefont {{Heetderks}}}, \bibinfo {author} {\bibfnamefont
  {K.}~\bibnamefont {{Heitmann}}}, \bibinfo {author} {\bibfnamefont {W.~A.}\
  \bibnamefont {{Hellwing}}}, \bibinfo {author} {\bibfnamefont {D.~A.}\
  \bibnamefont {{Herrera}}}, \bibinfo {author} {\bibfnamefont {S.}~\bibnamefont
  {{Ho}}}, \bibinfo {author} {\bibfnamefont {S.}~\bibnamefont {{Holland}}},
  \bibinfo {author} {\bibfnamefont {K.}~\bibnamefont {{Honscheid}}}, \bibinfo
  {author} {\bibfnamefont {E.}~\bibnamefont {{Huff}}}, \bibinfo {author}
  {\bibfnamefont {T.~A.}\ \bibnamefont {{Hutchinson}}}, \bibinfo {author}
  {\bibfnamefont {D.}~\bibnamefont {{Huterer}}}, \bibinfo {author}
  {\bibfnamefont {H.~S.}\ \bibnamefont {{Hwang}}}, \bibinfo {author}
  {\bibfnamefont {J.~M.}\ \bibnamefont {{Illa Laguna}}}, \bibinfo {author}
  {\bibfnamefont {Y.}~\bibnamefont {{Ishikawa}}}, \bibinfo {author}
  {\bibfnamefont {D.}~\bibnamefont {{Jacobs}}}, \bibinfo {author}
  {\bibfnamefont {N.}~\bibnamefont {{Jeffrey}}}, \bibinfo {author}
  {\bibfnamefont {P.}~\bibnamefont {{Jelinsky}}}, \bibinfo {author}
  {\bibfnamefont {E.}~\bibnamefont {{Jennings}}}, \bibinfo {author}
  {\bibfnamefont {L.}~\bibnamefont {{Jiang}}}, \bibinfo {author} {\bibfnamefont
  {J.}~\bibnamefont {{Jimenez}}}, \bibinfo {author} {\bibfnamefont
  {J.}~\bibnamefont {{Johnson}}}, \bibinfo {author} {\bibfnamefont
  {R.}~\bibnamefont {{Joyce}}}, \bibinfo {author} {\bibfnamefont
  {E.}~\bibnamefont {{Jullo}}}, \bibinfo {author} {\bibfnamefont
  {S.}~\bibnamefont {{Juneau}}}, \bibinfo {author} {\bibfnamefont
  {S.}~\bibnamefont {{Kama}}}, \bibinfo {author} {\bibfnamefont
  {A.}~\bibnamefont {{Karcher}}}, \bibinfo {author} {\bibfnamefont
  {S.}~\bibnamefont {{Karkar}}}, \bibinfo {author} {\bibfnamefont
  {R.}~\bibnamefont {{Kehoe}}}, \bibinfo {author} {\bibfnamefont
  {N.}~\bibnamefont {{Kennamer}}}, \bibinfo {author} {\bibfnamefont
  {S.}~\bibnamefont {{Kent}}}, \bibinfo {author} {\bibfnamefont
  {M.}~\bibnamefont {{Kilbinger}}}, \bibinfo {author} {\bibfnamefont {A.~G.}\
  \bibnamefont {{Kim}}}, \bibinfo {author} {\bibfnamefont {D.}~\bibnamefont
  {{Kirkby}}}, \bibinfo {author} {\bibfnamefont {T.}~\bibnamefont {{Kisner}}},
  \bibinfo {author} {\bibfnamefont {E.}~\bibnamefont {{Kitanidis}}}, \bibinfo
  {author} {\bibfnamefont {J.-P.}\ \bibnamefont {{Kneib}}}, \bibinfo {author}
  {\bibfnamefont {S.}~\bibnamefont {{Koposov}}}, \bibinfo {author}
  {\bibfnamefont {E.}~\bibnamefont {{Kovacs}}}, \bibinfo {author}
  {\bibfnamefont {K.}~\bibnamefont {{Koyama}}}, \bibinfo {author}
  {\bibfnamefont {A.}~\bibnamefont {{Kremin}}}, \bibinfo {author}
  {\bibfnamefont {R.}~\bibnamefont {{Kron}}}, \bibinfo {author} {\bibfnamefont
  {L.}~\bibnamefont {{Kronig}}}, \bibinfo {author} {\bibfnamefont
  {A.}~\bibnamefont {{Kueter-Young}}}, \bibinfo {author} {\bibfnamefont
  {C.~G.}\ \bibnamefont {{Lacey}}}, \bibinfo {author} {\bibfnamefont
  {R.}~\bibnamefont {{Lafever}}}, \bibinfo {author} {\bibfnamefont
  {O.}~\bibnamefont {{Lahav}}}, \bibinfo {author} {\bibfnamefont
  {A.}~\bibnamefont {{Lambert}}}, \bibinfo {author} {\bibfnamefont
  {M.}~\bibnamefont {{Lampton}}}, \bibinfo {author} {\bibfnamefont
  {M.}~\bibnamefont {{Landriau}}}, \bibinfo {author} {\bibfnamefont
  {D.}~\bibnamefont {{Lang}}}, \bibinfo {author} {\bibfnamefont {T.~R.}\
  \bibnamefont {{Lauer}}}, \bibinfo {author} {\bibfnamefont {J.-M.}\
  \bibnamefont {{Le Goff}}}, \bibinfo {author} {\bibfnamefont {L.}~\bibnamefont
  {{Le Guillou}}}, \bibinfo {author} {\bibfnamefont {A.}~\bibnamefont {{Le Van
  Suu}}}, \bibinfo {author} {\bibfnamefont {J.~H.}\ \bibnamefont {{Lee}}},
  \bibinfo {author} {\bibfnamefont {S.-J.}\ \bibnamefont {{Lee}}}, \bibinfo
  {author} {\bibfnamefont {D.}~\bibnamefont {{Leitner}}}, \bibinfo {author}
  {\bibfnamefont {M.}~\bibnamefont {{Lesser}}}, \bibinfo {author}
  {\bibfnamefont {M.~E.}\ \bibnamefont {{Levi}}}, \bibinfo {author}
  {\bibfnamefont {B.}~\bibnamefont {{L'Huillier}}}, \bibinfo {author}
  {\bibfnamefont {B.}~\bibnamefont {{Li}}}, \bibinfo {author} {\bibfnamefont
  {M.}~\bibnamefont {{Liang}}}, \bibinfo {author} {\bibfnamefont
  {H.}~\bibnamefont {{Lin}}}, \bibinfo {author} {\bibfnamefont
  {E.}~\bibnamefont {{Linder}}}, \bibinfo {author} {\bibfnamefont {S.~R.}\
  \bibnamefont {{Loebman}}}, \bibinfo {author} {\bibfnamefont {Z.}~\bibnamefont
  {{Luki{\'c}}}}, \bibinfo {author} {\bibfnamefont {J.}~\bibnamefont {{Ma}}},
  \bibinfo {author} {\bibfnamefont {N.}~\bibnamefont {{MacCrann}}}, \bibinfo
  {author} {\bibfnamefont {C.}~\bibnamefont {{Magneville}}}, \bibinfo {author}
  {\bibfnamefont {L.}~\bibnamefont {{Makarem}}}, \bibinfo {author}
  {\bibfnamefont {M.}~\bibnamefont {{Manera}}}, \bibinfo {author}
  {\bibfnamefont {C.~J.}\ \bibnamefont {{Manser}}}, \bibinfo {author}
  {\bibfnamefont {R.}~\bibnamefont {{Marshall}}}, \bibinfo {author}
  {\bibfnamefont {P.}~\bibnamefont {{Martini}}}, \bibinfo {author}
  {\bibfnamefont {R.}~\bibnamefont {{Massey}}}, \bibinfo {author}
  {\bibfnamefont {T.}~\bibnamefont {{Matheson}}}, \bibinfo {author}
  {\bibfnamefont {J.}~\bibnamefont {{McCauley}}}, \bibinfo {author}
  {\bibfnamefont {P.}~\bibnamefont {{McDonald}}}, \bibinfo {author}
  {\bibfnamefont {I.~D.}\ \bibnamefont {{McGreer}}}, \bibinfo {author}
  {\bibfnamefont {A.}~\bibnamefont {{Meisner}}}, \bibinfo {author}
  {\bibfnamefont {N.}~\bibnamefont {{Metcalfe}}}, \bibinfo {author}
  {\bibfnamefont {T.~N.}\ \bibnamefont {{Miller}}}, \bibinfo {author}
  {\bibfnamefont {R.}~\bibnamefont {{Miquel}}}, \bibinfo {author}
  {\bibfnamefont {J.}~\bibnamefont {{Moustakas}}}, \bibinfo {author}
  {\bibfnamefont {A.}~\bibnamefont {{Myers}}}, \bibinfo {author} {\bibfnamefont
  {M.}~\bibnamefont {{Naik}}}, \bibinfo {author} {\bibfnamefont {J.~A.}\
  \bibnamefont {{Newman}}}, \bibinfo {author} {\bibfnamefont {R.~C.}\
  \bibnamefont {{Nichol}}}, \bibinfo {author} {\bibfnamefont {A.}~\bibnamefont
  {{Nicola}}}, \bibinfo {author} {\bibfnamefont {L.}~\bibnamefont {{Nicolati da
  Costa}}}, \bibinfo {author} {\bibfnamefont {J.}~\bibnamefont {{Nie}}},
  \bibinfo {author} {\bibfnamefont {G.}~\bibnamefont {{Niz}}}, \bibinfo
  {author} {\bibfnamefont {P.}~\bibnamefont {{Norberg}}}, \bibinfo {author}
  {\bibfnamefont {B.}~\bibnamefont {{Nord}}}, \bibinfo {author} {\bibfnamefont
  {D.}~\bibnamefont {{Norman}}}, \bibinfo {author} {\bibfnamefont
  {P.}~\bibnamefont {{Nugent}}}, \bibinfo {author} {\bibfnamefont
  {T.}~\bibnamefont {{O'Brien}}}, \bibinfo {author} {\bibfnamefont
  {M.}~\bibnamefont {{Oh}}}, \bibinfo {author} {\bibfnamefont {K.~A.~G.}\
  \bibnamefont {{Olsen}}}, \bibinfo {author} {\bibfnamefont {C.}~\bibnamefont
  {{Padilla}}}, \bibinfo {author} {\bibfnamefont {H.}~\bibnamefont
  {{Padmanabhan}}}, \bibinfo {author} {\bibfnamefont {N.}~\bibnamefont
  {{Padmanabhan}}}, \bibinfo {author} {\bibfnamefont {N.}~\bibnamefont
  {{Palanque-Delabrouille}}}, \bibinfo {author} {\bibfnamefont
  {A.}~\bibnamefont {{Palmese}}}, \bibinfo {author} {\bibfnamefont
  {D.}~\bibnamefont {{Pappalardo}}}, \bibinfo {author} {\bibfnamefont
  {I.}~\bibnamefont {{P{\^a}ris}}}, \bibinfo {author} {\bibfnamefont
  {C.}~\bibnamefont {{Park}}}, \bibinfo {author} {\bibfnamefont
  {A.}~\bibnamefont {{Patej}}}, \bibinfo {author} {\bibfnamefont {J.~A.}\
  \bibnamefont {{Peacock}}}, \bibinfo {author} {\bibfnamefont {H.~V.}\
  \bibnamefont {{Peiris}}}, \bibinfo {author} {\bibfnamefont {X.}~\bibnamefont
  {{Peng}}}, \bibinfo {author} {\bibfnamefont {W.~J.}\ \bibnamefont
  {{Percival}}}, \bibinfo {author} {\bibfnamefont {S.}~\bibnamefont
  {{Perruchot}}}, \bibinfo {author} {\bibfnamefont {M.~M.}\ \bibnamefont
  {{Pieri}}}, \bibinfo {author} {\bibfnamefont {R.}~\bibnamefont {{Pogge}}},
  \bibinfo {author} {\bibfnamefont {J.~E.}\ \bibnamefont {{Pollack}}}, \bibinfo
  {author} {\bibfnamefont {C.}~\bibnamefont {{Poppett}}}, \bibinfo {author}
  {\bibfnamefont {F.}~\bibnamefont {{Prada}}}, \bibinfo {author} {\bibfnamefont
  {A.}~\bibnamefont {{Prakash}}}, \bibinfo {author} {\bibfnamefont {R.~G.}\
  \bibnamefont {{Probst}}}, \bibinfo {author} {\bibfnamefont {D.}~\bibnamefont
  {{Rabinowitz}}}, \bibinfo {author} {\bibfnamefont {A.}~\bibnamefont
  {{Raichoor}}}, \bibinfo {author} {\bibfnamefont {C.~H.}\ \bibnamefont
  {{Ree}}}, \bibinfo {author} {\bibfnamefont {A.}~\bibnamefont {{Refregier}}},
  \bibinfo {author} {\bibfnamefont {X.}~\bibnamefont {{Regal}}}, \bibinfo
  {author} {\bibfnamefont {B.}~\bibnamefont {{Reid}}}, \bibinfo {author}
  {\bibfnamefont {K.}~\bibnamefont {{Reil}}}, \bibinfo {author} {\bibfnamefont
  {M.}~\bibnamefont {{Rezaie}}}, \bibinfo {author} {\bibfnamefont {C.~M.}\
  \bibnamefont {{Rockosi}}}, \bibinfo {author} {\bibfnamefont {N.}~\bibnamefont
  {{Roe}}}, \bibinfo {author} {\bibfnamefont {S.}~\bibnamefont {{Ronayette}}},
  \bibinfo {author} {\bibfnamefont {A.}~\bibnamefont {{Roodman}}}, \bibinfo
  {author} {\bibfnamefont {A.~J.}\ \bibnamefont {{Ross}}}, \bibinfo {author}
  {\bibfnamefont {N.~P.}\ \bibnamefont {{Ross}}}, \bibinfo {author}
  {\bibfnamefont {G.}~\bibnamefont {{Rossi}}}, \bibinfo {author} {\bibfnamefont
  {E.}~\bibnamefont {{Rozo}}}, \bibinfo {author} {\bibfnamefont
  {V.}~\bibnamefont {{Ruhlmann-Kleider}}}, \bibinfo {author} {\bibfnamefont
  {E.~S.}\ \bibnamefont {{Rykoff}}}, \bibinfo {author} {\bibfnamefont
  {C.}~\bibnamefont {{Sabiu}}}, \bibinfo {author} {\bibfnamefont
  {L.}~\bibnamefont {{Samushia}}}, \bibinfo {author} {\bibfnamefont
  {E.}~\bibnamefont {{Sanchez}}}, \bibinfo {author} {\bibfnamefont
  {J.}~\bibnamefont {{Sanchez}}}, \bibinfo {author} {\bibfnamefont {D.~J.}\
  \bibnamefont {{Schlegel}}}, \bibinfo {author} {\bibfnamefont
  {M.}~\bibnamefont {{Schneider}}}, \bibinfo {author} {\bibfnamefont
  {M.}~\bibnamefont {{Schubnell}}}, \bibinfo {author} {\bibfnamefont
  {A.}~\bibnamefont {{Secroun}}}, \bibinfo {author} {\bibfnamefont
  {U.}~\bibnamefont {{Seljak}}}, \bibinfo {author} {\bibfnamefont {H.-J.}\
  \bibnamefont {{Seo}}}, \bibinfo {author} {\bibfnamefont {S.}~\bibnamefont
  {{Serrano}}}, \bibinfo {author} {\bibfnamefont {A.}~\bibnamefont
  {{Shafieloo}}}, \bibinfo {author} {\bibfnamefont {H.}~\bibnamefont {{Shan}}},
  \bibinfo {author} {\bibfnamefont {R.}~\bibnamefont {{Sharples}}}, \bibinfo
  {author} {\bibfnamefont {M.~J.}\ \bibnamefont {{Sholl}}}, \bibinfo {author}
  {\bibfnamefont {W.~V.}\ \bibnamefont {{Shourt}}}, \bibinfo {author}
  {\bibfnamefont {J.~H.}\ \bibnamefont {{Silber}}}, \bibinfo {author}
  {\bibfnamefont {D.~R.}\ \bibnamefont {{Silva}}}, \bibinfo {author}
  {\bibfnamefont {M.~M.}\ \bibnamefont {{Sirk}}}, \bibinfo {author}
  {\bibfnamefont {A.}~\bibnamefont {{Slosar}}}, \bibinfo {author}
  {\bibfnamefont {A.}~\bibnamefont {{Smith}}}, \bibinfo {author} {\bibfnamefont
  {G.~F.}\ \bibnamefont {{Smoot}}}, \bibinfo {author} {\bibfnamefont
  {D.}~\bibnamefont {{Som}}}, \bibinfo {author} {\bibfnamefont {Y.-S.}\
  \bibnamefont {{Song}}}, \bibinfo {author} {\bibfnamefont {D.}~\bibnamefont
  {{Sprayberry}}}, \bibinfo {author} {\bibfnamefont {R.}~\bibnamefont
  {{Staten}}}, \bibinfo {author} {\bibfnamefont {A.}~\bibnamefont
  {{Stefanik}}}, \bibinfo {author} {\bibfnamefont {G.}~\bibnamefont {{Tarle}}},
  \bibinfo {author} {\bibfnamefont {S.}~\bibnamefont {{Sien Tie}}}, \bibinfo
  {author} {\bibfnamefont {J.~L.}\ \bibnamefont {{Tinker}}}, \bibinfo {author}
  {\bibfnamefont {R.}~\bibnamefont {{Tojeiro}}}, \bibinfo {author}
  {\bibfnamefont {F.}~\bibnamefont {{Valdes}}}, \bibinfo {author}
  {\bibfnamefont {O.}~\bibnamefont {{Valenzuela}}}, \bibinfo {author}
  {\bibfnamefont {M.}~\bibnamefont {{Valluri}}}, \bibinfo {author}
  {\bibfnamefont {M.}~\bibnamefont {{Vargas-Magana}}}, \bibinfo {author}
  {\bibfnamefont {L.}~\bibnamefont {{Verde}}}, \bibinfo {author} {\bibfnamefont
  {A.~R.}\ \bibnamefont {{Walker}}}, \bibinfo {author} {\bibfnamefont
  {J.}~\bibnamefont {{Wang}}}, \bibinfo {author} {\bibfnamefont
  {Y.}~\bibnamefont {{Wang}}}, \bibinfo {author} {\bibfnamefont {B.~A.}\
  \bibnamefont {{Weaver}}}, \bibinfo {author} {\bibfnamefont {C.}~\bibnamefont
  {{Weaverdyck}}}, \bibinfo {author} {\bibfnamefont {R.~H.}\ \bibnamefont
  {{Wechsler}}}, \bibinfo {author} {\bibfnamefont {D.~H.}\ \bibnamefont
  {{Weinberg}}}, \bibinfo {author} {\bibfnamefont {M.}~\bibnamefont {{White}}},
  \bibinfo {author} {\bibfnamefont {Q.}~\bibnamefont {{Yang}}}, \bibinfo
  {author} {\bibfnamefont {C.}~\bibnamefont {{Yeche}}}, \bibinfo {author}
  {\bibfnamefont {T.}~\bibnamefont {{Zhang}}}, \bibinfo {author} {\bibfnamefont
  {G.-B.}\ \bibnamefont {{Zhao}}}, \bibinfo {author} {\bibfnamefont
  {Y.}~\bibnamefont {{Zheng}}}, \bibinfo {author} {\bibfnamefont
  {X.}~\bibnamefont {{Zhou}}}, \bibinfo {author} {\bibfnamefont
  {Z.}~\bibnamefont {{Zhou}}}, \bibinfo {author} {\bibfnamefont
  {Y.}~\bibnamefont {{Zhu}}}, \bibinfo {author} {\bibfnamefont
  {H.}~\bibnamefont {{Zou}}},\ and\ \bibinfo {author} {\bibfnamefont
  {Y.}~\bibnamefont {{Zu}}},\ }\bibfield  {title} {\bibinfo {title} {{The DESI
  Experiment Part I: Science,Targeting, and Survey Design}},\ }\href
  {https://doi.org/10.48550/arXiv.1611.00036} {\bibfield  {journal} {\bibinfo
  {journal} {arXiv e-prints}\ ,\ \bibinfo {eid} {arXiv:1611.00036}} (\bibinfo
  {year} {2016})},\ \Eprint {https://arxiv.org/abs/1611.00036}
  {arXiv:1611.00036 [astro-ph.IM]} \BibitemShut {NoStop}%
\bibitem [{\citenamefont {{Laureijs}}\ \emph {et~al.}(2011)\citenamefont
  {{Laureijs}}, \citenamefont {{Amiaux}}, \citenamefont {{Arduini}},
  \citenamefont {{Augu{\`e}res}}, \citenamefont {{Brinchmann}}, \citenamefont
  {{Cole}}, \citenamefont {{Cropper}}, \citenamefont {{Dabin}}, \citenamefont
  {{Duvet}}, \citenamefont {{Ealet}}, \citenamefont {{Garilli}}, \citenamefont
  {{Gondoin}}, \citenamefont {{Guzzo}}, \citenamefont {{Hoar}}, \citenamefont
  {{Hoekstra}}, \citenamefont {{Holmes}}, \citenamefont {{Kitching}},
  \citenamefont {{Maciaszek}}, \citenamefont {{Mellier}}, \citenamefont
  {{Pasian}}, \citenamefont {{Percival}}, \citenamefont {{Rhodes}},
  \citenamefont {{Saavedra Criado}}, \citenamefont {{Sauvage}}, \citenamefont
  {{Scaramella}}, \citenamefont {{Valenziano}}, \citenamefont {{Warren}},
  \citenamefont {{Bender}}, \citenamefont {{Castander}}, \citenamefont
  {{Cimatti}}, \citenamefont {{Le F{\`e}vre}}, \citenamefont {{Kurki-Suonio}},
  \citenamefont {{Levi}}, \citenamefont {{Lilje}}, \citenamefont {{Meylan}},
  \citenamefont {{Nichol}}, \citenamefont {{Pedersen}}, \citenamefont {{Popa}},
  \citenamefont {{Rebolo Lopez}}, \citenamefont {{Rix}}, \citenamefont
  {{Rottgering}}, \citenamefont {{Zeilinger}}, \citenamefont {{Grupp}},
  \citenamefont {{Hudelot}}, \citenamefont {{Massey}}, \citenamefont
  {{Meneghetti}}, \citenamefont {{Miller}}, \citenamefont {{Paltani}},
  \citenamefont {{Paulin-Henriksson}}, \citenamefont {{Pires}}, \citenamefont
  {{Saxton}}, \citenamefont {{Schrabback}}, \citenamefont {{Seidel}},
  \citenamefont {{Walsh}}, \citenamefont {{Aghanim}}, \citenamefont
  {{Amendola}}, \citenamefont {{Bartlett}}, \citenamefont {{Baccigalupi}},
  \citenamefont {{Beaulieu}}, \citenamefont {{Benabed}}, \citenamefont
  {{Cuby}}, \citenamefont {{Elbaz}}, \citenamefont {{Fosalba}}, \citenamefont
  {{Gavazzi}}, \citenamefont {{Helmi}}, \citenamefont {{Hook}}, \citenamefont
  {{Irwin}}, \citenamefont {{Kneib}}, \citenamefont {{Kunz}}, \citenamefont
  {{Mannucci}}, \citenamefont {{Moscardini}}, \citenamefont {{Tao}},
  \citenamefont {{Teyssier}}, \citenamefont {{Weller}}, \citenamefont
  {{Zamorani}}, \citenamefont {{Zapatero Osorio}}, \citenamefont {{Boulade}},
  \citenamefont {{Foumond}}, \citenamefont {{Di Giorgio}}, \citenamefont
  {{Guttridge}}, \citenamefont {{James}}, \citenamefont {{Kemp}}, \citenamefont
  {{Martignac}}, \citenamefont {{Spencer}}, \citenamefont {{Walton}},
  \citenamefont {{Bl{\"u}mchen}}, \citenamefont {{Bonoli}}, \citenamefont
  {{Bortoletto}}, \citenamefont {{Cerna}}, \citenamefont {{Corcione}},
  \citenamefont {{Fabron}}, \citenamefont {{Jahnke}}, \citenamefont {{Ligori}},
  \citenamefont {{Madrid}}, \citenamefont {{Martin}}, \citenamefont
  {{Morgante}}, \citenamefont {{Pamplona}}, \citenamefont {{Prieto}},
  \citenamefont {{Riva}}, \citenamefont {{Toledo}}, \citenamefont
  {{Trifoglio}}, \citenamefont {{Zerbi}}, \citenamefont {{Abdalla}},
  \citenamefont {{Douspis}}, \citenamefont {{Grenet}}, \citenamefont
  {{Borgani}}, \citenamefont {{Bouwens}}, \citenamefont {{Courbin}},
  \citenamefont {{Delouis}}, \citenamefont {{Dubath}}, \citenamefont
  {{Fontana}}, \citenamefont {{Frailis}}, \citenamefont {{Grazian}},
  \citenamefont {{Koppenh{\"o}fer}}, \citenamefont {{Mansutti}}, \citenamefont
  {{Melchior}}, \citenamefont {{Mignoli}}, \citenamefont {{Mohr}},
  \citenamefont {{Neissner}}, \citenamefont {{Noddle}}, \citenamefont
  {{Poncet}}, \citenamefont {{Scodeggio}}, \citenamefont {{Serrano}},
  \citenamefont {{Shane}}, \citenamefont {{Starck}}, \citenamefont {{Surace}},
  \citenamefont {{Taylor}}, \citenamefont {{Verdoes-Kleijn}}, \citenamefont
  {{Vuerli}}, \citenamefont {{Williams}}, \citenamefont {{Zacchei}},
  \citenamefont {{Altieri}}, \citenamefont {{Escudero Sanz}}, \citenamefont
  {{Kohley}}, \citenamefont {{Oosterbroek}}, \citenamefont {{Astier}},
  \citenamefont {{Bacon}}, \citenamefont {{Bardelli}}, \citenamefont {{Baugh}},
  \citenamefont {{Bellagamba}}, \citenamefont {{Benoist}}, \citenamefont
  {{Bianchi}}, \citenamefont {{Biviano}}, \citenamefont {{Branchini}},
  \citenamefont {{Carbone}}, \citenamefont {{Cardone}}, \citenamefont
  {{Clements}}, \citenamefont {{Colombi}}, \citenamefont {{Conselice}},
  \citenamefont {{Cresci}}, \citenamefont {{Deacon}}, \citenamefont {{Dunlop}},
  \citenamefont {{Fedeli}}, \citenamefont {{Fontanot}}, \citenamefont
  {{Franzetti}}, \citenamefont {{Giocoli}}, \citenamefont {{Garcia-Bellido}},
  \citenamefont {{Gow}}, \citenamefont {{Heavens}}, \citenamefont {{Hewett}},
  \citenamefont {{Heymans}}, \citenamefont {{Holland}}, \citenamefont
  {{Huang}}, \citenamefont {{Ilbert}}, \citenamefont {{Joachimi}},
  \citenamefont {{Jennins}}, \citenamefont {{Kerins}}, \citenamefont
  {{Kiessling}}, \citenamefont {{Kirk}}, \citenamefont {{Kotak}}, \citenamefont
  {{Krause}}, \citenamefont {{Lahav}}, \citenamefont {{van Leeuwen}},
  \citenamefont {{Lesgourgues}}, \citenamefont {{Lombardi}}, \citenamefont
  {{Magliocchetti}}, \citenamefont {{Maguire}}, \citenamefont {{Majerotto}},
  \citenamefont {{Maoli}}, \citenamefont {{Marulli}}, \citenamefont
  {{Maurogordato}}, \citenamefont {{McCracken}}, \citenamefont {{McLure}},
  \citenamefont {{Melchiorri}}, \citenamefont {{Merson}}, \citenamefont
  {{Moresco}}, \citenamefont {{Nonino}}, \citenamefont {{Norberg}},
  \citenamefont {{Peacock}}, \citenamefont {{Pello}}, \citenamefont {{Penny}},
  \citenamefont {{Pettorino}}, \citenamefont {{Di Porto}}, \citenamefont
  {{Pozzetti}}, \citenamefont {{Quercellini}}, \citenamefont {{Radovich}},
  \citenamefont {{Rassat}}, \citenamefont {{Roche}}, \citenamefont
  {{Ronayette}}, \citenamefont {{Rossetti}}, \citenamefont {{Sartoris}},
  \citenamefont {{Schneider}}, \citenamefont {{Semboloni}}, \citenamefont
  {{Serjeant}}, \citenamefont {{Simpson}}, \citenamefont {{Skordis}},
  \citenamefont {{Smadja}}, \citenamefont {{Smartt}}, \citenamefont {{Spano}},
  \citenamefont {{Spiro}}, \citenamefont {{Sullivan}}, \citenamefont
  {{Tilquin}}, \citenamefont {{Trotta}}, \citenamefont {{Verde}}, \citenamefont
  {{Wang}}, \citenamefont {{Williger}}, \citenamefont {{Zhao}}, \citenamefont
  {{Zoubian}},\ and\ \citenamefont {{Zucca}}}]{2011arXiv1110.3193L}%
  \BibitemOpen
  \bibfield  {author} {\bibinfo {author} {\bibfnamefont {R.}~\bibnamefont
  {{Laureijs}}}, \bibinfo {author} {\bibfnamefont {J.}~\bibnamefont
  {{Amiaux}}}, \bibinfo {author} {\bibfnamefont {S.}~\bibnamefont {{Arduini}}},
  \bibinfo {author} {\bibfnamefont {J.~L.}\ \bibnamefont {{Augu{\`e}res}}},
  \bibinfo {author} {\bibfnamefont {J.}~\bibnamefont {{Brinchmann}}}, \bibinfo
  {author} {\bibfnamefont {R.}~\bibnamefont {{Cole}}}, \bibinfo {author}
  {\bibfnamefont {M.}~\bibnamefont {{Cropper}}}, \bibinfo {author}
  {\bibfnamefont {C.}~\bibnamefont {{Dabin}}}, \bibinfo {author} {\bibfnamefont
  {L.}~\bibnamefont {{Duvet}}}, \bibinfo {author} {\bibfnamefont
  {A.}~\bibnamefont {{Ealet}}}, \bibinfo {author} {\bibfnamefont
  {B.}~\bibnamefont {{Garilli}}}, \bibinfo {author} {\bibfnamefont
  {P.}~\bibnamefont {{Gondoin}}}, \bibinfo {author} {\bibfnamefont
  {L.}~\bibnamefont {{Guzzo}}}, \bibinfo {author} {\bibfnamefont
  {J.}~\bibnamefont {{Hoar}}}, \bibinfo {author} {\bibfnamefont
  {H.}~\bibnamefont {{Hoekstra}}}, \bibinfo {author} {\bibfnamefont
  {R.}~\bibnamefont {{Holmes}}}, \bibinfo {author} {\bibfnamefont
  {T.}~\bibnamefont {{Kitching}}}, \bibinfo {author} {\bibfnamefont
  {T.}~\bibnamefont {{Maciaszek}}}, \bibinfo {author} {\bibfnamefont
  {Y.}~\bibnamefont {{Mellier}}}, \bibinfo {author} {\bibfnamefont
  {F.}~\bibnamefont {{Pasian}}}, \bibinfo {author} {\bibfnamefont
  {W.}~\bibnamefont {{Percival}}}, \bibinfo {author} {\bibfnamefont
  {J.}~\bibnamefont {{Rhodes}}}, \bibinfo {author} {\bibfnamefont
  {G.}~\bibnamefont {{Saavedra Criado}}}, \bibinfo {author} {\bibfnamefont
  {M.}~\bibnamefont {{Sauvage}}}, \bibinfo {author} {\bibfnamefont
  {R.}~\bibnamefont {{Scaramella}}}, \bibinfo {author} {\bibfnamefont
  {L.}~\bibnamefont {{Valenziano}}}, \bibinfo {author} {\bibfnamefont
  {S.}~\bibnamefont {{Warren}}}, \bibinfo {author} {\bibfnamefont
  {R.}~\bibnamefont {{Bender}}}, \bibinfo {author} {\bibfnamefont
  {F.}~\bibnamefont {{Castander}}}, \bibinfo {author} {\bibfnamefont
  {A.}~\bibnamefont {{Cimatti}}}, \bibinfo {author} {\bibfnamefont
  {O.}~\bibnamefont {{Le F{\`e}vre}}}, \bibinfo {author} {\bibfnamefont
  {H.}~\bibnamefont {{Kurki-Suonio}}}, \bibinfo {author} {\bibfnamefont
  {M.}~\bibnamefont {{Levi}}}, \bibinfo {author} {\bibfnamefont
  {P.}~\bibnamefont {{Lilje}}}, \bibinfo {author} {\bibfnamefont
  {G.}~\bibnamefont {{Meylan}}}, \bibinfo {author} {\bibfnamefont
  {R.}~\bibnamefont {{Nichol}}}, \bibinfo {author} {\bibfnamefont
  {K.}~\bibnamefont {{Pedersen}}}, \bibinfo {author} {\bibfnamefont
  {V.}~\bibnamefont {{Popa}}}, \bibinfo {author} {\bibfnamefont
  {R.}~\bibnamefont {{Rebolo Lopez}}}, \bibinfo {author} {\bibfnamefont
  {H.~W.}\ \bibnamefont {{Rix}}}, \bibinfo {author} {\bibfnamefont
  {H.}~\bibnamefont {{Rottgering}}}, \bibinfo {author} {\bibfnamefont
  {W.}~\bibnamefont {{Zeilinger}}}, \bibinfo {author} {\bibfnamefont
  {F.}~\bibnamefont {{Grupp}}}, \bibinfo {author} {\bibfnamefont
  {P.}~\bibnamefont {{Hudelot}}}, \bibinfo {author} {\bibfnamefont
  {R.}~\bibnamefont {{Massey}}}, \bibinfo {author} {\bibfnamefont
  {M.}~\bibnamefont {{Meneghetti}}}, \bibinfo {author} {\bibfnamefont
  {L.}~\bibnamefont {{Miller}}}, \bibinfo {author} {\bibfnamefont
  {S.}~\bibnamefont {{Paltani}}}, \bibinfo {author} {\bibfnamefont
  {S.}~\bibnamefont {{Paulin-Henriksson}}}, \bibinfo {author} {\bibfnamefont
  {S.}~\bibnamefont {{Pires}}}, \bibinfo {author} {\bibfnamefont
  {C.}~\bibnamefont {{Saxton}}}, \bibinfo {author} {\bibfnamefont
  {T.}~\bibnamefont {{Schrabback}}}, \bibinfo {author} {\bibfnamefont
  {G.}~\bibnamefont {{Seidel}}}, \bibinfo {author} {\bibfnamefont
  {J.}~\bibnamefont {{Walsh}}}, \bibinfo {author} {\bibfnamefont
  {N.}~\bibnamefont {{Aghanim}}}, \bibinfo {author} {\bibfnamefont
  {L.}~\bibnamefont {{Amendola}}}, \bibinfo {author} {\bibfnamefont
  {J.}~\bibnamefont {{Bartlett}}}, \bibinfo {author} {\bibfnamefont
  {C.}~\bibnamefont {{Baccigalupi}}}, \bibinfo {author} {\bibfnamefont {J.~P.}\
  \bibnamefont {{Beaulieu}}}, \bibinfo {author} {\bibfnamefont
  {K.}~\bibnamefont {{Benabed}}}, \bibinfo {author} {\bibfnamefont {J.~G.}\
  \bibnamefont {{Cuby}}}, \bibinfo {author} {\bibfnamefont {D.}~\bibnamefont
  {{Elbaz}}}, \bibinfo {author} {\bibfnamefont {P.}~\bibnamefont {{Fosalba}}},
  \bibinfo {author} {\bibfnamefont {G.}~\bibnamefont {{Gavazzi}}}, \bibinfo
  {author} {\bibfnamefont {A.}~\bibnamefont {{Helmi}}}, \bibinfo {author}
  {\bibfnamefont {I.}~\bibnamefont {{Hook}}}, \bibinfo {author} {\bibfnamefont
  {M.}~\bibnamefont {{Irwin}}}, \bibinfo {author} {\bibfnamefont {J.~P.}\
  \bibnamefont {{Kneib}}}, \bibinfo {author} {\bibfnamefont {M.}~\bibnamefont
  {{Kunz}}}, \bibinfo {author} {\bibfnamefont {F.}~\bibnamefont {{Mannucci}}},
  \bibinfo {author} {\bibfnamefont {L.}~\bibnamefont {{Moscardini}}}, \bibinfo
  {author} {\bibfnamefont {C.}~\bibnamefont {{Tao}}}, \bibinfo {author}
  {\bibfnamefont {R.}~\bibnamefont {{Teyssier}}}, \bibinfo {author}
  {\bibfnamefont {J.}~\bibnamefont {{Weller}}}, \bibinfo {author}
  {\bibfnamefont {G.}~\bibnamefont {{Zamorani}}}, \bibinfo {author}
  {\bibfnamefont {M.~R.}\ \bibnamefont {{Zapatero Osorio}}}, \bibinfo {author}
  {\bibfnamefont {O.}~\bibnamefont {{Boulade}}}, \bibinfo {author}
  {\bibfnamefont {J.~J.}\ \bibnamefont {{Foumond}}}, \bibinfo {author}
  {\bibfnamefont {A.}~\bibnamefont {{Di Giorgio}}}, \bibinfo {author}
  {\bibfnamefont {P.}~\bibnamefont {{Guttridge}}}, \bibinfo {author}
  {\bibfnamefont {A.}~\bibnamefont {{James}}}, \bibinfo {author} {\bibfnamefont
  {M.}~\bibnamefont {{Kemp}}}, \bibinfo {author} {\bibfnamefont
  {J.}~\bibnamefont {{Martignac}}}, \bibinfo {author} {\bibfnamefont
  {A.}~\bibnamefont {{Spencer}}}, \bibinfo {author} {\bibfnamefont
  {D.}~\bibnamefont {{Walton}}}, \bibinfo {author} {\bibfnamefont
  {T.}~\bibnamefont {{Bl{\"u}mchen}}}, \bibinfo {author} {\bibfnamefont
  {C.}~\bibnamefont {{Bonoli}}}, \bibinfo {author} {\bibfnamefont
  {F.}~\bibnamefont {{Bortoletto}}}, \bibinfo {author} {\bibfnamefont
  {C.}~\bibnamefont {{Cerna}}}, \bibinfo {author} {\bibfnamefont
  {L.}~\bibnamefont {{Corcione}}}, \bibinfo {author} {\bibfnamefont
  {C.}~\bibnamefont {{Fabron}}}, \bibinfo {author} {\bibfnamefont
  {K.}~\bibnamefont {{Jahnke}}}, \bibinfo {author} {\bibfnamefont
  {S.}~\bibnamefont {{Ligori}}}, \bibinfo {author} {\bibfnamefont
  {F.}~\bibnamefont {{Madrid}}}, \bibinfo {author} {\bibfnamefont
  {L.}~\bibnamefont {{Martin}}}, \bibinfo {author} {\bibfnamefont
  {G.}~\bibnamefont {{Morgante}}}, \bibinfo {author} {\bibfnamefont
  {T.}~\bibnamefont {{Pamplona}}}, \bibinfo {author} {\bibfnamefont
  {E.}~\bibnamefont {{Prieto}}}, \bibinfo {author} {\bibfnamefont
  {M.}~\bibnamefont {{Riva}}}, \bibinfo {author} {\bibfnamefont
  {R.}~\bibnamefont {{Toledo}}}, \bibinfo {author} {\bibfnamefont
  {M.}~\bibnamefont {{Trifoglio}}}, \bibinfo {author} {\bibfnamefont
  {F.}~\bibnamefont {{Zerbi}}}, \bibinfo {author} {\bibfnamefont
  {F.}~\bibnamefont {{Abdalla}}}, \bibinfo {author} {\bibfnamefont
  {M.}~\bibnamefont {{Douspis}}}, \bibinfo {author} {\bibfnamefont
  {C.}~\bibnamefont {{Grenet}}}, \bibinfo {author} {\bibfnamefont
  {S.}~\bibnamefont {{Borgani}}}, \bibinfo {author} {\bibfnamefont
  {R.}~\bibnamefont {{Bouwens}}}, \bibinfo {author} {\bibfnamefont
  {F.}~\bibnamefont {{Courbin}}}, \bibinfo {author} {\bibfnamefont {J.~M.}\
  \bibnamefont {{Delouis}}}, \bibinfo {author} {\bibfnamefont {P.}~\bibnamefont
  {{Dubath}}}, \bibinfo {author} {\bibfnamefont {A.}~\bibnamefont {{Fontana}}},
  \bibinfo {author} {\bibfnamefont {M.}~\bibnamefont {{Frailis}}}, \bibinfo
  {author} {\bibfnamefont {A.}~\bibnamefont {{Grazian}}}, \bibinfo {author}
  {\bibfnamefont {J.}~\bibnamefont {{Koppenh{\"o}fer}}}, \bibinfo {author}
  {\bibfnamefont {O.}~\bibnamefont {{Mansutti}}}, \bibinfo {author}
  {\bibfnamefont {M.}~\bibnamefont {{Melchior}}}, \bibinfo {author}
  {\bibfnamefont {M.}~\bibnamefont {{Mignoli}}}, \bibinfo {author}
  {\bibfnamefont {J.}~\bibnamefont {{Mohr}}}, \bibinfo {author} {\bibfnamefont
  {C.}~\bibnamefont {{Neissner}}}, \bibinfo {author} {\bibfnamefont
  {K.}~\bibnamefont {{Noddle}}}, \bibinfo {author} {\bibfnamefont
  {M.}~\bibnamefont {{Poncet}}}, \bibinfo {author} {\bibfnamefont
  {M.}~\bibnamefont {{Scodeggio}}}, \bibinfo {author} {\bibfnamefont
  {S.}~\bibnamefont {{Serrano}}}, \bibinfo {author} {\bibfnamefont
  {N.}~\bibnamefont {{Shane}}}, \bibinfo {author} {\bibfnamefont {J.~L.}\
  \bibnamefont {{Starck}}}, \bibinfo {author} {\bibfnamefont {C.}~\bibnamefont
  {{Surace}}}, \bibinfo {author} {\bibfnamefont {A.}~\bibnamefont {{Taylor}}},
  \bibinfo {author} {\bibfnamefont {G.}~\bibnamefont {{Verdoes-Kleijn}}},
  \bibinfo {author} {\bibfnamefont {C.}~\bibnamefont {{Vuerli}}}, \bibinfo
  {author} {\bibfnamefont {O.~R.}\ \bibnamefont {{Williams}}}, \bibinfo
  {author} {\bibfnamefont {A.}~\bibnamefont {{Zacchei}}}, \bibinfo {author}
  {\bibfnamefont {B.}~\bibnamefont {{Altieri}}}, \bibinfo {author}
  {\bibfnamefont {I.}~\bibnamefont {{Escudero Sanz}}}, \bibinfo {author}
  {\bibfnamefont {R.}~\bibnamefont {{Kohley}}}, \bibinfo {author}
  {\bibfnamefont {T.}~\bibnamefont {{Oosterbroek}}}, \bibinfo {author}
  {\bibfnamefont {P.}~\bibnamefont {{Astier}}}, \bibinfo {author}
  {\bibfnamefont {D.}~\bibnamefont {{Bacon}}}, \bibinfo {author} {\bibfnamefont
  {S.}~\bibnamefont {{Bardelli}}}, \bibinfo {author} {\bibfnamefont
  {C.}~\bibnamefont {{Baugh}}}, \bibinfo {author} {\bibfnamefont
  {F.}~\bibnamefont {{Bellagamba}}}, \bibinfo {author} {\bibfnamefont
  {C.}~\bibnamefont {{Benoist}}}, \bibinfo {author} {\bibfnamefont
  {D.}~\bibnamefont {{Bianchi}}}, \bibinfo {author} {\bibfnamefont
  {A.}~\bibnamefont {{Biviano}}}, \bibinfo {author} {\bibfnamefont
  {E.}~\bibnamefont {{Branchini}}}, \bibinfo {author} {\bibfnamefont
  {C.}~\bibnamefont {{Carbone}}}, \bibinfo {author} {\bibfnamefont
  {V.}~\bibnamefont {{Cardone}}}, \bibinfo {author} {\bibfnamefont
  {D.}~\bibnamefont {{Clements}}}, \bibinfo {author} {\bibfnamefont
  {S.}~\bibnamefont {{Colombi}}}, \bibinfo {author} {\bibfnamefont
  {C.}~\bibnamefont {{Conselice}}}, \bibinfo {author} {\bibfnamefont
  {G.}~\bibnamefont {{Cresci}}}, \bibinfo {author} {\bibfnamefont
  {N.}~\bibnamefont {{Deacon}}}, \bibinfo {author} {\bibfnamefont
  {J.}~\bibnamefont {{Dunlop}}}, \bibinfo {author} {\bibfnamefont
  {C.}~\bibnamefont {{Fedeli}}}, \bibinfo {author} {\bibfnamefont
  {F.}~\bibnamefont {{Fontanot}}}, \bibinfo {author} {\bibfnamefont
  {P.}~\bibnamefont {{Franzetti}}}, \bibinfo {author} {\bibfnamefont
  {C.}~\bibnamefont {{Giocoli}}}, \bibinfo {author} {\bibfnamefont
  {J.}~\bibnamefont {{Garcia-Bellido}}}, \bibinfo {author} {\bibfnamefont
  {J.}~\bibnamefont {{Gow}}}, \bibinfo {author} {\bibfnamefont
  {A.}~\bibnamefont {{Heavens}}}, \bibinfo {author} {\bibfnamefont
  {P.}~\bibnamefont {{Hewett}}}, \bibinfo {author} {\bibfnamefont
  {C.}~\bibnamefont {{Heymans}}}, \bibinfo {author} {\bibfnamefont
  {A.}~\bibnamefont {{Holland}}}, \bibinfo {author} {\bibfnamefont
  {Z.}~\bibnamefont {{Huang}}}, \bibinfo {author} {\bibfnamefont
  {O.}~\bibnamefont {{Ilbert}}}, \bibinfo {author} {\bibfnamefont
  {B.}~\bibnamefont {{Joachimi}}}, \bibinfo {author} {\bibfnamefont
  {E.}~\bibnamefont {{Jennins}}}, \bibinfo {author} {\bibfnamefont
  {E.}~\bibnamefont {{Kerins}}}, \bibinfo {author} {\bibfnamefont
  {A.}~\bibnamefont {{Kiessling}}}, \bibinfo {author} {\bibfnamefont
  {D.}~\bibnamefont {{Kirk}}}, \bibinfo {author} {\bibfnamefont
  {R.}~\bibnamefont {{Kotak}}}, \bibinfo {author} {\bibfnamefont
  {O.}~\bibnamefont {{Krause}}}, \bibinfo {author} {\bibfnamefont
  {O.}~\bibnamefont {{Lahav}}}, \bibinfo {author} {\bibfnamefont
  {F.}~\bibnamefont {{van Leeuwen}}}, \bibinfo {author} {\bibfnamefont
  {J.}~\bibnamefont {{Lesgourgues}}}, \bibinfo {author} {\bibfnamefont
  {M.}~\bibnamefont {{Lombardi}}}, \bibinfo {author} {\bibfnamefont
  {M.}~\bibnamefont {{Magliocchetti}}}, \bibinfo {author} {\bibfnamefont
  {K.}~\bibnamefont {{Maguire}}}, \bibinfo {author} {\bibfnamefont
  {E.}~\bibnamefont {{Majerotto}}}, \bibinfo {author} {\bibfnamefont
  {R.}~\bibnamefont {{Maoli}}}, \bibinfo {author} {\bibfnamefont
  {F.}~\bibnamefont {{Marulli}}}, \bibinfo {author} {\bibfnamefont
  {S.}~\bibnamefont {{Maurogordato}}}, \bibinfo {author} {\bibfnamefont
  {H.}~\bibnamefont {{McCracken}}}, \bibinfo {author} {\bibfnamefont
  {R.}~\bibnamefont {{McLure}}}, \bibinfo {author} {\bibfnamefont
  {A.}~\bibnamefont {{Melchiorri}}}, \bibinfo {author} {\bibfnamefont
  {A.}~\bibnamefont {{Merson}}}, \bibinfo {author} {\bibfnamefont
  {M.}~\bibnamefont {{Moresco}}}, \bibinfo {author} {\bibfnamefont
  {M.}~\bibnamefont {{Nonino}}}, \bibinfo {author} {\bibfnamefont
  {P.}~\bibnamefont {{Norberg}}}, \bibinfo {author} {\bibfnamefont
  {J.}~\bibnamefont {{Peacock}}}, \bibinfo {author} {\bibfnamefont
  {R.}~\bibnamefont {{Pello}}}, \bibinfo {author} {\bibfnamefont
  {M.}~\bibnamefont {{Penny}}}, \bibinfo {author} {\bibfnamefont
  {V.}~\bibnamefont {{Pettorino}}}, \bibinfo {author} {\bibfnamefont
  {C.}~\bibnamefont {{Di Porto}}}, \bibinfo {author} {\bibfnamefont
  {L.}~\bibnamefont {{Pozzetti}}}, \bibinfo {author} {\bibfnamefont
  {C.}~\bibnamefont {{Quercellini}}}, \bibinfo {author} {\bibfnamefont
  {M.}~\bibnamefont {{Radovich}}}, \bibinfo {author} {\bibfnamefont
  {A.}~\bibnamefont {{Rassat}}}, \bibinfo {author} {\bibfnamefont
  {N.}~\bibnamefont {{Roche}}}, \bibinfo {author} {\bibfnamefont
  {S.}~\bibnamefont {{Ronayette}}}, \bibinfo {author} {\bibfnamefont
  {E.}~\bibnamefont {{Rossetti}}}, \bibinfo {author} {\bibfnamefont
  {B.}~\bibnamefont {{Sartoris}}}, \bibinfo {author} {\bibfnamefont
  {P.}~\bibnamefont {{Schneider}}}, \bibinfo {author} {\bibfnamefont
  {E.}~\bibnamefont {{Semboloni}}}, \bibinfo {author} {\bibfnamefont
  {S.}~\bibnamefont {{Serjeant}}}, \bibinfo {author} {\bibfnamefont
  {F.}~\bibnamefont {{Simpson}}}, \bibinfo {author} {\bibfnamefont
  {C.}~\bibnamefont {{Skordis}}}, \bibinfo {author} {\bibfnamefont
  {G.}~\bibnamefont {{Smadja}}}, \bibinfo {author} {\bibfnamefont
  {S.}~\bibnamefont {{Smartt}}}, \bibinfo {author} {\bibfnamefont
  {P.}~\bibnamefont {{Spano}}}, \bibinfo {author} {\bibfnamefont
  {S.}~\bibnamefont {{Spiro}}}, \bibinfo {author} {\bibfnamefont
  {M.}~\bibnamefont {{Sullivan}}}, \bibinfo {author} {\bibfnamefont
  {A.}~\bibnamefont {{Tilquin}}}, \bibinfo {author} {\bibfnamefont
  {R.}~\bibnamefont {{Trotta}}}, \bibinfo {author} {\bibfnamefont
  {L.}~\bibnamefont {{Verde}}}, \bibinfo {author} {\bibfnamefont
  {Y.}~\bibnamefont {{Wang}}}, \bibinfo {author} {\bibfnamefont
  {G.}~\bibnamefont {{Williger}}}, \bibinfo {author} {\bibfnamefont
  {G.}~\bibnamefont {{Zhao}}}, \bibinfo {author} {\bibfnamefont
  {J.}~\bibnamefont {{Zoubian}}},\ and\ \bibinfo {author} {\bibfnamefont
  {E.}~\bibnamefont {{Zucca}}},\ }\bibfield  {title} {\bibinfo {title} {{Euclid
  Definition Study Report}},\ }\href {https://doi.org/10.48550/arXiv.1110.3193}
  {\bibfield  {journal} {\bibinfo  {journal} {arXiv e-prints}\ ,\ \bibinfo
  {eid} {arXiv:1110.3193}} (\bibinfo {year} {2011})},\ \Eprint
  {https://arxiv.org/abs/1110.3193} {arXiv:1110.3193 [astro-ph.CO]}
  \BibitemShut {NoStop}%
\bibitem [{\citenamefont {{LSST Science Collaboration}}\ \emph
  {et~al.}(2009)\citenamefont {{LSST Science Collaboration}}, \citenamefont
  {{Abell}}, \citenamefont {{Allison}}, \citenamefont {{Anderson}},
  \citenamefont {{Andrew}}, \citenamefont {{Angel}}, \citenamefont {{Armus}},
  \citenamefont {{Arnett}}, \citenamefont {{Asztalos}}, \citenamefont
  {{Axelrod}}, \citenamefont {{Bailey}}, \citenamefont {{Ballantyne}},
  \citenamefont {{Bankert}}, \citenamefont {{Barkhouse}}, \citenamefont
  {{Barr}}, \citenamefont {{Barrientos}}, \citenamefont {{Barth}},
  \citenamefont {{Bartlett}}, \citenamefont {{Becker}}, \citenamefont
  {{Becla}}, \citenamefont {{Beers}}, \citenamefont {{Bernstein}},
  \citenamefont {{Biswas}}, \citenamefont {{Blanton}}, \citenamefont {{Bloom}},
  \citenamefont {{Bochanski}}, \citenamefont {{Boeshaar}}, \citenamefont
  {{Borne}}, \citenamefont {{Bradac}}, \citenamefont {{Brandt}}, \citenamefont
  {{Bridge}}, \citenamefont {{Brown}}, \citenamefont {{Brunner}}, \citenamefont
  {{Bullock}}, \citenamefont {{Burgasser}}, \citenamefont {{Burge}},
  \citenamefont {{Burke}}, \citenamefont {{Cargile}}, \citenamefont
  {{Chandrasekharan}}, \citenamefont {{Chartas}}, \citenamefont {{Chesley}},
  \citenamefont {{Chu}}, \citenamefont {{Cinabro}}, \citenamefont {{Claire}},
  \citenamefont {{Claver}}, \citenamefont {{Clowe}}, \citenamefont
  {{Connolly}}, \citenamefont {{Cook}}, \citenamefont {{Cooke}}, \citenamefont
  {{Cooray}}, \citenamefont {{Covey}}, \citenamefont {{Culliton}},
  \citenamefont {{de Jong}}, \citenamefont {{de Vries}}, \citenamefont
  {{Debattista}}, \citenamefont {{Delgado}}, \citenamefont {{Dell'Antonio}},
  \citenamefont {{Dhital}}, \citenamefont {{Di Stefano}}, \citenamefont
  {{Dickinson}}, \citenamefont {{Dilday}}, \citenamefont {{Djorgovski}},
  \citenamefont {{Dobler}}, \citenamefont {{Donalek}}, \citenamefont
  {{Dubois-Felsmann}}, \citenamefont {{Durech}}, \citenamefont {{Eliasdottir}},
  \citenamefont {{Eracleous}}, \citenamefont {{Eyer}}, \citenamefont {{Falco}},
  \citenamefont {{Fan}}, \citenamefont {{Fassnacht}}, \citenamefont
  {{Ferguson}}, \citenamefont {{Fernandez}}, \citenamefont {{Fields}},
  \citenamefont {{Finkbeiner}}, \citenamefont {{Figueroa}}, \citenamefont
  {{Fox}}, \citenamefont {{Francke}}, \citenamefont {{Frank}}, \citenamefont
  {{Frieman}}, \citenamefont {{Fromenteau}}, \citenamefont {{Furqan}},
  \citenamefont {{Galaz}}, \citenamefont {{Gal-Yam}}, \citenamefont
  {{Garnavich}}, \citenamefont {{Gawiser}}, \citenamefont {{Geary}},
  \citenamefont {{Gee}}, \citenamefont {{Gibson}}, \citenamefont {{Gilmore}},
  \citenamefont {{Grace}}, \citenamefont {{Green}}, \citenamefont {{Gressler}},
  \citenamefont {{Grillmair}}, \citenamefont {{Habib}}, \citenamefont
  {{Haggerty}}, \citenamefont {{Hamuy}}, \citenamefont {{Harris}},
  \citenamefont {{Hawley}}, \citenamefont {{Heavens}}, \citenamefont {{Hebb}},
  \citenamefont {{Henry}}, \citenamefont {{Hileman}}, \citenamefont {{Hilton}},
  \citenamefont {{Hoadley}}, \citenamefont {{Holberg}}, \citenamefont
  {{Holman}}, \citenamefont {{Howell}}, \citenamefont {{Infante}},
  \citenamefont {{Ivezic}}, \citenamefont {{Jacoby}}, \citenamefont {{Jain}},
  \citenamefont {{R}}, \citenamefont {{Jedicke}}, \citenamefont {{Jee}},
  \citenamefont {{Garrett Jernigan}}, \citenamefont {{Jha}}, \citenamefont
  {{Johnston}}, \citenamefont {{Jones}}, \citenamefont {{Juric}}, \citenamefont
  {{Kaasalainen}}, \citenamefont {{Styliani}}, \citenamefont {{Kafka}},
  \citenamefont {{Kahn}}, \citenamefont {{Kaib}}, \citenamefont {{Kalirai}},
  \citenamefont {{Kantor}}, \citenamefont {{Kasliwal}}, \citenamefont
  {{Keeton}}, \citenamefont {{Kessler}}, \citenamefont {{Knezevic}},
  \citenamefont {{Kowalski}}, \citenamefont {{Krabbendam}}, \citenamefont
  {{Krughoff}}, \citenamefont {{Kulkarni}}, \citenamefont {{Kuhlman}},
  \citenamefont {{Lacy}}, \citenamefont {{Lepine}}, \citenamefont {{Liang}},
  \citenamefont {{Lien}}, \citenamefont {{Lira}}, \citenamefont {{Long}},
  \citenamefont {{Lorenz}}, \citenamefont {{Lotz}}, \citenamefont {{Lupton}},
  \citenamefont {{Lutz}}, \citenamefont {{Macri}}, \citenamefont {{Mahabal}},
  \citenamefont {{Mandelbaum}}, \citenamefont {{Marshall}}, \citenamefont
  {{May}}, \citenamefont {{McGehee}}, \citenamefont {{Meadows}}, \citenamefont
  {{Meert}}, \citenamefont {{Milani}}, \citenamefont {{Miller}}, \citenamefont
  {{Miller}}, \citenamefont {{Mills}}, \citenamefont {{Minniti}}, \citenamefont
  {{Monet}}, \citenamefont {{Mukadam}}, \citenamefont {{Nakar}}, \citenamefont
  {{Neill}}, \citenamefont {{Newman}}, \citenamefont {{Nikolaev}},
  \citenamefont {{Nordby}}, \citenamefont {{O'Connor}}, \citenamefont
  {{Oguri}}, \citenamefont {{Oliver}}, \citenamefont {{Olivier}}, \citenamefont
  {{Olsen}}, \citenamefont {{Olsen}}, \citenamefont {{Olszewski}},
  \citenamefont {{Oluseyi}}, \citenamefont {{Padilla}}, \citenamefont
  {{Parker}}, \citenamefont {{Pepper}}, \citenamefont {{Peterson}},
  \citenamefont {{Petry}}, \citenamefont {{Pinto}}, \citenamefont {{Pizagno}},
  \citenamefont {{Popescu}}, \citenamefont {{Prsa}}, \citenamefont {{Radcka}},
  \citenamefont {{Raddick}}, \citenamefont {{Rasmussen}}, \citenamefont
  {{Rau}}, \citenamefont {{Rho}}, \citenamefont {{Rhoads}}, \citenamefont
  {{Richards}}, \citenamefont {{Ridgway}}, \citenamefont {{Robertson}},
  \citenamefont {{Roskar}}, \citenamefont {{Saha}}, \citenamefont
  {{Sarajedini}}, \citenamefont {{Scannapieco}}, \citenamefont {{Schalk}},
  \citenamefont {{Schindler}}, \citenamefont {{Schmidt}}, \citenamefont
  {{Schmidt}}, \citenamefont {{Schneider}}, \citenamefont {{Schumacher}},
  \citenamefont {{Scranton}}, \citenamefont {{Sebag}}, \citenamefont
  {{Seppala}}, \citenamefont {{Shemmer}}, \citenamefont {{Simon}},
  \citenamefont {{Sivertz}}, \citenamefont {{Smith}}, \citenamefont {{Allyn
  Smith}}, \citenamefont {{Smith}}, \citenamefont {{Spitz}}, \citenamefont
  {{Stanford}}, \citenamefont {{Stassun}}, \citenamefont {{Strader}},
  \citenamefont {{Strauss}}, \citenamefont {{Stubbs}}, \citenamefont
  {{Sweeney}}, \citenamefont {{Szalay}}, \citenamefont {{Szkody}},
  \citenamefont {{Takada}}, \citenamefont {{Thorman}}, \citenamefont
  {{Trilling}}, \citenamefont {{Trimble}}, \citenamefont {{Tyson}},
  \citenamefont {{Van Berg}}, \citenamefont {{Vanden Berk}}, \citenamefont
  {{VanderPlas}}, \citenamefont {{Verde}}, \citenamefont {{Vrsnak}},
  \citenamefont {{Walkowicz}}, \citenamefont {{Wandelt}}, \citenamefont
  {{Wang}}, \citenamefont {{Wang}}, \citenamefont {{Warner}}, \citenamefont
  {{Wechsler}}, \citenamefont {{West}}, \citenamefont {{Wiecha}}, \citenamefont
  {{Williams}}, \citenamefont {{Willman}}, \citenamefont {{Wittman}},
  \citenamefont {{Wolff}}, \citenamefont {{Wood-Vasey}}, \citenamefont
  {{Wozniak}}, \citenamefont {{Young}}, \citenamefont {{Zentner}},\ and\
  \citenamefont {{Zhan}}}]{2009arXiv0912.0201L}%
  \BibitemOpen
  \bibfield  {author} {\bibinfo {author} {\bibnamefont {{LSST Science
  Collaboration}}}, \bibinfo {author} {\bibfnamefont {P.~A.}\ \bibnamefont
  {{Abell}}}, \bibinfo {author} {\bibfnamefont {J.}~\bibnamefont {{Allison}}},
  \bibinfo {author} {\bibfnamefont {S.~F.}\ \bibnamefont {{Anderson}}},
  \bibinfo {author} {\bibfnamefont {J.~R.}\ \bibnamefont {{Andrew}}}, \bibinfo
  {author} {\bibfnamefont {J.~R.~P.}\ \bibnamefont {{Angel}}}, \bibinfo
  {author} {\bibfnamefont {L.}~\bibnamefont {{Armus}}}, \bibinfo {author}
  {\bibfnamefont {D.}~\bibnamefont {{Arnett}}}, \bibinfo {author}
  {\bibfnamefont {S.~J.}\ \bibnamefont {{Asztalos}}}, \bibinfo {author}
  {\bibfnamefont {T.~S.}\ \bibnamefont {{Axelrod}}}, \bibinfo {author}
  {\bibfnamefont {S.}~\bibnamefont {{Bailey}}}, \bibinfo {author}
  {\bibfnamefont {D.~R.}\ \bibnamefont {{Ballantyne}}}, \bibinfo {author}
  {\bibfnamefont {J.~R.}\ \bibnamefont {{Bankert}}}, \bibinfo {author}
  {\bibfnamefont {W.~A.}\ \bibnamefont {{Barkhouse}}}, \bibinfo {author}
  {\bibfnamefont {J.~D.}\ \bibnamefont {{Barr}}}, \bibinfo {author}
  {\bibfnamefont {L.~F.}\ \bibnamefont {{Barrientos}}}, \bibinfo {author}
  {\bibfnamefont {A.~J.}\ \bibnamefont {{Barth}}}, \bibinfo {author}
  {\bibfnamefont {J.~G.}\ \bibnamefont {{Bartlett}}}, \bibinfo {author}
  {\bibfnamefont {A.~C.}\ \bibnamefont {{Becker}}}, \bibinfo {author}
  {\bibfnamefont {J.}~\bibnamefont {{Becla}}}, \bibinfo {author} {\bibfnamefont
  {T.~C.}\ \bibnamefont {{Beers}}}, \bibinfo {author} {\bibfnamefont {J.~P.}\
  \bibnamefont {{Bernstein}}}, \bibinfo {author} {\bibfnamefont
  {R.}~\bibnamefont {{Biswas}}}, \bibinfo {author} {\bibfnamefont {M.~R.}\
  \bibnamefont {{Blanton}}}, \bibinfo {author} {\bibfnamefont {J.~S.}\
  \bibnamefont {{Bloom}}}, \bibinfo {author} {\bibfnamefont {J.~J.}\
  \bibnamefont {{Bochanski}}}, \bibinfo {author} {\bibfnamefont
  {P.}~\bibnamefont {{Boeshaar}}}, \bibinfo {author} {\bibfnamefont {K.~D.}\
  \bibnamefont {{Borne}}}, \bibinfo {author} {\bibfnamefont {M.}~\bibnamefont
  {{Bradac}}}, \bibinfo {author} {\bibfnamefont {W.~N.}\ \bibnamefont
  {{Brandt}}}, \bibinfo {author} {\bibfnamefont {C.~R.}\ \bibnamefont
  {{Bridge}}}, \bibinfo {author} {\bibfnamefont {M.~E.}\ \bibnamefont
  {{Brown}}}, \bibinfo {author} {\bibfnamefont {R.~J.}\ \bibnamefont
  {{Brunner}}}, \bibinfo {author} {\bibfnamefont {J.~S.}\ \bibnamefont
  {{Bullock}}}, \bibinfo {author} {\bibfnamefont {A.~J.}\ \bibnamefont
  {{Burgasser}}}, \bibinfo {author} {\bibfnamefont {J.~H.}\ \bibnamefont
  {{Burge}}}, \bibinfo {author} {\bibfnamefont {D.~L.}\ \bibnamefont
  {{Burke}}}, \bibinfo {author} {\bibfnamefont {P.~A.}\ \bibnamefont
  {{Cargile}}}, \bibinfo {author} {\bibfnamefont {S.}~\bibnamefont
  {{Chandrasekharan}}}, \bibinfo {author} {\bibfnamefont {G.}~\bibnamefont
  {{Chartas}}}, \bibinfo {author} {\bibfnamefont {S.~R.}\ \bibnamefont
  {{Chesley}}}, \bibinfo {author} {\bibfnamefont {Y.-H.}\ \bibnamefont
  {{Chu}}}, \bibinfo {author} {\bibfnamefont {D.}~\bibnamefont {{Cinabro}}},
  \bibinfo {author} {\bibfnamefont {M.~W.}\ \bibnamefont {{Claire}}}, \bibinfo
  {author} {\bibfnamefont {C.~F.}\ \bibnamefont {{Claver}}}, \bibinfo {author}
  {\bibfnamefont {D.}~\bibnamefont {{Clowe}}}, \bibinfo {author} {\bibfnamefont
  {A.~J.}\ \bibnamefont {{Connolly}}}, \bibinfo {author} {\bibfnamefont
  {K.~H.}\ \bibnamefont {{Cook}}}, \bibinfo {author} {\bibfnamefont
  {J.}~\bibnamefont {{Cooke}}}, \bibinfo {author} {\bibfnamefont
  {A.}~\bibnamefont {{Cooray}}}, \bibinfo {author} {\bibfnamefont {K.~R.}\
  \bibnamefont {{Covey}}}, \bibinfo {author} {\bibfnamefont {C.~S.}\
  \bibnamefont {{Culliton}}}, \bibinfo {author} {\bibfnamefont
  {R.}~\bibnamefont {{de Jong}}}, \bibinfo {author} {\bibfnamefont {W.~H.}\
  \bibnamefont {{de Vries}}}, \bibinfo {author} {\bibfnamefont {V.~P.}\
  \bibnamefont {{Debattista}}}, \bibinfo {author} {\bibfnamefont
  {F.}~\bibnamefont {{Delgado}}}, \bibinfo {author} {\bibfnamefont {I.~P.}\
  \bibnamefont {{Dell'Antonio}}}, \bibinfo {author} {\bibfnamefont
  {S.}~\bibnamefont {{Dhital}}}, \bibinfo {author} {\bibfnamefont
  {R.}~\bibnamefont {{Di Stefano}}}, \bibinfo {author} {\bibfnamefont
  {M.}~\bibnamefont {{Dickinson}}}, \bibinfo {author} {\bibfnamefont
  {B.}~\bibnamefont {{Dilday}}}, \bibinfo {author} {\bibfnamefont {S.~G.}\
  \bibnamefont {{Djorgovski}}}, \bibinfo {author} {\bibfnamefont
  {G.}~\bibnamefont {{Dobler}}}, \bibinfo {author} {\bibfnamefont
  {C.}~\bibnamefont {{Donalek}}}, \bibinfo {author} {\bibfnamefont
  {G.}~\bibnamefont {{Dubois-Felsmann}}}, \bibinfo {author} {\bibfnamefont
  {J.}~\bibnamefont {{Durech}}}, \bibinfo {author} {\bibfnamefont
  {A.}~\bibnamefont {{Eliasdottir}}}, \bibinfo {author} {\bibfnamefont
  {M.}~\bibnamefont {{Eracleous}}}, \bibinfo {author} {\bibfnamefont
  {L.}~\bibnamefont {{Eyer}}}, \bibinfo {author} {\bibfnamefont {E.~E.}\
  \bibnamefont {{Falco}}}, \bibinfo {author} {\bibfnamefont {X.}~\bibnamefont
  {{Fan}}}, \bibinfo {author} {\bibfnamefont {C.~D.}\ \bibnamefont
  {{Fassnacht}}}, \bibinfo {author} {\bibfnamefont {H.~C.}\ \bibnamefont
  {{Ferguson}}}, \bibinfo {author} {\bibfnamefont {Y.~R.}\ \bibnamefont
  {{Fernandez}}}, \bibinfo {author} {\bibfnamefont {B.~D.}\ \bibnamefont
  {{Fields}}}, \bibinfo {author} {\bibfnamefont {D.}~\bibnamefont
  {{Finkbeiner}}}, \bibinfo {author} {\bibfnamefont {E.~E.}\ \bibnamefont
  {{Figueroa}}}, \bibinfo {author} {\bibfnamefont {D.~B.}\ \bibnamefont
  {{Fox}}}, \bibinfo {author} {\bibfnamefont {H.}~\bibnamefont {{Francke}}},
  \bibinfo {author} {\bibfnamefont {J.~S.}\ \bibnamefont {{Frank}}}, \bibinfo
  {author} {\bibfnamefont {J.}~\bibnamefont {{Frieman}}}, \bibinfo {author}
  {\bibfnamefont {S.}~\bibnamefont {{Fromenteau}}}, \bibinfo {author}
  {\bibfnamefont {M.}~\bibnamefont {{Furqan}}}, \bibinfo {author}
  {\bibfnamefont {G.}~\bibnamefont {{Galaz}}}, \bibinfo {author} {\bibfnamefont
  {A.}~\bibnamefont {{Gal-Yam}}}, \bibinfo {author} {\bibfnamefont
  {P.}~\bibnamefont {{Garnavich}}}, \bibinfo {author} {\bibfnamefont
  {E.}~\bibnamefont {{Gawiser}}}, \bibinfo {author} {\bibfnamefont
  {J.}~\bibnamefont {{Geary}}}, \bibinfo {author} {\bibfnamefont
  {P.}~\bibnamefont {{Gee}}}, \bibinfo {author} {\bibfnamefont {R.~R.}\
  \bibnamefont {{Gibson}}}, \bibinfo {author} {\bibfnamefont {K.}~\bibnamefont
  {{Gilmore}}}, \bibinfo {author} {\bibfnamefont {E.~A.}\ \bibnamefont
  {{Grace}}}, \bibinfo {author} {\bibfnamefont {R.~F.}\ \bibnamefont
  {{Green}}}, \bibinfo {author} {\bibfnamefont {W.~J.}\ \bibnamefont
  {{Gressler}}}, \bibinfo {author} {\bibfnamefont {C.~J.}\ \bibnamefont
  {{Grillmair}}}, \bibinfo {author} {\bibfnamefont {S.}~\bibnamefont
  {{Habib}}}, \bibinfo {author} {\bibfnamefont {J.~S.}\ \bibnamefont
  {{Haggerty}}}, \bibinfo {author} {\bibfnamefont {M.}~\bibnamefont {{Hamuy}}},
  \bibinfo {author} {\bibfnamefont {A.~W.}\ \bibnamefont {{Harris}}}, \bibinfo
  {author} {\bibfnamefont {S.~L.}\ \bibnamefont {{Hawley}}}, \bibinfo {author}
  {\bibfnamefont {A.~F.}\ \bibnamefont {{Heavens}}}, \bibinfo {author}
  {\bibfnamefont {L.}~\bibnamefont {{Hebb}}}, \bibinfo {author} {\bibfnamefont
  {T.~J.}\ \bibnamefont {{Henry}}}, \bibinfo {author} {\bibfnamefont
  {E.}~\bibnamefont {{Hileman}}}, \bibinfo {author} {\bibfnamefont {E.~J.}\
  \bibnamefont {{Hilton}}}, \bibinfo {author} {\bibfnamefont {K.}~\bibnamefont
  {{Hoadley}}}, \bibinfo {author} {\bibfnamefont {J.~B.}\ \bibnamefont
  {{Holberg}}}, \bibinfo {author} {\bibfnamefont {M.~J.}\ \bibnamefont
  {{Holman}}}, \bibinfo {author} {\bibfnamefont {S.~B.}\ \bibnamefont
  {{Howell}}}, \bibinfo {author} {\bibfnamefont {L.}~\bibnamefont {{Infante}}},
  \bibinfo {author} {\bibfnamefont {Z.}~\bibnamefont {{Ivezic}}}, \bibinfo
  {author} {\bibfnamefont {S.~H.}\ \bibnamefont {{Jacoby}}}, \bibinfo {author}
  {\bibfnamefont {B.}~\bibnamefont {{Jain}}}, \bibinfo {author} {\bibnamefont
  {{R}}}, \bibinfo {author} {\bibnamefont {{Jedicke}}}, \bibinfo {author}
  {\bibfnamefont {M.~J.}\ \bibnamefont {{Jee}}}, \bibinfo {author}
  {\bibfnamefont {J.}~\bibnamefont {{Garrett Jernigan}}}, \bibinfo {author}
  {\bibfnamefont {S.~W.}\ \bibnamefont {{Jha}}}, \bibinfo {author}
  {\bibfnamefont {K.~V.}\ \bibnamefont {{Johnston}}}, \bibinfo {author}
  {\bibfnamefont {R.~L.}\ \bibnamefont {{Jones}}}, \bibinfo {author}
  {\bibfnamefont {M.}~\bibnamefont {{Juric}}}, \bibinfo {author} {\bibfnamefont
  {M.}~\bibnamefont {{Kaasalainen}}}, \bibinfo {author} {\bibnamefont
  {{Styliani}}}, \bibinfo {author} {\bibnamefont {{Kafka}}}, \bibinfo {author}
  {\bibfnamefont {S.~M.}\ \bibnamefont {{Kahn}}}, \bibinfo {author}
  {\bibfnamefont {N.~A.}\ \bibnamefont {{Kaib}}}, \bibinfo {author}
  {\bibfnamefont {J.}~\bibnamefont {{Kalirai}}}, \bibinfo {author}
  {\bibfnamefont {J.}~\bibnamefont {{Kantor}}}, \bibinfo {author}
  {\bibfnamefont {M.~M.}\ \bibnamefont {{Kasliwal}}}, \bibinfo {author}
  {\bibfnamefont {C.~R.}\ \bibnamefont {{Keeton}}}, \bibinfo {author}
  {\bibfnamefont {R.}~\bibnamefont {{Kessler}}}, \bibinfo {author}
  {\bibfnamefont {Z.}~\bibnamefont {{Knezevic}}}, \bibinfo {author}
  {\bibfnamefont {A.}~\bibnamefont {{Kowalski}}}, \bibinfo {author}
  {\bibfnamefont {V.~L.}\ \bibnamefont {{Krabbendam}}}, \bibinfo {author}
  {\bibfnamefont {K.~S.}\ \bibnamefont {{Krughoff}}}, \bibinfo {author}
  {\bibfnamefont {S.}~\bibnamefont {{Kulkarni}}}, \bibinfo {author}
  {\bibfnamefont {S.}~\bibnamefont {{Kuhlman}}}, \bibinfo {author}
  {\bibfnamefont {M.}~\bibnamefont {{Lacy}}}, \bibinfo {author} {\bibfnamefont
  {S.}~\bibnamefont {{Lepine}}}, \bibinfo {author} {\bibfnamefont
  {M.}~\bibnamefont {{Liang}}}, \bibinfo {author} {\bibfnamefont
  {A.}~\bibnamefont {{Lien}}}, \bibinfo {author} {\bibfnamefont
  {P.}~\bibnamefont {{Lira}}}, \bibinfo {author} {\bibfnamefont {K.~S.}\
  \bibnamefont {{Long}}}, \bibinfo {author} {\bibfnamefont {S.}~\bibnamefont
  {{Lorenz}}}, \bibinfo {author} {\bibfnamefont {J.~M.}\ \bibnamefont
  {{Lotz}}}, \bibinfo {author} {\bibfnamefont {R.~H.}\ \bibnamefont
  {{Lupton}}}, \bibinfo {author} {\bibfnamefont {J.}~\bibnamefont {{Lutz}}},
  \bibinfo {author} {\bibfnamefont {L.~M.}\ \bibnamefont {{Macri}}}, \bibinfo
  {author} {\bibfnamefont {A.~A.}\ \bibnamefont {{Mahabal}}}, \bibinfo {author}
  {\bibfnamefont {R.}~\bibnamefont {{Mandelbaum}}}, \bibinfo {author}
  {\bibfnamefont {P.}~\bibnamefont {{Marshall}}}, \bibinfo {author}
  {\bibfnamefont {M.}~\bibnamefont {{May}}}, \bibinfo {author} {\bibfnamefont
  {P.~M.}\ \bibnamefont {{McGehee}}}, \bibinfo {author} {\bibfnamefont {B.~T.}\
  \bibnamefont {{Meadows}}}, \bibinfo {author} {\bibfnamefont {A.}~\bibnamefont
  {{Meert}}}, \bibinfo {author} {\bibfnamefont {A.}~\bibnamefont {{Milani}}},
  \bibinfo {author} {\bibfnamefont {C.~J.}\ \bibnamefont {{Miller}}}, \bibinfo
  {author} {\bibfnamefont {M.}~\bibnamefont {{Miller}}}, \bibinfo {author}
  {\bibfnamefont {D.}~\bibnamefont {{Mills}}}, \bibinfo {author} {\bibfnamefont
  {D.}~\bibnamefont {{Minniti}}}, \bibinfo {author} {\bibfnamefont
  {D.}~\bibnamefont {{Monet}}}, \bibinfo {author} {\bibfnamefont {A.~S.}\
  \bibnamefont {{Mukadam}}}, \bibinfo {author} {\bibfnamefont {E.}~\bibnamefont
  {{Nakar}}}, \bibinfo {author} {\bibfnamefont {D.~R.}\ \bibnamefont
  {{Neill}}}, \bibinfo {author} {\bibfnamefont {J.~A.}\ \bibnamefont
  {{Newman}}}, \bibinfo {author} {\bibfnamefont {S.}~\bibnamefont
  {{Nikolaev}}}, \bibinfo {author} {\bibfnamefont {M.}~\bibnamefont
  {{Nordby}}}, \bibinfo {author} {\bibfnamefont {P.}~\bibnamefont
  {{O'Connor}}}, \bibinfo {author} {\bibfnamefont {M.}~\bibnamefont {{Oguri}}},
  \bibinfo {author} {\bibfnamefont {J.}~\bibnamefont {{Oliver}}}, \bibinfo
  {author} {\bibfnamefont {S.~S.}\ \bibnamefont {{Olivier}}}, \bibinfo {author}
  {\bibfnamefont {J.~K.}\ \bibnamefont {{Olsen}}}, \bibinfo {author}
  {\bibfnamefont {K.}~\bibnamefont {{Olsen}}}, \bibinfo {author} {\bibfnamefont
  {E.~W.}\ \bibnamefont {{Olszewski}}}, \bibinfo {author} {\bibfnamefont
  {H.}~\bibnamefont {{Oluseyi}}}, \bibinfo {author} {\bibfnamefont {N.~D.}\
  \bibnamefont {{Padilla}}}, \bibinfo {author} {\bibfnamefont {A.}~\bibnamefont
  {{Parker}}}, \bibinfo {author} {\bibfnamefont {J.}~\bibnamefont {{Pepper}}},
  \bibinfo {author} {\bibfnamefont {J.~R.}\ \bibnamefont {{Peterson}}},
  \bibinfo {author} {\bibfnamefont {C.}~\bibnamefont {{Petry}}}, \bibinfo
  {author} {\bibfnamefont {P.~A.}\ \bibnamefont {{Pinto}}}, \bibinfo {author}
  {\bibfnamefont {J.~L.}\ \bibnamefont {{Pizagno}}}, \bibinfo {author}
  {\bibfnamefont {B.}~\bibnamefont {{Popescu}}}, \bibinfo {author}
  {\bibfnamefont {A.}~\bibnamefont {{Prsa}}}, \bibinfo {author} {\bibfnamefont
  {V.}~\bibnamefont {{Radcka}}}, \bibinfo {author} {\bibfnamefont {M.~J.}\
  \bibnamefont {{Raddick}}}, \bibinfo {author} {\bibfnamefont {A.}~\bibnamefont
  {{Rasmussen}}}, \bibinfo {author} {\bibfnamefont {A.}~\bibnamefont {{Rau}}},
  \bibinfo {author} {\bibfnamefont {J.}~\bibnamefont {{Rho}}}, \bibinfo
  {author} {\bibfnamefont {J.~E.}\ \bibnamefont {{Rhoads}}}, \bibinfo {author}
  {\bibfnamefont {G.~T.}\ \bibnamefont {{Richards}}}, \bibinfo {author}
  {\bibfnamefont {S.~T.}\ \bibnamefont {{Ridgway}}}, \bibinfo {author}
  {\bibfnamefont {B.~E.}\ \bibnamefont {{Robertson}}}, \bibinfo {author}
  {\bibfnamefont {R.}~\bibnamefont {{Roskar}}}, \bibinfo {author}
  {\bibfnamefont {A.}~\bibnamefont {{Saha}}}, \bibinfo {author} {\bibfnamefont
  {A.}~\bibnamefont {{Sarajedini}}}, \bibinfo {author} {\bibfnamefont
  {E.}~\bibnamefont {{Scannapieco}}}, \bibinfo {author} {\bibfnamefont
  {T.}~\bibnamefont {{Schalk}}}, \bibinfo {author} {\bibfnamefont
  {R.}~\bibnamefont {{Schindler}}}, \bibinfo {author} {\bibfnamefont
  {S.}~\bibnamefont {{Schmidt}}}, \bibinfo {author} {\bibfnamefont
  {S.}~\bibnamefont {{Schmidt}}}, \bibinfo {author} {\bibfnamefont {D.~P.}\
  \bibnamefont {{Schneider}}}, \bibinfo {author} {\bibfnamefont
  {G.}~\bibnamefont {{Schumacher}}}, \bibinfo {author} {\bibfnamefont
  {R.}~\bibnamefont {{Scranton}}}, \bibinfo {author} {\bibfnamefont
  {J.}~\bibnamefont {{Sebag}}}, \bibinfo {author} {\bibfnamefont {L.~G.}\
  \bibnamefont {{Seppala}}}, \bibinfo {author} {\bibfnamefont {O.}~\bibnamefont
  {{Shemmer}}}, \bibinfo {author} {\bibfnamefont {J.~D.}\ \bibnamefont
  {{Simon}}}, \bibinfo {author} {\bibfnamefont {M.}~\bibnamefont {{Sivertz}}},
  \bibinfo {author} {\bibfnamefont {H.~A.}\ \bibnamefont {{Smith}}}, \bibinfo
  {author} {\bibfnamefont {J.}~\bibnamefont {{Allyn Smith}}}, \bibinfo {author}
  {\bibfnamefont {N.}~\bibnamefont {{Smith}}}, \bibinfo {author} {\bibfnamefont
  {A.~H.}\ \bibnamefont {{Spitz}}}, \bibinfo {author} {\bibfnamefont
  {A.}~\bibnamefont {{Stanford}}}, \bibinfo {author} {\bibfnamefont {K.~G.}\
  \bibnamefont {{Stassun}}}, \bibinfo {author} {\bibfnamefont {J.}~\bibnamefont
  {{Strader}}}, \bibinfo {author} {\bibfnamefont {M.~A.}\ \bibnamefont
  {{Strauss}}}, \bibinfo {author} {\bibfnamefont {C.~W.}\ \bibnamefont
  {{Stubbs}}}, \bibinfo {author} {\bibfnamefont {D.~W.}\ \bibnamefont
  {{Sweeney}}}, \bibinfo {author} {\bibfnamefont {A.}~\bibnamefont {{Szalay}}},
  \bibinfo {author} {\bibfnamefont {P.}~\bibnamefont {{Szkody}}}, \bibinfo
  {author} {\bibfnamefont {M.}~\bibnamefont {{Takada}}}, \bibinfo {author}
  {\bibfnamefont {P.}~\bibnamefont {{Thorman}}}, \bibinfo {author}
  {\bibfnamefont {D.~E.}\ \bibnamefont {{Trilling}}}, \bibinfo {author}
  {\bibfnamefont {V.}~\bibnamefont {{Trimble}}}, \bibinfo {author}
  {\bibfnamefont {A.}~\bibnamefont {{Tyson}}}, \bibinfo {author} {\bibfnamefont
  {R.}~\bibnamefont {{Van Berg}}}, \bibinfo {author} {\bibfnamefont
  {D.}~\bibnamefont {{Vanden Berk}}}, \bibinfo {author} {\bibfnamefont
  {J.}~\bibnamefont {{VanderPlas}}}, \bibinfo {author} {\bibfnamefont
  {L.}~\bibnamefont {{Verde}}}, \bibinfo {author} {\bibfnamefont
  {B.}~\bibnamefont {{Vrsnak}}}, \bibinfo {author} {\bibfnamefont {L.~M.}\
  \bibnamefont {{Walkowicz}}}, \bibinfo {author} {\bibfnamefont {B.~D.}\
  \bibnamefont {{Wandelt}}}, \bibinfo {author} {\bibfnamefont {S.}~\bibnamefont
  {{Wang}}}, \bibinfo {author} {\bibfnamefont {Y.}~\bibnamefont {{Wang}}},
  \bibinfo {author} {\bibfnamefont {M.}~\bibnamefont {{Warner}}}, \bibinfo
  {author} {\bibfnamefont {R.~H.}\ \bibnamefont {{Wechsler}}}, \bibinfo
  {author} {\bibfnamefont {A.~A.}\ \bibnamefont {{West}}}, \bibinfo {author}
  {\bibfnamefont {O.}~\bibnamefont {{Wiecha}}}, \bibinfo {author}
  {\bibfnamefont {B.~F.}\ \bibnamefont {{Williams}}}, \bibinfo {author}
  {\bibfnamefont {B.}~\bibnamefont {{Willman}}}, \bibinfo {author}
  {\bibfnamefont {D.}~\bibnamefont {{Wittman}}}, \bibinfo {author}
  {\bibfnamefont {S.~C.}\ \bibnamefont {{Wolff}}}, \bibinfo {author}
  {\bibfnamefont {W.~M.}\ \bibnamefont {{Wood-Vasey}}}, \bibinfo {author}
  {\bibfnamefont {P.}~\bibnamefont {{Wozniak}}}, \bibinfo {author}
  {\bibfnamefont {P.}~\bibnamefont {{Young}}}, \bibinfo {author} {\bibfnamefont
  {A.}~\bibnamefont {{Zentner}}},\ and\ \bibinfo {author} {\bibfnamefont
  {H.}~\bibnamefont {{Zhan}}},\ }\bibfield  {title} {\bibinfo {title} {{LSST
  Science Book, Version 2.0}},\ }\href
  {https://doi.org/10.48550/arXiv.0912.0201} {\bibfield  {journal} {\bibinfo
  {journal} {arXiv e-prints}\ ,\ \bibinfo {eid} {arXiv:0912.0201}} (\bibinfo
  {year} {2009})},\ \Eprint {https://arxiv.org/abs/0912.0201} {arXiv:0912.0201
  [astro-ph.IM]} \BibitemShut {NoStop}%
\bibitem [{\citenamefont {{Spergel}}\ \emph {et~al.}(2015)\citenamefont
  {{Spergel}}, \citenamefont {{Gehrels}}, \citenamefont {{Baltay}},
  \citenamefont {{Bennett}}, \citenamefont {{Breckinridge}}, \citenamefont
  {{Donahue}}, \citenamefont {{Dressler}}, \citenamefont {{Gaudi}},
  \citenamefont {{Greene}}, \citenamefont {{Guyon}}, \citenamefont {{Hirata}},
  \citenamefont {{Kalirai}}, \citenamefont {{Kasdin}}, \citenamefont
  {{Macintosh}}, \citenamefont {{Moos}}, \citenamefont {{Perlmutter}},
  \citenamefont {{Postman}}, \citenamefont {{Rauscher}}, \citenamefont
  {{Rhodes}}, \citenamefont {{Wang}}, \citenamefont {{Weinberg}}, \citenamefont
  {{Benford}}, \citenamefont {{Hudson}}, \citenamefont {{Jeong}}, \citenamefont
  {{Mellier}}, \citenamefont {{Traub}}, \citenamefont {{Yamada}}, \citenamefont
  {{Capak}}, \citenamefont {{Colbert}}, \citenamefont {{Masters}},
  \citenamefont {{Penny}}, \citenamefont {{Savransky}}, \citenamefont
  {{Stern}}, \citenamefont {{Zimmerman}}, \citenamefont {{Barry}},
  \citenamefont {{Bartusek}}, \citenamefont {{Carpenter}}, \citenamefont
  {{Cheng}}, \citenamefont {{Content}}, \citenamefont {{Dekens}}, \citenamefont
  {{Demers}}, \citenamefont {{Grady}}, \citenamefont {{Jackson}}, \citenamefont
  {{Kuan}}, \citenamefont {{Kruk}}, \citenamefont {{Melton}}, \citenamefont
  {{Nemati}}, \citenamefont {{Parvin}}, \citenamefont {{Poberezhskiy}},
  \citenamefont {{Peddie}}, \citenamefont {{Ruffa}}, \citenamefont {{Wallace}},
  \citenamefont {{Whipple}}, \citenamefont {{Wollack}},\ and\ \citenamefont
  {{Zhao}}}]{2015arXiv150303757S}%
  \BibitemOpen
  \bibfield  {author} {\bibinfo {author} {\bibfnamefont {D.}~\bibnamefont
  {{Spergel}}}, \bibinfo {author} {\bibfnamefont {N.}~\bibnamefont
  {{Gehrels}}}, \bibinfo {author} {\bibfnamefont {C.}~\bibnamefont {{Baltay}}},
  \bibinfo {author} {\bibfnamefont {D.}~\bibnamefont {{Bennett}}}, \bibinfo
  {author} {\bibfnamefont {J.}~\bibnamefont {{Breckinridge}}}, \bibinfo
  {author} {\bibfnamefont {M.}~\bibnamefont {{Donahue}}}, \bibinfo {author}
  {\bibfnamefont {A.}~\bibnamefont {{Dressler}}}, \bibinfo {author}
  {\bibfnamefont {B.~S.}\ \bibnamefont {{Gaudi}}}, \bibinfo {author}
  {\bibfnamefont {T.}~\bibnamefont {{Greene}}}, \bibinfo {author}
  {\bibfnamefont {O.}~\bibnamefont {{Guyon}}}, \bibinfo {author} {\bibfnamefont
  {C.}~\bibnamefont {{Hirata}}}, \bibinfo {author} {\bibfnamefont
  {J.}~\bibnamefont {{Kalirai}}}, \bibinfo {author} {\bibfnamefont {N.~J.}\
  \bibnamefont {{Kasdin}}}, \bibinfo {author} {\bibfnamefont {B.}~\bibnamefont
  {{Macintosh}}}, \bibinfo {author} {\bibfnamefont {W.}~\bibnamefont {{Moos}}},
  \bibinfo {author} {\bibfnamefont {S.}~\bibnamefont {{Perlmutter}}}, \bibinfo
  {author} {\bibfnamefont {M.}~\bibnamefont {{Postman}}}, \bibinfo {author}
  {\bibfnamefont {B.}~\bibnamefont {{Rauscher}}}, \bibinfo {author}
  {\bibfnamefont {J.}~\bibnamefont {{Rhodes}}}, \bibinfo {author}
  {\bibfnamefont {Y.}~\bibnamefont {{Wang}}}, \bibinfo {author} {\bibfnamefont
  {D.}~\bibnamefont {{Weinberg}}}, \bibinfo {author} {\bibfnamefont
  {D.}~\bibnamefont {{Benford}}}, \bibinfo {author} {\bibfnamefont
  {M.}~\bibnamefont {{Hudson}}}, \bibinfo {author} {\bibfnamefont {W.~S.}\
  \bibnamefont {{Jeong}}}, \bibinfo {author} {\bibfnamefont {Y.}~\bibnamefont
  {{Mellier}}}, \bibinfo {author} {\bibfnamefont {W.}~\bibnamefont {{Traub}}},
  \bibinfo {author} {\bibfnamefont {T.}~\bibnamefont {{Yamada}}}, \bibinfo
  {author} {\bibfnamefont {P.}~\bibnamefont {{Capak}}}, \bibinfo {author}
  {\bibfnamefont {J.}~\bibnamefont {{Colbert}}}, \bibinfo {author}
  {\bibfnamefont {D.}~\bibnamefont {{Masters}}}, \bibinfo {author}
  {\bibfnamefont {M.}~\bibnamefont {{Penny}}}, \bibinfo {author} {\bibfnamefont
  {D.}~\bibnamefont {{Savransky}}}, \bibinfo {author} {\bibfnamefont
  {D.}~\bibnamefont {{Stern}}}, \bibinfo {author} {\bibfnamefont
  {N.}~\bibnamefont {{Zimmerman}}}, \bibinfo {author} {\bibfnamefont
  {R.}~\bibnamefont {{Barry}}}, \bibinfo {author} {\bibfnamefont
  {L.}~\bibnamefont {{Bartusek}}}, \bibinfo {author} {\bibfnamefont
  {K.}~\bibnamefont {{Carpenter}}}, \bibinfo {author} {\bibfnamefont
  {E.}~\bibnamefont {{Cheng}}}, \bibinfo {author} {\bibfnamefont
  {D.}~\bibnamefont {{Content}}}, \bibinfo {author} {\bibfnamefont
  {F.}~\bibnamefont {{Dekens}}}, \bibinfo {author} {\bibfnamefont
  {R.}~\bibnamefont {{Demers}}}, \bibinfo {author} {\bibfnamefont
  {K.}~\bibnamefont {{Grady}}}, \bibinfo {author} {\bibfnamefont
  {C.}~\bibnamefont {{Jackson}}}, \bibinfo {author} {\bibfnamefont
  {G.}~\bibnamefont {{Kuan}}}, \bibinfo {author} {\bibfnamefont
  {J.}~\bibnamefont {{Kruk}}}, \bibinfo {author} {\bibfnamefont
  {M.}~\bibnamefont {{Melton}}}, \bibinfo {author} {\bibfnamefont
  {B.}~\bibnamefont {{Nemati}}}, \bibinfo {author} {\bibfnamefont
  {B.}~\bibnamefont {{Parvin}}}, \bibinfo {author} {\bibfnamefont
  {I.}~\bibnamefont {{Poberezhskiy}}}, \bibinfo {author} {\bibfnamefont
  {C.}~\bibnamefont {{Peddie}}}, \bibinfo {author} {\bibfnamefont
  {J.}~\bibnamefont {{Ruffa}}}, \bibinfo {author} {\bibfnamefont {J.~K.}\
  \bibnamefont {{Wallace}}}, \bibinfo {author} {\bibfnamefont {A.}~\bibnamefont
  {{Whipple}}}, \bibinfo {author} {\bibfnamefont {E.}~\bibnamefont
  {{Wollack}}},\ and\ \bibinfo {author} {\bibfnamefont {F.}~\bibnamefont
  {{Zhao}}},\ }\bibfield  {title} {\bibinfo {title} {{Wide-Field InfrarRed
  Survey Telescope-Astrophysics Focused Telescope Assets WFIRST-AFTA 2015
  Report}},\ }\href {https://doi.org/10.48550/arXiv.1503.03757} {\bibfield
  {journal} {\bibinfo  {journal} {arXiv e-prints}\ ,\ \bibinfo {eid}
  {arXiv:1503.03757}} (\bibinfo {year} {2015})},\ \Eprint
  {https://arxiv.org/abs/1503.03757} {arXiv:1503.03757 [astro-ph.IM]}
  \BibitemShut {NoStop}%
\bibitem [{\citenamefont {{Eifler}}\ \emph {et~al.}(2021)\citenamefont
  {{Eifler}}, \citenamefont {{Miyatake}}, \citenamefont {{Krause}},
  \citenamefont {{Heinrich}}, \citenamefont {{Miranda}}, \citenamefont
  {{Hirata}}, \citenamefont {{Xu}}, \citenamefont {{Hemmati}}, \citenamefont
  {{Simet}}, \citenamefont {{Capak}}, \citenamefont {{Choi}}, \citenamefont
  {{Dor{\'e}}}, \citenamefont {{Doux}}, \citenamefont {{Fang}}, \citenamefont
  {{Hounsell}}, \citenamefont {{Huff}}, \citenamefont {{Huang}}, \citenamefont
  {{Jarvis}}, \citenamefont {{Kruk}}, \citenamefont {{Masters}}, \citenamefont
  {{Rozo}}, \citenamefont {{Scolnic}}, \citenamefont {{Spergel}}, \citenamefont
  {{Troxel}}, \citenamefont {{von der Linden}}, \citenamefont {{Wang}},
  \citenamefont {{Weinberg}}, \citenamefont {{Wenzl}},\ and\ \citenamefont
  {{Wu}}}]{2021MNRAS.507.1746E}%
  \BibitemOpen
  \bibfield  {author} {\bibinfo {author} {\bibfnamefont {T.}~\bibnamefont
  {{Eifler}}}, \bibinfo {author} {\bibfnamefont {H.}~\bibnamefont
  {{Miyatake}}}, \bibinfo {author} {\bibfnamefont {E.}~\bibnamefont
  {{Krause}}}, \bibinfo {author} {\bibfnamefont {C.}~\bibnamefont
  {{Heinrich}}}, \bibinfo {author} {\bibfnamefont {V.}~\bibnamefont
  {{Miranda}}}, \bibinfo {author} {\bibfnamefont {C.}~\bibnamefont {{Hirata}}},
  \bibinfo {author} {\bibfnamefont {J.}~\bibnamefont {{Xu}}}, \bibinfo {author}
  {\bibfnamefont {S.}~\bibnamefont {{Hemmati}}}, \bibinfo {author}
  {\bibfnamefont {M.}~\bibnamefont {{Simet}}}, \bibinfo {author} {\bibfnamefont
  {P.}~\bibnamefont {{Capak}}}, \bibinfo {author} {\bibfnamefont
  {A.}~\bibnamefont {{Choi}}}, \bibinfo {author} {\bibfnamefont
  {O.}~\bibnamefont {{Dor{\'e}}}}, \bibinfo {author} {\bibfnamefont
  {C.}~\bibnamefont {{Doux}}}, \bibinfo {author} {\bibfnamefont
  {X.}~\bibnamefont {{Fang}}}, \bibinfo {author} {\bibfnamefont
  {R.}~\bibnamefont {{Hounsell}}}, \bibinfo {author} {\bibfnamefont
  {E.}~\bibnamefont {{Huff}}}, \bibinfo {author} {\bibfnamefont {H.-J.}\
  \bibnamefont {{Huang}}}, \bibinfo {author} {\bibfnamefont {M.}~\bibnamefont
  {{Jarvis}}}, \bibinfo {author} {\bibfnamefont {J.}~\bibnamefont {{Kruk}}},
  \bibinfo {author} {\bibfnamefont {D.}~\bibnamefont {{Masters}}}, \bibinfo
  {author} {\bibfnamefont {E.}~\bibnamefont {{Rozo}}}, \bibinfo {author}
  {\bibfnamefont {D.}~\bibnamefont {{Scolnic}}}, \bibinfo {author}
  {\bibfnamefont {D.~N.}\ \bibnamefont {{Spergel}}}, \bibinfo {author}
  {\bibfnamefont {M.}~\bibnamefont {{Troxel}}}, \bibinfo {author}
  {\bibfnamefont {A.}~\bibnamefont {{von der Linden}}}, \bibinfo {author}
  {\bibfnamefont {Y.}~\bibnamefont {{Wang}}}, \bibinfo {author} {\bibfnamefont
  {D.~H.}\ \bibnamefont {{Weinberg}}}, \bibinfo {author} {\bibfnamefont
  {L.}~\bibnamefont {{Wenzl}}},\ and\ \bibinfo {author} {\bibfnamefont {H.-Y.}\
  \bibnamefont {{Wu}}},\ }\bibfield  {title} {\bibinfo {title} {{Cosmology with
  the Roman Space Telescope - multiprobe strategies}},\ }\href
  {https://doi.org/10.1093/mnras/stab1762} {\bibfield  {journal} {\bibinfo
  {journal} {\mnras}\ }\textbf {\bibinfo {volume} {507}},\ \bibinfo {pages}
  {1746} (\bibinfo {year} {2021})},\ \Eprint {https://arxiv.org/abs/2004.05271}
  {arXiv:2004.05271 [astro-ph.CO]} \BibitemShut {NoStop}%
\bibitem [{\citenamefont {{Eisenstein}}\ \emph {et~al.}(2005)\citenamefont
  {{Eisenstein}}, \citenamefont {{Zehavi}}, \citenamefont {{Hogg}},
  \citenamefont {{Scoccimarro}}, \citenamefont {{Blanton}}, \citenamefont
  {{Nichol}}, \citenamefont {{Scranton}}, \citenamefont {{Seo}}, \citenamefont
  {{Tegmark}}, \citenamefont {{Zheng}}, \citenamefont {{Anderson}},
  \citenamefont {{Annis}}, \citenamefont {{Bahcall}}, \citenamefont
  {{Brinkmann}}, \citenamefont {{Burles}}, \citenamefont {{Castander}},
  \citenamefont {{Connolly}}, \citenamefont {{Csabai}}, \citenamefont {{Doi}},
  \citenamefont {{Fukugita}}, \citenamefont {{Frieman}}, \citenamefont
  {{Glazebrook}}, \citenamefont {{Gunn}}, \citenamefont {{Hendry}},
  \citenamefont {{Hennessy}}, \citenamefont {{Ivezi{\'c}}}, \citenamefont
  {{Kent}}, \citenamefont {{Knapp}}, \citenamefont {{Lin}}, \citenamefont
  {{Loh}}, \citenamefont {{Lupton}}, \citenamefont {{Margon}}, \citenamefont
  {{McKay}}, \citenamefont {{Meiksin}}, \citenamefont {{Munn}}, \citenamefont
  {{Pope}}, \citenamefont {{Richmond}}, \citenamefont {{Schlegel}},
  \citenamefont {{Schneider}}, \citenamefont {{Shimasaku}}, \citenamefont
  {{Stoughton}}, \citenamefont {{Strauss}}, \citenamefont {{SubbaRao}},
  \citenamefont {{Szalay}}, \citenamefont {{Szapudi}}, \citenamefont
  {{Tucker}}, \citenamefont {{Yanny}},\ and\ \citenamefont
  {{York}}}]{2005ApJ...633..560E}%
  \BibitemOpen
  \bibfield  {author} {\bibinfo {author} {\bibfnamefont {D.~J.}\ \bibnamefont
  {{Eisenstein}}}, \bibinfo {author} {\bibfnamefont {I.}~\bibnamefont
  {{Zehavi}}}, \bibinfo {author} {\bibfnamefont {D.~W.}\ \bibnamefont
  {{Hogg}}}, \bibinfo {author} {\bibfnamefont {R.}~\bibnamefont
  {{Scoccimarro}}}, \bibinfo {author} {\bibfnamefont {M.~R.}\ \bibnamefont
  {{Blanton}}}, \bibinfo {author} {\bibfnamefont {R.~C.}\ \bibnamefont
  {{Nichol}}}, \bibinfo {author} {\bibfnamefont {R.}~\bibnamefont
  {{Scranton}}}, \bibinfo {author} {\bibfnamefont {H.-J.}\ \bibnamefont
  {{Seo}}}, \bibinfo {author} {\bibfnamefont {M.}~\bibnamefont {{Tegmark}}},
  \bibinfo {author} {\bibfnamefont {Z.}~\bibnamefont {{Zheng}}}, \bibinfo
  {author} {\bibfnamefont {S.~F.}\ \bibnamefont {{Anderson}}}, \bibinfo
  {author} {\bibfnamefont {J.}~\bibnamefont {{Annis}}}, \bibinfo {author}
  {\bibfnamefont {N.}~\bibnamefont {{Bahcall}}}, \bibinfo {author}
  {\bibfnamefont {J.}~\bibnamefont {{Brinkmann}}}, \bibinfo {author}
  {\bibfnamefont {S.}~\bibnamefont {{Burles}}}, \bibinfo {author}
  {\bibfnamefont {F.~J.}\ \bibnamefont {{Castander}}}, \bibinfo {author}
  {\bibfnamefont {A.}~\bibnamefont {{Connolly}}}, \bibinfo {author}
  {\bibfnamefont {I.}~\bibnamefont {{Csabai}}}, \bibinfo {author}
  {\bibfnamefont {M.}~\bibnamefont {{Doi}}}, \bibinfo {author} {\bibfnamefont
  {M.}~\bibnamefont {{Fukugita}}}, \bibinfo {author} {\bibfnamefont {J.~A.}\
  \bibnamefont {{Frieman}}}, \bibinfo {author} {\bibfnamefont {K.}~\bibnamefont
  {{Glazebrook}}}, \bibinfo {author} {\bibfnamefont {J.~E.}\ \bibnamefont
  {{Gunn}}}, \bibinfo {author} {\bibfnamefont {J.~S.}\ \bibnamefont
  {{Hendry}}}, \bibinfo {author} {\bibfnamefont {G.}~\bibnamefont
  {{Hennessy}}}, \bibinfo {author} {\bibfnamefont {Z.}~\bibnamefont
  {{Ivezi{\'c}}}}, \bibinfo {author} {\bibfnamefont {S.}~\bibnamefont
  {{Kent}}}, \bibinfo {author} {\bibfnamefont {G.~R.}\ \bibnamefont {{Knapp}}},
  \bibinfo {author} {\bibfnamefont {H.}~\bibnamefont {{Lin}}}, \bibinfo
  {author} {\bibfnamefont {Y.-S.}\ \bibnamefont {{Loh}}}, \bibinfo {author}
  {\bibfnamefont {R.~H.}\ \bibnamefont {{Lupton}}}, \bibinfo {author}
  {\bibfnamefont {B.}~\bibnamefont {{Margon}}}, \bibinfo {author}
  {\bibfnamefont {T.~A.}\ \bibnamefont {{McKay}}}, \bibinfo {author}
  {\bibfnamefont {A.}~\bibnamefont {{Meiksin}}}, \bibinfo {author}
  {\bibfnamefont {J.~A.}\ \bibnamefont {{Munn}}}, \bibinfo {author}
  {\bibfnamefont {A.}~\bibnamefont {{Pope}}}, \bibinfo {author} {\bibfnamefont
  {M.~W.}\ \bibnamefont {{Richmond}}}, \bibinfo {author} {\bibfnamefont
  {D.}~\bibnamefont {{Schlegel}}}, \bibinfo {author} {\bibfnamefont {D.~P.}\
  \bibnamefont {{Schneider}}}, \bibinfo {author} {\bibfnamefont
  {K.}~\bibnamefont {{Shimasaku}}}, \bibinfo {author} {\bibfnamefont
  {C.}~\bibnamefont {{Stoughton}}}, \bibinfo {author} {\bibfnamefont {M.~A.}\
  \bibnamefont {{Strauss}}}, \bibinfo {author} {\bibfnamefont {M.}~\bibnamefont
  {{SubbaRao}}}, \bibinfo {author} {\bibfnamefont {A.~S.}\ \bibnamefont
  {{Szalay}}}, \bibinfo {author} {\bibfnamefont {I.}~\bibnamefont {{Szapudi}}},
  \bibinfo {author} {\bibfnamefont {D.~L.}\ \bibnamefont {{Tucker}}}, \bibinfo
  {author} {\bibfnamefont {B.}~\bibnamefont {{Yanny}}},\ and\ \bibinfo {author}
  {\bibfnamefont {D.~G.}\ \bibnamefont {{York}}},\ }\bibfield  {title}
  {\bibinfo {title} {{Detection of the Baryon Acoustic Peak in the Large-Scale
  Correlation Function of SDSS Luminous Red Galaxies}},\ }\href
  {https://doi.org/10.1086/466512} {\bibfield  {journal} {\bibinfo  {journal}
  {\apj}\ }\textbf {\bibinfo {volume} {633}},\ \bibinfo {pages} {560} (\bibinfo
  {year} {2005})},\ \Eprint {https://arxiv.org/abs/astro-ph/0501171}
  {arXiv:astro-ph/0501171 [astro-ph]} \BibitemShut {NoStop}%
\bibitem [{\citenamefont {{Peacock}}\ and\ \citenamefont
  {{Smith}}(2000)}]{2000MNRAS.318.1144P}%
  \BibitemOpen
  \bibfield  {author} {\bibinfo {author} {\bibfnamefont {J.~A.}\ \bibnamefont
  {{Peacock}}}\ and\ \bibinfo {author} {\bibfnamefont {R.~E.}\ \bibnamefont
  {{Smith}}},\ }\bibfield  {title} {\bibinfo {title} {{Halo occupation numbers
  and galaxy bias}},\ }\href {https://doi.org/10.1046/j.1365-8711.2000.03779.x}
  {\bibfield  {journal} {\bibinfo  {journal} {\mnras}\ }\textbf {\bibinfo
  {volume} {318}},\ \bibinfo {pages} {1144} (\bibinfo {year} {2000})},\ \Eprint
  {https://arxiv.org/abs/astro-ph/0005010} {arXiv:astro-ph/0005010 [astro-ph]}
  \BibitemShut {NoStop}%
\bibitem [{\citenamefont {{Seljak}}(2000)}]{2000MNRAS.318..203S}%
  \BibitemOpen
  \bibfield  {author} {\bibinfo {author} {\bibfnamefont {U.}~\bibnamefont
  {{Seljak}}},\ }\bibfield  {title} {\bibinfo {title} {{Analytic model for
  galaxy and dark matter clustering}},\ }\href
  {https://doi.org/10.1046/j.1365-8711.2000.03715.x} {\bibfield  {journal}
  {\bibinfo  {journal} {\mnras}\ }\textbf {\bibinfo {volume} {318}},\ \bibinfo
  {pages} {203} (\bibinfo {year} {2000})},\ \Eprint
  {https://arxiv.org/abs/astro-ph/0001493} {arXiv:astro-ph/0001493 [astro-ph]}
  \BibitemShut {NoStop}%
\bibitem [{\citenamefont {{Benson}}\ \emph {et~al.}(2000)\citenamefont
  {{Benson}}, \citenamefont {{Cole}}, \citenamefont {{Frenk}}, \citenamefont
  {{Baugh}},\ and\ \citenamefont {{Lacey}}}]{2000MNRAS.311..793B}%
  \BibitemOpen
  \bibfield  {author} {\bibinfo {author} {\bibfnamefont {A.~J.}\ \bibnamefont
  {{Benson}}}, \bibinfo {author} {\bibfnamefont {S.}~\bibnamefont {{Cole}}},
  \bibinfo {author} {\bibfnamefont {C.~S.}\ \bibnamefont {{Frenk}}}, \bibinfo
  {author} {\bibfnamefont {C.~M.}\ \bibnamefont {{Baugh}}},\ and\ \bibinfo
  {author} {\bibfnamefont {C.~G.}\ \bibnamefont {{Lacey}}},\ }\bibfield
  {title} {\bibinfo {title} {{The nature of galaxy bias and clustering}},\
  }\href {https://doi.org/10.1046/j.1365-8711.2000.03101.x} {\bibfield
  {journal} {\bibinfo  {journal} {\mnras}\ }\textbf {\bibinfo {volume} {311}},\
  \bibinfo {pages} {793} (\bibinfo {year} {2000})},\ \Eprint
  {https://arxiv.org/abs/astro-ph/9903343} {arXiv:astro-ph/9903343 [astro-ph]}
  \BibitemShut {NoStop}%
\bibitem [{\citenamefont {{White}}\ \emph {et~al.}(2001)\citenamefont
  {{White}}, \citenamefont {{Hernquist}},\ and\ \citenamefont
  {{Springel}}}]{2001ApJ...550L.129W}%
  \BibitemOpen
  \bibfield  {author} {\bibinfo {author} {\bibfnamefont {M.}~\bibnamefont
  {{White}}}, \bibinfo {author} {\bibfnamefont {L.}~\bibnamefont
  {{Hernquist}}},\ and\ \bibinfo {author} {\bibfnamefont {V.}~\bibnamefont
  {{Springel}}},\ }\bibfield  {title} {\bibinfo {title} {{The Halo Model and
  Numerical Simulations}},\ }\href {https://doi.org/10.1086/319644} {\bibfield
  {journal} {\bibinfo  {journal} {\apjl}\ }\textbf {\bibinfo {volume} {550}},\
  \bibinfo {pages} {L129} (\bibinfo {year} {2001})},\ \Eprint
  {https://arxiv.org/abs/astro-ph/0012518} {arXiv:astro-ph/0012518 [astro-ph]}
  \BibitemShut {NoStop}%
\bibitem [{\citenamefont {{Berlind}}\ and\ \citenamefont
  {{Weinberg}}(2002)}]{2002ApJ...575..587B}%
  \BibitemOpen
  \bibfield  {author} {\bibinfo {author} {\bibfnamefont {A.~A.}\ \bibnamefont
  {{Berlind}}}\ and\ \bibinfo {author} {\bibfnamefont {D.~H.}\ \bibnamefont
  {{Weinberg}}},\ }\bibfield  {title} {\bibinfo {title} {{The Halo Occupation
  Distribution: Toward an Empirical Determination of the Relation between
  Galaxies and Mass}},\ }\href {https://doi.org/10.1086/341469} {\bibfield
  {journal} {\bibinfo  {journal} {\apj}\ }\textbf {\bibinfo {volume} {575}},\
  \bibinfo {pages} {587} (\bibinfo {year} {2002})},\ \Eprint
  {https://arxiv.org/abs/astro-ph/0109001} {arXiv:astro-ph/0109001 [astro-ph]}
  \BibitemShut {NoStop}%
\bibitem [{\citenamefont {{Cooray}}\ and\ \citenamefont
  {{Sheth}}(2002)}]{2002PhR...372....1C}%
  \BibitemOpen
  \bibfield  {author} {\bibinfo {author} {\bibfnamefont {A.}~\bibnamefont
  {{Cooray}}}\ and\ \bibinfo {author} {\bibfnamefont {R.}~\bibnamefont
  {{Sheth}}},\ }\bibfield  {title} {\bibinfo {title} {{Halo models of large
  scale structure}},\ }\href {https://doi.org/10.1016/S0370-1573(02)00276-4}
  {\bibfield  {journal} {\bibinfo  {journal} {\physrep}\ }\textbf {\bibinfo
  {volume} {372}},\ \bibinfo {pages} {1} (\bibinfo {year} {2002})},\ \Eprint
  {https://arxiv.org/abs/astro-ph/0206508} {arXiv:astro-ph/0206508 [astro-ph]}
  \BibitemShut {NoStop}%
\bibitem [{\citenamefont {{Yang}}\ \emph {et~al.}(2009)\citenamefont {{Yang}},
  \citenamefont {{Mo}},\ and\ \citenamefont {{van den
  Bosch}}}]{2009ApJ...695..900Y}%
  \BibitemOpen
  \bibfield  {author} {\bibinfo {author} {\bibfnamefont {X.}~\bibnamefont
  {{Yang}}}, \bibinfo {author} {\bibfnamefont {H.~J.}\ \bibnamefont {{Mo}}},\
  and\ \bibinfo {author} {\bibfnamefont {F.~C.}\ \bibnamefont {{van den
  Bosch}}},\ }\bibfield  {title} {\bibinfo {title} {{Galaxy Groups in the SDSS
  DR4. III. The Luminosity and Stellar Mass Functions}},\ }\href
  {https://doi.org/10.1088/0004-637X/695/2/900} {\bibfield  {journal} {\bibinfo
   {journal} {\apj}\ }\textbf {\bibinfo {volume} {695}},\ \bibinfo {pages}
  {900} (\bibinfo {year} {2009})},\ \Eprint {https://arxiv.org/abs/0808.0539}
  {arXiv:0808.0539 [astro-ph]} \BibitemShut {NoStop}%
\bibitem [{\citenamefont {{Yang}}\ \emph {et~al.}(2004)\citenamefont {{Yang}},
  \citenamefont {{Mo}}, \citenamefont {{Jing}}, \citenamefont {{van den
  Bosch}},\ and\ \citenamefont {{Chu}}}]{2004MNRAS.350.1153Y}%
  \BibitemOpen
  \bibfield  {author} {\bibinfo {author} {\bibfnamefont {X.}~\bibnamefont
  {{Yang}}}, \bibinfo {author} {\bibfnamefont {H.~J.}\ \bibnamefont {{Mo}}},
  \bibinfo {author} {\bibfnamefont {Y.~P.}\ \bibnamefont {{Jing}}}, \bibinfo
  {author} {\bibfnamefont {F.~C.}\ \bibnamefont {{van den Bosch}}},\ and\
  \bibinfo {author} {\bibfnamefont {Y.}~\bibnamefont {{Chu}}},\ }\bibfield
  {title} {\bibinfo {title} {{Populating dark matter haloes with galaxies:
  comparing the 2dFGRS with mock galaxy redshift surveys}},\ }\href
  {https://doi.org/10.1111/j.1365-2966.2004.07744.x} {\bibfield  {journal}
  {\bibinfo  {journal} {\mnras}\ }\textbf {\bibinfo {volume} {350}},\ \bibinfo
  {pages} {1153} (\bibinfo {year} {2004})},\ \Eprint
  {https://arxiv.org/abs/astro-ph/0303524} {arXiv:astro-ph/0303524 [astro-ph]}
  \BibitemShut {NoStop}%
\bibitem [{\citenamefont {{Zheng}}\ \emph {et~al.}(2005)\citenamefont
  {{Zheng}}, \citenamefont {{Berlind}}, \citenamefont {{Weinberg}},
  \citenamefont {{Benson}}, \citenamefont {{Baugh}}, \citenamefont {{Cole}},
  \citenamefont {{Dav{\'e}}}, \citenamefont {{Frenk}}, \citenamefont {{Katz}},\
  and\ \citenamefont {{Lacey}}}]{2005ApJ...633..791Z}%
  \BibitemOpen
  \bibfield  {author} {\bibinfo {author} {\bibfnamefont {Z.}~\bibnamefont
  {{Zheng}}}, \bibinfo {author} {\bibfnamefont {A.~A.}\ \bibnamefont
  {{Berlind}}}, \bibinfo {author} {\bibfnamefont {D.~H.}\ \bibnamefont
  {{Weinberg}}}, \bibinfo {author} {\bibfnamefont {A.~J.}\ \bibnamefont
  {{Benson}}}, \bibinfo {author} {\bibfnamefont {C.~M.}\ \bibnamefont
  {{Baugh}}}, \bibinfo {author} {\bibfnamefont {S.}~\bibnamefont {{Cole}}},
  \bibinfo {author} {\bibfnamefont {R.}~\bibnamefont {{Dav{\'e}}}}, \bibinfo
  {author} {\bibfnamefont {C.~S.}\ \bibnamefont {{Frenk}}}, \bibinfo {author}
  {\bibfnamefont {N.}~\bibnamefont {{Katz}}},\ and\ \bibinfo {author}
  {\bibfnamefont {C.~G.}\ \bibnamefont {{Lacey}}},\ }\bibfield  {title}
  {\bibinfo {title} {{Theoretical Models of the Halo Occupation Distribution:
  Separating Central and Satellite Galaxies}},\ }\href
  {https://doi.org/10.1086/466510} {\bibfield  {journal} {\bibinfo  {journal}
  {\apj}\ }\textbf {\bibinfo {volume} {633}},\ \bibinfo {pages} {791} (\bibinfo
  {year} {2005})},\ \Eprint {https://arxiv.org/abs/astro-ph/0408564}
  {arXiv:astro-ph/0408564 [astro-ph]} \BibitemShut {NoStop}%
\bibitem [{\citenamefont {{Conroy}}\ \emph {et~al.}(2006)\citenamefont
  {{Conroy}}, \citenamefont {{Wechsler}},\ and\ \citenamefont
  {{Kravtsov}}}]{2006ApJ...647..201C}%
  \BibitemOpen
  \bibfield  {author} {\bibinfo {author} {\bibfnamefont {C.}~\bibnamefont
  {{Conroy}}}, \bibinfo {author} {\bibfnamefont {R.~H.}\ \bibnamefont
  {{Wechsler}}},\ and\ \bibinfo {author} {\bibfnamefont {A.~V.}\ \bibnamefont
  {{Kravtsov}}},\ }\bibfield  {title} {\bibinfo {title} {{Modeling
  Luminosity-dependent Galaxy Clustering through Cosmic Time}},\ }\href
  {https://doi.org/10.1086/503602} {\bibfield  {journal} {\bibinfo  {journal}
  {\apj}\ }\textbf {\bibinfo {volume} {647}},\ \bibinfo {pages} {201} (\bibinfo
  {year} {2006})},\ \Eprint {https://arxiv.org/abs/astro-ph/0512234}
  {arXiv:astro-ph/0512234 [astro-ph]} \BibitemShut {NoStop}%
\bibitem [{\citenamefont {{Behroozi}}\ \emph {et~al.}(2010)\citenamefont
  {{Behroozi}}, \citenamefont {{Conroy}},\ and\ \citenamefont
  {{Wechsler}}}]{2010ApJ...717..379B}%
  \BibitemOpen
  \bibfield  {author} {\bibinfo {author} {\bibfnamefont {P.~S.}\ \bibnamefont
  {{Behroozi}}}, \bibinfo {author} {\bibfnamefont {C.}~\bibnamefont
  {{Conroy}}},\ and\ \bibinfo {author} {\bibfnamefont {R.~H.}\ \bibnamefont
  {{Wechsler}}},\ }\bibfield  {title} {\bibinfo {title} {{A Comprehensive
  Analysis of Uncertainties Affecting the Stellar Mass-Halo Mass Relation for 0
  < z < 4}},\ }\href {https://doi.org/10.1088/0004-637X/717/1/379} {\bibfield
  {journal} {\bibinfo  {journal} {\apj}\ }\textbf {\bibinfo {volume} {717}},\
  \bibinfo {pages} {379} (\bibinfo {year} {2010})},\ \Eprint
  {https://arxiv.org/abs/1001.0015} {arXiv:1001.0015 [astro-ph.CO]}
  \BibitemShut {NoStop}%
\bibitem [{\citenamefont {{Reddick}}\ \emph {et~al.}(2014)\citenamefont
  {{Reddick}}, \citenamefont {{Tinker}}, \citenamefont {{Wechsler}},\ and\
  \citenamefont {{Lu}}}]{2014ApJ...783..118R}%
  \BibitemOpen
  \bibfield  {author} {\bibinfo {author} {\bibfnamefont {R.~M.}\ \bibnamefont
  {{Reddick}}}, \bibinfo {author} {\bibfnamefont {J.~L.}\ \bibnamefont
  {{Tinker}}}, \bibinfo {author} {\bibfnamefont {R.~H.}\ \bibnamefont
  {{Wechsler}}},\ and\ \bibinfo {author} {\bibfnamefont {Y.}~\bibnamefont
  {{Lu}}},\ }\bibfield  {title} {\bibinfo {title} {{Cosmological Constraints
  from Galaxy Clustering and the Mass-to-number Ratio of Galaxy Clusters:
  Marginalizing over the Physics of Galaxy Formation}},\ }\href
  {https://doi.org/10.1088/0004-637X/783/2/118} {\bibfield  {journal} {\bibinfo
   {journal} {\apj}\ }\textbf {\bibinfo {volume} {783}},\ \bibinfo {eid} {118}
  (\bibinfo {year} {2014})},\ \Eprint {https://arxiv.org/abs/1306.4686}
  {arXiv:1306.4686 [astro-ph.CO]} \BibitemShut {NoStop}%
\bibitem [{\citenamefont {{Guo}}\ \emph {et~al.}(2016)\citenamefont {{Guo}},
  \citenamefont {{Zheng}}, \citenamefont {{Behroozi}}, \citenamefont
  {{Zehavi}}, \citenamefont {{Chuang}}, \citenamefont {{Comparat}},
  \citenamefont {{Favole}}, \citenamefont {{Gottloeber}}, \citenamefont
  {{Klypin}}, \citenamefont {{Prada}}, \citenamefont {{Rodr{\'\i}guez-Torres}},
  \citenamefont {{Weinberg}},\ and\ \citenamefont
  {{Yepes}}}]{2016MNRAS.459.3040G}%
  \BibitemOpen
  \bibfield  {author} {\bibinfo {author} {\bibfnamefont {H.}~\bibnamefont
  {{Guo}}}, \bibinfo {author} {\bibfnamefont {Z.}~\bibnamefont {{Zheng}}},
  \bibinfo {author} {\bibfnamefont {P.~S.}\ \bibnamefont {{Behroozi}}},
  \bibinfo {author} {\bibfnamefont {I.}~\bibnamefont {{Zehavi}}}, \bibinfo
  {author} {\bibfnamefont {C.-H.}\ \bibnamefont {{Chuang}}}, \bibinfo {author}
  {\bibfnamefont {J.}~\bibnamefont {{Comparat}}}, \bibinfo {author}
  {\bibfnamefont {G.}~\bibnamefont {{Favole}}}, \bibinfo {author}
  {\bibfnamefont {S.}~\bibnamefont {{Gottloeber}}}, \bibinfo {author}
  {\bibfnamefont {A.}~\bibnamefont {{Klypin}}}, \bibinfo {author}
  {\bibfnamefont {F.}~\bibnamefont {{Prada}}}, \bibinfo {author} {\bibfnamefont
  {S.~A.}\ \bibnamefont {{Rodr{\'\i}guez-Torres}}}, \bibinfo {author}
  {\bibfnamefont {D.~H.}\ \bibnamefont {{Weinberg}}},\ and\ \bibinfo {author}
  {\bibfnamefont {G.}~\bibnamefont {{Yepes}}},\ }\bibfield  {title} {\bibinfo
  {title} {{Modelling galaxy clustering: halo occupation distribution versus
  subhalo matching}},\ }\href {https://doi.org/10.1093/mnras/stw845} {\bibfield
   {journal} {\bibinfo  {journal} {\mnras}\ }\textbf {\bibinfo {volume}
  {459}},\ \bibinfo {pages} {3040} (\bibinfo {year} {2016})},\ \Eprint
  {https://arxiv.org/abs/1508.07012} {arXiv:1508.07012 [astro-ph.CO]}
  \BibitemShut {NoStop}%
\bibitem [{\citenamefont {{Chaves-Montero}}\ \emph {et~al.}(2016)\citenamefont
  {{Chaves-Montero}}, \citenamefont {{Angulo}}, \citenamefont {{Schaye}},
  \citenamefont {{Schaller}}, \citenamefont {{Crain}}, \citenamefont
  {{Furlong}},\ and\ \citenamefont {{Theuns}}}]{2016MNRAS.460.3100C}%
  \BibitemOpen
  \bibfield  {author} {\bibinfo {author} {\bibfnamefont {J.}~\bibnamefont
  {{Chaves-Montero}}}, \bibinfo {author} {\bibfnamefont {R.~E.}\ \bibnamefont
  {{Angulo}}}, \bibinfo {author} {\bibfnamefont {J.}~\bibnamefont {{Schaye}}},
  \bibinfo {author} {\bibfnamefont {M.}~\bibnamefont {{Schaller}}}, \bibinfo
  {author} {\bibfnamefont {R.~A.}\ \bibnamefont {{Crain}}}, \bibinfo {author}
  {\bibfnamefont {M.}~\bibnamefont {{Furlong}}},\ and\ \bibinfo {author}
  {\bibfnamefont {T.}~\bibnamefont {{Theuns}}},\ }\bibfield  {title} {\bibinfo
  {title} {{Subhalo abundance matching and assembly bias in the EAGLE
  simulation}},\ }\href {https://doi.org/10.1093/mnras/stw1225} {\bibfield
  {journal} {\bibinfo  {journal} {\mnras}\ }\textbf {\bibinfo {volume} {460}},\
  \bibinfo {pages} {3100} (\bibinfo {year} {2016})},\ \Eprint
  {https://arxiv.org/abs/1507.01948} {arXiv:1507.01948 [astro-ph.GA]}
  \BibitemShut {NoStop}%
\bibitem [{\citenamefont {{Favole}}\ \emph {et~al.}(2022)\citenamefont
  {{Favole}}, \citenamefont {{Montero-Dorta}}, \citenamefont {{Artale}},
  \citenamefont {{Contreras}}, \citenamefont {{Zehavi}},\ and\ \citenamefont
  {{Xu}}}]{2022MNRAS.509.1614F}%
  \BibitemOpen
  \bibfield  {author} {\bibinfo {author} {\bibfnamefont {G.}~\bibnamefont
  {{Favole}}}, \bibinfo {author} {\bibfnamefont {A.~D.}\ \bibnamefont
  {{Montero-Dorta}}}, \bibinfo {author} {\bibfnamefont {M.~C.}\ \bibnamefont
  {{Artale}}}, \bibinfo {author} {\bibfnamefont {S.}~\bibnamefont
  {{Contreras}}}, \bibinfo {author} {\bibfnamefont {I.}~\bibnamefont
  {{Zehavi}}},\ and\ \bibinfo {author} {\bibfnamefont {X.}~\bibnamefont
  {{Xu}}},\ }\bibfield  {title} {\bibinfo {title} {{Subhalo abundance matching
  through the lens of a hydrodynamical simulation}},\ }\href
  {https://doi.org/10.1093/mnras/stab3006} {\bibfield  {journal} {\bibinfo
  {journal} {\mnras}\ }\textbf {\bibinfo {volume} {509}},\ \bibinfo {pages}
  {1614} (\bibinfo {year} {2022})},\ \Eprint {https://arxiv.org/abs/2101.10733}
  {arXiv:2101.10733 [astro-ph.GA]} \BibitemShut {NoStop}%
\bibitem [{\citenamefont {{Norberg}}\ \emph {et~al.}(2001)\citenamefont
  {{Norberg}}, \citenamefont {{Baugh}}, \citenamefont {{Hawkins}},
  \citenamefont {{Maddox}}, \citenamefont {{Peacock}}, \citenamefont {{Cole}},
  \citenamefont {{Frenk}}, \citenamefont {{Bland-Hawthorn}}, \citenamefont
  {{Bridges}}, \citenamefont {{Cannon}}, \citenamefont {{Colless}},
  \citenamefont {{Collins}}, \citenamefont {{Couch}}, \citenamefont {{Dalton}},
  \citenamefont {{De Propris}}, \citenamefont {{Driver}}, \citenamefont
  {{Efstathiou}}, \citenamefont {{Ellis}}, \citenamefont {{Glazebrook}},
  \citenamefont {{Jackson}}, \citenamefont {{Lahav}}, \citenamefont {{Lewis}},
  \citenamefont {{Lumsden}}, \citenamefont {{Madgwick}}, \citenamefont
  {{Peterson}}, \citenamefont {{Sutherland}},\ and\ \citenamefont
  {{Taylor}}}]{Norberg}%
  \BibitemOpen
  \bibfield  {author} {\bibinfo {author} {\bibfnamefont {P.}~\bibnamefont
  {{Norberg}}}, \bibinfo {author} {\bibfnamefont {C.~M.}\ \bibnamefont
  {{Baugh}}}, \bibinfo {author} {\bibfnamefont {E.}~\bibnamefont {{Hawkins}}},
  \bibinfo {author} {\bibfnamefont {S.}~\bibnamefont {{Maddox}}}, \bibinfo
  {author} {\bibfnamefont {J.~A.}\ \bibnamefont {{Peacock}}}, \bibinfo {author}
  {\bibfnamefont {S.}~\bibnamefont {{Cole}}}, \bibinfo {author} {\bibfnamefont
  {C.~S.}\ \bibnamefont {{Frenk}}}, \bibinfo {author} {\bibfnamefont
  {J.}~\bibnamefont {{Bland-Hawthorn}}}, \bibinfo {author} {\bibfnamefont
  {T.}~\bibnamefont {{Bridges}}}, \bibinfo {author} {\bibfnamefont
  {R.}~\bibnamefont {{Cannon}}}, \bibinfo {author} {\bibfnamefont
  {M.}~\bibnamefont {{Colless}}}, \bibinfo {author} {\bibfnamefont
  {C.}~\bibnamefont {{Collins}}}, \bibinfo {author} {\bibfnamefont
  {W.}~\bibnamefont {{Couch}}}, \bibinfo {author} {\bibfnamefont
  {G.}~\bibnamefont {{Dalton}}}, \bibinfo {author} {\bibfnamefont
  {R.}~\bibnamefont {{De Propris}}}, \bibinfo {author} {\bibfnamefont {S.~P.}\
  \bibnamefont {{Driver}}}, \bibinfo {author} {\bibfnamefont {G.}~\bibnamefont
  {{Efstathiou}}}, \bibinfo {author} {\bibfnamefont {R.~S.}\ \bibnamefont
  {{Ellis}}}, \bibinfo {author} {\bibfnamefont {K.}~\bibnamefont
  {{Glazebrook}}}, \bibinfo {author} {\bibfnamefont {C.}~\bibnamefont
  {{Jackson}}}, \bibinfo {author} {\bibfnamefont {O.}~\bibnamefont {{Lahav}}},
  \bibinfo {author} {\bibfnamefont {I.}~\bibnamefont {{Lewis}}}, \bibinfo
  {author} {\bibfnamefont {S.}~\bibnamefont {{Lumsden}}}, \bibinfo {author}
  {\bibfnamefont {D.}~\bibnamefont {{Madgwick}}}, \bibinfo {author}
  {\bibfnamefont {B.~A.}\ \bibnamefont {{Peterson}}}, \bibinfo {author}
  {\bibfnamefont {W.}~\bibnamefont {{Sutherland}}},\ and\ \bibinfo {author}
  {\bibfnamefont {K.}~\bibnamefont {{Taylor}}},\ }\bibfield  {title} {\bibinfo
  {title} {{The 2dF Galaxy Redshift Survey: luminosity dependence of galaxy
  clustering}},\ }\href {https://doi.org/10.1046/j.1365-8711.2001.04839.x}
  {\bibfield  {journal} {\bibinfo  {journal} {\mnras}\ }\textbf {\bibinfo
  {volume} {328}},\ \bibinfo {pages} {64} (\bibinfo {year} {2001})},\ \Eprint
  {https://arxiv.org/abs/astro-ph/0105500} {astro-ph/0105500} \BibitemShut
  {NoStop}%
\bibitem [{\citenamefont {{Zehavi}}\ \emph
  {et~al.}(2002{\natexlab{a}})\citenamefont {{Zehavi}}, \citenamefont
  {{Blanton}}, \citenamefont {{Frieman}}, \citenamefont {{Weinberg}},
  \citenamefont {{Mo}}, \citenamefont {{Strauss}}, \citenamefont {{Anderson}},
  \citenamefont {{Annis}}, \citenamefont {{Bahcall}}, \citenamefont
  {{Bernardi}}, \citenamefont {{Briggs}}, \citenamefont {{Brinkmann}},
  \citenamefont {{Burles}}, \citenamefont {{Carey}}, \citenamefont
  {{Castander}}, \citenamefont {{Connolly}}, \citenamefont {{Csabai}},
  \citenamefont {{Dalcanton}}, \citenamefont {{Dodelson}}, \citenamefont
  {{Doi}}, \citenamefont {{Eisenstein}}, \citenamefont {{Evans}}, \citenamefont
  {{Finkbeiner}}, \citenamefont {{Friedman}}, \citenamefont {{Fukugita}},
  \citenamefont {{Gunn}}, \citenamefont {{Hennessy}}, \citenamefont
  {{Hindsley}}, \citenamefont {{Ivezi{\'c}}}, \citenamefont {{Kent}},
  \citenamefont {{Knapp}}, \citenamefont {{Kron}}, \citenamefont {{Kunszt}},
  \citenamefont {{Lamb}}, \citenamefont {{Leger}}, \citenamefont {{Long}},
  \citenamefont {{Loveday}}, \citenamefont {{Lupton}}, \citenamefont {{McKay}},
  \citenamefont {{Meiksin}}, \citenamefont {{Merrelli}}, \citenamefont
  {{Munn}}, \citenamefont {{Narayanan}}, \citenamefont {{Newcomb}},
  \citenamefont {{Nichol}}, \citenamefont {{Owen}}, \citenamefont {{Peoples}},
  \citenamefont {{Pope}}, \citenamefont {{Rockosi}}, \citenamefont
  {{Schlegel}}, \citenamefont {{Schneider}}, \citenamefont {{Scoccimarro}},
  \citenamefont {{Sheth}}, \citenamefont {{Siegmund}}, \citenamefont {{Smee}},
  \citenamefont {{Snir}}, \citenamefont {{Stebbins}}, \citenamefont
  {{Stoughton}}, \citenamefont {{SubbaRao}}, \citenamefont {{Szalay}},
  \citenamefont {{Szapudi}}, \citenamefont {{Tegmark}}, \citenamefont
  {{Tucker}}, \citenamefont {{Uomoto}}, \citenamefont {{Vanden Berk}},
  \citenamefont {{Vogeley}}, \citenamefont {{Waddell}}, \citenamefont
  {{Yanny}},\ and\ \citenamefont {{York}}}]{Zehavi}%
  \BibitemOpen
  \bibfield  {author} {\bibinfo {author} {\bibfnamefont {I.}~\bibnamefont
  {{Zehavi}}}, \bibinfo {author} {\bibfnamefont {M.~R.}\ \bibnamefont
  {{Blanton}}}, \bibinfo {author} {\bibfnamefont {J.~A.}\ \bibnamefont
  {{Frieman}}}, \bibinfo {author} {\bibfnamefont {D.~H.}\ \bibnamefont
  {{Weinberg}}}, \bibinfo {author} {\bibfnamefont {H.~J.}\ \bibnamefont
  {{Mo}}}, \bibinfo {author} {\bibfnamefont {M.~A.}\ \bibnamefont {{Strauss}}},
  \bibinfo {author} {\bibfnamefont {S.~F.}\ \bibnamefont {{Anderson}}},
  \bibinfo {author} {\bibfnamefont {J.}~\bibnamefont {{Annis}}}, \bibinfo
  {author} {\bibfnamefont {N.~A.}\ \bibnamefont {{Bahcall}}}, \bibinfo {author}
  {\bibfnamefont {M.}~\bibnamefont {{Bernardi}}}, \bibinfo {author}
  {\bibfnamefont {J.~W.}\ \bibnamefont {{Briggs}}}, \bibinfo {author}
  {\bibfnamefont {J.}~\bibnamefont {{Brinkmann}}}, \bibinfo {author}
  {\bibfnamefont {S.}~\bibnamefont {{Burles}}}, \bibinfo {author}
  {\bibfnamefont {L.}~\bibnamefont {{Carey}}}, \bibinfo {author} {\bibfnamefont
  {F.~J.}\ \bibnamefont {{Castander}}}, \bibinfo {author} {\bibfnamefont
  {A.~J.}\ \bibnamefont {{Connolly}}}, \bibinfo {author} {\bibfnamefont
  {I.}~\bibnamefont {{Csabai}}}, \bibinfo {author} {\bibfnamefont {J.~J.}\
  \bibnamefont {{Dalcanton}}}, \bibinfo {author} {\bibfnamefont
  {S.}~\bibnamefont {{Dodelson}}}, \bibinfo {author} {\bibfnamefont
  {M.}~\bibnamefont {{Doi}}}, \bibinfo {author} {\bibfnamefont
  {D.}~\bibnamefont {{Eisenstein}}}, \bibinfo {author} {\bibfnamefont {M.~L.}\
  \bibnamefont {{Evans}}}, \bibinfo {author} {\bibfnamefont {D.~P.}\
  \bibnamefont {{Finkbeiner}}}, \bibinfo {author} {\bibfnamefont
  {S.}~\bibnamefont {{Friedman}}}, \bibinfo {author} {\bibfnamefont
  {M.}~\bibnamefont {{Fukugita}}}, \bibinfo {author} {\bibfnamefont {J.~E.}\
  \bibnamefont {{Gunn}}}, \bibinfo {author} {\bibfnamefont {G.~S.}\
  \bibnamefont {{Hennessy}}}, \bibinfo {author} {\bibfnamefont {R.~B.}\
  \bibnamefont {{Hindsley}}}, \bibinfo {author} {\bibfnamefont
  {{\v{Z}}.}~\bibnamefont {{Ivezi{\'c}}}}, \bibinfo {author} {\bibfnamefont
  {S.}~\bibnamefont {{Kent}}}, \bibinfo {author} {\bibfnamefont {G.~R.}\
  \bibnamefont {{Knapp}}}, \bibinfo {author} {\bibfnamefont {R.}~\bibnamefont
  {{Kron}}}, \bibinfo {author} {\bibfnamefont {P.}~\bibnamefont {{Kunszt}}},
  \bibinfo {author} {\bibfnamefont {D.~Q.}\ \bibnamefont {{Lamb}}}, \bibinfo
  {author} {\bibfnamefont {R.~F.}\ \bibnamefont {{Leger}}}, \bibinfo {author}
  {\bibfnamefont {D.~C.}\ \bibnamefont {{Long}}}, \bibinfo {author}
  {\bibfnamefont {J.}~\bibnamefont {{Loveday}}}, \bibinfo {author}
  {\bibfnamefont {R.~H.}\ \bibnamefont {{Lupton}}}, \bibinfo {author}
  {\bibfnamefont {T.}~\bibnamefont {{McKay}}}, \bibinfo {author} {\bibfnamefont
  {A.}~\bibnamefont {{Meiksin}}}, \bibinfo {author} {\bibfnamefont
  {A.}~\bibnamefont {{Merrelli}}}, \bibinfo {author} {\bibfnamefont {J.~A.}\
  \bibnamefont {{Munn}}}, \bibinfo {author} {\bibfnamefont {V.}~\bibnamefont
  {{Narayanan}}}, \bibinfo {author} {\bibfnamefont {M.}~\bibnamefont
  {{Newcomb}}}, \bibinfo {author} {\bibfnamefont {R.~C.}\ \bibnamefont
  {{Nichol}}}, \bibinfo {author} {\bibfnamefont {R.}~\bibnamefont {{Owen}}},
  \bibinfo {author} {\bibfnamefont {J.}~\bibnamefont {{Peoples}}}, \bibinfo
  {author} {\bibfnamefont {A.}~\bibnamefont {{Pope}}}, \bibinfo {author}
  {\bibfnamefont {C.~M.}\ \bibnamefont {{Rockosi}}}, \bibinfo {author}
  {\bibfnamefont {D.}~\bibnamefont {{Schlegel}}}, \bibinfo {author}
  {\bibfnamefont {D.~P.}\ \bibnamefont {{Schneider}}}, \bibinfo {author}
  {\bibfnamefont {R.}~\bibnamefont {{Scoccimarro}}}, \bibinfo {author}
  {\bibfnamefont {R.~K.}\ \bibnamefont {{Sheth}}}, \bibinfo {author}
  {\bibfnamefont {W.}~\bibnamefont {{Siegmund}}}, \bibinfo {author}
  {\bibfnamefont {S.}~\bibnamefont {{Smee}}}, \bibinfo {author} {\bibfnamefont
  {Y.}~\bibnamefont {{Snir}}}, \bibinfo {author} {\bibfnamefont
  {A.}~\bibnamefont {{Stebbins}}}, \bibinfo {author} {\bibfnamefont
  {C.}~\bibnamefont {{Stoughton}}}, \bibinfo {author} {\bibfnamefont
  {M.}~\bibnamefont {{SubbaRao}}}, \bibinfo {author} {\bibfnamefont {A.~S.}\
  \bibnamefont {{Szalay}}}, \bibinfo {author} {\bibfnamefont {I.}~\bibnamefont
  {{Szapudi}}}, \bibinfo {author} {\bibfnamefont {M.}~\bibnamefont
  {{Tegmark}}}, \bibinfo {author} {\bibfnamefont {D.~L.}\ \bibnamefont
  {{Tucker}}}, \bibinfo {author} {\bibfnamefont {A.}~\bibnamefont {{Uomoto}}},
  \bibinfo {author} {\bibfnamefont {D.}~\bibnamefont {{Vanden Berk}}}, \bibinfo
  {author} {\bibfnamefont {M.~S.}\ \bibnamefont {{Vogeley}}}, \bibinfo {author}
  {\bibfnamefont {P.}~\bibnamefont {{Waddell}}}, \bibinfo {author}
  {\bibfnamefont {B.}~\bibnamefont {{Yanny}}},\ and\ \bibinfo {author}
  {\bibfnamefont {D.~G.}\ \bibnamefont {{York}}},\ }\bibfield  {title}
  {\bibinfo {title} {{Galaxy Clustering in Early Sloan Digital Sky Survey
  Redshift Data}},\ }\href {https://doi.org/10.1086/339893} {\bibfield
  {journal} {\bibinfo  {journal} {\apj}\ }\textbf {\bibinfo {volume} {571}},\
  \bibinfo {pages} {172} (\bibinfo {year} {2002}{\natexlab{a}})},\ \Eprint
  {https://arxiv.org/abs/astro-ph/0106476} {arXiv:astro-ph/0106476 [astro-ph]}
  \BibitemShut {NoStop}%
\bibitem [{\citenamefont {{Leauthaud}}\ \emph {et~al.}(2017)\citenamefont
  {{Leauthaud}}, \citenamefont {{Saito}}, \citenamefont {{Hilbert}},
  \citenamefont {{Barreira}}, \citenamefont {{More}}, \citenamefont {{White}},
  \citenamefont {{Alam}}, \citenamefont {{Behroozi}}, \citenamefont {{Bundy}},
  \citenamefont {{Coupon}}, \citenamefont {{Erben}}, \citenamefont {{Heymans}},
  \citenamefont {{Hildebrandt}}, \citenamefont {{Mandelbaum}}, \citenamefont
  {{Miller}}, \citenamefont {{Moraes}}, \citenamefont {{Pereira}},
  \citenamefont {{Rodr{\'\i}guez-Torres}}, \citenamefont {{Schmidt}},
  \citenamefont {{Shan}}, \citenamefont {{Viel}},\ and\ \citenamefont
  {{Villaescusa-Navarro}}}]{2017MNRAS.467.3024L}%
  \BibitemOpen
  \bibfield  {author} {\bibinfo {author} {\bibfnamefont {A.}~\bibnamefont
  {{Leauthaud}}}, \bibinfo {author} {\bibfnamefont {S.}~\bibnamefont
  {{Saito}}}, \bibinfo {author} {\bibfnamefont {S.}~\bibnamefont {{Hilbert}}},
  \bibinfo {author} {\bibfnamefont {A.}~\bibnamefont {{Barreira}}}, \bibinfo
  {author} {\bibfnamefont {S.}~\bibnamefont {{More}}}, \bibinfo {author}
  {\bibfnamefont {M.}~\bibnamefont {{White}}}, \bibinfo {author} {\bibfnamefont
  {S.}~\bibnamefont {{Alam}}}, \bibinfo {author} {\bibfnamefont
  {P.}~\bibnamefont {{Behroozi}}}, \bibinfo {author} {\bibfnamefont
  {K.}~\bibnamefont {{Bundy}}}, \bibinfo {author} {\bibfnamefont
  {J.}~\bibnamefont {{Coupon}}}, \bibinfo {author} {\bibfnamefont
  {T.}~\bibnamefont {{Erben}}}, \bibinfo {author} {\bibfnamefont
  {C.}~\bibnamefont {{Heymans}}}, \bibinfo {author} {\bibfnamefont
  {H.}~\bibnamefont {{Hildebrandt}}}, \bibinfo {author} {\bibfnamefont
  {R.}~\bibnamefont {{Mandelbaum}}}, \bibinfo {author} {\bibfnamefont
  {L.}~\bibnamefont {{Miller}}}, \bibinfo {author} {\bibfnamefont
  {B.}~\bibnamefont {{Moraes}}}, \bibinfo {author} {\bibfnamefont {M.~E.~S.}\
  \bibnamefont {{Pereira}}}, \bibinfo {author} {\bibfnamefont {S.~A.}\
  \bibnamefont {{Rodr{\'\i}guez-Torres}}}, \bibinfo {author} {\bibfnamefont
  {F.}~\bibnamefont {{Schmidt}}}, \bibinfo {author} {\bibfnamefont {H.-Y.}\
  \bibnamefont {{Shan}}}, \bibinfo {author} {\bibfnamefont {M.}~\bibnamefont
  {{Viel}}},\ and\ \bibinfo {author} {\bibfnamefont {F.}~\bibnamefont
  {{Villaescusa-Navarro}}},\ }\bibfield  {title} {\bibinfo {title} {{Lensing is
  low: cosmology, galaxy formation or new physics?}},\ }\href
  {https://doi.org/10.1093/mnras/stx258} {\bibfield  {journal} {\bibinfo
  {journal} {\mnras}\ }\textbf {\bibinfo {volume} {467}},\ \bibinfo {pages}
  {3024} (\bibinfo {year} {2017})},\ \Eprint {https://arxiv.org/abs/1611.08606}
  {arXiv:1611.08606 [astro-ph.CO]} \BibitemShut {NoStop}%
\bibitem [{\citenamefont {{Zehavi}}\ \emph {et~al.}(2019)\citenamefont
  {{Zehavi}}, \citenamefont {{Kerby}}, \citenamefont {{Contreras}},
  \citenamefont {{Jim{\'e}nez}}, \citenamefont {{Padilla}},\ and\ \citenamefont
  {{Baugh}}}]{2019ApJ...887...17Z}%
  \BibitemOpen
  \bibfield  {author} {\bibinfo {author} {\bibfnamefont {I.}~\bibnamefont
  {{Zehavi}}}, \bibinfo {author} {\bibfnamefont {S.~E.}\ \bibnamefont
  {{Kerby}}}, \bibinfo {author} {\bibfnamefont {S.}~\bibnamefont
  {{Contreras}}}, \bibinfo {author} {\bibfnamefont {E.}~\bibnamefont
  {{Jim{\'e}nez}}}, \bibinfo {author} {\bibfnamefont {N.}~\bibnamefont
  {{Padilla}}},\ and\ \bibinfo {author} {\bibfnamefont {C.~M.}\ \bibnamefont
  {{Baugh}}},\ }\bibfield  {title} {\bibinfo {title} {{On the Prospect of Using
  the Maximum Circular Velocity of Halos to Encapsulate Assembly Bias in the
  Galaxy-Halo Connection}},\ }\href {https://doi.org/10.3847/1538-4357/ab4d4d}
  {\bibfield  {journal} {\bibinfo  {journal} {\apj}\ }\textbf {\bibinfo
  {volume} {887}},\ \bibinfo {eid} {17} (\bibinfo {year} {2019})},\ \Eprint
  {https://arxiv.org/abs/1907.05424} {arXiv:1907.05424 [astro-ph.GA]}
  \BibitemShut {NoStop}%
\bibitem [{\citenamefont {{Hadzhiyska}}\ \emph {et~al.}(2020)\citenamefont
  {{Hadzhiyska}}, \citenamefont {{Bose}}, \citenamefont {{Eisenstein}},
  \citenamefont {{Hernquist}},\ and\ \citenamefont
  {{Spergel}}}]{2020MNRAS.493.5506H}%
  \BibitemOpen
  \bibfield  {author} {\bibinfo {author} {\bibfnamefont {B.}~\bibnamefont
  {{Hadzhiyska}}}, \bibinfo {author} {\bibfnamefont {S.}~\bibnamefont
  {{Bose}}}, \bibinfo {author} {\bibfnamefont {D.}~\bibnamefont
  {{Eisenstein}}}, \bibinfo {author} {\bibfnamefont {L.}~\bibnamefont
  {{Hernquist}}},\ and\ \bibinfo {author} {\bibfnamefont {D.~N.}\ \bibnamefont
  {{Spergel}}},\ }\bibfield  {title} {\bibinfo {title} {{Limitations to the
  `basic' HOD model and beyond}},\ }\href
  {https://doi.org/10.1093/mnras/staa623} {\bibfield  {journal} {\bibinfo
  {journal} {\mnras}\ }\textbf {\bibinfo {volume} {493}},\ \bibinfo {pages}
  {5506} (\bibinfo {year} {2020})},\ \Eprint {https://arxiv.org/abs/1911.02610}
  {arXiv:1911.02610 [astro-ph.CO]} \BibitemShut {NoStop}%
\bibitem [{\citenamefont {{Contreras}}\ \emph {et~al.}(2021)\citenamefont
  {{Contreras}}, \citenamefont {{Angulo}},\ and\ \citenamefont
  {{Zennaro}}}]{2021MNRAS.508..175C}%
  \BibitemOpen
  \bibfield  {author} {\bibinfo {author} {\bibfnamefont {S.}~\bibnamefont
  {{Contreras}}}, \bibinfo {author} {\bibfnamefont {R.~E.}\ \bibnamefont
  {{Angulo}}},\ and\ \bibinfo {author} {\bibfnamefont {M.}~\bibnamefont
  {{Zennaro}}},\ }\bibfield  {title} {\bibinfo {title} {{A flexible subhalo
  abundance matching model for galaxy clustering in redshift space}},\ }\href
  {https://doi.org/10.1093/mnras/stab2560} {\bibfield  {journal} {\bibinfo
  {journal} {\mnras}\ }\textbf {\bibinfo {volume} {508}},\ \bibinfo {pages}
  {175} (\bibinfo {year} {2021})},\ \Eprint {https://arxiv.org/abs/2012.06596}
  {arXiv:2012.06596 [astro-ph.CO]} \BibitemShut {NoStop}%
\bibitem [{\citenamefont {{Wechsler}}\ and\ \citenamefont
  {{Tinker}}(2018)}]{2018ARA&A..56..435W}%
  \BibitemOpen
  \bibfield  {author} {\bibinfo {author} {\bibfnamefont {R.~H.}\ \bibnamefont
  {{Wechsler}}}\ and\ \bibinfo {author} {\bibfnamefont {J.~L.}\ \bibnamefont
  {{Tinker}}},\ }\bibfield  {title} {\bibinfo {title} {{The Connection Between
  Galaxies and Their Dark Matter Halos}},\ }\href
  {https://doi.org/10.1146/annurev-astro-081817-051756} {\bibfield  {journal}
  {\bibinfo  {journal} {\araa}\ }\textbf {\bibinfo {volume} {56}},\ \bibinfo
  {pages} {435} (\bibinfo {year} {2018})},\ \Eprint
  {https://arxiv.org/abs/1804.03097} {arXiv:1804.03097 [astro-ph.GA]}
  \BibitemShut {NoStop}%
\bibitem [{\citenamefont {{Angulo}}\ and\ \citenamefont
  {{Hahn}}(2022)}]{2022LRCA....8....1A}%
  \BibitemOpen
  \bibfield  {author} {\bibinfo {author} {\bibfnamefont {R.~E.}\ \bibnamefont
  {{Angulo}}}\ and\ \bibinfo {author} {\bibfnamefont {O.}~\bibnamefont
  {{Hahn}}},\ }\bibfield  {title} {\bibinfo {title} {{Large-scale dark matter
  simulations}},\ }\href {https://doi.org/10.1007/s41115-021-00013-z}
  {\bibfield  {journal} {\bibinfo  {journal} {Living Reviews in Computational
  Astrophysics}\ }\textbf {\bibinfo {volume} {8}},\ \bibinfo {eid} {1}
  (\bibinfo {year} {2022})},\ \Eprint {https://arxiv.org/abs/2112.05165}
  {arXiv:2112.05165 [astro-ph.CO]} \BibitemShut {NoStop}%
\bibitem [{\citenamefont {{Gao}}\ \emph {et~al.}(2005)\citenamefont {{Gao}},
  \citenamefont {{Springel}},\ and\ \citenamefont {{White}}}]{Gao2005}%
  \BibitemOpen
  \bibfield  {author} {\bibinfo {author} {\bibfnamefont {L.}~\bibnamefont
  {{Gao}}}, \bibinfo {author} {\bibfnamefont {V.}~\bibnamefont {{Springel}}},\
  and\ \bibinfo {author} {\bibfnamefont {S.~D.~M.}\ \bibnamefont {{White}}},\
  }\bibfield  {title} {\bibinfo {title} {{The age dependence of halo
  clustering}},\ }\href {https://doi.org/10.1111/j.1745-3933.2005.00084.x}
  {\bibfield  {journal} {\bibinfo  {journal} {\mnras}\ }\textbf {\bibinfo
  {volume} {363}},\ \bibinfo {pages} {L66} (\bibinfo {year} {2005})},\ \Eprint
  {https://arxiv.org/abs/astro-ph/0506510} {arXiv:astro-ph/0506510 [astro-ph]}
  \BibitemShut {NoStop}%
\bibitem [{\citenamefont {{Croton}}\ \emph {et~al.}(2007)\citenamefont
  {{Croton}}, \citenamefont {{Gao}},\ and\ \citenamefont
  {{White}}}]{2007MNRAS.374.1303C}%
  \BibitemOpen
  \bibfield  {author} {\bibinfo {author} {\bibfnamefont {D.~J.}\ \bibnamefont
  {{Croton}}}, \bibinfo {author} {\bibfnamefont {L.}~\bibnamefont {{Gao}}},\
  and\ \bibinfo {author} {\bibfnamefont {S.~D.~M.}\ \bibnamefont {{White}}},\
  }\bibfield  {title} {\bibinfo {title} {{Halo assembly bias and its effects on
  galaxy clustering}},\ }\href
  {https://doi.org/10.1111/j.1365-2966.2006.11230.x} {\bibfield  {journal}
  {\bibinfo  {journal} {\mnras}\ }\textbf {\bibinfo {volume} {374}},\ \bibinfo
  {pages} {1303} (\bibinfo {year} {2007})},\ \Eprint
  {https://arxiv.org/abs/astro-ph/0605636} {arXiv:astro-ph/0605636 [astro-ph]}
  \BibitemShut {NoStop}%
\bibitem [{\citenamefont {{Hadzhiyska}}\ \emph
  {et~al.}(2021{\natexlab{a}})\citenamefont {{Hadzhiyska}}, \citenamefont
  {{Bose}}, \citenamefont {{Eisenstein}},\ and\ \citenamefont
  {{Hernquist}}}]{2021MNRAS.501.1603H}%
  \BibitemOpen
  \bibfield  {author} {\bibinfo {author} {\bibfnamefont {B.}~\bibnamefont
  {{Hadzhiyska}}}, \bibinfo {author} {\bibfnamefont {S.}~\bibnamefont
  {{Bose}}}, \bibinfo {author} {\bibfnamefont {D.}~\bibnamefont
  {{Eisenstein}}},\ and\ \bibinfo {author} {\bibfnamefont {L.}~\bibnamefont
  {{Hernquist}}},\ }\bibfield  {title} {\bibinfo {title} {{Extensions to models
  of the galaxy-halo connection}},\ }\href
  {https://doi.org/10.1093/mnras/staa3776} {\bibfield  {journal} {\bibinfo
  {journal} {\mnras}\ }\textbf {\bibinfo {volume} {501}},\ \bibinfo {pages}
  {1603} (\bibinfo {year} {2021}{\natexlab{a}})},\ \Eprint
  {https://arxiv.org/abs/2008.04913} {arXiv:2008.04913 [astro-ph.CO]}
  \BibitemShut {NoStop}%
\bibitem [{\citenamefont {{Vogelsberger}}\ \emph
  {et~al.}(2014{\natexlab{a}})\citenamefont {{Vogelsberger}}, \citenamefont
  {{Genel}}, \citenamefont {{Springel}}, \citenamefont {{Torrey}},
  \citenamefont {{Sijacki}}, \citenamefont {{Xu}}, \citenamefont {{Snyder}},
  \citenamefont {{Nelson}},\ and\ \citenamefont
  {{Hernquist}}}]{2014MNRAS.444.1518V}%
  \BibitemOpen
  \bibfield  {author} {\bibinfo {author} {\bibfnamefont {M.}~\bibnamefont
  {{Vogelsberger}}}, \bibinfo {author} {\bibfnamefont {S.}~\bibnamefont
  {{Genel}}}, \bibinfo {author} {\bibfnamefont {V.}~\bibnamefont {{Springel}}},
  \bibinfo {author} {\bibfnamefont {P.}~\bibnamefont {{Torrey}}}, \bibinfo
  {author} {\bibfnamefont {D.}~\bibnamefont {{Sijacki}}}, \bibinfo {author}
  {\bibfnamefont {D.}~\bibnamefont {{Xu}}}, \bibinfo {author} {\bibfnamefont
  {G.}~\bibnamefont {{Snyder}}}, \bibinfo {author} {\bibfnamefont
  {D.}~\bibnamefont {{Nelson}}},\ and\ \bibinfo {author} {\bibfnamefont
  {L.}~\bibnamefont {{Hernquist}}},\ }\bibfield  {title} {\bibinfo {title}
  {{Introducing the Illustris Project: simulating the coevolution of dark and
  visible matter in the Universe}},\ }\href
  {https://doi.org/10.1093/mnras/stu1536} {\bibfield  {journal} {\bibinfo
  {journal} {\mnras}\ }\textbf {\bibinfo {volume} {444}},\ \bibinfo {pages}
  {1518} (\bibinfo {year} {2014}{\natexlab{a}})},\ \Eprint
  {https://arxiv.org/abs/1405.2921} {arXiv:1405.2921 [astro-ph.CO]}
  \BibitemShut {NoStop}%
\bibitem [{\citenamefont {{Vogelsberger}}\ \emph
  {et~al.}(2014{\natexlab{b}})\citenamefont {{Vogelsberger}}, \citenamefont
  {{Genel}}, \citenamefont {{Springel}}, \citenamefont {{Torrey}},
  \citenamefont {{Sijacki}}, \citenamefont {{Xu}}, \citenamefont {{Snyder}},
  \citenamefont {{Bird}}, \citenamefont {{Nelson}},\ and\ \citenamefont
  {{Hernquist}}}]{2014Natur.509..177V}%
  \BibitemOpen
  \bibfield  {author} {\bibinfo {author} {\bibfnamefont {M.}~\bibnamefont
  {{Vogelsberger}}}, \bibinfo {author} {\bibfnamefont {S.}~\bibnamefont
  {{Genel}}}, \bibinfo {author} {\bibfnamefont {V.}~\bibnamefont {{Springel}}},
  \bibinfo {author} {\bibfnamefont {P.}~\bibnamefont {{Torrey}}}, \bibinfo
  {author} {\bibfnamefont {D.}~\bibnamefont {{Sijacki}}}, \bibinfo {author}
  {\bibfnamefont {D.}~\bibnamefont {{Xu}}}, \bibinfo {author} {\bibfnamefont
  {G.}~\bibnamefont {{Snyder}}}, \bibinfo {author} {\bibfnamefont
  {S.}~\bibnamefont {{Bird}}}, \bibinfo {author} {\bibfnamefont
  {D.}~\bibnamefont {{Nelson}}},\ and\ \bibinfo {author} {\bibfnamefont
  {L.}~\bibnamefont {{Hernquist}}},\ }\bibfield  {title} {\bibinfo {title}
  {{Properties of galaxies reproduced by a hydrodynamic simulation}},\ }\href
  {https://doi.org/10.1038/nature13316} {\bibfield  {journal} {\bibinfo
  {journal} {\nat}\ }\textbf {\bibinfo {volume} {509}},\ \bibinfo {pages} {177}
  (\bibinfo {year} {2014}{\natexlab{b}})},\ \Eprint
  {https://arxiv.org/abs/1405.1418} {arXiv:1405.1418 [astro-ph.CO]}
  \BibitemShut {NoStop}%
\bibitem [{\citenamefont {{Genel}}\ \emph {et~al.}(2014)\citenamefont
  {{Genel}}, \citenamefont {{Vogelsberger}}, \citenamefont {{Springel}},
  \citenamefont {{Sijacki}}, \citenamefont {{Nelson}}, \citenamefont
  {{Snyder}}, \citenamefont {{Rodriguez-Gomez}}, \citenamefont {{Torrey}},\
  and\ \citenamefont {{Hernquist}}}]{2014MNRAS.445..175G}%
  \BibitemOpen
  \bibfield  {author} {\bibinfo {author} {\bibfnamefont {S.}~\bibnamefont
  {{Genel}}}, \bibinfo {author} {\bibfnamefont {M.}~\bibnamefont
  {{Vogelsberger}}}, \bibinfo {author} {\bibfnamefont {V.}~\bibnamefont
  {{Springel}}}, \bibinfo {author} {\bibfnamefont {D.}~\bibnamefont
  {{Sijacki}}}, \bibinfo {author} {\bibfnamefont {D.}~\bibnamefont {{Nelson}}},
  \bibinfo {author} {\bibfnamefont {G.}~\bibnamefont {{Snyder}}}, \bibinfo
  {author} {\bibfnamefont {V.}~\bibnamefont {{Rodriguez-Gomez}}}, \bibinfo
  {author} {\bibfnamefont {P.}~\bibnamefont {{Torrey}}},\ and\ \bibinfo
  {author} {\bibfnamefont {L.}~\bibnamefont {{Hernquist}}},\ }\bibfield
  {title} {\bibinfo {title} {{Introducing the Illustris project: the evolution
  of galaxy populations across cosmic time}},\ }\href
  {https://doi.org/10.1093/mnras/stu1654} {\bibfield  {journal} {\bibinfo
  {journal} {\mnras}\ }\textbf {\bibinfo {volume} {445}},\ \bibinfo {pages}
  {175} (\bibinfo {year} {2014})},\ \Eprint {https://arxiv.org/abs/1405.3749}
  {arXiv:1405.3749 [astro-ph.CO]} \BibitemShut {NoStop}%
\bibitem [{\citenamefont {{Schaye}}\ \emph {et~al.}(2015)\citenamefont
  {{Schaye}}, \citenamefont {{Crain}}, \citenamefont {{Bower}}, \citenamefont
  {{Furlong}}, \citenamefont {{Schaller}}, \citenamefont {{Theuns}},
  \citenamefont {{Dalla Vecchia}}, \citenamefont {{Frenk}}, \citenamefont
  {{McCarthy}}, \citenamefont {{Helly}}, \citenamefont {{Jenkins}},
  \citenamefont {{Rosas-Guevara}}, \citenamefont {{White}}, \citenamefont
  {{Baes}}, \citenamefont {{Booth}}, \citenamefont {{Camps}}, \citenamefont
  {{Navarro}}, \citenamefont {{Qu}}, \citenamefont {{Rahmati}}, \citenamefont
  {{Sawala}}, \citenamefont {{Thomas}},\ and\ \citenamefont
  {{Trayford}}}]{2015MNRAS.446..521S}%
  \BibitemOpen
  \bibfield  {author} {\bibinfo {author} {\bibfnamefont {J.}~\bibnamefont
  {{Schaye}}}, \bibinfo {author} {\bibfnamefont {R.~A.}\ \bibnamefont
  {{Crain}}}, \bibinfo {author} {\bibfnamefont {R.~G.}\ \bibnamefont
  {{Bower}}}, \bibinfo {author} {\bibfnamefont {M.}~\bibnamefont {{Furlong}}},
  \bibinfo {author} {\bibfnamefont {M.}~\bibnamefont {{Schaller}}}, \bibinfo
  {author} {\bibfnamefont {T.}~\bibnamefont {{Theuns}}}, \bibinfo {author}
  {\bibfnamefont {C.}~\bibnamefont {{Dalla Vecchia}}}, \bibinfo {author}
  {\bibfnamefont {C.~S.}\ \bibnamefont {{Frenk}}}, \bibinfo {author}
  {\bibfnamefont {I.~G.}\ \bibnamefont {{McCarthy}}}, \bibinfo {author}
  {\bibfnamefont {J.~C.}\ \bibnamefont {{Helly}}}, \bibinfo {author}
  {\bibfnamefont {A.}~\bibnamefont {{Jenkins}}}, \bibinfo {author}
  {\bibfnamefont {Y.~M.}\ \bibnamefont {{Rosas-Guevara}}}, \bibinfo {author}
  {\bibfnamefont {S.~D.~M.}\ \bibnamefont {{White}}}, \bibinfo {author}
  {\bibfnamefont {M.}~\bibnamefont {{Baes}}}, \bibinfo {author} {\bibfnamefont
  {C.~M.}\ \bibnamefont {{Booth}}}, \bibinfo {author} {\bibfnamefont
  {P.}~\bibnamefont {{Camps}}}, \bibinfo {author} {\bibfnamefont {J.~F.}\
  \bibnamefont {{Navarro}}}, \bibinfo {author} {\bibfnamefont {Y.}~\bibnamefont
  {{Qu}}}, \bibinfo {author} {\bibfnamefont {A.}~\bibnamefont {{Rahmati}}},
  \bibinfo {author} {\bibfnamefont {T.}~\bibnamefont {{Sawala}}}, \bibinfo
  {author} {\bibfnamefont {P.~A.}\ \bibnamefont {{Thomas}}},\ and\ \bibinfo
  {author} {\bibfnamefont {J.}~\bibnamefont {{Trayford}}},\ }\bibfield  {title}
  {\bibinfo {title} {{The EAGLE project: simulating the evolution and assembly
  of galaxies and their environments}},\ }\href
  {https://doi.org/10.1093/mnras/stu2058} {\bibfield  {journal} {\bibinfo
  {journal} {\mnras}\ }\textbf {\bibinfo {volume} {446}},\ \bibinfo {pages}
  {521} (\bibinfo {year} {2015})},\ \Eprint {https://arxiv.org/abs/1407.7040}
  {arXiv:1407.7040 [astro-ph.GA]} \BibitemShut {NoStop}%
\bibitem [{\citenamefont {{Nelson}}\ \emph
  {et~al.}(2019{\natexlab{a}})\citenamefont {{Nelson}}, \citenamefont
  {{Springel}}, \citenamefont {{Pillepich}}, \citenamefont {{Rodriguez-Gomez}},
  \citenamefont {{Torrey}}, \citenamefont {{Genel}}, \citenamefont
  {{Vogelsberger}}, \citenamefont {{Pakmor}}, \citenamefont {{Marinacci}},
  \citenamefont {{Weinberger}}, \citenamefont {{Kelley}}, \citenamefont
  {{Lovell}}, \citenamefont {{Diemer}},\ and\ \citenamefont
  {{Hernquist}}}]{2019ComAC...6....2N}%
  \BibitemOpen
  \bibfield  {author} {\bibinfo {author} {\bibfnamefont {D.}~\bibnamefont
  {{Nelson}}}, \bibinfo {author} {\bibfnamefont {V.}~\bibnamefont
  {{Springel}}}, \bibinfo {author} {\bibfnamefont {A.}~\bibnamefont
  {{Pillepich}}}, \bibinfo {author} {\bibfnamefont {V.}~\bibnamefont
  {{Rodriguez-Gomez}}}, \bibinfo {author} {\bibfnamefont {P.}~\bibnamefont
  {{Torrey}}}, \bibinfo {author} {\bibfnamefont {S.}~\bibnamefont {{Genel}}},
  \bibinfo {author} {\bibfnamefont {M.}~\bibnamefont {{Vogelsberger}}},
  \bibinfo {author} {\bibfnamefont {R.}~\bibnamefont {{Pakmor}}}, \bibinfo
  {author} {\bibfnamefont {F.}~\bibnamefont {{Marinacci}}}, \bibinfo {author}
  {\bibfnamefont {R.}~\bibnamefont {{Weinberger}}}, \bibinfo {author}
  {\bibfnamefont {L.}~\bibnamefont {{Kelley}}}, \bibinfo {author}
  {\bibfnamefont {M.}~\bibnamefont {{Lovell}}}, \bibinfo {author}
  {\bibfnamefont {B.}~\bibnamefont {{Diemer}}},\ and\ \bibinfo {author}
  {\bibfnamefont {L.}~\bibnamefont {{Hernquist}}},\ }\bibfield  {title}
  {\bibinfo {title} {{The IllustrisTNG simulations: public data release}},\
  }\href {https://doi.org/10.1186/s40668-019-0028-x} {\bibfield  {journal}
  {\bibinfo  {journal} {Computational Astrophysics and Cosmology}\ }\textbf
  {\bibinfo {volume} {6}},\ \bibinfo {eid} {2} (\bibinfo {year}
  {2019}{\natexlab{a}})},\ \Eprint {https://arxiv.org/abs/1812.05609}
  {arXiv:1812.05609 [astro-ph.GA]} \BibitemShut {NoStop}%
\bibitem [{\citenamefont {{Kannan}}\ \emph {et~al.}(2022)\citenamefont
  {{Kannan}}, \citenamefont {{Garaldi}}, \citenamefont {{Smith}}, \citenamefont
  {{Pakmor}}, \citenamefont {{Springel}}, \citenamefont {{Vogelsberger}},\ and\
  \citenamefont {{Hernquist}}}]{2022MNRAS.511.4005K}%
  \BibitemOpen
  \bibfield  {author} {\bibinfo {author} {\bibfnamefont {R.}~\bibnamefont
  {{Kannan}}}, \bibinfo {author} {\bibfnamefont {E.}~\bibnamefont {{Garaldi}}},
  \bibinfo {author} {\bibfnamefont {A.}~\bibnamefont {{Smith}}}, \bibinfo
  {author} {\bibfnamefont {R.}~\bibnamefont {{Pakmor}}}, \bibinfo {author}
  {\bibfnamefont {V.}~\bibnamefont {{Springel}}}, \bibinfo {author}
  {\bibfnamefont {M.}~\bibnamefont {{Vogelsberger}}},\ and\ \bibinfo {author}
  {\bibfnamefont {L.}~\bibnamefont {{Hernquist}}},\ }\bibfield  {title}
  {\bibinfo {title} {{Introducing the THESAN project:
  radiation-magnetohydrodynamic simulations of the epoch of reionization}},\
  }\href {https://doi.org/10.1093/mnras/stab3710} {\bibfield  {journal}
  {\bibinfo  {journal} {\mnras}\ }\textbf {\bibinfo {volume} {511}},\ \bibinfo
  {pages} {4005} (\bibinfo {year} {2022})},\ \Eprint
  {https://arxiv.org/abs/2110.00584} {arXiv:2110.00584 [astro-ph.GA]}
  \BibitemShut {NoStop}%
\bibitem [{\citenamefont {{Shapley}}\ \emph {et~al.}(2003)\citenamefont
  {{Shapley}}, \citenamefont {{Steidel}}, \citenamefont {{Pettini}},\ and\
  \citenamefont {{Adelberger}}}]{2003ApJ...588...65S}%
  \BibitemOpen
  \bibfield  {author} {\bibinfo {author} {\bibfnamefont {A.~E.}\ \bibnamefont
  {{Shapley}}}, \bibinfo {author} {\bibfnamefont {C.~C.}\ \bibnamefont
  {{Steidel}}}, \bibinfo {author} {\bibfnamefont {M.}~\bibnamefont
  {{Pettini}}},\ and\ \bibinfo {author} {\bibfnamefont {K.~L.}\ \bibnamefont
  {{Adelberger}}},\ }\bibfield  {title} {\bibinfo {title} {{Rest-Frame
  Ultraviolet Spectra of z\raisebox{-0.5ex}\textasciitilde3 Lyman Break
  Galaxies}},\ }\href {https://doi.org/10.1086/373922} {\bibfield  {journal}
  {\bibinfo  {journal} {\apj}\ }\textbf {\bibinfo {volume} {588}},\ \bibinfo
  {pages} {65} (\bibinfo {year} {2003})},\ \Eprint
  {https://arxiv.org/abs/astro-ph/0301230} {arXiv:astro-ph/0301230 [astro-ph]}
  \BibitemShut {NoStop}%
\bibitem [{\citenamefont {{Ouchi}}\ \emph {et~al.}(2005)\citenamefont
  {{Ouchi}}, \citenamefont {{Shimasaku}}, \citenamefont {{Akiyama}},
  \citenamefont {{Sekiguchi}}, \citenamefont {{Furusawa}}, \citenamefont
  {{Okamura}}, \citenamefont {{Kashikawa}}, \citenamefont {{Iye}},
  \citenamefont {{Kodama}}, \citenamefont {{Saito}}, \citenamefont {{Sasaki}},
  \citenamefont {{Simpson}}, \citenamefont {{Takata}}, \citenamefont
  {{Yamada}}, \citenamefont {{Yamanoi}}, \citenamefont {{Yoshida}},\ and\
  \citenamefont {{Yoshida}}}]{2005ApJ...620L...1O}%
  \BibitemOpen
  \bibfield  {author} {\bibinfo {author} {\bibfnamefont {M.}~\bibnamefont
  {{Ouchi}}}, \bibinfo {author} {\bibfnamefont {K.}~\bibnamefont
  {{Shimasaku}}}, \bibinfo {author} {\bibfnamefont {M.}~\bibnamefont
  {{Akiyama}}}, \bibinfo {author} {\bibfnamefont {K.}~\bibnamefont
  {{Sekiguchi}}}, \bibinfo {author} {\bibfnamefont {H.}~\bibnamefont
  {{Furusawa}}}, \bibinfo {author} {\bibfnamefont {S.}~\bibnamefont
  {{Okamura}}}, \bibinfo {author} {\bibfnamefont {N.}~\bibnamefont
  {{Kashikawa}}}, \bibinfo {author} {\bibfnamefont {M.}~\bibnamefont {{Iye}}},
  \bibinfo {author} {\bibfnamefont {T.}~\bibnamefont {{Kodama}}}, \bibinfo
  {author} {\bibfnamefont {T.}~\bibnamefont {{Saito}}}, \bibinfo {author}
  {\bibfnamefont {T.}~\bibnamefont {{Sasaki}}}, \bibinfo {author}
  {\bibfnamefont {C.}~\bibnamefont {{Simpson}}}, \bibinfo {author}
  {\bibfnamefont {T.}~\bibnamefont {{Takata}}}, \bibinfo {author}
  {\bibfnamefont {T.}~\bibnamefont {{Yamada}}}, \bibinfo {author}
  {\bibfnamefont {H.}~\bibnamefont {{Yamanoi}}}, \bibinfo {author}
  {\bibfnamefont {M.}~\bibnamefont {{Yoshida}}},\ and\ \bibinfo {author}
  {\bibfnamefont {M.}~\bibnamefont {{Yoshida}}},\ }\bibfield  {title} {\bibinfo
  {title} {{The Discovery of Primeval Large-Scale Structures with Forming
  Clusters at Redshift 6}},\ }\href {https://doi.org/10.1086/428499} {\bibfield
   {journal} {\bibinfo  {journal} {\apjl}\ }\textbf {\bibinfo {volume} {620}},\
  \bibinfo {pages} {L1} (\bibinfo {year} {2005})},\ \Eprint
  {https://arxiv.org/abs/astro-ph/0412648} {arXiv:astro-ph/0412648 [astro-ph]}
  \BibitemShut {NoStop}%
\bibitem [{\citenamefont {{Hu}}\ and\ \citenamefont
  {{Cowie}}(2006)}]{2006Natur.440.1145H}%
  \BibitemOpen
  \bibfield  {author} {\bibinfo {author} {\bibfnamefont {E.~M.}\ \bibnamefont
  {{Hu}}}\ and\ \bibinfo {author} {\bibfnamefont {L.~L.}\ \bibnamefont
  {{Cowie}}},\ }\bibfield  {title} {\bibinfo {title} {{High-redshift galaxy
  populations}},\ }\href {https://doi.org/10.1038/nature04806} {\bibfield
  {journal} {\bibinfo  {journal} {\nat}\ }\textbf {\bibinfo {volume} {440}},\
  \bibinfo {pages} {1145} (\bibinfo {year} {2006})}\BibitemShut {NoStop}%
\bibitem [{\citenamefont {{Hern{\'a}ndez-Aguayo}}\ \emph
  {et~al.}(2023)\citenamefont {{Hern{\'a}ndez-Aguayo}}, \citenamefont
  {{Springel}}, \citenamefont {{Pakmor}}, \citenamefont {{Barrera}},
  \citenamefont {{Ferlito}}, \citenamefont {{White}}, \citenamefont
  {{Hernquist}}, \citenamefont {{Hadzhiyska}}, \citenamefont {{Delgado}},
  \citenamefont {{Kannan}}, \citenamefont {{Bose}},\ and\ \citenamefont
  {{Frenk}}}]{Aguayo2022}%
  \BibitemOpen
  \bibfield  {author} {\bibinfo {author} {\bibfnamefont {C.}~\bibnamefont
  {{Hern{\'a}ndez-Aguayo}}}, \bibinfo {author} {\bibfnamefont {V.}~\bibnamefont
  {{Springel}}}, \bibinfo {author} {\bibfnamefont {R.}~\bibnamefont
  {{Pakmor}}}, \bibinfo {author} {\bibfnamefont {M.}~\bibnamefont {{Barrera}}},
  \bibinfo {author} {\bibfnamefont {F.}~\bibnamefont {{Ferlito}}}, \bibinfo
  {author} {\bibfnamefont {S.~D.~M.}\ \bibnamefont {{White}}}, \bibinfo
  {author} {\bibfnamefont {L.}~\bibnamefont {{Hernquist}}}, \bibinfo {author}
  {\bibfnamefont {B.}~\bibnamefont {{Hadzhiyska}}}, \bibinfo {author}
  {\bibfnamefont {A.~M.}\ \bibnamefont {{Delgado}}}, \bibinfo {author}
  {\bibfnamefont {R.}~\bibnamefont {{Kannan}}}, \bibinfo {author}
  {\bibfnamefont {S.}~\bibnamefont {{Bose}}},\ and\ \bibinfo {author}
  {\bibfnamefont {C.}~\bibnamefont {{Frenk}}},\ }\bibfield  {title} {\bibinfo
  {title} {{The MillenniumTNG Project: high-precision predictions for matter
  clustering and halo statistics}},\ }\href
  {https://doi.org/10.1093/mnras/stad1657} {\bibfield  {journal} {\bibinfo
  {journal} {\mnras}\ }\textbf {\bibinfo {volume} {524}},\ \bibinfo {pages}
  {2556} (\bibinfo {year} {2023})},\ \Eprint {https://arxiv.org/abs/2210.10059}
  {arXiv:2210.10059 [astro-ph.CO]} \BibitemShut {NoStop}%
\bibitem [{\citenamefont {{Pakmor}}\ \emph {et~al.}(2023)\citenamefont
  {{Pakmor}}, \citenamefont {{Springel}}, \citenamefont {{Coles}},
  \citenamefont {{Guillet}}, \citenamefont {{Pfrommer}}, \citenamefont
  {{Bose}}, \citenamefont {{Barrera}}, \citenamefont {{Delgado}}, \citenamefont
  {{Ferlito}}, \citenamefont {{Frenk}}, \citenamefont {{Hadzhiyska}},
  \citenamefont {{Hern{\'a}ndez-Aguayo}}, \citenamefont {{Hernquist}},
  \citenamefont {{Kannan}},\ and\ \citenamefont {{White}}}]{Pakmor2022}%
  \BibitemOpen
  \bibfield  {author} {\bibinfo {author} {\bibfnamefont {R.}~\bibnamefont
  {{Pakmor}}}, \bibinfo {author} {\bibfnamefont {V.}~\bibnamefont
  {{Springel}}}, \bibinfo {author} {\bibfnamefont {J.~P.}\ \bibnamefont
  {{Coles}}}, \bibinfo {author} {\bibfnamefont {T.}~\bibnamefont {{Guillet}}},
  \bibinfo {author} {\bibfnamefont {C.}~\bibnamefont {{Pfrommer}}}, \bibinfo
  {author} {\bibfnamefont {S.}~\bibnamefont {{Bose}}}, \bibinfo {author}
  {\bibfnamefont {M.}~\bibnamefont {{Barrera}}}, \bibinfo {author}
  {\bibfnamefont {A.~M.}\ \bibnamefont {{Delgado}}}, \bibinfo {author}
  {\bibfnamefont {F.}~\bibnamefont {{Ferlito}}}, \bibinfo {author}
  {\bibfnamefont {C.}~\bibnamefont {{Frenk}}}, \bibinfo {author} {\bibfnamefont
  {B.}~\bibnamefont {{Hadzhiyska}}}, \bibinfo {author} {\bibfnamefont
  {C.}~\bibnamefont {{Hern{\'a}ndez-Aguayo}}}, \bibinfo {author} {\bibfnamefont
  {L.}~\bibnamefont {{Hernquist}}}, \bibinfo {author} {\bibfnamefont
  {R.}~\bibnamefont {{Kannan}}},\ and\ \bibinfo {author} {\bibfnamefont
  {S.~D.~M.}\ \bibnamefont {{White}}},\ }\bibfield  {title} {\bibinfo {title}
  {{The MillenniumTNG Project: the hydrodynamical full physics simulation and a
  first look at its galaxy clusters}},\ }\href
  {https://doi.org/10.1093/mnras/stac3620} {\bibfield  {journal} {\bibinfo
  {journal} {\mnras}\ }\textbf {\bibinfo {volume} {524}},\ \bibinfo {pages}
  {2539} (\bibinfo {year} {2023})},\ \Eprint {https://arxiv.org/abs/2210.10060}
  {arXiv:2210.10060 [astro-ph.CO]} \BibitemShut {NoStop}%
\bibitem [{\citenamefont {{Kannan}}\ \emph {et~al.}(2023)\citenamefont
  {{Kannan}}, \citenamefont {{Springel}}, \citenamefont {{Hernquist}},
  \citenamefont {{Pakmor}}, \citenamefont {{Delgado}}, \citenamefont
  {{Hadzhiyska}}, \citenamefont {{Hern{\'a}ndez-Aguayo}}, \citenamefont
  {{Barrera}}, \citenamefont {{Ferlito}}, \citenamefont {{Bose}}, \citenamefont
  {{White}}, \citenamefont {{Frenk}}, \citenamefont {{Smith}},\ and\
  \citenamefont {{Garaldi}}}]{Kannan2022}%
  \BibitemOpen
  \bibfield  {author} {\bibinfo {author} {\bibfnamefont {R.}~\bibnamefont
  {{Kannan}}}, \bibinfo {author} {\bibfnamefont {V.}~\bibnamefont
  {{Springel}}}, \bibinfo {author} {\bibfnamefont {L.}~\bibnamefont
  {{Hernquist}}}, \bibinfo {author} {\bibfnamefont {R.}~\bibnamefont
  {{Pakmor}}}, \bibinfo {author} {\bibfnamefont {A.~M.}\ \bibnamefont
  {{Delgado}}}, \bibinfo {author} {\bibfnamefont {B.}~\bibnamefont
  {{Hadzhiyska}}}, \bibinfo {author} {\bibfnamefont {C.}~\bibnamefont
  {{Hern{\'a}ndez-Aguayo}}}, \bibinfo {author} {\bibfnamefont {M.}~\bibnamefont
  {{Barrera}}}, \bibinfo {author} {\bibfnamefont {F.}~\bibnamefont
  {{Ferlito}}}, \bibinfo {author} {\bibfnamefont {S.}~\bibnamefont {{Bose}}},
  \bibinfo {author} {\bibfnamefont {S.~D.~M.}\ \bibnamefont {{White}}},
  \bibinfo {author} {\bibfnamefont {C.}~\bibnamefont {{Frenk}}}, \bibinfo
  {author} {\bibfnamefont {A.}~\bibnamefont {{Smith}}},\ and\ \bibinfo {author}
  {\bibfnamefont {E.}~\bibnamefont {{Garaldi}}},\ }\bibfield  {title} {\bibinfo
  {title} {{The MillenniumTNG project: the galaxy population at z
  {\ensuremath{\geq}} 8}},\ }\href {https://doi.org/10.1093/mnras/stac3743}
  {\bibfield  {journal} {\bibinfo  {journal} {\mnras}\ }\textbf {\bibinfo
  {volume} {524}},\ \bibinfo {pages} {2594} (\bibinfo {year} {2023})},\ \Eprint
  {https://arxiv.org/abs/2210.10066} {arXiv:2210.10066 [astro-ph.GA]}
  \BibitemShut {NoStop}%
\bibitem [{\citenamefont {{Ferlito}}\ \emph {et~al.}(2023)\citenamefont
  {{Ferlito}}, \citenamefont {{Springel}}, \citenamefont {{Davies}},
  \citenamefont {{Hern{\'a}ndez-Aguayo}}, \citenamefont {{Pakmor}},
  \citenamefont {{Barrera}}, \citenamefont {{White}}, \citenamefont
  {{Delgado}}, \citenamefont {{Hadzhiyska}}, \citenamefont {{Hernquist}},
  \citenamefont {{Kannan}}, \citenamefont {{Bose}},\ and\ \citenamefont
  {{Frenk}}}]{Ferlito2022}%
  \BibitemOpen
  \bibfield  {author} {\bibinfo {author} {\bibfnamefont {F.}~\bibnamefont
  {{Ferlito}}}, \bibinfo {author} {\bibfnamefont {V.}~\bibnamefont
  {{Springel}}}, \bibinfo {author} {\bibfnamefont {C.~T.}\ \bibnamefont
  {{Davies}}}, \bibinfo {author} {\bibfnamefont {C.}~\bibnamefont
  {{Hern{\'a}ndez-Aguayo}}}, \bibinfo {author} {\bibfnamefont {R.}~\bibnamefont
  {{Pakmor}}}, \bibinfo {author} {\bibfnamefont {M.}~\bibnamefont {{Barrera}}},
  \bibinfo {author} {\bibfnamefont {S.~D.~M.}\ \bibnamefont {{White}}},
  \bibinfo {author} {\bibfnamefont {A.~M.}\ \bibnamefont {{Delgado}}}, \bibinfo
  {author} {\bibfnamefont {B.}~\bibnamefont {{Hadzhiyska}}}, \bibinfo {author}
  {\bibfnamefont {L.}~\bibnamefont {{Hernquist}}}, \bibinfo {author}
  {\bibfnamefont {R.}~\bibnamefont {{Kannan}}}, \bibinfo {author}
  {\bibfnamefont {S.}~\bibnamefont {{Bose}}},\ and\ \bibinfo {author}
  {\bibfnamefont {C.}~\bibnamefont {{Frenk}}},\ }\bibfield  {title} {\bibinfo
  {title} {{The MillenniumTNG Project: the impact of baryons and massive
  neutrinos on high-resolution weak gravitational lensing convergence maps}},\
  }\href {https://doi.org/10.1093/mnras/stad2205} {\bibfield  {journal}
  {\bibinfo  {journal} {\mnras}\ }\textbf {\bibinfo {volume} {524}},\ \bibinfo
  {pages} {5591} (\bibinfo {year} {2023})},\ \Eprint
  {https://arxiv.org/abs/2304.12338} {arXiv:2304.12338 [astro-ph.CO]}
  \BibitemShut {NoStop}%
\bibitem [{\citenamefont {{Delgado}}\ \emph {et~al.}(2023)\citenamefont
  {{Delgado}}, \citenamefont {{Angl{\'e}s-Alc{\'a}zar}}, \citenamefont
  {{Thiele}}, \citenamefont {{Pandey}}, \citenamefont {{Lehman}}, \citenamefont
  {{Somerville}}, \citenamefont {{Ntampaka}}, \citenamefont {{Genel}},
  \citenamefont {{Villaescusa-Navarro}},\ and\ \citenamefont
  {{Hernquist}}}]{Delgado2022}%
  \BibitemOpen
  \bibfield  {author} {\bibinfo {author} {\bibfnamefont {A.~M.}\ \bibnamefont
  {{Delgado}}}, \bibinfo {author} {\bibfnamefont {D.}~\bibnamefont
  {{Angl{\'e}s-Alc{\'a}zar}}}, \bibinfo {author} {\bibfnamefont
  {L.}~\bibnamefont {{Thiele}}}, \bibinfo {author} {\bibfnamefont
  {S.}~\bibnamefont {{Pandey}}}, \bibinfo {author} {\bibfnamefont
  {K.}~\bibnamefont {{Lehman}}}, \bibinfo {author} {\bibfnamefont {R.~S.}\
  \bibnamefont {{Somerville}}}, \bibinfo {author} {\bibfnamefont
  {M.}~\bibnamefont {{Ntampaka}}}, \bibinfo {author} {\bibfnamefont
  {S.}~\bibnamefont {{Genel}}}, \bibinfo {author} {\bibfnamefont
  {F.}~\bibnamefont {{Villaescusa-Navarro}}},\ and\ \bibinfo {author}
  {\bibfnamefont {L.}~\bibnamefont {{Hernquist}}},\ }\bibfield  {title}
  {\bibinfo {title} {{Predicting the impact of feedback on matter clustering
  with machine learning in CAMELS}},\ }\bibfield  {journal} {\bibinfo
  {journal} {\mnras}\ }\href {https://doi.org/10.1093/mnras/stad2992}
  {10.1093/mnras/stad2992} (\bibinfo {year} {2023}),\ \Eprint
  {https://arxiv.org/abs/2301.02231} {arXiv:2301.02231 [astro-ph.GA]}
  \BibitemShut {NoStop}%
\bibitem [{\citenamefont {{Bose}}\ \emph {et~al.}(2023)\citenamefont {{Bose}},
  \citenamefont {{Hadzhiyska}}, \citenamefont {{Barrera}}, \citenamefont
  {{Delgado}}, \citenamefont {{Ferlito}}, \citenamefont {{Frenk}},
  \citenamefont {{Hern{\'a}ndez-Aguayo}}, \citenamefont {{Hernquist}},
  \citenamefont {{Kannan}}, \citenamefont {{Pakmor}}, \citenamefont
  {{Springel}},\ and\ \citenamefont {{White}}}]{Bose2022}%
  \BibitemOpen
  \bibfield  {author} {\bibinfo {author} {\bibfnamefont {S.}~\bibnamefont
  {{Bose}}}, \bibinfo {author} {\bibfnamefont {B.}~\bibnamefont
  {{Hadzhiyska}}}, \bibinfo {author} {\bibfnamefont {M.}~\bibnamefont
  {{Barrera}}}, \bibinfo {author} {\bibfnamefont {A.~M.}\ \bibnamefont
  {{Delgado}}}, \bibinfo {author} {\bibfnamefont {F.}~\bibnamefont
  {{Ferlito}}}, \bibinfo {author} {\bibfnamefont {C.}~\bibnamefont {{Frenk}}},
  \bibinfo {author} {\bibfnamefont {C.}~\bibnamefont {{Hern{\'a}ndez-Aguayo}}},
  \bibinfo {author} {\bibfnamefont {L.}~\bibnamefont {{Hernquist}}}, \bibinfo
  {author} {\bibfnamefont {R.}~\bibnamefont {{Kannan}}}, \bibinfo {author}
  {\bibfnamefont {R.}~\bibnamefont {{Pakmor}}}, \bibinfo {author}
  {\bibfnamefont {V.}~\bibnamefont {{Springel}}},\ and\ \bibinfo {author}
  {\bibfnamefont {S.~D.~M.}\ \bibnamefont {{White}}},\ }\bibfield  {title}
  {\bibinfo {title} {{The MillenniumTNG Project: the large-scale clustering of
  galaxies}},\ }\href {https://doi.org/10.1093/mnras/stad1097} {\bibfield
  {journal} {\bibinfo  {journal} {\mnras}\ }\textbf {\bibinfo {volume} {524}},\
  \bibinfo {pages} {2579} (\bibinfo {year} {2023})},\ \Eprint
  {https://arxiv.org/abs/2210.10065} {arXiv:2210.10065 [astro-ph.CO]}
  \BibitemShut {NoStop}%
\bibitem [{\citenamefont {{Contreras}}\ \emph {et~al.}(2023)\citenamefont
  {{Contreras}}, \citenamefont {{Angulo}}, \citenamefont {{Springel}},
  \citenamefont {{White}}, \citenamefont {{Hadzhiyska}}, \citenamefont
  {{Hernquist}}, \citenamefont {{Pakmor}}, \citenamefont {{Kannan}},
  \citenamefont {{Hern{\'a}ndez-Aguayo}}, \citenamefont {{Barrera}},
  \citenamefont {{Ferlito}}, \citenamefont {{Delgado}}, \citenamefont
  {{Bose}},\ and\ \citenamefont {{Frenk}}}]{Contreras2022}%
  \BibitemOpen
  \bibfield  {author} {\bibinfo {author} {\bibfnamefont {S.}~\bibnamefont
  {{Contreras}}}, \bibinfo {author} {\bibfnamefont {R.~E.}\ \bibnamefont
  {{Angulo}}}, \bibinfo {author} {\bibfnamefont {V.}~\bibnamefont
  {{Springel}}}, \bibinfo {author} {\bibfnamefont {S.~D.~M.}\ \bibnamefont
  {{White}}}, \bibinfo {author} {\bibfnamefont {B.}~\bibnamefont
  {{Hadzhiyska}}}, \bibinfo {author} {\bibfnamefont {L.}~\bibnamefont
  {{Hernquist}}}, \bibinfo {author} {\bibfnamefont {R.}~\bibnamefont
  {{Pakmor}}}, \bibinfo {author} {\bibfnamefont {R.}~\bibnamefont {{Kannan}}},
  \bibinfo {author} {\bibfnamefont {C.}~\bibnamefont {{Hern{\'a}ndez-Aguayo}}},
  \bibinfo {author} {\bibfnamefont {M.}~\bibnamefont {{Barrera}}}, \bibinfo
  {author} {\bibfnamefont {F.}~\bibnamefont {{Ferlito}}}, \bibinfo {author}
  {\bibfnamefont {A.~M.}\ \bibnamefont {{Delgado}}}, \bibinfo {author}
  {\bibfnamefont {S.}~\bibnamefont {{Bose}}},\ and\ \bibinfo {author}
  {\bibfnamefont {C.}~\bibnamefont {{Frenk}}},\ }\bibfield  {title} {\bibinfo
  {title} {{The MillenniumTNG Project: inferring cosmology from galaxy
  clustering with accelerated N-body scaling and subhalo abundance matching}},\
  }\href {https://doi.org/10.1093/mnras/stac3699} {\bibfield  {journal}
  {\bibinfo  {journal} {\mnras}\ }\textbf {\bibinfo {volume} {524}},\ \bibinfo
  {pages} {2489} (\bibinfo {year} {2023})},\ \Eprint
  {https://arxiv.org/abs/2210.10075} {arXiv:2210.10075 [astro-ph.GA]}
  \BibitemShut {NoStop}%
\bibitem [{\citenamefont {{Hadzhiyska}}\ \emph
  {et~al.}(2023{\natexlab{a}})\citenamefont {{Hadzhiyska}}, \citenamefont
  {{Eisenstein}}, \citenamefont {{Hernquist}}, \citenamefont {{Pakmor}},
  \citenamefont {{Bose}}, \citenamefont {{Delgado}}, \citenamefont
  {{Contreras}}, \citenamefont {{Kannan}}, \citenamefont {{White}},
  \citenamefont {{Springel}}, \citenamefont {{Frenk}}, \citenamefont
  {{Hern{\'a}ndez-Aguayo}}, \citenamefont {{Barrera}},\ and\ \citenamefont
  {{Monica}}}]{2023MNRAS.524.2507H}%
  \BibitemOpen
  \bibfield  {author} {\bibinfo {author} {\bibfnamefont {B.}~\bibnamefont
  {{Hadzhiyska}}}, \bibinfo {author} {\bibfnamefont {D.}~\bibnamefont
  {{Eisenstein}}}, \bibinfo {author} {\bibfnamefont {L.}~\bibnamefont
  {{Hernquist}}}, \bibinfo {author} {\bibfnamefont {R.}~\bibnamefont
  {{Pakmor}}}, \bibinfo {author} {\bibfnamefont {S.}~\bibnamefont {{Bose}}},
  \bibinfo {author} {\bibfnamefont {A.~M.}\ \bibnamefont {{Delgado}}}, \bibinfo
  {author} {\bibfnamefont {S.}~\bibnamefont {{Contreras}}}, \bibinfo {author}
  {\bibfnamefont {R.}~\bibnamefont {{Kannan}}}, \bibinfo {author}
  {\bibfnamefont {S.~D.~M.}\ \bibnamefont {{White}}}, \bibinfo {author}
  {\bibfnamefont {V.}~\bibnamefont {{Springel}}}, \bibinfo {author}
  {\bibfnamefont {C.}~\bibnamefont {{Frenk}}}, \bibinfo {author} {\bibfnamefont
  {C.}~\bibnamefont {{Hern{\'a}ndez-Aguayo}}}, \bibinfo {author} {\bibfnamefont
  {F.~F.}\ \bibnamefont {{Barrera}}},\ and\ \bibinfo {author} {\bibnamefont
  {{Monica}}},\ }\bibfield  {title} {\bibinfo {title} {{The MillenniumTNG
  Project: an improved two-halo model for the galaxy-halo connection of red and
  blue galaxies}},\ }\href {https://doi.org/10.1093/mnras/stad731} {\bibfield
  {journal} {\bibinfo  {journal} {\mnras}\ }\textbf {\bibinfo {volume} {524}},\
  \bibinfo {pages} {2507} (\bibinfo {year} {2023}{\natexlab{a}})},\ \Eprint
  {https://arxiv.org/abs/2210.10072} {arXiv:2210.10072 [astro-ph.CO]}
  \BibitemShut {NoStop}%
\bibitem [{\citenamefont {{Hadzhiyska}}\ \emph
  {et~al.}(2023{\natexlab{b}})\citenamefont {{Hadzhiyska}}, \citenamefont
  {{Hernquist}}, \citenamefont {{Eisenstein}}, \citenamefont {{Delgado}},
  \citenamefont {{Bose}}, \citenamefont {{Kannan}}, \citenamefont {{Pakmor}},
  \citenamefont {{Springel}}, \citenamefont {{Contreras}}, \citenamefont
  {{Barrera}}, \citenamefont {{Ferlito}}, \citenamefont
  {{Hern{\'a}ndez-Aguayo}}, \citenamefont {{White}},\ and\ \citenamefont
  {{Frenk}}}]{2023MNRAS.524.2524H}%
  \BibitemOpen
  \bibfield  {author} {\bibinfo {author} {\bibfnamefont {B.}~\bibnamefont
  {{Hadzhiyska}}}, \bibinfo {author} {\bibfnamefont {L.}~\bibnamefont
  {{Hernquist}}}, \bibinfo {author} {\bibfnamefont {D.}~\bibnamefont
  {{Eisenstein}}}, \bibinfo {author} {\bibfnamefont {A.~M.}\ \bibnamefont
  {{Delgado}}}, \bibinfo {author} {\bibfnamefont {S.}~\bibnamefont {{Bose}}},
  \bibinfo {author} {\bibfnamefont {R.}~\bibnamefont {{Kannan}}}, \bibinfo
  {author} {\bibfnamefont {R.}~\bibnamefont {{Pakmor}}}, \bibinfo {author}
  {\bibfnamefont {V.}~\bibnamefont {{Springel}}}, \bibinfo {author}
  {\bibfnamefont {S.}~\bibnamefont {{Contreras}}}, \bibinfo {author}
  {\bibfnamefont {M.}~\bibnamefont {{Barrera}}}, \bibinfo {author}
  {\bibfnamefont {F.}~\bibnamefont {{Ferlito}}}, \bibinfo {author}
  {\bibfnamefont {C.}~\bibnamefont {{Hern{\'a}ndez-Aguayo}}}, \bibinfo {author}
  {\bibfnamefont {S.~D.~M.}\ \bibnamefont {{White}}},\ and\ \bibinfo {author}
  {\bibfnamefont {C.}~\bibnamefont {{Frenk}}},\ }\bibfield  {title} {\bibinfo
  {title} {{The MillenniumTNG Project: refining the one-halo model of red and
  blue galaxies at different redshifts}},\ }\href
  {https://doi.org/10.1093/mnras/stad279} {\bibfield  {journal} {\bibinfo
  {journal} {\mnras}\ }\textbf {\bibinfo {volume} {524}},\ \bibinfo {pages}
  {2524} (\bibinfo {year} {2023}{\natexlab{b}})},\ \Eprint
  {https://arxiv.org/abs/2210.10068} {arXiv:2210.10068 [astro-ph.CO]}
  \BibitemShut {NoStop}%
\bibitem [{\citenamefont {{Barrera}}\ \emph {et~al.}(2023)\citenamefont
  {{Barrera}}, \citenamefont {{Springel}}, \citenamefont {{White}},
  \citenamefont {{Hern{\'a}ndez-Aguayo}}, \citenamefont {{Hernquist}},
  \citenamefont {{Frenk}}, \citenamefont {{Pakmor}}, \citenamefont {{Ferlito}},
  \citenamefont {{Hadzhiyska}}, \citenamefont {{Delgado}}, \citenamefont
  {{Kannan}},\ and\ \citenamefont {{Bose}}}]{Barrera2022}%
  \BibitemOpen
  \bibfield  {author} {\bibinfo {author} {\bibfnamefont {M.}~\bibnamefont
  {{Barrera}}}, \bibinfo {author} {\bibfnamefont {V.}~\bibnamefont
  {{Springel}}}, \bibinfo {author} {\bibfnamefont {S.~D.~M.}\ \bibnamefont
  {{White}}}, \bibinfo {author} {\bibfnamefont {C.}~\bibnamefont
  {{Hern{\'a}ndez-Aguayo}}}, \bibinfo {author} {\bibfnamefont {L.}~\bibnamefont
  {{Hernquist}}}, \bibinfo {author} {\bibfnamefont {C.}~\bibnamefont
  {{Frenk}}}, \bibinfo {author} {\bibfnamefont {R.}~\bibnamefont {{Pakmor}}},
  \bibinfo {author} {\bibfnamefont {F.}~\bibnamefont {{Ferlito}}}, \bibinfo
  {author} {\bibfnamefont {B.}~\bibnamefont {{Hadzhiyska}}}, \bibinfo {author}
  {\bibfnamefont {A.~M.}\ \bibnamefont {{Delgado}}}, \bibinfo {author}
  {\bibfnamefont {R.}~\bibnamefont {{Kannan}}},\ and\ \bibinfo {author}
  {\bibfnamefont {S.}~\bibnamefont {{Bose}}},\ }\bibfield  {title} {\bibinfo
  {title} {{The MillenniumTNG Project: semi-analytic galaxy formation models on
  the past lightcone}},\ }\href {https://doi.org/10.1093/mnras/stad2688}
  {\bibfield  {journal} {\bibinfo  {journal} {\mnras}\ }\textbf {\bibinfo
  {volume} {525}},\ \bibinfo {pages} {6312} (\bibinfo {year} {2023})},\ \Eprint
  {https://arxiv.org/abs/2210.10419} {arXiv:2210.10419 [astro-ph.CO]}
  \BibitemShut {NoStop}%
\bibitem [{\citenamefont {{Weinberger}}\ \emph {et~al.}(2017)\citenamefont
  {{Weinberger}}, \citenamefont {{Springel}}, \citenamefont {{Hernquist}},
  \citenamefont {{Pillepich}}, \citenamefont {{Marinacci}}, \citenamefont
  {{Pakmor}}, \citenamefont {{Nelson}}, \citenamefont {{Genel}}, \citenamefont
  {{Vogelsberger}}, \citenamefont {{Naiman}},\ and\ \citenamefont
  {{Torrey}}}]{2017MNRAS.465.3291W}%
  \BibitemOpen
  \bibfield  {author} {\bibinfo {author} {\bibfnamefont {R.}~\bibnamefont
  {{Weinberger}}}, \bibinfo {author} {\bibfnamefont {V.}~\bibnamefont
  {{Springel}}}, \bibinfo {author} {\bibfnamefont {L.}~\bibnamefont
  {{Hernquist}}}, \bibinfo {author} {\bibfnamefont {A.}~\bibnamefont
  {{Pillepich}}}, \bibinfo {author} {\bibfnamefont {F.}~\bibnamefont
  {{Marinacci}}}, \bibinfo {author} {\bibfnamefont {R.}~\bibnamefont
  {{Pakmor}}}, \bibinfo {author} {\bibfnamefont {D.}~\bibnamefont {{Nelson}}},
  \bibinfo {author} {\bibfnamefont {S.}~\bibnamefont {{Genel}}}, \bibinfo
  {author} {\bibfnamefont {M.}~\bibnamefont {{Vogelsberger}}}, \bibinfo
  {author} {\bibfnamefont {J.}~\bibnamefont {{Naiman}}},\ and\ \bibinfo
  {author} {\bibfnamefont {P.}~\bibnamefont {{Torrey}}},\ }\bibfield  {title}
  {\bibinfo {title} {{Simulating galaxy formation with black hole driven
  thermal and kinetic feedback}},\ }\href
  {https://doi.org/10.1093/mnras/stw2944} {\bibfield  {journal} {\bibinfo
  {journal} {\mnras}\ }\textbf {\bibinfo {volume} {465}},\ \bibinfo {pages}
  {3291} (\bibinfo {year} {2017})},\ \Eprint {https://arxiv.org/abs/1607.03486}
  {arXiv:1607.03486 [astro-ph.GA]} \BibitemShut {NoStop}%
\bibitem [{\citenamefont {{Pillepich}}\ \emph
  {et~al.}(2018{\natexlab{a}})\citenamefont {{Pillepich}}, \citenamefont
  {{Springel}}, \citenamefont {{Nelson}}, \citenamefont {{Genel}},
  \citenamefont {{Naiman}}, \citenamefont {{Pakmor}}, \citenamefont
  {{Hernquist}}, \citenamefont {{Torrey}}, \citenamefont {{Vogelsberger}},
  \citenamefont {{Weinberger}},\ and\ \citenamefont
  {{Marinacci}}}]{2018MNRAS.473.4077P}%
  \BibitemOpen
  \bibfield  {author} {\bibinfo {author} {\bibfnamefont {A.}~\bibnamefont
  {{Pillepich}}}, \bibinfo {author} {\bibfnamefont {V.}~\bibnamefont
  {{Springel}}}, \bibinfo {author} {\bibfnamefont {D.}~\bibnamefont
  {{Nelson}}}, \bibinfo {author} {\bibfnamefont {S.}~\bibnamefont {{Genel}}},
  \bibinfo {author} {\bibfnamefont {J.}~\bibnamefont {{Naiman}}}, \bibinfo
  {author} {\bibfnamefont {R.}~\bibnamefont {{Pakmor}}}, \bibinfo {author}
  {\bibfnamefont {L.}~\bibnamefont {{Hernquist}}}, \bibinfo {author}
  {\bibfnamefont {P.}~\bibnamefont {{Torrey}}}, \bibinfo {author}
  {\bibfnamefont {M.}~\bibnamefont {{Vogelsberger}}}, \bibinfo {author}
  {\bibfnamefont {R.}~\bibnamefont {{Weinberger}}},\ and\ \bibinfo {author}
  {\bibfnamefont {F.}~\bibnamefont {{Marinacci}}},\ }\bibfield  {title}
  {\bibinfo {title} {{Simulating galaxy formation with the IllustrisTNG
  model}},\ }\href {https://doi.org/10.1093/mnras/stx2656} {\bibfield
  {journal} {\bibinfo  {journal} {\mnras}\ }\textbf {\bibinfo {volume} {473}},\
  \bibinfo {pages} {4077} (\bibinfo {year} {2018}{\natexlab{a}})},\ \Eprint
  {https://arxiv.org/abs/1703.02970} {arXiv:1703.02970 [astro-ph.GA]}
  \BibitemShut {NoStop}%
\bibitem [{\citenamefont {{Pillepich}}\ \emph
  {et~al.}(2018{\natexlab{b}})\citenamefont {{Pillepich}}, \citenamefont
  {{Nelson}}, \citenamefont {{Hernquist}}, \citenamefont {{Springel}},
  \citenamefont {{Pakmor}}, \citenamefont {{Torrey}}, \citenamefont
  {{Weinberger}}, \citenamefont {{Genel}}, \citenamefont {{Naiman}},
  \citenamefont {{Marinacci}},\ and\ \citenamefont
  {{Vogelsberger}}}]{2018MNRAS.475..648P}%
  \BibitemOpen
  \bibfield  {author} {\bibinfo {author} {\bibfnamefont {A.}~\bibnamefont
  {{Pillepich}}}, \bibinfo {author} {\bibfnamefont {D.}~\bibnamefont
  {{Nelson}}}, \bibinfo {author} {\bibfnamefont {L.}~\bibnamefont
  {{Hernquist}}}, \bibinfo {author} {\bibfnamefont {V.}~\bibnamefont
  {{Springel}}}, \bibinfo {author} {\bibfnamefont {R.}~\bibnamefont
  {{Pakmor}}}, \bibinfo {author} {\bibfnamefont {P.}~\bibnamefont {{Torrey}}},
  \bibinfo {author} {\bibfnamefont {R.}~\bibnamefont {{Weinberger}}}, \bibinfo
  {author} {\bibfnamefont {S.}~\bibnamefont {{Genel}}}, \bibinfo {author}
  {\bibfnamefont {J.~P.}\ \bibnamefont {{Naiman}}}, \bibinfo {author}
  {\bibfnamefont {F.}~\bibnamefont {{Marinacci}}},\ and\ \bibinfo {author}
  {\bibfnamefont {M.}~\bibnamefont {{Vogelsberger}}},\ }\bibfield  {title}
  {\bibinfo {title} {{First results from the IllustrisTNG simulations: the
  stellar mass content of groups and clusters of galaxies}},\ }\href
  {https://doi.org/10.1093/mnras/stx3112} {\bibfield  {journal} {\bibinfo
  {journal} {\mnras}\ }\textbf {\bibinfo {volume} {475}},\ \bibinfo {pages}
  {648} (\bibinfo {year} {2018}{\natexlab{b}})},\ \Eprint
  {https://arxiv.org/abs/1707.03406} {arXiv:1707.03406 [astro-ph.GA]}
  \BibitemShut {NoStop}%
\bibitem [{\citenamefont {{Nelson}}\ \emph {et~al.}(2018)\citenamefont
  {{Nelson}}, \citenamefont {{Pillepich}}, \citenamefont {{Springel}},
  \citenamefont {{Weinberger}}, \citenamefont {{Hernquist}}, \citenamefont
  {{Pakmor}}, \citenamefont {{Genel}}, \citenamefont {{Torrey}}, \citenamefont
  {{Vogelsberger}}, \citenamefont {{Kauffmann}}, \citenamefont {{Marinacci}},\
  and\ \citenamefont {{Naiman}}}]{2018MNRAS.475..624N}%
  \BibitemOpen
  \bibfield  {author} {\bibinfo {author} {\bibfnamefont {D.}~\bibnamefont
  {{Nelson}}}, \bibinfo {author} {\bibfnamefont {A.}~\bibnamefont
  {{Pillepich}}}, \bibinfo {author} {\bibfnamefont {V.}~\bibnamefont
  {{Springel}}}, \bibinfo {author} {\bibfnamefont {R.}~\bibnamefont
  {{Weinberger}}}, \bibinfo {author} {\bibfnamefont {L.}~\bibnamefont
  {{Hernquist}}}, \bibinfo {author} {\bibfnamefont {R.}~\bibnamefont
  {{Pakmor}}}, \bibinfo {author} {\bibfnamefont {S.}~\bibnamefont {{Genel}}},
  \bibinfo {author} {\bibfnamefont {P.}~\bibnamefont {{Torrey}}}, \bibinfo
  {author} {\bibfnamefont {M.}~\bibnamefont {{Vogelsberger}}}, \bibinfo
  {author} {\bibfnamefont {G.}~\bibnamefont {{Kauffmann}}}, \bibinfo {author}
  {\bibfnamefont {F.}~\bibnamefont {{Marinacci}}},\ and\ \bibinfo {author}
  {\bibfnamefont {J.}~\bibnamefont {{Naiman}}},\ }\bibfield  {title} {\bibinfo
  {title} {{First results from the IllustrisTNG simulations: the galaxy colour
  bimodality}},\ }\href {https://doi.org/10.1093/mnras/stx3040} {\bibfield
  {journal} {\bibinfo  {journal} {\mnras}\ }\textbf {\bibinfo {volume} {475}},\
  \bibinfo {pages} {624} (\bibinfo {year} {2018})},\ \Eprint
  {https://arxiv.org/abs/1707.03395} {arXiv:1707.03395 [astro-ph.GA]}
  \BibitemShut {NoStop}%
\bibitem [{\citenamefont {{Naiman}}\ \emph {et~al.}(2018)\citenamefont
  {{Naiman}}, \citenamefont {{Pillepich}}, \citenamefont {{Springel}},
  \citenamefont {{Ramirez-Ruiz}}, \citenamefont {{Torrey}}, \citenamefont
  {{Vogelsberger}}, \citenamefont {{Pakmor}}, \citenamefont {{Nelson}},
  \citenamefont {{Marinacci}}, \citenamefont {{Hernquist}}, \citenamefont
  {{Weinberger}},\ and\ \citenamefont {{Genel}}}]{2018MNRAS.477.1206N}%
  \BibitemOpen
  \bibfield  {author} {\bibinfo {author} {\bibfnamefont {J.~P.}\ \bibnamefont
  {{Naiman}}}, \bibinfo {author} {\bibfnamefont {A.}~\bibnamefont
  {{Pillepich}}}, \bibinfo {author} {\bibfnamefont {V.}~\bibnamefont
  {{Springel}}}, \bibinfo {author} {\bibfnamefont {E.}~\bibnamefont
  {{Ramirez-Ruiz}}}, \bibinfo {author} {\bibfnamefont {P.}~\bibnamefont
  {{Torrey}}}, \bibinfo {author} {\bibfnamefont {M.}~\bibnamefont
  {{Vogelsberger}}}, \bibinfo {author} {\bibfnamefont {R.}~\bibnamefont
  {{Pakmor}}}, \bibinfo {author} {\bibfnamefont {D.}~\bibnamefont {{Nelson}}},
  \bibinfo {author} {\bibfnamefont {F.}~\bibnamefont {{Marinacci}}}, \bibinfo
  {author} {\bibfnamefont {L.}~\bibnamefont {{Hernquist}}}, \bibinfo {author}
  {\bibfnamefont {R.}~\bibnamefont {{Weinberger}}},\ and\ \bibinfo {author}
  {\bibfnamefont {S.}~\bibnamefont {{Genel}}},\ }\bibfield  {title} {\bibinfo
  {title} {{First results from the IllustrisTNG simulations: a tale of two
  elements - chemical evolution of magnesium and europium}},\ }\href
  {https://doi.org/10.1093/mnras/sty618} {\bibfield  {journal} {\bibinfo
  {journal} {\mnras}\ }\textbf {\bibinfo {volume} {477}},\ \bibinfo {pages}
  {1206} (\bibinfo {year} {2018})},\ \Eprint {https://arxiv.org/abs/1707.03401}
  {arXiv:1707.03401 [astro-ph.GA]} \BibitemShut {NoStop}%
\bibitem [{\citenamefont {{Marinacci}}\ \emph {et~al.}(2018)\citenamefont
  {{Marinacci}}, \citenamefont {{Vogelsberger}}, \citenamefont {{Pakmor}},
  \citenamefont {{Torrey}}, \citenamefont {{Springel}}, \citenamefont
  {{Hernquist}}, \citenamefont {{Nelson}}, \citenamefont {{Weinberger}},
  \citenamefont {{Pillepich}}, \citenamefont {{Naiman}},\ and\ \citenamefont
  {{Genel}}}]{2018MNRAS.480.5113M}%
  \BibitemOpen
  \bibfield  {author} {\bibinfo {author} {\bibfnamefont {F.}~\bibnamefont
  {{Marinacci}}}, \bibinfo {author} {\bibfnamefont {M.}~\bibnamefont
  {{Vogelsberger}}}, \bibinfo {author} {\bibfnamefont {R.}~\bibnamefont
  {{Pakmor}}}, \bibinfo {author} {\bibfnamefont {P.}~\bibnamefont {{Torrey}}},
  \bibinfo {author} {\bibfnamefont {V.}~\bibnamefont {{Springel}}}, \bibinfo
  {author} {\bibfnamefont {L.}~\bibnamefont {{Hernquist}}}, \bibinfo {author}
  {\bibfnamefont {D.}~\bibnamefont {{Nelson}}}, \bibinfo {author}
  {\bibfnamefont {R.}~\bibnamefont {{Weinberger}}}, \bibinfo {author}
  {\bibfnamefont {A.}~\bibnamefont {{Pillepich}}}, \bibinfo {author}
  {\bibfnamefont {J.}~\bibnamefont {{Naiman}}},\ and\ \bibinfo {author}
  {\bibfnamefont {S.}~\bibnamefont {{Genel}}},\ }\bibfield  {title} {\bibinfo
  {title} {{First results from the IllustrisTNG simulations: radio haloes and
  magnetic fields}},\ }\href {https://doi.org/10.1093/mnras/sty2206} {\bibfield
   {journal} {\bibinfo  {journal} {\mnras}\ }\textbf {\bibinfo {volume}
  {480}},\ \bibinfo {pages} {5113} (\bibinfo {year} {2018})},\ \Eprint
  {https://arxiv.org/abs/1707.03396} {arXiv:1707.03396 [astro-ph.CO]}
  \BibitemShut {NoStop}%
\bibitem [{\citenamefont {{Springel}}\ \emph {et~al.}(2018)\citenamefont
  {{Springel}}, \citenamefont {{Pakmor}}, \citenamefont {{Pillepich}},
  \citenamefont {{Weinberger}}, \citenamefont {{Nelson}}, \citenamefont
  {{Hernquist}}, \citenamefont {{Vogelsberger}}, \citenamefont {{Genel}},
  \citenamefont {{Torrey}}, \citenamefont {{Marinacci}},\ and\ \citenamefont
  {{Naiman}}}]{2018MNRAS.475..676S}%
  \BibitemOpen
  \bibfield  {author} {\bibinfo {author} {\bibfnamefont {V.}~\bibnamefont
  {{Springel}}}, \bibinfo {author} {\bibfnamefont {R.}~\bibnamefont
  {{Pakmor}}}, \bibinfo {author} {\bibfnamefont {A.}~\bibnamefont
  {{Pillepich}}}, \bibinfo {author} {\bibfnamefont {R.}~\bibnamefont
  {{Weinberger}}}, \bibinfo {author} {\bibfnamefont {D.}~\bibnamefont
  {{Nelson}}}, \bibinfo {author} {\bibfnamefont {L.}~\bibnamefont
  {{Hernquist}}}, \bibinfo {author} {\bibfnamefont {M.}~\bibnamefont
  {{Vogelsberger}}}, \bibinfo {author} {\bibfnamefont {S.}~\bibnamefont
  {{Genel}}}, \bibinfo {author} {\bibfnamefont {P.}~\bibnamefont {{Torrey}}},
  \bibinfo {author} {\bibfnamefont {F.}~\bibnamefont {{Marinacci}}},\ and\
  \bibinfo {author} {\bibfnamefont {J.}~\bibnamefont {{Naiman}}},\ }\bibfield
  {title} {\bibinfo {title} {{First results from the IllustrisTNG simulations:
  matter and galaxy clustering}},\ }\href
  {https://doi.org/10.1093/mnras/stx3304} {\bibfield  {journal} {\bibinfo
  {journal} {\mnras}\ }\textbf {\bibinfo {volume} {475}},\ \bibinfo {pages}
  {676} (\bibinfo {year} {2018})},\ \Eprint {https://arxiv.org/abs/1707.03397}
  {arXiv:1707.03397 [astro-ph.GA]} \BibitemShut {NoStop}%
\bibitem [{\citenamefont {{Nelson}}\ \emph
  {et~al.}(2019{\natexlab{b}})\citenamefont {{Nelson}}, \citenamefont
  {{Pillepich}}, \citenamefont {{Springel}}, \citenamefont {{Pakmor}},
  \citenamefont {{Weinberger}}, \citenamefont {{Genel}}, \citenamefont
  {{Torrey}}, \citenamefont {{Vogelsberger}}, \citenamefont {{Marinacci}},\
  and\ \citenamefont {{Hernquist}}}]{2019MNRAS.490.3234N}%
  \BibitemOpen
  \bibfield  {author} {\bibinfo {author} {\bibfnamefont {D.}~\bibnamefont
  {{Nelson}}}, \bibinfo {author} {\bibfnamefont {A.}~\bibnamefont
  {{Pillepich}}}, \bibinfo {author} {\bibfnamefont {V.}~\bibnamefont
  {{Springel}}}, \bibinfo {author} {\bibfnamefont {R.}~\bibnamefont
  {{Pakmor}}}, \bibinfo {author} {\bibfnamefont {R.}~\bibnamefont
  {{Weinberger}}}, \bibinfo {author} {\bibfnamefont {S.}~\bibnamefont
  {{Genel}}}, \bibinfo {author} {\bibfnamefont {P.}~\bibnamefont {{Torrey}}},
  \bibinfo {author} {\bibfnamefont {M.}~\bibnamefont {{Vogelsberger}}},
  \bibinfo {author} {\bibfnamefont {F.}~\bibnamefont {{Marinacci}}},\ and\
  \bibinfo {author} {\bibfnamefont {L.}~\bibnamefont {{Hernquist}}},\
  }\bibfield  {title} {\bibinfo {title} {{First results from the TNG50
  simulation: galactic outflows driven by supernovae and black hole
  feedback}},\ }\href {https://doi.org/10.1093/mnras/stz2306} {\bibfield
  {journal} {\bibinfo  {journal} {\mnras}\ }\textbf {\bibinfo {volume} {490}},\
  \bibinfo {pages} {3234} (\bibinfo {year} {2019}{\natexlab{b}})},\ \Eprint
  {https://arxiv.org/abs/1902.05554} {arXiv:1902.05554 [astro-ph.GA]}
  \BibitemShut {NoStop}%
\bibitem [{\citenamefont {{Pillepich}}\ \emph {et~al.}(2019)\citenamefont
  {{Pillepich}}, \citenamefont {{Nelson}}, \citenamefont {{Springel}},
  \citenamefont {{Pakmor}}, \citenamefont {{Torrey}}, \citenamefont
  {{Weinberger}}, \citenamefont {{Vogelsberger}}, \citenamefont {{Marinacci}},
  \citenamefont {{Genel}}, \citenamefont {{van der Wel}},\ and\ \citenamefont
  {{Hernquist}}}]{2019MNRAS.490.3196P}%
  \BibitemOpen
  \bibfield  {author} {\bibinfo {author} {\bibfnamefont {A.}~\bibnamefont
  {{Pillepich}}}, \bibinfo {author} {\bibfnamefont {D.}~\bibnamefont
  {{Nelson}}}, \bibinfo {author} {\bibfnamefont {V.}~\bibnamefont
  {{Springel}}}, \bibinfo {author} {\bibfnamefont {R.}~\bibnamefont
  {{Pakmor}}}, \bibinfo {author} {\bibfnamefont {P.}~\bibnamefont {{Torrey}}},
  \bibinfo {author} {\bibfnamefont {R.}~\bibnamefont {{Weinberger}}}, \bibinfo
  {author} {\bibfnamefont {M.}~\bibnamefont {{Vogelsberger}}}, \bibinfo
  {author} {\bibfnamefont {F.}~\bibnamefont {{Marinacci}}}, \bibinfo {author}
  {\bibfnamefont {S.}~\bibnamefont {{Genel}}}, \bibinfo {author} {\bibfnamefont
  {A.}~\bibnamefont {{van der Wel}}},\ and\ \bibinfo {author} {\bibfnamefont
  {L.}~\bibnamefont {{Hernquist}}},\ }\bibfield  {title} {\bibinfo {title}
  {{First results from the TNG50 simulation: the evolution of stellar and
  gaseous discs across cosmic time}},\ }\href
  {https://doi.org/10.1093/mnras/stz2338} {\bibfield  {journal} {\bibinfo
  {journal} {\mnras}\ }\textbf {\bibinfo {volume} {490}},\ \bibinfo {pages}
  {3196} (\bibinfo {year} {2019})},\ \Eprint {https://arxiv.org/abs/1902.05553}
  {arXiv:1902.05553 [astro-ph.GA]} \BibitemShut {NoStop}%
\bibitem [{\citenamefont {{Springel}}(2010)}]{2010MNRAS.401..791S}%
  \BibitemOpen
  \bibfield  {author} {\bibinfo {author} {\bibfnamefont {V.}~\bibnamefont
  {{Springel}}},\ }\bibfield  {title} {\bibinfo {title} {{E pur si muove:
  Galilean-invariant cosmological hydrodynamical simulations on a moving
  mesh}},\ }\href {https://doi.org/10.1111/j.1365-2966.2009.15715.x} {\bibfield
   {journal} {\bibinfo  {journal} {\mnras}\ }\textbf {\bibinfo {volume}
  {401}},\ \bibinfo {pages} {791} (\bibinfo {year} {2010})},\ \Eprint
  {https://arxiv.org/abs/0901.4107} {arXiv:0901.4107 [astro-ph.CO]}
  \BibitemShut {NoStop}%
\bibitem [{\citenamefont {{Davis}}\ \emph {et~al.}(1985)\citenamefont
  {{Davis}}, \citenamefont {{Efstathiou}}, \citenamefont {{Frenk}},\ and\
  \citenamefont {{White}}}]{1985ApJ...292..371D}%
  \BibitemOpen
  \bibfield  {author} {\bibinfo {author} {\bibfnamefont {M.}~\bibnamefont
  {{Davis}}}, \bibinfo {author} {\bibfnamefont {G.}~\bibnamefont
  {{Efstathiou}}}, \bibinfo {author} {\bibfnamefont {C.~S.}\ \bibnamefont
  {{Frenk}}},\ and\ \bibinfo {author} {\bibfnamefont {S.~D.~M.}\ \bibnamefont
  {{White}}},\ }\bibfield  {title} {\bibinfo {title} {{The evolution of
  large-scale structure in a universe dominated by cold dark matter}},\ }\href
  {https://doi.org/10.1086/163168} {\bibfield  {journal} {\bibinfo  {journal}
  {\apj}\ }\textbf {\bibinfo {volume} {292}},\ \bibinfo {pages} {371} (\bibinfo
  {year} {1985})}\BibitemShut {NoStop}%
\bibitem [{\citenamefont {{Springel}}\ \emph {et~al.}(2021)\citenamefont
  {{Springel}}, \citenamefont {{Pakmor}}, \citenamefont {{Zier}},\ and\
  \citenamefont {{Reinecke}}}]{Springel2021}%
  \BibitemOpen
  \bibfield  {author} {\bibinfo {author} {\bibfnamefont {V.}~\bibnamefont
  {{Springel}}}, \bibinfo {author} {\bibfnamefont {R.}~\bibnamefont
  {{Pakmor}}}, \bibinfo {author} {\bibfnamefont {O.}~\bibnamefont {{Zier}}},\
  and\ \bibinfo {author} {\bibfnamefont {M.}~\bibnamefont {{Reinecke}}},\
  }\bibfield  {title} {\bibinfo {title} {{Simulating cosmic structure formation
  with the GADGET-4 code}},\ }\href {https://doi.org/10.1093/mnras/stab1855}
  {\bibfield  {journal} {\bibinfo  {journal} {\mnras}\ }\textbf {\bibinfo
  {volume} {506}},\ \bibinfo {pages} {2871} (\bibinfo {year} {2021})},\ \Eprint
  {https://arxiv.org/abs/2010.03567} {arXiv:2010.03567 [astro-ph.IM]}
  \BibitemShut {NoStop}%
\bibitem [{\citenamefont {{Bryan}}\ and\ \citenamefont
  {{Norman}}(1998)}]{1998ApJ...495...80B}%
  \BibitemOpen
  \bibfield  {author} {\bibinfo {author} {\bibfnamefont {G.~L.}\ \bibnamefont
  {{Bryan}}}\ and\ \bibinfo {author} {\bibfnamefont {M.~L.}\ \bibnamefont
  {{Norman}}},\ }\bibfield  {title} {\bibinfo {title} {{Statistical Properties
  of X-Ray Clusters: Analytic and Numerical Comparisons}},\ }\href
  {https://doi.org/10.1086/305262} {\bibfield  {journal} {\bibinfo  {journal}
  {\apj}\ }\textbf {\bibinfo {volume} {495}},\ \bibinfo {pages} {80} (\bibinfo
  {year} {1998})},\ \Eprint {https://arxiv.org/abs/astro-ph/9710107}
  {arXiv:astro-ph/9710107 [astro-ph]} \BibitemShut {NoStop}%
\bibitem [{\citenamefont {Zheng}\ \emph {et~al.}(2005)\citenamefont {Zheng},
  \citenamefont {Berlind}, \citenamefont {Weinberg}, \citenamefont {Benson},
  \citenamefont {Baugh}, \citenamefont {Cole}, \citenamefont {Dave},
  \citenamefont {Frenk}, \citenamefont {Katz},\ and\ \citenamefont
  {Lacey}}]{Zheng:2004id}%
  \BibitemOpen
  \bibfield  {author} {\bibinfo {author} {\bibfnamefont {Z.}~\bibnamefont
  {Zheng}}, \bibinfo {author} {\bibfnamefont {A.~A.}\ \bibnamefont {Berlind}},
  \bibinfo {author} {\bibfnamefont {D.~H.}\ \bibnamefont {Weinberg}}, \bibinfo
  {author} {\bibfnamefont {A.~J.}\ \bibnamefont {Benson}}, \bibinfo {author}
  {\bibfnamefont {C.~M.}\ \bibnamefont {Baugh}}, \bibinfo {author}
  {\bibfnamefont {S.}~\bibnamefont {Cole}}, \bibinfo {author} {\bibfnamefont
  {R.}~\bibnamefont {Dave}}, \bibinfo {author} {\bibfnamefont {C.~S.}\
  \bibnamefont {Frenk}}, \bibinfo {author} {\bibfnamefont {N.}~\bibnamefont
  {Katz}},\ and\ \bibinfo {author} {\bibfnamefont {C.~G.}\ \bibnamefont
  {Lacey}},\ }\bibfield  {title} {\bibinfo {title} {{Theoretical models of the
  halo occupation distribution: Separating central and satellite galaxies}},\
  }\href {https://doi.org/10.1086/466510} {\bibfield  {journal} {\bibinfo
  {journal} {Astrophys. J.}\ }\textbf {\bibinfo {volume} {633}},\ \bibinfo
  {pages} {791} (\bibinfo {year} {2005})},\ \Eprint
  {https://arxiv.org/abs/astro-ph/0408564} {arXiv:astro-ph/0408564 [astro-ph]}
  \BibitemShut {NoStop}%
\bibitem [{\citenamefont {{Ouchi}}\ \emph {et~al.}(2020)\citenamefont
  {{Ouchi}}, \citenamefont {{Ono}},\ and\ \citenamefont
  {{Shibuya}}}]{2020ARA&A..58..617O}%
  \BibitemOpen
  \bibfield  {author} {\bibinfo {author} {\bibfnamefont {M.}~\bibnamefont
  {{Ouchi}}}, \bibinfo {author} {\bibfnamefont {Y.}~\bibnamefont {{Ono}}},\
  and\ \bibinfo {author} {\bibfnamefont {T.}~\bibnamefont {{Shibuya}}},\
  }\bibfield  {title} {\bibinfo {title} {{Observations of the
  Lyman-{\ensuremath{\alpha}} Universe}},\ }\href
  {https://doi.org/10.1146/annurev-astro-032620-021859} {\bibfield  {journal}
  {\bibinfo  {journal} {\araa}\ }\textbf {\bibinfo {volume} {58}},\ \bibinfo
  {pages} {617} (\bibinfo {year} {2020})},\ \Eprint
  {https://arxiv.org/abs/2012.07960} {arXiv:2012.07960 [astro-ph.GA]}
  \BibitemShut {NoStop}%
\bibitem [{\citenamefont {{Dijkstra}}\ and\ \citenamefont
  {{Westra}}(2010)}]{2010MNRAS.401.2343D}%
  \BibitemOpen
  \bibfield  {author} {\bibinfo {author} {\bibfnamefont {M.}~\bibnamefont
  {{Dijkstra}}}\ and\ \bibinfo {author} {\bibfnamefont {E.}~\bibnamefont
  {{Westra}}},\ }\bibfield  {title} {\bibinfo {title} {{Star formation
  indicators and line equivalent width in Ly{\ensuremath{\alpha}} galaxies}},\
  }\href {https://doi.org/10.1111/j.1365-2966.2009.15859.x} {\bibfield
  {journal} {\bibinfo  {journal} {\mnras}\ }\textbf {\bibinfo {volume} {401}},\
  \bibinfo {pages} {2343} (\bibinfo {year} {2010})},\ \Eprint
  {https://arxiv.org/abs/0911.1357} {arXiv:0911.1357 [astro-ph.CO]}
  \BibitemShut {NoStop}%
\bibitem [{\citenamefont {{Nagamine}}\ \emph {et~al.}(2010)\citenamefont
  {{Nagamine}}, \citenamefont {{Ouchi}}, \citenamefont {{Springel}},\ and\
  \citenamefont {{Hernquist}}}]{Nagamine}%
  \BibitemOpen
  \bibfield  {author} {\bibinfo {author} {\bibfnamefont {K.}~\bibnamefont
  {{Nagamine}}}, \bibinfo {author} {\bibfnamefont {M.}~\bibnamefont {{Ouchi}}},
  \bibinfo {author} {\bibfnamefont {V.}~\bibnamefont {{Springel}}},\ and\
  \bibinfo {author} {\bibfnamefont {L.}~\bibnamefont {{Hernquist}}},\
  }\bibfield  {title} {\bibinfo {title} {{Lyman-{\ensuremath{\alpha}} Emitters
  and Lyman-Break Galaxies at z = 3-6 in Cosmological SPH Simulations}},\
  }\href {https://doi.org/10.1093/pasj/62.6.1455} {\bibfield  {journal}
  {\bibinfo  {journal} {\pasj}\ }\textbf {\bibinfo {volume} {62}},\ \bibinfo
  {pages} {1455} (\bibinfo {year} {2010})},\ \Eprint
  {https://arxiv.org/abs/0802.0228} {arXiv:0802.0228 [astro-ph]} \BibitemShut
  {NoStop}%
\bibitem [{\citenamefont {{Inoue}}\ \emph {et~al.}(2014)\citenamefont
  {{Inoue}}, \citenamefont {{Shimizu}}, \citenamefont {{Iwata}},\ and\
  \citenamefont {{Tanaka}}}]{2014MNRAS.442.1805I}%
  \BibitemOpen
  \bibfield  {author} {\bibinfo {author} {\bibfnamefont {A.~K.}\ \bibnamefont
  {{Inoue}}}, \bibinfo {author} {\bibfnamefont {I.}~\bibnamefont {{Shimizu}}},
  \bibinfo {author} {\bibfnamefont {I.}~\bibnamefont {{Iwata}}},\ and\ \bibinfo
  {author} {\bibfnamefont {M.}~\bibnamefont {{Tanaka}}},\ }\bibfield  {title}
  {\bibinfo {title} {{An updated analytic model for attenuation by the
  intergalactic medium}},\ }\href {https://doi.org/10.1093/mnras/stu936}
  {\bibfield  {journal} {\bibinfo  {journal} {\mnras}\ }\textbf {\bibinfo
  {volume} {442}},\ \bibinfo {pages} {1805} (\bibinfo {year} {2014})},\ \Eprint
  {https://arxiv.org/abs/1402.0677} {arXiv:1402.0677 [astro-ph.CO]}
  \BibitemShut {NoStop}%
\bibitem [{\citenamefont {{Reid}}\ and\ \citenamefont
  {{White}}(2011)}]{2011MNRAS.417.1913R}%
  \BibitemOpen
  \bibfield  {author} {\bibinfo {author} {\bibfnamefont {B.~A.}\ \bibnamefont
  {{Reid}}}\ and\ \bibinfo {author} {\bibfnamefont {M.}~\bibnamefont
  {{White}}},\ }\bibfield  {title} {\bibinfo {title} {{Towards an accurate
  model of the redshift-space clustering of haloes in the quasi-linear
  regime}},\ }\href {https://doi.org/10.1111/j.1365-2966.2011.19379.x}
  {\bibfield  {journal} {\bibinfo  {journal} {\mnras}\ }\textbf {\bibinfo
  {volume} {417}},\ \bibinfo {pages} {1913} (\bibinfo {year} {2011})},\ \Eprint
  {https://arxiv.org/abs/1105.4165} {arXiv:1105.4165 [astro-ph.CO]}
  \BibitemShut {NoStop}%
\bibitem [{\citenamefont {{Gao}}\ and\ \citenamefont
  {{White}}(2007)}]{2007MNRAS.377L...5G}%
  \BibitemOpen
  \bibfield  {author} {\bibinfo {author} {\bibfnamefont {L.}~\bibnamefont
  {{Gao}}}\ and\ \bibinfo {author} {\bibfnamefont {S.~D.~M.}\ \bibnamefont
  {{White}}},\ }\bibfield  {title} {\bibinfo {title} {{Assembly bias in the
  clustering of dark matter haloes}},\ }\href
  {https://doi.org/10.1111/j.1745-3933.2007.00292.x} {\bibfield  {journal}
  {\bibinfo  {journal} {\mnras}\ }\textbf {\bibinfo {volume} {377}},\ \bibinfo
  {pages} {L5} (\bibinfo {year} {2007})},\ \Eprint
  {https://arxiv.org/abs/astro-ph/0611921} {arXiv:astro-ph/0611921 [astro-ph]}
  \BibitemShut {NoStop}%
\bibitem [{\citenamefont {{Zehavi}}\ \emph {et~al.}(2018)\citenamefont
  {{Zehavi}}, \citenamefont {{Contreras}}, \citenamefont {{Padilla}},
  \citenamefont {{Smith}}, \citenamefont {{Baugh}},\ and\ \citenamefont
  {{Norberg}}}]{2018ApJ...853...84Z}%
  \BibitemOpen
  \bibfield  {author} {\bibinfo {author} {\bibfnamefont {I.}~\bibnamefont
  {{Zehavi}}}, \bibinfo {author} {\bibfnamefont {S.}~\bibnamefont
  {{Contreras}}}, \bibinfo {author} {\bibfnamefont {N.}~\bibnamefont
  {{Padilla}}}, \bibinfo {author} {\bibfnamefont {N.~J.}\ \bibnamefont
  {{Smith}}}, \bibinfo {author} {\bibfnamefont {C.~M.}\ \bibnamefont
  {{Baugh}}},\ and\ \bibinfo {author} {\bibfnamefont {P.}~\bibnamefont
  {{Norberg}}},\ }\bibfield  {title} {\bibinfo {title} {{The Impact of Assembly
  Bias on the Galaxy Content of Dark Matter Halos}},\ }\href
  {https://doi.org/10.3847/1538-4357/aaa54a} {\bibfield  {journal} {\bibinfo
  {journal} {\apj}\ }\textbf {\bibinfo {volume} {853}},\ \bibinfo {eid} {84}
  (\bibinfo {year} {2018})},\ \Eprint {https://arxiv.org/abs/1706.07871}
  {arXiv:1706.07871 [astro-ph.GA]} \BibitemShut {NoStop}%
\bibitem [{\citenamefont {{Artale}}\ \emph {et~al.}(2018)\citenamefont
  {{Artale}}, \citenamefont {{Zehavi}}, \citenamefont {{Contreras}},\ and\
  \citenamefont {{Norberg}}}]{2018MNRAS.480.3978A}%
  \BibitemOpen
  \bibfield  {author} {\bibinfo {author} {\bibfnamefont {M.~C.}\ \bibnamefont
  {{Artale}}}, \bibinfo {author} {\bibfnamefont {I.}~\bibnamefont {{Zehavi}}},
  \bibinfo {author} {\bibfnamefont {S.}~\bibnamefont {{Contreras}}},\ and\
  \bibinfo {author} {\bibfnamefont {P.}~\bibnamefont {{Norberg}}},\ }\bibfield
  {title} {\bibinfo {title} {{The impact of assembly bias on the halo
  occupation in hydrodynamical simulations}},\ }\href
  {https://doi.org/10.1093/mnras/sty2110} {\bibfield  {journal} {\bibinfo
  {journal} {\mnras}\ }\textbf {\bibinfo {volume} {480}},\ \bibinfo {pages}
  {3978} (\bibinfo {year} {2018})},\ \Eprint {https://arxiv.org/abs/1805.06938}
  {arXiv:1805.06938 [astro-ph.GA]} \BibitemShut {NoStop}%
\bibitem [{\citenamefont {{Yuan}}\ \emph {et~al.}(2022)\citenamefont {{Yuan}},
  \citenamefont {{Garrison}}, \citenamefont {{Hadzhiyska}}, \citenamefont
  {{Bose}},\ and\ \citenamefont {{Eisenstein}}}]{2022MNRAS.510.3301Y}%
  \BibitemOpen
  \bibfield  {author} {\bibinfo {author} {\bibfnamefont {S.}~\bibnamefont
  {{Yuan}}}, \bibinfo {author} {\bibfnamefont {L.~H.}\ \bibnamefont
  {{Garrison}}}, \bibinfo {author} {\bibfnamefont {B.}~\bibnamefont
  {{Hadzhiyska}}}, \bibinfo {author} {\bibfnamefont {S.}~\bibnamefont
  {{Bose}}},\ and\ \bibinfo {author} {\bibfnamefont {D.~J.}\ \bibnamefont
  {{Eisenstein}}},\ }\bibfield  {title} {\bibinfo {title} {{ABACUSHOD: a highly
  efficient extended multitracer HOD framework and its application to BOSS and
  eBOSS data}},\ }\href {https://doi.org/10.1093/mnras/stab3355} {\bibfield
  {journal} {\bibinfo  {journal} {\mnras}\ }\textbf {\bibinfo {volume} {510}},\
  \bibinfo {pages} {3301} (\bibinfo {year} {2022})},\ \Eprint
  {https://arxiv.org/abs/2110.11412} {arXiv:2110.11412 [astro-ph.CO]}
  \BibitemShut {NoStop}%
\bibitem [{\citenamefont {{Maksimova}}\ \emph {et~al.}(2021)\citenamefont
  {{Maksimova}}, \citenamefont {{Garrison}}, \citenamefont {{Eisenstein}},
  \citenamefont {{Hadzhiyska}}, \citenamefont {{Bose}},\ and\ \citenamefont
  {{Satterthwaite}}}]{2021MNRAS.508.4017M}%
  \BibitemOpen
  \bibfield  {author} {\bibinfo {author} {\bibfnamefont {N.~A.}\ \bibnamefont
  {{Maksimova}}}, \bibinfo {author} {\bibfnamefont {L.~H.}\ \bibnamefont
  {{Garrison}}}, \bibinfo {author} {\bibfnamefont {D.~J.}\ \bibnamefont
  {{Eisenstein}}}, \bibinfo {author} {\bibfnamefont {B.}~\bibnamefont
  {{Hadzhiyska}}}, \bibinfo {author} {\bibfnamefont {S.}~\bibnamefont
  {{Bose}}},\ and\ \bibinfo {author} {\bibfnamefont {T.~P.}\ \bibnamefont
  {{Satterthwaite}}},\ }\bibfield  {title} {\bibinfo {title} {{ABACUSSUMMIT: a
  massive set of high-accuracy, high-resolution N-body simulations}},\ }\href
  {https://doi.org/10.1093/mnras/stab2484} {\bibfield  {journal} {\bibinfo
  {journal} {\mnras}\ }\textbf {\bibinfo {volume} {508}},\ \bibinfo {pages}
  {4017} (\bibinfo {year} {2021})},\ \Eprint {https://arxiv.org/abs/2110.11398}
  {arXiv:2110.11398 [astro-ph.CO]} \BibitemShut {NoStop}%
\bibitem [{Note1()}]{Note1}%
  \BibitemOpen
  \bibinfo {note} {A caveat in that statement is the dependence of halo
  properties on the halo finding algorithm and simulation
  resolution}\BibitemShut {NoStop}%
\bibitem [{\citenamefont {{Hadzhiyska}}\ \emph
  {et~al.}(2021{\natexlab{b}})\citenamefont {{Hadzhiyska}}, \citenamefont
  {{Tacchella}}, \citenamefont {{Bose}},\ and\ \citenamefont
  {{Eisenstein}}}]{2021MNRAS.502.3599H}%
  \BibitemOpen
  \bibfield  {author} {\bibinfo {author} {\bibfnamefont {B.}~\bibnamefont
  {{Hadzhiyska}}}, \bibinfo {author} {\bibfnamefont {S.}~\bibnamefont
  {{Tacchella}}}, \bibinfo {author} {\bibfnamefont {S.}~\bibnamefont
  {{Bose}}},\ and\ \bibinfo {author} {\bibfnamefont {D.~J.}\ \bibnamefont
  {{Eisenstein}}},\ }\bibfield  {title} {\bibinfo {title} {{The galaxy-halo
  connection of emission-line galaxies in IllustrisTNG}},\ }\href
  {https://doi.org/10.1093/mnras/stab243} {\bibfield  {journal} {\bibinfo
  {journal} {\mnras}\ }\textbf {\bibinfo {volume} {502}},\ \bibinfo {pages}
  {3599} (\bibinfo {year} {2021}{\natexlab{b}})},\ \Eprint
  {https://arxiv.org/abs/2011.05331} {arXiv:2011.05331 [astro-ph.GA]}
  \BibitemShut {NoStop}%
\bibitem [{\citenamefont {{More}}\ \emph {et~al.}(2011)\citenamefont {{More}},
  \citenamefont {{Kravtsov}}, \citenamefont {{Dalal}},\ and\ \citenamefont
  {{Gottl{\"o}ber}}}]{2011ApJS..195....4M}%
  \BibitemOpen
  \bibfield  {author} {\bibinfo {author} {\bibfnamefont {S.}~\bibnamefont
  {{More}}}, \bibinfo {author} {\bibfnamefont {A.~V.}\ \bibnamefont
  {{Kravtsov}}}, \bibinfo {author} {\bibfnamefont {N.}~\bibnamefont
  {{Dalal}}},\ and\ \bibinfo {author} {\bibfnamefont {S.}~\bibnamefont
  {{Gottl{\"o}ber}}},\ }\bibfield  {title} {\bibinfo {title} {{The Overdensity
  and Masses of the Friends-of-friends Halos and Universality of Halo Mass
  Function}},\ }\href {https://doi.org/10.1088/0067-0049/195/1/4} {\bibfield
  {journal} {\bibinfo  {journal} {\apjs}\ }\textbf {\bibinfo {volume} {195}},\
  \bibinfo {eid} {4} (\bibinfo {year} {2011})},\ \Eprint
  {https://arxiv.org/abs/1103.0005} {arXiv:1103.0005 [astro-ph.CO]}
  \BibitemShut {NoStop}%
\bibitem [{\citenamefont {{Gawiser}}\ \emph {et~al.}(2007)\citenamefont
  {{Gawiser}}, \citenamefont {{Francke}}, \citenamefont {{Lai}}, \citenamefont
  {{Schawinski}}, \citenamefont {{Gronwall}}, \citenamefont {{Ciardullo}},
  \citenamefont {{Quadri}}, \citenamefont {{Orsi}}, \citenamefont
  {{Barrientos}}, \citenamefont {{Blanc}}, \citenamefont {{Fazio}},
  \citenamefont {{Feldmeier}}, \citenamefont {{Huang}}, \citenamefont
  {{Infante}}, \citenamefont {{Lira}}, \citenamefont {{Padilla}}, \citenamefont
  {{Taylor}}, \citenamefont {{Treister}}, \citenamefont {{Urry}}, \citenamefont
  {{van Dokkum}},\ and\ \citenamefont {{Virani}}}]{Gawiser}%
  \BibitemOpen
  \bibfield  {author} {\bibinfo {author} {\bibfnamefont {E.}~\bibnamefont
  {{Gawiser}}}, \bibinfo {author} {\bibfnamefont {H.}~\bibnamefont
  {{Francke}}}, \bibinfo {author} {\bibfnamefont {K.}~\bibnamefont {{Lai}}},
  \bibinfo {author} {\bibfnamefont {K.}~\bibnamefont {{Schawinski}}}, \bibinfo
  {author} {\bibfnamefont {C.}~\bibnamefont {{Gronwall}}}, \bibinfo {author}
  {\bibfnamefont {R.}~\bibnamefont {{Ciardullo}}}, \bibinfo {author}
  {\bibfnamefont {R.}~\bibnamefont {{Quadri}}}, \bibinfo {author}
  {\bibfnamefont {A.}~\bibnamefont {{Orsi}}}, \bibinfo {author} {\bibfnamefont
  {L.~F.}\ \bibnamefont {{Barrientos}}}, \bibinfo {author} {\bibfnamefont
  {G.~A.}\ \bibnamefont {{Blanc}}}, \bibinfo {author} {\bibfnamefont
  {G.}~\bibnamefont {{Fazio}}}, \bibinfo {author} {\bibfnamefont {J.~J.}\
  \bibnamefont {{Feldmeier}}}, \bibinfo {author} {\bibfnamefont {J.-s.}\
  \bibnamefont {{Huang}}}, \bibinfo {author} {\bibfnamefont {L.}~\bibnamefont
  {{Infante}}}, \bibinfo {author} {\bibfnamefont {P.}~\bibnamefont {{Lira}}},
  \bibinfo {author} {\bibfnamefont {N.}~\bibnamefont {{Padilla}}}, \bibinfo
  {author} {\bibfnamefont {E.~N.}\ \bibnamefont {{Taylor}}}, \bibinfo {author}
  {\bibfnamefont {E.}~\bibnamefont {{Treister}}}, \bibinfo {author}
  {\bibfnamefont {C.~M.}\ \bibnamefont {{Urry}}}, \bibinfo {author}
  {\bibfnamefont {P.~G.}\ \bibnamefont {{van Dokkum}}},\ and\ \bibinfo {author}
  {\bibfnamefont {S.~N.}\ \bibnamefont {{Virani}}},\ }\bibfield  {title}
  {\bibinfo {title} {{Ly{\ensuremath{\alpha}}-Emitting Galaxies at z = 3.1: L*
  Progenitors Experiencing Rapid Star Formation}},\ }\href
  {https://doi.org/10.1086/522955} {\bibfield  {journal} {\bibinfo  {journal}
  {\apj}\ }\textbf {\bibinfo {volume} {671}},\ \bibinfo {pages} {278} (\bibinfo
  {year} {2007})},\ \Eprint {https://arxiv.org/abs/0710.2697} {arXiv:0710.2697
  [astro-ph]} \BibitemShut {NoStop}%
\bibitem [{\citenamefont {{Guaita}}\ \emph {et~al.}(2010)\citenamefont
  {{Guaita}}, \citenamefont {{Gawiser}}, \citenamefont {{Padilla}},
  \citenamefont {{Francke}}, \citenamefont {{Bond}}, \citenamefont
  {{Gronwall}}, \citenamefont {{Ciardullo}}, \citenamefont {{Feldmeier}},
  \citenamefont {{Sinawa}}, \citenamefont {{Blanc}},\ and\ \citenamefont
  {{Virani}}}]{Guaita}%
  \BibitemOpen
  \bibfield  {author} {\bibinfo {author} {\bibfnamefont {L.}~\bibnamefont
  {{Guaita}}}, \bibinfo {author} {\bibfnamefont {E.}~\bibnamefont {{Gawiser}}},
  \bibinfo {author} {\bibfnamefont {N.}~\bibnamefont {{Padilla}}}, \bibinfo
  {author} {\bibfnamefont {H.}~\bibnamefont {{Francke}}}, \bibinfo {author}
  {\bibfnamefont {N.~A.}\ \bibnamefont {{Bond}}}, \bibinfo {author}
  {\bibfnamefont {C.}~\bibnamefont {{Gronwall}}}, \bibinfo {author}
  {\bibfnamefont {R.}~\bibnamefont {{Ciardullo}}}, \bibinfo {author}
  {\bibfnamefont {J.~J.}\ \bibnamefont {{Feldmeier}}}, \bibinfo {author}
  {\bibfnamefont {S.}~\bibnamefont {{Sinawa}}}, \bibinfo {author}
  {\bibfnamefont {G.~A.}\ \bibnamefont {{Blanc}}},\ and\ \bibinfo {author}
  {\bibfnamefont {S.}~\bibnamefont {{Virani}}},\ }\bibfield  {title} {\bibinfo
  {title} {{Ly{\ensuremath{\alpha}}-emitting Galaxies at z = 2.1 in ECDF-S:
  Building Blocks of Typical Present-day Galaxies?}},\ }\href
  {https://doi.org/10.1088/0004-637X/714/1/255} {\bibfield  {journal} {\bibinfo
   {journal} {\apj}\ }\textbf {\bibinfo {volume} {714}},\ \bibinfo {pages}
  {255} (\bibinfo {year} {2010})},\ \Eprint {https://arxiv.org/abs/0910.2244}
  {arXiv:0910.2244 [astro-ph.CO]} \BibitemShut {NoStop}%
\bibitem [{\citenamefont {{Ouchi}}\ \emph {et~al.}(2010)\citenamefont
  {{Ouchi}}, \citenamefont {{Shimasaku}}, \citenamefont {{Furusawa}},
  \citenamefont {{Saito}}, \citenamefont {{Yoshida}}, \citenamefont
  {{Akiyama}}, \citenamefont {{Ono}}, \citenamefont {{Yamada}}, \citenamefont
  {{Ota}}, \citenamefont {{Kashikawa}}, \citenamefont {{Iye}}, \citenamefont
  {{Kodama}}, \citenamefont {{Okamura}}, \citenamefont {{Simpson}},\ and\
  \citenamefont {{Yoshida}}}]{Ouchi}%
  \BibitemOpen
  \bibfield  {author} {\bibinfo {author} {\bibfnamefont {M.}~\bibnamefont
  {{Ouchi}}}, \bibinfo {author} {\bibfnamefont {K.}~\bibnamefont
  {{Shimasaku}}}, \bibinfo {author} {\bibfnamefont {H.}~\bibnamefont
  {{Furusawa}}}, \bibinfo {author} {\bibfnamefont {T.}~\bibnamefont {{Saito}}},
  \bibinfo {author} {\bibfnamefont {M.}~\bibnamefont {{Yoshida}}}, \bibinfo
  {author} {\bibfnamefont {M.}~\bibnamefont {{Akiyama}}}, \bibinfo {author}
  {\bibfnamefont {Y.}~\bibnamefont {{Ono}}}, \bibinfo {author} {\bibfnamefont
  {T.}~\bibnamefont {{Yamada}}}, \bibinfo {author} {\bibfnamefont
  {K.}~\bibnamefont {{Ota}}}, \bibinfo {author} {\bibfnamefont
  {N.}~\bibnamefont {{Kashikawa}}}, \bibinfo {author} {\bibfnamefont
  {M.}~\bibnamefont {{Iye}}}, \bibinfo {author} {\bibfnamefont
  {T.}~\bibnamefont {{Kodama}}}, \bibinfo {author} {\bibfnamefont
  {S.}~\bibnamefont {{Okamura}}}, \bibinfo {author} {\bibfnamefont
  {C.}~\bibnamefont {{Simpson}}},\ and\ \bibinfo {author} {\bibfnamefont
  {M.}~\bibnamefont {{Yoshida}}},\ }\bibfield  {title} {\bibinfo {title}
  {{Statistics of 207 Ly{\ensuremath{\alpha}} Emitters at a Redshift Near 7:
  Constraints on Reionization and Galaxy Formation Models}},\ }\href
  {https://doi.org/10.1088/0004-637X/723/1/869} {\bibfield  {journal} {\bibinfo
   {journal} {\apj}\ }\textbf {\bibinfo {volume} {723}},\ \bibinfo {pages}
  {869} (\bibinfo {year} {2010})},\ \Eprint {https://arxiv.org/abs/1007.2961}
  {arXiv:1007.2961 [astro-ph.CO]} \BibitemShut {NoStop}%
\bibitem [{\citenamefont {{Bielby}}\ \emph {et~al.}(2016)\citenamefont
  {{Bielby}}, \citenamefont {{Tummuangpak}}, \citenamefont {{Shanks}},
  \citenamefont {{Francke}}, \citenamefont {{Crighton}}, \citenamefont
  {{Ba{\~n}ados}}, \citenamefont {{Gonz{\'a}lez-L{\'o}pez}}, \citenamefont
  {{Infante}},\ and\ \citenamefont {{Orsi}}}]{Bielby}%
  \BibitemOpen
  \bibfield  {author} {\bibinfo {author} {\bibfnamefont {R.~M.}\ \bibnamefont
  {{Bielby}}}, \bibinfo {author} {\bibfnamefont {P.}~\bibnamefont
  {{Tummuangpak}}}, \bibinfo {author} {\bibfnamefont {T.}~\bibnamefont
  {{Shanks}}}, \bibinfo {author} {\bibfnamefont {H.}~\bibnamefont {{Francke}}},
  \bibinfo {author} {\bibfnamefont {N.~H.~M.}\ \bibnamefont {{Crighton}}},
  \bibinfo {author} {\bibfnamefont {E.}~\bibnamefont {{Ba{\~n}ados}}}, \bibinfo
  {author} {\bibfnamefont {J.}~\bibnamefont {{Gonz{\'a}lez-L{\'o}pez}}},
  \bibinfo {author} {\bibfnamefont {L.}~\bibnamefont {{Infante}}},\ and\
  \bibinfo {author} {\bibfnamefont {A.}~\bibnamefont {{Orsi}}},\ }\bibfield
  {title} {\bibinfo {title} {{The VLT LBG redshift survey - V. Characterizing
  the z = 3.1 Lyman {\ensuremath{\alpha}} emitter population}},\ }\href
  {https://doi.org/10.1093/mnras/stv2914} {\bibfield  {journal} {\bibinfo
  {journal} {\mnras}\ }\textbf {\bibinfo {volume} {456}},\ \bibinfo {pages}
  {4061} (\bibinfo {year} {2016})},\ \Eprint {https://arxiv.org/abs/1501.01215}
  {arXiv:1501.01215 [astro-ph.GA]} \BibitemShut {NoStop}%
\bibitem [{\citenamefont {{Hao}}\ \emph {et~al.}(2018)\citenamefont {{Hao}},
  \citenamefont {{Huang}}, \citenamefont {{Xia}}, \citenamefont {{Zheng}},
  \citenamefont {{Jiang}},\ and\ \citenamefont {{Li}}}]{Hao}%
  \BibitemOpen
  \bibfield  {author} {\bibinfo {author} {\bibfnamefont {C.-N.}\ \bibnamefont
  {{Hao}}}, \bibinfo {author} {\bibfnamefont {J.-S.}\ \bibnamefont {{Huang}}},
  \bibinfo {author} {\bibfnamefont {X.}~\bibnamefont {{Xia}}}, \bibinfo
  {author} {\bibfnamefont {X.}~\bibnamefont {{Zheng}}}, \bibinfo {author}
  {\bibfnamefont {C.}~\bibnamefont {{Jiang}}},\ and\ \bibinfo {author}
  {\bibfnamefont {C.}~\bibnamefont {{Li}}},\ }\bibfield  {title} {\bibinfo
  {title} {{A Deep Ly{\ensuremath{\alpha}} Survey in ECDF-S and COSMOS. I.
  General Properties of Ly{\ensuremath{\alpha}} Emitters at z
  {\ensuremath{\sim}} 2}},\ }\href {https://doi.org/10.3847/1538-4357/aad80b}
  {\bibfield  {journal} {\bibinfo  {journal} {\apj}\ }\textbf {\bibinfo
  {volume} {864}},\ \bibinfo {eid} {145} (\bibinfo {year} {2018})},\ \Eprint
  {https://arxiv.org/abs/1808.02704} {arXiv:1808.02704 [astro-ph.GA]}
  \BibitemShut {NoStop}%
\bibitem [{\citenamefont {{Khostovan}}\ \emph {et~al.}(2019)\citenamefont
  {{Khostovan}}, \citenamefont {{Sobral}}, \citenamefont {{Mobasher}},
  \citenamefont {{Matthee}}, \citenamefont {{Cochrane}}, \citenamefont
  {{Chartab}}, \citenamefont {{Jafariyazani}}, \citenamefont
  {{Paulino-Afonso}}, \citenamefont {{Santos}},\ and\ \citenamefont
  {{Calhau}}}]{Khostovan}%
  \BibitemOpen
  \bibfield  {author} {\bibinfo {author} {\bibfnamefont {A.~A.}\ \bibnamefont
  {{Khostovan}}}, \bibinfo {author} {\bibfnamefont {D.}~\bibnamefont
  {{Sobral}}}, \bibinfo {author} {\bibfnamefont {B.}~\bibnamefont
  {{Mobasher}}}, \bibinfo {author} {\bibfnamefont {J.}~\bibnamefont
  {{Matthee}}}, \bibinfo {author} {\bibfnamefont {R.~K.}\ \bibnamefont
  {{Cochrane}}}, \bibinfo {author} {\bibfnamefont {N.}~\bibnamefont
  {{Chartab}}}, \bibinfo {author} {\bibfnamefont {M.}~\bibnamefont
  {{Jafariyazani}}}, \bibinfo {author} {\bibfnamefont {A.}~\bibnamefont
  {{Paulino-Afonso}}}, \bibinfo {author} {\bibfnamefont {S.}~\bibnamefont
  {{Santos}}},\ and\ \bibinfo {author} {\bibfnamefont {J.}~\bibnamefont
  {{Calhau}}},\ }\bibfield  {title} {\bibinfo {title} {{The clustering of
  typical Ly {\ensuremath{\alpha}} emitters from z {\ensuremath{\sim}} 2.5-6:
  host halo masses depend on Ly {\ensuremath{\alpha}} and UV luminosities}},\
  }\href {https://doi.org/10.1093/mnras/stz2149} {\bibfield  {journal}
  {\bibinfo  {journal} {\mnras}\ }\textbf {\bibinfo {volume} {489}},\ \bibinfo
  {pages} {555} (\bibinfo {year} {2019})},\ \Eprint
  {https://arxiv.org/abs/1811.00556} {arXiv:1811.00556 [astro-ph.GA]}
  \BibitemShut {NoStop}%
\bibitem [{\citenamefont {{Ramakrishnan}}\ \emph {et~al.}(2023)\citenamefont
  {{Ramakrishnan}}, \citenamefont {{Moon}}, \citenamefont {{Im}}, \citenamefont
  {{Farooq}}, \citenamefont {{Lee}}, \citenamefont {{Gawiser}}, \citenamefont
  {{Yang}}, \citenamefont {{Park}}, \citenamefont {{Hwang}}, \citenamefont
  {{Valdes}}, \citenamefont {{Artale}}, \citenamefont {{Ciardullo}},
  \citenamefont {{Dey}}, \citenamefont {{Gronwall}}, \citenamefont {{Guaita}},
  \citenamefont {{Jeong}}, \citenamefont {{Padilla}}, \citenamefont {{Singh}},\
  and\ \citenamefont {{Zabludoff}}}]{2023ApJ...951..119R}%
  \BibitemOpen
  \bibfield  {author} {\bibinfo {author} {\bibfnamefont {V.}~\bibnamefont
  {{Ramakrishnan}}}, \bibinfo {author} {\bibfnamefont {B.}~\bibnamefont
  {{Moon}}}, \bibinfo {author} {\bibfnamefont {S.~H.}\ \bibnamefont {{Im}}},
  \bibinfo {author} {\bibfnamefont {R.}~\bibnamefont {{Farooq}}}, \bibinfo
  {author} {\bibfnamefont {K.-S.}\ \bibnamefont {{Lee}}}, \bibinfo {author}
  {\bibfnamefont {E.}~\bibnamefont {{Gawiser}}}, \bibinfo {author}
  {\bibfnamefont {Y.}~\bibnamefont {{Yang}}}, \bibinfo {author} {\bibfnamefont
  {C.}~\bibnamefont {{Park}}}, \bibinfo {author} {\bibfnamefont {H.~S.}\
  \bibnamefont {{Hwang}}}, \bibinfo {author} {\bibfnamefont {F.}~\bibnamefont
  {{Valdes}}}, \bibinfo {author} {\bibfnamefont {M.~C.}\ \bibnamefont
  {{Artale}}}, \bibinfo {author} {\bibfnamefont {R.}~\bibnamefont
  {{Ciardullo}}}, \bibinfo {author} {\bibfnamefont {A.}~\bibnamefont {{Dey}}},
  \bibinfo {author} {\bibfnamefont {C.}~\bibnamefont {{Gronwall}}}, \bibinfo
  {author} {\bibfnamefont {L.}~\bibnamefont {{Guaita}}}, \bibinfo {author}
  {\bibfnamefont {W.-S.}\ \bibnamefont {{Jeong}}}, \bibinfo {author}
  {\bibfnamefont {N.}~\bibnamefont {{Padilla}}}, \bibinfo {author}
  {\bibfnamefont {A.}~\bibnamefont {{Singh}}},\ and\ \bibinfo {author}
  {\bibfnamefont {A.}~\bibnamefont {{Zabludoff}}},\ }\bibfield  {title}
  {\bibinfo {title} {{ODIN: Where Do Ly{\ensuremath{\alpha}} Blobs Live?
  Contextualizing Blob Environments within Large-scale Structure}},\ }\href
  {https://doi.org/10.3847/1538-4357/acd341} {\bibfield  {journal} {\bibinfo
  {journal} {\apj}\ }\textbf {\bibinfo {volume} {951}},\ \bibinfo {eid} {119}
  (\bibinfo {year} {2023})},\ \Eprint {https://arxiv.org/abs/2302.07860}
  {arXiv:2302.07860 [astro-ph.GA]} \BibitemShut {NoStop}%
\bibitem [{\citenamefont {{Lee}}\ \emph {et~al.}(2024)\citenamefont {{Lee}},
  \citenamefont {{Gawiser}}, \citenamefont {{Park}}, \citenamefont {{Yang}},
  \citenamefont {{Valdes}}, \citenamefont {{Lang}}, \citenamefont
  {{Ramakrishnan}}, \citenamefont {{Moon}}, \citenamefont {{Firestone}},
  \citenamefont {{Appleby}}, \citenamefont {{Artale}}, \citenamefont
  {{Andrews}}, \citenamefont {{Bauer}}, \citenamefont {{Benda}}, \citenamefont
  {{Broussard}}, \citenamefont {{Chiang}}, \citenamefont {{Ciardullo}},
  \citenamefont {{Dey}}, \citenamefont {{Farooq}}, \citenamefont {{Gronwall}},
  \citenamefont {{Guaita}}, \citenamefont {{Huang}}, \citenamefont {{Hwang}},
  \citenamefont {{Im}}, \citenamefont {{Jeong}}, \citenamefont {{Karthikeyan}},
  \citenamefont {{Kim}}, \citenamefont {{Kim}}, \citenamefont {{Kumar}},
  \citenamefont {{Nagaraj}}, \citenamefont {{Nantais}}, \citenamefont
  {{Padilla}}, \citenamefont {{Park}}, \citenamefont {{Pope}}, \citenamefont
  {{Popescu}}, \citenamefont {{Schlegel}}, \citenamefont {{Seo}}, \citenamefont
  {{Singh}}, \citenamefont {{Song}}, \citenamefont {{Troncoso}}, \citenamefont
  {{Vivas}}, \citenamefont {{Zabludoff}},\ and\ \citenamefont
  {{Zenteno}}}]{2024ApJ...962...36L}%
  \BibitemOpen
  \bibfield  {author} {\bibinfo {author} {\bibfnamefont {K.-S.}\ \bibnamefont
  {{Lee}}}, \bibinfo {author} {\bibfnamefont {E.}~\bibnamefont {{Gawiser}}},
  \bibinfo {author} {\bibfnamefont {C.}~\bibnamefont {{Park}}}, \bibinfo
  {author} {\bibfnamefont {Y.}~\bibnamefont {{Yang}}}, \bibinfo {author}
  {\bibfnamefont {F.}~\bibnamefont {{Valdes}}}, \bibinfo {author}
  {\bibfnamefont {D.}~\bibnamefont {{Lang}}}, \bibinfo {author} {\bibfnamefont
  {V.}~\bibnamefont {{Ramakrishnan}}}, \bibinfo {author} {\bibfnamefont
  {B.}~\bibnamefont {{Moon}}}, \bibinfo {author} {\bibfnamefont
  {N.}~\bibnamefont {{Firestone}}}, \bibinfo {author} {\bibfnamefont
  {S.}~\bibnamefont {{Appleby}}}, \bibinfo {author} {\bibfnamefont {M.~C.}\
  \bibnamefont {{Artale}}}, \bibinfo {author} {\bibfnamefont {M.}~\bibnamefont
  {{Andrews}}}, \bibinfo {author} {\bibfnamefont {F.}~\bibnamefont {{Bauer}}},
  \bibinfo {author} {\bibfnamefont {B.}~\bibnamefont {{Benda}}}, \bibinfo
  {author} {\bibfnamefont {A.}~\bibnamefont {{Broussard}}}, \bibinfo {author}
  {\bibfnamefont {Y.-K.}\ \bibnamefont {{Chiang}}}, \bibinfo {author}
  {\bibfnamefont {R.}~\bibnamefont {{Ciardullo}}}, \bibinfo {author}
  {\bibfnamefont {A.}~\bibnamefont {{Dey}}}, \bibinfo {author} {\bibfnamefont
  {R.}~\bibnamefont {{Farooq}}}, \bibinfo {author} {\bibfnamefont
  {C.}~\bibnamefont {{Gronwall}}}, \bibinfo {author} {\bibfnamefont
  {L.}~\bibnamefont {{Guaita}}}, \bibinfo {author} {\bibfnamefont
  {Y.}~\bibnamefont {{Huang}}}, \bibinfo {author} {\bibfnamefont {H.~S.}\
  \bibnamefont {{Hwang}}}, \bibinfo {author} {\bibfnamefont {S.~H.}\
  \bibnamefont {{Im}}}, \bibinfo {author} {\bibfnamefont {W.-S.}\ \bibnamefont
  {{Jeong}}}, \bibinfo {author} {\bibfnamefont {S.}~\bibnamefont
  {{Karthikeyan}}}, \bibinfo {author} {\bibfnamefont {H.}~\bibnamefont
  {{Kim}}}, \bibinfo {author} {\bibfnamefont {S.}~\bibnamefont {{Kim}}},
  \bibinfo {author} {\bibfnamefont {A.}~\bibnamefont {{Kumar}}}, \bibinfo
  {author} {\bibfnamefont {G.~R.}\ \bibnamefont {{Nagaraj}}}, \bibinfo {author}
  {\bibfnamefont {J.}~\bibnamefont {{Nantais}}}, \bibinfo {author}
  {\bibfnamefont {N.}~\bibnamefont {{Padilla}}}, \bibinfo {author}
  {\bibfnamefont {J.}~\bibnamefont {{Park}}}, \bibinfo {author} {\bibfnamefont
  {A.}~\bibnamefont {{Pope}}}, \bibinfo {author} {\bibfnamefont
  {R.}~\bibnamefont {{Popescu}}}, \bibinfo {author} {\bibfnamefont
  {D.}~\bibnamefont {{Schlegel}}}, \bibinfo {author} {\bibfnamefont
  {E.}~\bibnamefont {{Seo}}}, \bibinfo {author} {\bibfnamefont
  {A.}~\bibnamefont {{Singh}}}, \bibinfo {author} {\bibfnamefont
  {H.}~\bibnamefont {{Song}}}, \bibinfo {author} {\bibfnamefont
  {P.}~\bibnamefont {{Troncoso}}}, \bibinfo {author} {\bibfnamefont {A.~K.}\
  \bibnamefont {{Vivas}}}, \bibinfo {author} {\bibfnamefont {A.}~\bibnamefont
  {{Zabludoff}}},\ and\ \bibinfo {author} {\bibfnamefont {A.}~\bibnamefont
  {{Zenteno}}},\ }\bibfield  {title} {\bibinfo {title} {{The
  One-hundred-deg$^{2}$ DECam Imaging in Narrowbands (ODIN): Survey Design and
  Science Goals}},\ }\href {https://doi.org/10.3847/1538-4357/ad165e}
  {\bibfield  {journal} {\bibinfo  {journal} {\apj}\ }\textbf {\bibinfo
  {volume} {962}},\ \bibinfo {eid} {36} (\bibinfo {year} {2024})},\ \Eprint
  {https://arxiv.org/abs/2309.10191} {arXiv:2309.10191 [astro-ph.GA]}
  \BibitemShut {NoStop}%
\bibitem [{\citenamefont {{McQuinn}}\ \emph {et~al.}(2007)\citenamefont
  {{McQuinn}}, \citenamefont {{Hernquist}}, \citenamefont {{Zaldarriaga}},\
  and\ \citenamefont {{Dutta}}}]{2007MNRAS.381...75M}%
  \BibitemOpen
  \bibfield  {author} {\bibinfo {author} {\bibfnamefont {M.}~\bibnamefont
  {{McQuinn}}}, \bibinfo {author} {\bibfnamefont {L.}~\bibnamefont
  {{Hernquist}}}, \bibinfo {author} {\bibfnamefont {M.}~\bibnamefont
  {{Zaldarriaga}}},\ and\ \bibinfo {author} {\bibfnamefont {S.}~\bibnamefont
  {{Dutta}}},\ }\bibfield  {title} {\bibinfo {title} {{Studying reionization
  with Ly{\ensuremath{\alpha}} emitters}},\ }\href
  {https://doi.org/10.1111/j.1365-2966.2007.12085.x} {\bibfield  {journal}
  {\bibinfo  {journal} {\mnras}\ }\textbf {\bibinfo {volume} {381}},\ \bibinfo
  {pages} {75} (\bibinfo {year} {2007})},\ \Eprint
  {https://arxiv.org/abs/0704.2239} {arXiv:0704.2239 [astro-ph]} \BibitemShut
  {NoStop}%
\bibitem [{\citenamefont {{Weinberger}}\ \emph {et~al.}(2018)\citenamefont
  {{Weinberger}}, \citenamefont {{Kulkarni}}, \citenamefont {{Haehnelt}},
  \citenamefont {{Choudhury}},\ and\ \citenamefont
  {{Puchwein}}}]{2018MNRAS.479.2564W}%
  \BibitemOpen
  \bibfield  {author} {\bibinfo {author} {\bibfnamefont {L.~H.}\ \bibnamefont
  {{Weinberger}}}, \bibinfo {author} {\bibfnamefont {G.}~\bibnamefont
  {{Kulkarni}}}, \bibinfo {author} {\bibfnamefont {M.~G.}\ \bibnamefont
  {{Haehnelt}}}, \bibinfo {author} {\bibfnamefont {T.~R.}\ \bibnamefont
  {{Choudhury}}},\ and\ \bibinfo {author} {\bibfnamefont {E.}~\bibnamefont
  {{Puchwein}}},\ }\bibfield  {title} {\bibinfo {title}
  {{Lyman-{\ensuremath{\alpha}} emitters gone missing: the different evolution
  of the bright and faint populations}},\ }\href
  {https://doi.org/10.1093/mnras/sty1563} {\bibfield  {journal} {\bibinfo
  {journal} {\mnras}\ }\textbf {\bibinfo {volume} {479}},\ \bibinfo {pages}
  {2564} (\bibinfo {year} {2018})},\ \Eprint {https://arxiv.org/abs/1803.03789}
  {arXiv:1803.03789 [astro-ph.GA]} \BibitemShut {NoStop}%
\bibitem [{\citenamefont {{Weinberger}}\ \emph {et~al.}(2019)\citenamefont
  {{Weinberger}}, \citenamefont {{Haehnelt}},\ and\ \citenamefont
  {{Kulkarni}}}]{2019MNRAS.485.1350W}%
  \BibitemOpen
  \bibfield  {author} {\bibinfo {author} {\bibfnamefont {L.~H.}\ \bibnamefont
  {{Weinberger}}}, \bibinfo {author} {\bibfnamefont {M.~G.}\ \bibnamefont
  {{Haehnelt}}},\ and\ \bibinfo {author} {\bibfnamefont {G.}~\bibnamefont
  {{Kulkarni}}},\ }\bibfield  {title} {\bibinfo {title} {{Modelling the
  observed luminosity function and clustering evolution of Ly
  {\ensuremath{\alpha}} emitters: growing evidence for late reionization}},\
  }\href {https://doi.org/10.1093/mnras/stz481} {\bibfield  {journal} {\bibinfo
   {journal} {\mnras}\ }\textbf {\bibinfo {volume} {485}},\ \bibinfo {pages}
  {1350} (\bibinfo {year} {2019})},\ \Eprint {https://arxiv.org/abs/1902.05077}
  {arXiv:1902.05077 [astro-ph.GA]} \BibitemShut {NoStop}%
\bibitem [{\citenamefont {{Garel}}\ \emph {et~al.}(2015)\citenamefont
  {{Garel}}, \citenamefont {{Blaizot}}, \citenamefont {{Guiderdoni}},
  \citenamefont {{Michel-Dansac}}, \citenamefont {{Hayes}},\ and\ \citenamefont
  {{Verhamme}}}]{Garel}%
  \BibitemOpen
  \bibfield  {author} {\bibinfo {author} {\bibfnamefont {T.}~\bibnamefont
  {{Garel}}}, \bibinfo {author} {\bibfnamefont {J.}~\bibnamefont {{Blaizot}}},
  \bibinfo {author} {\bibfnamefont {B.}~\bibnamefont {{Guiderdoni}}}, \bibinfo
  {author} {\bibfnamefont {L.}~\bibnamefont {{Michel-Dansac}}}, \bibinfo
  {author} {\bibfnamefont {M.}~\bibnamefont {{Hayes}}},\ and\ \bibinfo {author}
  {\bibfnamefont {A.}~\bibnamefont {{Verhamme}}},\ }\bibfield  {title}
  {\bibinfo {title} {{The UV, Lyman {\ensuremath{\alpha}}, and dark matter halo
  properties of high-redshift galaxies}},\ }\href
  {https://doi.org/10.1093/mnras/stv374} {\bibfield  {journal} {\bibinfo
  {journal} {\mnras}\ }\textbf {\bibinfo {volume} {450}},\ \bibinfo {pages}
  {1279} (\bibinfo {year} {2015})},\ \Eprint {https://arxiv.org/abs/1503.06635}
  {arXiv:1503.06635 [astro-ph.GA]} \BibitemShut {NoStop}%
\bibitem [{\citenamefont {{Zheng}}\ \emph {et~al.}(2011)\citenamefont
  {{Zheng}}, \citenamefont {{Cen}}, \citenamefont {{Trac}},\ and\ \citenamefont
  {{Miralda-Escud{\'e}}}}]{2011ApJ...726...38Z}%
  \BibitemOpen
  \bibfield  {author} {\bibinfo {author} {\bibfnamefont {Z.}~\bibnamefont
  {{Zheng}}}, \bibinfo {author} {\bibfnamefont {R.}~\bibnamefont {{Cen}}},
  \bibinfo {author} {\bibfnamefont {H.}~\bibnamefont {{Trac}}},\ and\ \bibinfo
  {author} {\bibfnamefont {J.}~\bibnamefont {{Miralda-Escud{\'e}}}},\
  }\bibfield  {title} {\bibinfo {title} {{Radiative Transfer Modeling of
  Ly{\ensuremath{\alpha}} Emitters. II. New Effects on Galaxy Clustering}},\
  }\href {https://doi.org/10.1088/0004-637X/726/1/38} {\bibfield  {journal}
  {\bibinfo  {journal} {\apj}\ }\textbf {\bibinfo {volume} {726}},\ \bibinfo
  {eid} {38} (\bibinfo {year} {2011})},\ \Eprint
  {https://arxiv.org/abs/1003.4990} {arXiv:1003.4990 [astro-ph.CO]}
  \BibitemShut {NoStop}%
\bibitem [{\citenamefont {{Behrens}}\ \emph {et~al.}(2018)\citenamefont
  {{Behrens}}, \citenamefont {{Byrohl}}, \citenamefont {{Saito}},\ and\
  \citenamefont {{Niemeyer}}}]{2018A&A...614A..31B}%
  \BibitemOpen
  \bibfield  {author} {\bibinfo {author} {\bibfnamefont {C.}~\bibnamefont
  {{Behrens}}}, \bibinfo {author} {\bibfnamefont {C.}~\bibnamefont {{Byrohl}}},
  \bibinfo {author} {\bibfnamefont {S.}~\bibnamefont {{Saito}}},\ and\ \bibinfo
  {author} {\bibfnamefont {J.~C.}\ \bibnamefont {{Niemeyer}}},\ }\bibfield
  {title} {\bibinfo {title} {{The impact of Lyman-{\ensuremath{\alpha}}
  radiative transfer on large-scale clustering in the Illustris simulation}},\
  }\href {https://doi.org/10.1051/0004-6361/201731783} {\bibfield  {journal}
  {\bibinfo  {journal} {\aap}\ }\textbf {\bibinfo {volume} {614}},\ \bibinfo
  {eid} {A31} (\bibinfo {year} {2018})},\ \Eprint
  {https://arxiv.org/abs/1710.06171} {arXiv:1710.06171 [astro-ph.GA]}
  \BibitemShut {NoStop}%
\bibitem [{\citenamefont {{Momose}}\ \emph {et~al.}(2021)\citenamefont
  {{Momose}}, \citenamefont {{Shimasaku}}, \citenamefont {{Nagamine}},
  \citenamefont {{Shimizu}}, \citenamefont {{Kashikawa}}, \citenamefont
  {{Ando}},\ and\ \citenamefont {{Kusakabe}}}]{2021ApJ...912L..24M}%
  \BibitemOpen
  \bibfield  {author} {\bibinfo {author} {\bibfnamefont {R.}~\bibnamefont
  {{Momose}}}, \bibinfo {author} {\bibfnamefont {K.}~\bibnamefont
  {{Shimasaku}}}, \bibinfo {author} {\bibfnamefont {K.}~\bibnamefont
  {{Nagamine}}}, \bibinfo {author} {\bibfnamefont {I.}~\bibnamefont
  {{Shimizu}}}, \bibinfo {author} {\bibfnamefont {N.}~\bibnamefont
  {{Kashikawa}}}, \bibinfo {author} {\bibfnamefont {M.}~\bibnamefont
  {{Ando}}},\ and\ \bibinfo {author} {\bibfnamefont {H.}~\bibnamefont
  {{Kusakabe}}},\ }\bibfield  {title} {\bibinfo {title} {{Catch Me if You Can:
  Biased Distribution of Ly{\ensuremath{\alpha}}-emitting Galaxies according to
  the Viewing Direction}},\ }\href {https://doi.org/10.3847/2041-8213/abf04c}
  {\bibfield  {journal} {\bibinfo  {journal} {\apjl}\ }\textbf {\bibinfo
  {volume} {912}},\ \bibinfo {eid} {L24} (\bibinfo {year} {2021})},\ \Eprint
  {https://arxiv.org/abs/2104.10580} {arXiv:2104.10580 [astro-ph.GA]}
  \BibitemShut {NoStop}%
\bibitem [{\citenamefont {{Ebina}}\ and\ \citenamefont
  {{White}}(2024)}]{2024arXiv240113166E}%
  \BibitemOpen
  \bibfield  {author} {\bibinfo {author} {\bibfnamefont {H.}~\bibnamefont
  {{Ebina}}}\ and\ \bibinfo {author} {\bibfnamefont {M.}~\bibnamefont
  {{White}}},\ }\bibfield  {title} {\bibinfo {title} {{Cosmology before noon
  with multiple galaxy populations}},\ }\href
  {https://doi.org/10.48550/arXiv.2401.13166} {\bibfield  {journal} {\bibinfo
  {journal} {arXiv e-prints}\ ,\ \bibinfo {eid} {arXiv:2401.13166}} (\bibinfo
  {year} {2024})},\ \Eprint {https://arxiv.org/abs/2401.13166}
  {arXiv:2401.13166 [astro-ph.CO]} \BibitemShut {NoStop}%
\bibitem [{\citenamefont {{Zehavi}}\ \emph
  {et~al.}(2002{\natexlab{b}})\citenamefont {{Zehavi}}, \citenamefont
  {{Blanton}}, \citenamefont {{Frieman}}, \citenamefont {{Weinberg}},
  \citenamefont {{Mo}}, \citenamefont {{Strauss}}, \citenamefont {{Anderson}},
  \citenamefont {{Annis}}, \citenamefont {{Bahcall}}, \citenamefont
  {{Bernardi}}, \citenamefont {{Briggs}}, \citenamefont {{Brinkmann}},
  \citenamefont {{Burles}}, \citenamefont {{Carey}}, \citenamefont
  {{Castander}}, \citenamefont {{Connolly}}, \citenamefont {{Csabai}},
  \citenamefont {{Dalcanton}}, \citenamefont {{Dodelson}}, \citenamefont
  {{Doi}}, \citenamefont {{Eisenstein}}, \citenamefont {{Evans}}, \citenamefont
  {{Finkbeiner}}, \citenamefont {{Friedman}}, \citenamefont {{Fukugita}},
  \citenamefont {{Gunn}}, \citenamefont {{Hennessy}}, \citenamefont
  {{Hindsley}}, \citenamefont {{Ivezi{\'c}}}, \citenamefont {{Kent}},
  \citenamefont {{Knapp}}, \citenamefont {{Kron}}, \citenamefont {{Kunszt}},
  \citenamefont {{Lamb}}, \citenamefont {{Leger}}, \citenamefont {{Long}},
  \citenamefont {{Loveday}}, \citenamefont {{Lupton}}, \citenamefont {{McKay}},
  \citenamefont {{Meiksin}}, \citenamefont {{Merrelli}}, \citenamefont
  {{Munn}}, \citenamefont {{Narayanan}}, \citenamefont {{Newcomb}},
  \citenamefont {{Nichol}}, \citenamefont {{Owen}}, \citenamefont {{Peoples}},
  \citenamefont {{Pope}}, \citenamefont {{Rockosi}}, \citenamefont
  {{Schlegel}}, \citenamefont {{Schneider}}, \citenamefont {{Scoccimarro}},
  \citenamefont {{Sheth}}, \citenamefont {{Siegmund}}, \citenamefont {{Smee}},
  \citenamefont {{Snir}}, \citenamefont {{Stebbins}}, \citenamefont
  {{Stoughton}}, \citenamefont {{SubbaRao}}, \citenamefont {{Szalay}},
  \citenamefont {{Szapudi}}, \citenamefont {{Tegmark}}, \citenamefont
  {{Tucker}}, \citenamefont {{Uomoto}}, \citenamefont {{Vanden Berk}},
  \citenamefont {{Vogeley}}, \citenamefont {{Waddell}}, \citenamefont
  {{Yanny}},\ and\ \citenamefont {{York}}}]{2002ApJ...571..172Z}%
  \BibitemOpen
  \bibfield  {author} {\bibinfo {author} {\bibfnamefont {I.}~\bibnamefont
  {{Zehavi}}}, \bibinfo {author} {\bibfnamefont {M.~R.}\ \bibnamefont
  {{Blanton}}}, \bibinfo {author} {\bibfnamefont {J.~A.}\ \bibnamefont
  {{Frieman}}}, \bibinfo {author} {\bibfnamefont {D.~H.}\ \bibnamefont
  {{Weinberg}}}, \bibinfo {author} {\bibfnamefont {H.~J.}\ \bibnamefont
  {{Mo}}}, \bibinfo {author} {\bibfnamefont {M.~A.}\ \bibnamefont {{Strauss}}},
  \bibinfo {author} {\bibfnamefont {S.~F.}\ \bibnamefont {{Anderson}}},
  \bibinfo {author} {\bibfnamefont {J.}~\bibnamefont {{Annis}}}, \bibinfo
  {author} {\bibfnamefont {N.~A.}\ \bibnamefont {{Bahcall}}}, \bibinfo {author}
  {\bibfnamefont {M.}~\bibnamefont {{Bernardi}}}, \bibinfo {author}
  {\bibfnamefont {J.~W.}\ \bibnamefont {{Briggs}}}, \bibinfo {author}
  {\bibfnamefont {J.}~\bibnamefont {{Brinkmann}}}, \bibinfo {author}
  {\bibfnamefont {S.}~\bibnamefont {{Burles}}}, \bibinfo {author}
  {\bibfnamefont {L.}~\bibnamefont {{Carey}}}, \bibinfo {author} {\bibfnamefont
  {F.~J.}\ \bibnamefont {{Castander}}}, \bibinfo {author} {\bibfnamefont
  {A.~J.}\ \bibnamefont {{Connolly}}}, \bibinfo {author} {\bibfnamefont
  {I.}~\bibnamefont {{Csabai}}}, \bibinfo {author} {\bibfnamefont {J.~J.}\
  \bibnamefont {{Dalcanton}}}, \bibinfo {author} {\bibfnamefont
  {S.}~\bibnamefont {{Dodelson}}}, \bibinfo {author} {\bibfnamefont
  {M.}~\bibnamefont {{Doi}}}, \bibinfo {author} {\bibfnamefont
  {D.}~\bibnamefont {{Eisenstein}}}, \bibinfo {author} {\bibfnamefont {M.~L.}\
  \bibnamefont {{Evans}}}, \bibinfo {author} {\bibfnamefont {D.~P.}\
  \bibnamefont {{Finkbeiner}}}, \bibinfo {author} {\bibfnamefont
  {S.}~\bibnamefont {{Friedman}}}, \bibinfo {author} {\bibfnamefont
  {M.}~\bibnamefont {{Fukugita}}}, \bibinfo {author} {\bibfnamefont {J.~E.}\
  \bibnamefont {{Gunn}}}, \bibinfo {author} {\bibfnamefont {G.~S.}\
  \bibnamefont {{Hennessy}}}, \bibinfo {author} {\bibfnamefont {R.~B.}\
  \bibnamefont {{Hindsley}}}, \bibinfo {author} {\bibfnamefont
  {{\v{Z}}.}~\bibnamefont {{Ivezi{\'c}}}}, \bibinfo {author} {\bibfnamefont
  {S.}~\bibnamefont {{Kent}}}, \bibinfo {author} {\bibfnamefont {G.~R.}\
  \bibnamefont {{Knapp}}}, \bibinfo {author} {\bibfnamefont {R.}~\bibnamefont
  {{Kron}}}, \bibinfo {author} {\bibfnamefont {P.}~\bibnamefont {{Kunszt}}},
  \bibinfo {author} {\bibfnamefont {D.~Q.}\ \bibnamefont {{Lamb}}}, \bibinfo
  {author} {\bibfnamefont {R.~F.}\ \bibnamefont {{Leger}}}, \bibinfo {author}
  {\bibfnamefont {D.~C.}\ \bibnamefont {{Long}}}, \bibinfo {author}
  {\bibfnamefont {J.}~\bibnamefont {{Loveday}}}, \bibinfo {author}
  {\bibfnamefont {R.~H.}\ \bibnamefont {{Lupton}}}, \bibinfo {author}
  {\bibfnamefont {T.}~\bibnamefont {{McKay}}}, \bibinfo {author} {\bibfnamefont
  {A.}~\bibnamefont {{Meiksin}}}, \bibinfo {author} {\bibfnamefont
  {A.}~\bibnamefont {{Merrelli}}}, \bibinfo {author} {\bibfnamefont {J.~A.}\
  \bibnamefont {{Munn}}}, \bibinfo {author} {\bibfnamefont {V.}~\bibnamefont
  {{Narayanan}}}, \bibinfo {author} {\bibfnamefont {M.}~\bibnamefont
  {{Newcomb}}}, \bibinfo {author} {\bibfnamefont {R.~C.}\ \bibnamefont
  {{Nichol}}}, \bibinfo {author} {\bibfnamefont {R.}~\bibnamefont {{Owen}}},
  \bibinfo {author} {\bibfnamefont {J.}~\bibnamefont {{Peoples}}}, \bibinfo
  {author} {\bibfnamefont {A.}~\bibnamefont {{Pope}}}, \bibinfo {author}
  {\bibfnamefont {C.~M.}\ \bibnamefont {{Rockosi}}}, \bibinfo {author}
  {\bibfnamefont {D.}~\bibnamefont {{Schlegel}}}, \bibinfo {author}
  {\bibfnamefont {D.~P.}\ \bibnamefont {{Schneider}}}, \bibinfo {author}
  {\bibfnamefont {R.}~\bibnamefont {{Scoccimarro}}}, \bibinfo {author}
  {\bibfnamefont {R.~K.}\ \bibnamefont {{Sheth}}}, \bibinfo {author}
  {\bibfnamefont {W.}~\bibnamefont {{Siegmund}}}, \bibinfo {author}
  {\bibfnamefont {S.}~\bibnamefont {{Smee}}}, \bibinfo {author} {\bibfnamefont
  {Y.}~\bibnamefont {{Snir}}}, \bibinfo {author} {\bibfnamefont
  {A.}~\bibnamefont {{Stebbins}}}, \bibinfo {author} {\bibfnamefont
  {C.}~\bibnamefont {{Stoughton}}}, \bibinfo {author} {\bibfnamefont
  {M.}~\bibnamefont {{SubbaRao}}}, \bibinfo {author} {\bibfnamefont {A.~S.}\
  \bibnamefont {{Szalay}}}, \bibinfo {author} {\bibfnamefont {I.}~\bibnamefont
  {{Szapudi}}}, \bibinfo {author} {\bibfnamefont {M.}~\bibnamefont
  {{Tegmark}}}, \bibinfo {author} {\bibfnamefont {D.~L.}\ \bibnamefont
  {{Tucker}}}, \bibinfo {author} {\bibfnamefont {A.}~\bibnamefont {{Uomoto}}},
  \bibinfo {author} {\bibfnamefont {D.}~\bibnamefont {{Vanden Berk}}}, \bibinfo
  {author} {\bibfnamefont {M.~S.}\ \bibnamefont {{Vogeley}}}, \bibinfo {author}
  {\bibfnamefont {P.}~\bibnamefont {{Waddell}}}, \bibinfo {author}
  {\bibfnamefont {B.}~\bibnamefont {{Yanny}}},\ and\ \bibinfo {author}
  {\bibfnamefont {D.~G.}\ \bibnamefont {{York}}},\ }\bibfield  {title}
  {\bibinfo {title} {{Galaxy Clustering in Early Sloan Digital Sky Survey
  Redshift Data}},\ }\href {https://doi.org/10.1086/339893} {\bibfield
  {journal} {\bibinfo  {journal} {\apj}\ }\textbf {\bibinfo {volume} {571}},\
  \bibinfo {pages} {172} (\bibinfo {year} {2002}{\natexlab{b}})},\ \Eprint
  {https://arxiv.org/abs/astro-ph/0106476} {arXiv:astro-ph/0106476 [astro-ph]}
  \BibitemShut {NoStop}%
\bibitem [{\citenamefont {{Baugh}}(2008)}]{2008RSPTA.366.4381B}%
  \BibitemOpen
  \bibfield  {author} {\bibinfo {author} {\bibfnamefont {C.~M.}\ \bibnamefont
  {{Baugh}}},\ }\bibfield  {title} {\bibinfo {title} {{Creating synthetic
  universes in a computer}},\ }\href {https://doi.org/10.1098/rsta.2008.0192}
  {\bibfield  {journal} {\bibinfo  {journal} {Philosophical Transactions of the
  Royal Society of London Series A}\ }\textbf {\bibinfo {volume} {366}},\
  \bibinfo {pages} {4381} (\bibinfo {year} {2008})}\BibitemShut {NoStop}%
\bibitem [{\citenamefont {Zheng}(2005)}]{feyn54}%
  \BibitemOpen
  \bibfield  {author} {\bibinfo {author} {\bibfnamefont {Z.}~\bibnamefont
  {Zheng}},\ }\href@noop {} {\bibfield  {journal} {\bibinfo  {journal}
  {Astrophys. J.}\ }\textbf {\bibinfo {volume} {633}},\ \bibinfo {pages} {38}
  (\bibinfo {year} {2005})}\BibitemShut {NoStop}%
\bibitem [{\citenamefont {Landy}\ and\ \citenamefont {Szalay}(1993)}]{epr}%
  \BibitemOpen
  \bibfield  {author} {\bibinfo {author} {\bibfnamefont {M.}~\bibnamefont
  {Landy}}\ and\ \bibinfo {author} {\bibfnamefont {A.}~\bibnamefont {Szalay}},\
  }\href@noop {} {\bibfield  {journal} {\bibinfo  {journal} {ApJ}\ }\textbf
  {\bibinfo {volume} {412}},\ \bibinfo {pages} {8} (\bibinfo {year}
  {1993})}\BibitemShut {NoStop}%
\bibitem [{\citenamefont {Dijkstra}(2009)}]{Berman1983}%
  \BibitemOpen
  \bibfield  {author} {\bibinfo {author} {\bibfnamefont {M.}~\bibnamefont
  {Dijkstra}},\ }\href@noop {} {\bibfield  {journal} {\bibinfo  {journal} {Mon.
  Not. R. Astron. Soc}\ }\textbf {\bibinfo {volume} {401}},\ \bibinfo {pages}
  {6} (\bibinfo {year} {2009})}\BibitemShut {NoStop}%
\bibitem [{\citenamefont {Bryan}\ and\ \citenamefont
  {Norman}(1998)}]{Davies1w998}%
  \BibitemOpen
  \bibfield  {author} {\bibinfo {author} {\bibfnamefont {G.}~\bibnamefont
  {Bryan}}\ and\ \bibinfo {author} {\bibfnamefont {M.}~\bibnamefont {Norman}},\
  }\href@noop {} {\bibfield  {journal} {\bibinfo  {journal} {Astrophys. J.}\
  }\textbf {\bibinfo {volume} {495}},\ \bibinfo {pages} {20} (\bibinfo {year}
  {1998})}\BibitemShut {NoStop}%
\end{thebibliography}%

\end{document}